\documentclass[10pt,letterpaper]{article}
\pdfoutput=1

\usepackage[margin=1in]{geometry}
\usepackage[utf8]{inputenc}
\usepackage[T1]{fontenc}
\usepackage[nopatch=footnote]{microtype}
\usepackage{amsfonts}
\usepackage{amssymb}
\usepackage{amsmath}
\usepackage{amsthm}
\usepackage{booktabs}
\usepackage[inline]{enumitem}
\usepackage{setspace}
\usepackage[hidelinks,bookmarks,bookmarksopen,bookmarksnumbered]{hyperref}
\usepackage[capitalize]{cleveref}
\usepackage{mathtools}
\usepackage{tikz}
\usepackage[font=small,labelfont=bf]{caption}
\usepackage{titlesec}
\usepackage{thmtools}
\usepackage{wrapfig}
\usepackage{tablefootnote}
\usepackage{thm-restate}
\usepackage{soul}
\usepackage[draft]{fixme}
\usepackage{framed}
\usepackage{tikz}
\usepackage{subcaption}
\usepackage[norefs,nocites,nomsgs]{refcheck}
\usepackage{xspace}

\usetikzlibrary{trees}
\usetikzlibrary{arrows}
\usetikzlibrary{arrows.meta}
\usetikzlibrary{decorations.pathreplacing}
\usetikzlibrary{positioning}
\usetikzlibrary{calc}

\input{settings}

\newcommand{\bigO}{\mathcal{O}}
\newcommand{\Oh}{\bigO}
\DeclareMathOperator{\polylog}{polylog}

\newcommand{\ceil}[1]{\left\lceil #1 \right\rceil}
\newcommand{\floor}[1]{\left\lfloor #1 \right\rfloor}
\newcommand{\dd}{\mathinner{.\,.}}
\newcommand{\probname}[1]{\text{\sc #1}}

\newcommand{\Pat}{P}
\newcommand{\Text}{T}

\newcommand{\Textlen}{n}
\newcommand{\Seqlen}{m}
\newcommand{\AlphabetSize}{\sigma}
\newcommand{\IntegerAlphabet}{[0 \dd \AlphabetSize)}
\newcommand{\BinaryAlphabet}{\{{\tt 0}, {\tt 1}\}}
\newcommand{\emptystring}{\varepsilon}

\newcommand{\Z}{\mathbb{Z}}
\newcommand{\Zz}{\Z_{\ge 0}}

\newcommand{\SA}[1]{\mathrm{SA}_{#1}}
\newcommand{\ISA}[1]{\mathrm{SA}^{-1}_{#1}}

\newcommand{\LCE}[3]{\mathrm{LCE}_{#1}(#2,#3)}
\newcommand{\lcp}[2]{\mathrm{lcp}(#1,#2)}
\newcommand{\per}[1]{\mathrm{per}(#1)}
\newcommand{\revstr}[1]{\overline{#1}}

\newcommand{\Val}[2]{\mathrm{val}_{#1}(#2)}
\newcommand{\BasicInt}[2]{\mathrm{int}_{#1}(#2)}
\newcommand{\AlphabetMap}[3]{\mathrm{map}_{#1,#2}(#3)}

\newcommand{\PackedRepresentation}[3]{\mathrm{packed}_{#1,#2}(#3)}
\newcommand{\PackedSeqRepresentation}[3]{\PackedRepresentation{#1}{#2}{#3}}
\newcommand{\srted}[2]{{#1}_{#2}^{\rm sorted}}

\newcommand{\OccTwo}[2]{\mathrm{Occ}(#1, #2)}
\newcommand{\RangeBegTwo}[2]{\mathrm{RangeBeg}(#1, #2)}
\newcommand{\RangeEndTwo}[2]{\mathrm{RangeEnd}(#1, #2)}

\newcommand{\Rank}[3]{\mathsf{rank}_{#1}(#2,#3)}
\newcommand{\SpecialRank}[2]{\mathsf{special\mbox{-}rank}_{#1}(#2)}
\newcommand{\Select}[3]{\mathsf{select}_{#1}(#2,#3)}

\newcommand{\PrefixRank}[3]{\mathsf{prefix\mbox{-}rank}_{#1}(#2,#3)}
\newcommand{\PrefixSpecialRank}[3]{\mathsf{prefix\mbox{-}special\mbox{-}rank}_{#1}(#2,#3)}
\newcommand{\PrefixSelect}[3]{\mathsf{prefix\mbox{-}select}_{#1}(#2,#3)}

\newcommand{\zero}{{\tt 0}}
\newcommand{\one}{{\tt 1}}

\hypersetup{pageanchor=false}
\begin{document}

\title{Constant-Time Inverse Suffix Array Queries in Compact Space\\ and Sublinear-Time Construction of Suffix Array Indexes}

\author{
  \large Dominik Kempa\thanks{Partially funded by the
  NSF CAREER Award 2337891.}\\[-0.3ex]
  \normalsize Department of Computer Science,\\[-0.3ex]
  \normalsize Stony Brook University,\\[-0.3ex]
  \normalsize Stony Brook, NY, USA\\[-0.3ex]
  \normalsize \texttt{kempa@cs.stonybrook.edu}
  \and
  \large Tomasz Kociumaka\\[-0.3ex]
  \normalsize Max Planck Institute for Informatics,\\[-0.3ex]
  \normalsize Saarland Informatics Campus,\\[-0.3ex]
  \normalsize Saarbrücken, Germany\\[-0.3ex]
  \normalsize \texttt{tomasz.kociumaka@mpi-inf.mpg.de}
}

\date{\vspace{-1.0cm}}
\maketitle

\begin{abstract}
  For a text $\Text\in[0\dd\AlphabetSize)^{\Textlen}$, where
  $2\leq\AlphabetSize\leq\Textlen$, the suffix array
  $\SA{\Text}$ lists the starting positions of the suffixes of $\Text$ in
  lexicographic order, whereas the inverse suffix array $\ISA{\Text}$ maps
  each text position to the lexicographic rank of the corresponding suffix.
  Access to either array is a basic operation in full-text indexes, with
  numerous applications
  in string processing, data compression, and sequence analysis.  Stored
  explicitly, either array occupies $\Theta(\Textlen\log\Textlen)$ bits, even
  though the underlying text can be stored in
  $\Theta(\Textlen\log\AlphabetSize)$ bits.  Since the introduction of the
  Compressed Suffix Array (Grossi and Vitter; STOC 2000) and the FM-Index (Ferragina and Manzini; FOCS 2000), a central goal has been to support
  these queries quickly in space proportional to the size of the text.  For every fixed
  $\epsilon>0$, compact representations using
  $\bigO(\Textlen\log\AlphabetSize)$ bits are known with query time
  $\bigO(\log^{\epsilon}_{\AlphabetSize}\Textlen)$.  Recently, Thankachan
  (FOCS 2026) reduced the inverse suffix array query time to
  $\bigO(\log\log\Textlen / \log \log\AlphabetSize)$ using
  $\bigO(\Textlen\log\AlphabetSize)$ bits, yet it remained open whether
  constant-time queries could be achieved in this space.

  We answer this question affirmatively with an
  $\bigO(\Textlen\log\AlphabetSize)$-bit data structure that answers inverse
  suffix array queries in $\bigO(1)$ time.  For binary texts, this establishes
  an unconditional asymptotic separation between inverse suffix array and
  suffix array access, for which every deterministic $\Oh(\Textlen)$-bit suffix array
  structure in the cell-probe model with $\Theta(\log\Textlen)$-bit cells has query time
  $\Omega\big(\tfrac{\log\log\Textlen}{\log\log\log\Textlen}\big)$.

  \medskip

  Efficiently constructing full-text indexes presents a second difficulty.
  The text and a compact index both occupy
  $\Theta(\Textlen \log \AlphabetSize /\log\Textlen)$ machine words, so a
  $\Theta(\Textlen)$-time construction is a factor of
  $\Theta(\log_{\AlphabetSize}\Textlen)$ slower than reading the input or writing the index.
  Previously, $o(\Textlen)$-time construction was known for only one
  space-efficient index supporting both suffix array and inverse suffix array
  queries.  Specifically, at SODA 2023 we gave a deterministic
  $\bigO(\Textlen\min(1,
  \log\AlphabetSize/\sqrt{\log\Textlen}))$-time construction of an
  $\bigO(\Textlen\log\AlphabetSize)$-bit index that answers both queries in
  $\bigO(\log^\epsilon \Textlen)$ time, for any fixed $\epsilon>0$.

  In this work, we provide deterministic algorithms that achieve the same
  $\bigO(\Textlen\min(1,\log\AlphabetSize/\sqrt{\log\Textlen}))$ construction
  time for the new optimal inverse-suffix-array structure and for two families of suffix-array
  structures that together interpolate among all state-of-the-art time--space
  tradeoffs.  For an integer parameter $B\geq2$, the first family uses
  $\bigO(\Textlen\log\AlphabetSize
  (1+\log_B\log_{\AlphabetSize}\Textlen))$ bits and has query time
  $\bigO(B(1+\log_B\log_{\AlphabetSize}\Textlen))$, whereas the second uses
  $\bigO(B\Textlen\log\AlphabetSize
  (1+\log_B\log_{\AlphabetSize}\Textlen))$ bits and has query time
  $\bigO(1+\log_B\log_{\AlphabetSize}\Textlen)$.
  Each construction has peak preprocessing space bounded by the
  corresponding~index~size~bound.

  For binary texts, the cell-probe lower bound shows that the second family
  attains the optimal time-space tradeoff when
  $B \ge (\log\log\Textlen)^{\Omega(1)}$.
  Furthermore, outside the slowest-query regimes, improving the construction
  time for binary texts to $o(\Textlen/\sqrt{\log\Textlen})$ would yield an
  equally fast Dictionary Matching algorithm, overcoming a major obstacle to
  faster algorithms for numerous problems in string processing and beyond.
\end{abstract}

\thispagestyle{empty}
\clearpage
\hypersetup{pageanchor=true}
\setcounter{page}{1}

\section{Introduction}\label{sec:intro}

Let $\Text=\Text[1]\cdots\Text[\Textlen]$ be a text of length $\Textlen$ over
an ordered alphabet of size $\AlphabetSize$.  For every
$j\in[1\dd\Textlen]$, the string $\Text[j\dd\Textlen]$ is the suffix of
$\Text$ starting at position $j$.  The \emph{suffix array}
$\SA{\Text}[1\dd\Textlen]$ is the permutation of $[1\dd\Textlen]$ that
orders these suffixes lexicographically: $\SA{\Text}[i]=j$ if
$\Text[j\dd\Textlen]$ is the $i$th suffix in this order.  The
\emph{inverse suffix array} $\ISA{\Text}[1\dd\Textlen]$ is the inverse
permutation, that is, $\ISA{\Text}[j]=i$ if and only if
$\SA{\Text}[i]=j$.  A suffix array query takes a rank $i$ and returns the
starting position $\SA{\Text}[i]$, whereas an inverse suffix array query takes
a text position $j$ and returns its suffix rank $\ISA{\Text}[j]$.

Suffix array and inverse suffix array queries are used in numerous
problems, including data
compression~\cite{bwt,OkanoharaS09,CrochemoreIS08,OhlebuschG11,GotoB13},
string analysis~\cite{AbouelhodaKO04,IlieS11,BartonHMP14,
Charalampopoulos21}, pattern matching and text
indexing~\cite{ManberM90,FerraginaM05,GrossiV05,CaceresPZ20,Gagie2020}, and
biological sequence
analysis~\cite{bwa,bowtie2,kurtz2004versatile,GonnellaK12,ilie2011hitec}.
More applications are described in recent
papers~\cite{Charalampopoulos20,MatsudaSST20,MunroNN20,NishimotoT21,breaking},
surveys~\cite{PuglisiST07,NavarroM07,Grossi11,NavarroIndexes}, and
textbooks~\cite{bwtbook,ennobook,navarrobook,MBCT2023}.

Stored explicitly, the suffix array and its inverse are permutations of
$[1\dd\Textlen]$.  Each uses $\Theta(\Textlen\log\Textlen)$ bits and supports
constant-time queries.  For integer alphabets $\IntegerAlphabet$ in the word RAM, both arrays can
be constructed in $\Oh(\Textlen)$ time~\cite{Farach-ColtonFM00,KoA03,KarkkainenSB06}.
The text itself, however, takes only $\Theta(\Textlen\log\AlphabetSize)$ bits, which is a
factor of $\Theta(\log_\AlphabetSize \Textlen)$ smaller.
This difference led to the \emph{Compressed
Suffix Array (CSA)} of Grossi and Vitter~\cite{GrossiV00,GrossiV05} and the \emph{FM-Index}
of Ferragina and Manzini~\cite{FerraginaM00,FerraginaM05}, both of which use
$\bigO(\Textlen\log\AlphabetSize)$ bits.  For every fixed $\epsilon>0$, these
compact structures answer suffix array and inverse suffix array queries
in $\bigO(\log^{\epsilon}_{\AlphabetSize}\Textlen)$ time.
Grossi and Vitter~\cite{GrossiV05} also obtained a suffix array using
  $\bigO(\Textlen\log\AlphabetSize
  (1+\log\log_{\AlphabetSize}\Textlen))$ bits with query time
$\bigO(1+\log\log_{\AlphabetSize}\Textlen)$, while Rao~\cite{Rao02} obtained constant
query time over a binary alphabet using
$\bigO(\Textlen\log^\epsilon\Textlen)$ bits.

Except for large alphabets $\AlphabetSize\geq\Textlen^{\Omega(1)}$, these compact indexes
use asymptotically optimal space but have superconstant query time.
Consequently, Grossi and Vitter~\cite{GrossiV00} explicitly asked whether
binary suffix-array access was possible in constant time using
$\bigO(\Textlen)$ bits.
A recent lower bound~\cite{SaPerfectEquiv} shows that the
answer is negative: for binary texts of length at most $N$, every deterministic
suffix array structure using $\bigO(N)$ bits
has worst-case query time
$\Omega(\log\log N/\log\log\log N)$ in the cell-probe model with $\Theta(\log N)$-bit cells.

Inverse suffix arrays are not covered by this query-time lower bound.
Thankachan~\cite{ISA26} very recently improved upon the suffix-array access tradeoffs and
obtained $\bigO(\log\log\Textlen / \log\log\AlphabetSize)$ query time in
$\bigO(\Textlen\log\AlphabetSize)$ bits.
He also posed the following question, already open for binary texts.

\par\medskip
\noindent\hspace*{.025\linewidth}%
\begin{tikzpicture}
  \node[
    draw=black!48,
    fill=black!2,
    line width=.35pt,
    rounded corners=1pt,
    inner xsep=1.25em,
    inner ysep=.85em,
    text width=.895\linewidth
  ] {
    \begin{minipage}{\linewidth}
      \raggedright
      \begin{enumerate}[
        label=\textbf{Question \arabic*.},
        leftmargin=*,
        labelsep=.55em,
        itemsep=.8ex,
        topsep=3pt,
        parsep=0pt
      ]
      \item Can \emph{inverse suffix array} queries be answered in $\Oh(1)$ time using $\bigO(\Textlen\log\AlphabetSize)$ bits?
      \end{enumerate}
    \end{minipage}
  };
\end{tikzpicture}
\vspace{-.2cm}

\subparagraph*{Construction Algorithms}
A compact data structure with fast queries need not make an index efficient
to use, because its construction is often the computational
bottleneck: a compact final representation is of limited help if building it
still requires temporarily storing the explicit
$\Theta(\Textlen\log\Textlen)$-bit suffix array or takes
$\Omega(\Textlen)$ time.  For a binary text, this gap
is already a factor of $\Omega(\log\Textlen)$ in both measures.  In the word
RAM with $\Theta(\log\Textlen)$-bit words, the packed input and an
$\bigO(\Textlen)$-bit index each occupy only
$\bigO(\Textlen/\log\Textlen)$ words.  A linear-time construction is a factor
of $\Theta(\log\Textlen)$ slower than the time needed to read or write this
many words, while the explicit suffix array uses
$\Theta(\Textlen\log\Textlen)$ bits rather than $\bigO(\Textlen)$ bits.

Unfortunately, Thankachan's inverse suffix array~\cite{ISA26} was presented without a construction algorithm, whereas the original compressed suffix array~\cite{GrossiV00,GrossiV05} was supplied only with an $\Oh(\Textlen\log\AlphabetSize)$-time construction using $\Oh(\Textlen\log \Textlen)$ bits of working space.
A long sequence of improvements~\cite{HonSS03,Belazzougui14,MunroNN17} resulted in a deterministic $\Oh(\Textlen)$-time CSA construction using the optimal $\Oh(\Textlen\log\AlphabetSize)$ bits of space.
The only $o(\Textlen)$-time construction, obtained in~\cite{breaking}, builds an
$\bigO(\Textlen\log\AlphabetSize)$-bit structure supporting both queries in
$\bigO(\log^\epsilon\Textlen)$ time for any fixed $\epsilon>0$.
The construction takes
$\bigO(\Textlen\min(1,
\log\AlphabetSize/\sqrt{\log\Textlen}))$ time and uses
$\bigO(\Textlen\log\AlphabetSize)$ bits of working space.
This leaves the following questions:
\par\medskip
\noindent\hspace*{.025\linewidth}%
\begin{tikzpicture}
  \node[
    draw=black!48,
    fill=black!2,
    line width=.35pt,
    rounded corners=1pt,
    inner xsep=1.25em,
    inner ysep=.85em,
    text width=.895\linewidth
  ] {
    \begin{minipage}{\linewidth}
      \raggedright
      Can the following be constructed in $o(\Textlen)$ time for sufficiently small alphabet size $\AlphabetSize\ge 2$?
      \begin{enumerate}[
        resume,
        label=\textbf{Question \arabic*.},
        leftmargin=*,
        labelsep=.55em,
        itemsep=.8ex,
        topsep=3pt,
        parsep=0pt
      ]      \setcounter{enumi}{1}
      \item \emph{Suffix array} structures achieving all compressed suffix array (CSA) tradeoffs.
      \item \emph{Inverse suffix array} structures improving upon the CSA tradeoffs (e.g.,~\cite{ISA26}).
      \end{enumerate}
    \end{minipage}
  };
\end{tikzpicture}
\par\smallskip
\pagebreak

\paragraph{Our Results}
We answer all three questions affirmatively with deterministic constructions.

\medskip
\noindent
\ul{Inverse suffix arrays.}\quad
Our first result is a new data structure for inverse suffix array queries.  It
is the first compact structure with constant query time for every
alphabet size.  Whenever
$\log\AlphabetSize=o(\sqrt{\log\Textlen})$, including every constant
alphabet, its construction also takes sublinear time and uses
$\bigO(\Textlen\log\AlphabetSize)$ bits of peak preprocessing space.
For larger $\AlphabetSize$, the construction runs in $\bigO(\Textlen)$ time and
uses $\bigO(\Textlen \log \AlphabetSize)$ bits of space.

\begin{theorem}[Constant-time compact inverse suffix arrays]
  \label{th:intro-inverse-suffix-array}
  Let $2\leq\AlphabetSize\leq\Textlen$.
  For every text $\Text\in[0\dd\AlphabetSize)^{\Textlen}$, there is an $\bigO(\Textlen\log\AlphabetSize)$-bit data structure that answers inverse
  suffix array queries in $\bigO(1)$ time.
  It can be constructed from the packed representation of $\Text$ in
  $\bigO(\Textlen\min(1,
  \log\AlphabetSize/\sqrt{\log\Textlen}))$ time using
  $\bigO(\Textlen\log\AlphabetSize)$ bits~of~space.
\end{theorem}

The index size, query time, and peak preprocessing space are simultaneously
optimal.  The index size matches the $\Omega(\Textlen\log\AlphabetSize)$-bit
lower bound established in~\cite{SaPerfectEquiv}, query time cannot be
subconstant, and peak preprocessing space must accommodate the final index.

For binary texts, \cref{th:intro-inverse-suffix-array} gives an
$\bigO(\Textlen)$-bit structure with constant query time, constructible in
$\bigO(\Textlen/\sqrt{\log\Textlen})$ time using
$\bigO(\Textlen)$ bits of peak preprocessing space.
As discussed at the end of the introduction, faster construction would yield
a faster Dictionary Matching algorithm, overcoming a major obstacle to faster
algorithms for many problems in string algorithms and beyond~\cite{hierarchy}.

Together with the cell-probe lower bound for suffix array
queries~\cite{SaPerfectEquiv} (see
\cref{th:suffix-array-cell-probe-lower-bound}), this yields an unconditional
separation among deterministic compact data structures: for binary texts, inverse suffix array queries take
constant time, whereas suffix array queries require
$\Omega(\log\log\Textlen/\log\log\log\Textlen)$ time.

\medskip
\noindent
\ul{Suffix arrays.}\quad
For suffix arrays, we give two families parameterized by an integer $B\geq2$;
for small $\AlphabetSize$, both have sublinear-time construction and optimal
preprocessing space.

\begin{theorem}[Suffix array tradeoffs, simplified]
  \label{th:intro-suffix-array}
  Let $2\leq\AlphabetSize\leq\Textlen$, and let $B\geq2$ be an integer.  Given
  the packed representation of a text
  $\Text\in[0\dd\AlphabetSize)^{\Textlen}$, one can construct the following
  suffix array representations:
  \begin{enumerate}[label=(\alph*),ref=(\alph*)]
  \item\label{it:intro-sa-first-family} one using
    $\bigO(\Textlen\log\AlphabetSize
      (1+\log_B\log_{\AlphabetSize}\Textlen))$ bits and answering queries in
    $\bigO(B(1+\log_B\log_{\AlphabetSize}\Textlen))$ time,
  \item\label{it:intro-sa-second-family} one using
    $\bigO(B\Textlen\log\AlphabetSize
      (1+\log_B\log_{\AlphabetSize}\Textlen))$ bits and answering queries in
    $\bigO(1+\log_B\log_{\AlphabetSize}\Textlen)$ time.
  \end{enumerate}
  Both constructions take
  $\bigO(\Textlen\min(1,
  \log\AlphabetSize/\sqrt{\log\Textlen}))$ time, and their
  peak preprocessing space satisfies the same asymptotic bound as the index
  size.
\end{theorem}

For binary texts, the query time in tradeoff~\ref{it:intro-sa-second-family} matches the deterministic
cell-probe lower bound from~\cite{SaPerfectEquiv} whenever $B \ge (\log\log\Textlen)^{\Omega(1)}$.
  For example, setting $B=\lceil(\log\log\Textlen)^\delta\rceil$ for a constant $\delta>0$ gives an $\bigO(\Textlen(\log\log\Textlen)^{1+\delta}/
  \log\log\log\Textlen)$-bit index with query time
  $\bigO(\log\log\Textlen/\log\log\log\Textlen)$.  Setting
  $B=\lceil\log^\epsilon\Textlen\rceil$, where $\epsilon\in(0,1)$, gives an
  $\bigO(\Textlen\log^\epsilon\Textlen)$-bit index with $\bigO(1)$ query time.
  In both examples, the query time is
  optimal for deterministic data structures using the respective
  index-size bounds. These bounds match the general tradeoff of
Rao~\cite{Rao02}. Our result, however, extends both tradeoffs to all
alphabet sizes and gives sublinear-time constructions with optimal
preprocessing space for small alphabets.
The following three choices are particularly useful:
\begin{itemize}
\item Setting $B=2$ in tradeoff~\ref{it:intro-sa-first-family} or~\ref{it:intro-sa-second-family} gives a structure using
  $\bigO(\Textlen\log\AlphabetSize
  (1+\log\log_{\AlphabetSize}\Textlen))$ bits and answering queries in
  $\bigO(1+\log\log_{\AlphabetSize}\Textlen)$ time.  It matches the first
  Grossi--Vitter structure~\cite[Thm.~2(i)]{GrossiV05} in space and query time,
  but has faster preprocessing and peak preprocessing space matching the
  index-size bound.  For $\AlphabetSize=2$, the resulting bounds are
  $\bigO(\Textlen\log\log\Textlen)$ bits of index space,
  $\bigO(\log\log\Textlen)$ query time,
  $\bigO(\Textlen/\sqrt{\log\Textlen})$ preprocessing time, and
  $\bigO(\Textlen\log\log\Textlen)$ bits of peak preprocessing space.
\item Setting
  $B=\max(2,\lceil\log^{\epsilon}_{\AlphabetSize}\Textlen\rceil)$, where
$\epsilon\in(0,1)$ is a constant, in
  tradeoff~\ref{it:intro-sa-first-family} gives optimal
  $\bigO(\Textlen\log\AlphabetSize)$-bit size and
  $\bigO(\log^{\epsilon}_{\AlphabetSize}\Textlen)$ query time.  The index size and
  query time match Grossi--Vitter~\cite[Thm.~2(ii)]{GrossiV05}.  It improves
  the $\bigO(\log^\epsilon\Textlen)$ query time from~\cite{breaking} and has
  faster preprocessing and peak preprocessing space matching the index size.
\item Setting $B=\max(2,\lceil\log^{\epsilon}_{\AlphabetSize}\Textlen\rceil)$, where
$\epsilon\in(0,1)$ is a constant, in
  tradeoff~\ref{it:intro-sa-second-family} gives constant query time using
  $\bigO(\Textlen\log\AlphabetSize
  \log^{\epsilon}_{\AlphabetSize}\Textlen)$ bits.  Our construction extends
  Rao's binary-alphabet tradeoff~\cite[Thm.~5]{Rao02} to every alphabet, with
  sublinear-time preprocessing for small alphabets and peak preprocessing space
  matching the index size.  For $\AlphabetSize=2$, the resulting index uses
  $\bigO(\Textlen\log^\epsilon\Textlen)$ bits, has
  $\bigO(1)$ query time,
  $\bigO(\Textlen/\sqrt{\log\Textlen})$ preprocessing time, and
  $\bigO(\Textlen\log^\epsilon\Textlen)$ bits of peak preprocessing space.
\end{itemize}

All three structures are constructed in
$\bigO(\Textlen\min(1,
\log\AlphabetSize/\sqrt{\log\Textlen}))$ time, and each has peak preprocessing
space satisfying its index-size bound.  The tradeoffs are also
listed in \cref{tab:intro-tradeoffs}.

\begin{table}[t!]
  \centering
  \caption{A list of tradeoffs for suffix array and inverse suffix array
    access.  Here, $B\geq2$ is an integer and
    $\epsilon\in(0,1)$ is an arbitrary constant.  The two binary-alphabet
    tradeoffs reported in~\cite[p.~308]{Rao02} are stated as using
    $\bigO(\Textlen t)$ bits with query time
    $\bigO(t(\log\Textlen)^{1/t})$, and using
    $\bigO(\Textlen t(\log\Textlen)^{1/t})$ bits with query time
    $\bigO(t)$, where $1 \le t\le \log\log\Textlen$.  Under the substitution
    $B=(\log\Textlen)^{1/t}$, or equivalently
    $t=\log_B\log\Textlen$, these become the two forms displayed below, and
    the range of $t$ corresponds to $2 \le B\le \log\Textlen$.  Integer
    rounding and endpoint choices affect only constant factors.  The final
    three suffix array rows for this paper are obtained from the two
    parameterized rows by taking
    $B=2$ and
    $B=\max(2,\lceil\log^{\epsilon}_{\AlphabetSize}\Textlen\rceil)$.  In every
    row reporting a preprocessing-time bound, one can also achieve
    $\bigO(\Textlen)$.  Both space columns are measured in bits.  A dash
    means that the cited result does not provide a construction
    algorithm.}
  \label{tab:intro-tradeoffs}
  \scriptsize
  \setlength{\tabcolsep}{1.25pt}
  \renewcommand{\arraystretch}{1.12}
  \begin{tabular*}{\textwidth}{@{\extracolsep{\fill}}llllll@{}}
    \toprule
    Problem & Index space (in bits) & Query time & Prep. time & Prep. space (in bits) & Reference \\
    \midrule
    \multicolumn{6}{c}{\emph{Previous results}} \\
    \midrule
    SA/ISA & $\bigO(\Textlen\log\Textlen)$ & $\bigO(1)$
      & $\bigO(\Textlen)$ & $\bigO(\Textlen\log\Textlen)$
      & \cite{Farach-ColtonFM00} \\
    SA
      & $\bigO(\Textlen\log\AlphabetSize
          (1+\log\log_{\AlphabetSize}\Textlen))$
      & $\bigO(1+\log\log_{\AlphabetSize}\Textlen)$
      & $\bigO(\Textlen\log\AlphabetSize)$
      & $\bigO(\Textlen\log\Textlen)$
      & \cite[Thm.~2(i)]{GrossiV05} \\
    SA & $\bigO(\Textlen\log\AlphabetSize)$
      & $\bigO(\log^{\epsilon}_{\AlphabetSize}\Textlen)$
      & $\bigO(\Textlen\log\AlphabetSize)$
      & $\bigO(\Textlen\log\Textlen)$
      & \cite[Thm.~2(ii)]{GrossiV05} \\
    SA ($\AlphabetSize=2$)
      & $\bigO(\Textlen\log_B\log\Textlen)$
      & $\bigO(B\log_B\log\Textlen)$
      & -- & -- & \cite[p.~308]{Rao02} \\
    SA ($\AlphabetSize=2$)
      & $\bigO(B\Textlen\log_B\log\Textlen)$
      & $\bigO(\log_B\log\Textlen)$
      & -- & -- & \cite[Thm.~5]{Rao02} \\
    SA/ISA & $\bigO(\Textlen\log\AlphabetSize)$
      & $\bigO(\log^\epsilon\Textlen)$
      & $\bigO(\Textlen\log\AlphabetSize/\sqrt{\log\Textlen})$
      & $\bigO(\Textlen\log\AlphabetSize)$ & \cite{breaking} \\
    ISA & $\bigO(\Textlen\log\AlphabetSize)$
      & $\bigO(\log\log\Textlen / \log \log \AlphabetSize)$ & -- & -- & \cite{ISA26} \\
    \midrule
    \multicolumn{6}{c}{\emph{This paper}} \\
    \midrule
    ISA & $\bigO(\Textlen\log\AlphabetSize)$
      & $\bigO(1)$
      & $\bigO(\Textlen\log\AlphabetSize/\sqrt{\log\Textlen})$
      & $\bigO(\Textlen\log\AlphabetSize)$
      & \cref{th:inverse-suffix-array-optimal} \\
    SA
      & $\bigO(\Textlen\log\AlphabetSize
          (1+\log_B\log_{\AlphabetSize}\Textlen))$
      & $\bigO(B(1+\log_B\log_{\AlphabetSize}\Textlen))$
      & $\bigO(\Textlen\log\AlphabetSize/\sqrt{\log\Textlen})$
      & $\bigO(\Textlen\log\AlphabetSize
          (1+\log_B\log_{\AlphabetSize}\Textlen))$
      & \cref{th:suffix-array-parameterized-tradeoffs} \\
    SA
      & $\bigO(B\Textlen\log\AlphabetSize
          (1+\log_B\log_{\AlphabetSize}\Textlen))$
      & $\bigO(1+\log_B\log_{\AlphabetSize}\Textlen)$
      & $\bigO(\Textlen\log\AlphabetSize/\sqrt{\log\Textlen})$
      & $\bigO(B\Textlen\log\AlphabetSize
          (1+\log_B\log_{\AlphabetSize}\Textlen))$
      & \cref{th:suffix-array-parameterized-tradeoffs} \\
    \addlinespace[.35ex]\midrule
    SA
      & $\bigO(\Textlen\log\AlphabetSize
          (1+\log\log_{\AlphabetSize}\Textlen))$
      & $\bigO(1+\log\log_{\AlphabetSize}\Textlen)$
      & $\bigO(\Textlen\log\AlphabetSize/\sqrt{\log\Textlen})$
      & $\bigO(\Textlen\log\AlphabetSize
          (1+\log\log_{\AlphabetSize}\Textlen))$
      & \cref{th:suffix-array-tradeoffs} \\
    SA & $\bigO(\Textlen\log\AlphabetSize)$
      & $\bigO(\log^{\epsilon}_{\AlphabetSize}\Textlen)$
      & $\bigO(\Textlen\log\AlphabetSize/\sqrt{\log\Textlen})$
      & $\bigO(\Textlen\log\AlphabetSize)$
      & \cref{th:suffix-array-tradeoffs} \\
    SA
      & $\bigO(\Textlen\log\AlphabetSize
          \log^{\epsilon}_{\AlphabetSize}\Textlen)$
      & $\bigO(1)$
      & $\bigO(\Textlen\log\AlphabetSize/\sqrt{\log\Textlen})$
      & $\bigO(\Textlen\log\AlphabetSize
          \log^{\epsilon}_{\AlphabetSize}\Textlen)$
      & \cref{th:suffix-array-tradeoffs} \\
    \bottomrule
\end{tabular*}
\end{table}

\pagebreak
\smallskip
\noindent
\ul{Construction-Time Optimality.}\quad
For every constant alphabet, all our construction time bounds become
$\bigO(\Textlen/\sqrt{\log\Textlen})$.
We next discuss the conditional optimality of this bound for binary texts
under the hardness of Dictionary Matching.  Given an $\Oh(N)$-bit text and
$\Oh(N/\log N)$ patterns of $\Oh(\log N)$ bits each, Dictionary Matching asks
whether any pattern occurs in the text; see
\cref{sec:overview-lower,sec:conditional-lower-bounds-dictionary-matching}.
The fastest known Dictionary Matching algorithm runs in
$\Oh(N/\sqrt{\log N})$ time.  As proved in~\cite{hierarchy},
state-of-the-art algorithms for many problems, such as Burrows--Wheeler
transform construction, inversion counting, and batched orthogonal range
queries, are optimal unless Dictionary Matching can be
solved faster.
We prove analogous conditional
optimality results for constructing data structures that support efficient
access to (inverse) suffix arrays.
\begin{itemize}
\item \emph{Inverse suffix arrays.}
  One of the reductions in~\cite{hierarchy} shows that every instance of Dictionary Matching can be reduced to a single batch of $\Oh(N/\log N)$ inverse suffix array queries to an $\Oh(N)$-bit text that can be constructed in $\Oh(N/\log N)$ time.
  Consequently, a data structure that answers inverse suffix array queries in
  $o(\sqrt{\log \Textlen})$ time and can be constructed in
  $o(\Textlen/\sqrt{\log \Textlen})$ time would yield an
  $o(N/\sqrt{\log N})$-time algorithm for Dictionary Matching.
  In particular, our compact constant-query-time structure has conditionally
  optimal preprocessing time.
\item \emph{Suffix arrays.}
  We develop a new reduction that solves every instance of Dictionary Matching
  using $\Oh(\log\log N)$ batches of suffix array queries on its
  $\Oh(N)$-bit input text.  Each batch contains $\Oh(N/\log N)$ queries, and
  the reduction performs $\Oh(N\log\log N/\log N)$ additional work.
  Consequently, a data structure that answers suffix array queries in
  $o(\sqrt{\log \Textlen}/\log \log \Textlen)$ time and can be constructed in
  $o(\Textlen/\sqrt{\log \Textlen})$ time would yield an
  $o(N/\sqrt{\log N})$-time algorithm for Dictionary Matching.
  Thus, under the Dictionary Matching premise, the
  $\bigO(\Textlen/\sqrt{\log\Textlen})$ preprocessing time is optimal among
  deterministic constructions for tradeoff~\ref{it:intro-sa-first-family}
  whenever
  $B=o(\sqrt{\log\Textlen}/\log\log\Textlen)$.  For
  tradeoff~\ref{it:intro-sa-second-family}, it is optimal for every~$B$.
  Accordingly, this preprocessing bound is conditionally optimal for the
  first displayed specialization, for the second when
  $\epsilon\in(0,1/2)$, and for the third when $\epsilon\in(0,1)$.
\end{itemize}

\paragraph{Organization}
\Cref{sec:overview} summarizes the equivalences, main results, and conditional
lower bounds.  \Cref{sec:prelim} introduces the remaining notation and standard
ingredients.
\Cref{sec:tools,sec:packed-permuting} develop the packed-string and routing
tools used by our constructions.
\Cref{sec:prefix-select,sec:prefix-special-rank} give the prefix-select
and prefix-special-rank structures and derive the suffix array and inverse
suffix array results.  Finally, \cref{sec:conditional-lower-bounds} gives the
conditional preprocessing-time lower bounds.

\section{Technical Overview}\label{sec:overview}

The starting point of our constructions is a pair of equivalences established
in~\cite{SaPerfectEquiv}, building upon~\cite{breaking,PrefixEquiv}:
suffix-array access is equivalent
to prefix select, whereas inverse-suffix-array access is equivalent to prefix
special rank.  We first define the underlying sequence queries and their prefix
variants, and then state the precise complexity guarantees of the equivalences.

\paragraph*{Notation and Model.}

All logarithms are binary.  For integers $a,b$, we write
$[a\dd b]\coloneqq \{k\in\mathbb{Z}:a\leq k\leq b\}$ and
$[a\dd b)\coloneqq \{k\in\mathbb{Z}:a\leq k<b\}$.
We work in the standard word RAM model~\cite{Hagerup98} with $w$-bit machine
words.  When inputs have length at most $N$, we assume $w=\Theta(\log N)$.
Usual arithmetic and bitwise operations on machine words take $\bigO(1)$ time.
The \emph{packed representation} of a length-$m$ string over
$[0\dd\AlphabetSize)$ is a standard representation of strings over integer
alphabet in the word RAM model; it uses $\lceil\log\AlphabetSize\rceil$ bits per
character and $\bigO(\ceil{m\log\AlphabetSize/w})$ words in total.
Consequently, for $w=\Theta(\log N)$, a length-$m$ string can be read in
$\bigO(\ceil{m\log\AlphabetSize/\log N})$ time.
A sequence
of equal-length strings is represented by packing their concatenation.
Unqualified space bounds are
measured in words, whereas bounds stated in bits use that unit explicitly.
Preprocessing space always means peak space and includes the input and output.

\subsection{Query Problems and Main Results}
\label{sec:overview-query-problems-and-results}

\subsubsection{Rank, Select, and Special Rank Queries}

We begin with three standard operations on a string.  Rank counts the
occurrences of a specified character, select locates a specified occurrence,
and special rank returns the rank of the character at the queried position.

\begin{definition}[Rank, select, and special-rank queries]\label{def:rank-select-to}
  Let $S \in \Sigma^{m}$ be a length-$m$ string over an alphabet~$\Sigma$.
  \begin{description}[style=sameline,itemsep=1ex]
  \item[Rank query:]
    For every $a \in \Sigma$ and $j \in [0 \dd m]$, we define
    $\Rank{S}{j}{a} \coloneqq  |\{i \in [1\dd j]: S[i] = a\}|$.
  \item[Special rank query:]
    For every $j \in [1 \dd m]$, we define
    $\SpecialRank{S}{j} \coloneqq  \Rank{S}{j}{S[j]}$.
  \item[Select query:]
    For every $a \in \Sigma$ and $r \in \Z_{\geq 1}$, we define
    $\Select{S}{r}{a}$ as the $r$th smallest element of
    $\{i \in [1 \dd m] : S[i] = a\}$ if $r \leq \Rank{S}{m}{a}$, and as
    $\infty$ otherwise.
  \end{description}
\end{definition}

For a fixed character, rank and select are inverse operations on its
occurrences.

\begin{example}\label{ex:rank-and-select-queries-to}
  For $S = \texttt{abbaabab}$, we have
  $\Rank{S}{5}{\texttt{a}}=3$, $\SpecialRank{S}{3}=2$,
  $\Select{S}{4}{\texttt{a}}=7$, and
  $\Select{S}{5}{\texttt{b}}=\infty$.
\end{example}

The queries of \cref{def:rank-select-to} are well understood.  For
$w=\Theta(\log m)$ and $\AlphabetSize=|\Sigma|$, select can be answered in
$\Oh(1)$ time using
$\Oh(m\log\AlphabetSize)$ bits,
whereas rank takes the optimal
$\Oh(\log(2+\frac{\log\AlphabetSize}{\log\log m}))$ time~\cite{BelazzouguiN15}.
Special rank, often called \emph{partial rank}, can be answered in
$\Oh(1)$ time using an auxiliary data structure of
$\Oh(m(1+\log\log\AlphabetSize))$ bits, provided that $S[j]$ is supplied
with the query $\SpecialRank{S}{j}$~\cite{MunroNN17}.  Together with the
packed string itself, the total space is $\Oh(m\log\AlphabetSize)$ bits.
These constant-time select and
special-rank structures admit deterministic $\Oh(m)$-time constructions
using $\Oh(m\log\AlphabetSize)$ bits of peak
space~\cite{BelazzouguiCKM20,MunroNN17}.

Faster construction is possible from a packed input.  Belazzougui and
Puglisi~\cite{BelazzouguiP16} construct Elias--Fano representations, which
split the positions in each character-occurrence list into high and low
parts; Gao, He, and Nekrich~\cite{Gao0N20} later state the resulting
primitive explicitly.  The resulting $\Oh(m\log\AlphabetSize)$-bit structure
supports select in $\Oh(1)$ time and rank in $\Oh(\log\log m)$ time, and it can
be constructed from the packed representation of $S$ in
$\Oh(\AlphabetSize+m\log^2\AlphabetSize/\log m)$ time.  For
$\AlphabetSize\leq\log^{\Oh(1)}m$, the rank time can be further reduced to
$\Oh(1)$ within the same construction bound~\cite{Gao0N20}.

\subsubsection{Prefix Rank, Prefix Select, and Prefix Special Rank Queries}

The prefix variants of rank, select, and special-rank queries are defined on a
sequence of strings.  They replace
equality to a given character with the relation ``has a given string as a
prefix.''

\begin{definition}[Prefix queries]\label{def:prefix-range-queries-to}
  Let $W \in (\Sigma^{*})^m$ be a sequence of strings over an alphabet
  $\Sigma$ of size $\AlphabetSize$.
  \begin{description}[style=sameline,itemsep=1ex]
  \item[Prefix rank query:]
    For every string $X \in \Sigma^{*}$ and position $j \in [0 \dd m]$, we define
    $\PrefixRank{W}{j}{X} \coloneqq  |\{i \in [1 \dd j]: X\text{ is a prefix of }W[i]\}|$.
  \item[Prefix special rank query:]
    For every $j \in [1 \dd m]$ and $p \in [0 \dd |W[j]|]$, we define
    $\PrefixSpecialRank{W}{j}{p} \coloneqq \allowbreak
      \PrefixRank{W}{j}{W[j][1 \dd p]}$.
  \item[Prefix select query:]
    For every string $X \in \Sigma^{*}$ and rank $r \in \Z_{\geq 1}$, we
    define $\PrefixSelect{W}{r}{X}$ as the $r$th smallest index
    $i\in[1\dd m]$ such that $X$ is a prefix of $W[i]$, or as $\infty$ if
    fewer than $r$ such indices exist.
  \end{description}
\end{definition}

\begin{example}\label{ex:prefix-range-queries-to}
  Suppose that $W$ consists, in order, of
  \texttt{caba}, \texttt{baba}, \texttt{abba}, \texttt{bbab},
  \texttt{baaa}, \texttt{aabb}, \texttt{bbaa}, \texttt{abab}, and
  \texttt{bbba}.  Exactly $W[4]$, $W[7]$, and $W[9]$ have prefix
  $\texttt{bb}$.  Hence, $\PrefixRank{W}{8}{\texttt{bb}}=2$,
  $\PrefixSpecialRank{W}{7}{2}=2$, $\PrefixSelect{W}{3}{\texttt{bb}}=9$, and
  $\PrefixSelect{W}{4}{\texttt{bb}}=\infty$.
\end{example}

In our results, $W$ consists of strings of a common length $\ell$
over the integer alphabet $\IntegerAlphabet$.  Its packed representation uses
$\Oh(m\ell\log\AlphabetSize)$ bits.  We further assume that
$\AlphabetSize^\ell\leq m$ so that $\ell\log\AlphabetSize\leq\log m$ and, in particular,
each string $W[i]$ and each query string $X$ fits in a single machine word.
We write $\emptystring$ for the empty string; thus,
$W[j][1\dd0]\coloneqq \emptystring$.
For $X\in[0\dd\AlphabetSize)^p$, we write
$\Val{\AlphabetSize}{X}\coloneqq \sum_{i=1}^p X[i]\AlphabetSize^{p-i}$ for its value
as a base-$\AlphabetSize$ integer.

\subsubsection{The Equivalence Theorem}

Earlier reductions based on the queries of
\cref{def:prefix-range-queries-to} applied only to binary strings and
lost an additive term of $\Oh(\log\log\Textlen)$ in query time
\cite{breaking,PrefixEquiv}.  The contributions of~\cite{SaPerfectEquiv}
remove both limitations and prove the following complexity-preserving
equivalences.

\begin{theorem}[{Complexity-preserving equivalences~\cite[Theorems 1.2 and 1.3]{SaPerfectEquiv}}]
  \label{th:overview-prefix-query-equivalences}
  Let $\AlphabetSize,N\in\Z_{\geq2}$ satisfy $\AlphabetSize\leq N$, and
  consider the word RAM with word size $w=c\log N$, where $c\geq2$ is a
  constant.\footnote{As usual in the word RAM model, we assume
  $w=\Theta(\log N)$. We write $w=c\log N$ only to make the hidden
  constant explicit. Choosing $c\geq2$ keeps the proofs clean by ensuring
  that the relevant packed strings fit in one machine word.}  For the prefix problems,
  consider sequences
  $W[1\dd m]$ of $m\geq\AlphabetSize$ strings of length
  $\ell=\lfloor\log_{\AlphabetSize}m\rfloor$, and let their input length be
  $m\ell$.  For the suffix-array problems, let the input length of a packed
  text $\Text\in[0\dd\AlphabetSize)^\Textlen$, where
  $\Textlen\geq\AlphabetSize$, be $\Textlen$.

  Consider either the pair (prefix select, suffix-array access) or the pair
  (prefix special rank, inverse-suffix-array access).
  Suppose that one problem in the pair admits, for every valid packed input of
  length at most $N$, a data structure with size
  $S(\AlphabetSize,N)$ bits, preprocessing time $P_t(\AlphabetSize,N)$,
  peak preprocessing space $P_s(\AlphabetSize,N)$ bits, and query time
  $Q(\AlphabetSize,N)$.
  Then, there exists $N'=\Theta(N)$ such that the other
  problem admits, for every valid packed input of length at most $N'$, a data
  structure with the respective bounds $\bigO(S(\AlphabetSize,N))$ bits,
  $\bigO(P_t(\AlphabetSize,N))$, $\bigO(P_s(\AlphabetSize,N))$ bits, and
  $\bigO(Q(\AlphabetSize,N))$.  The implication holds in both directions.
\end{theorem}

Note that \cref{th:overview-prefix-query-equivalences} is an equivalence
between families of indexing problems, not a pointwise
identification of a particular sequence $W$ with a particular text $\Text$.
Its only losses are a constant factor in the maximum input length and constant
factors in the four complexity measures.  For the upper bounds in this paper,
we use the directions from the prefix problems to the suffix-array problems.
Thus, our task is to construct data structures for prefix select and prefix
special rank.  The next subsubsection states the resulting bounds.  We use the
reductions behind \cref{th:overview-prefix-query-equivalences} as black boxes;
their proofs are given in~\cite{SaPerfectEquiv}.

\subsubsection{Our Prefix-Query Results}

The key technical contribution of this paper is a new pair of tradeoffs for
prefix select and a constant-time compact data structure for prefix special
rank.  By \cref{th:overview-prefix-query-equivalences}, these results transfer
directly to suffix-array and inverse-suffix-array access, proving
\cref{th:intro-suffix-array} and \cref{th:intro-inverse-suffix-array},
respectively.  We
state the results in their more general form, for arbitrary integer parameters
$m,\ell,\AlphabetSize$ satisfying
$2\le \AlphabetSize^\ell\leq m$, rather than immediately specializing to
$\ell=\lfloor\log_{\AlphabetSize}m\rfloor$ as required by the equivalence
theorem.

\begin{theorem}[Prefix-select tradeoffs, simplified]
  \label{th:overview-prefix-select-tradeoffs}
  Let $m,\AlphabetSize,\ell\in\Z_{\geq2}$ satisfy
  $2\le \AlphabetSize^\ell\leq m$, and consider the word RAM with word size
  $\Theta(\log m)$.  For every $B\in[2\dd \ell]$, the following two
  prefix-select data structures~exist:
  \begin{itemize}
  \item The first has index size and peak preprocessing space
    $\bigO(m\ell\log\AlphabetSize\log_B\ell)$ bits and query time
    $\bigO(B\log_B\ell)$.
  \item The second has index size and peak preprocessing space
    $\bigO(Bm\ell\log\AlphabetSize\log_B\ell)$ bits and query time
    $\bigO(\log_B\ell)$.
  \end{itemize}
  Both data structures can be constructed deterministically in time
  \[
    \bigO\left(
      m\min\left\{
        \ell,
        \frac{\ell\log\AlphabetSize}{\sqrt{\log m}},
        \frac{(\ell\log\AlphabetSize)^2}{\log m}
      \right\}
    \right).
  \]
\end{theorem}

\begin{theorem}[Prefix special rank, simplified]
  \label{th:overview-prefix-special-rank}
  Let $m,\AlphabetSize,\ell\in\Z_{\geq1}$ satisfy
  $2\le \AlphabetSize^\ell\leq m$, and consider the word RAM with word size
  $\Theta(\log m)$.  There exists a prefix-special-rank data structure with
  index size and peak preprocessing space
  $\bigO(m\ell\log\AlphabetSize)$ bits and query time $\bigO(1)$.  It can be
  constructed deterministically in time
  \[
    \bigO\left(
      m\min\left\{
        \ell,
        \frac{\ell\log\AlphabetSize}{\sqrt{\log m}},
        \frac{(\ell\log\AlphabetSize)^2}{\log m}
      \right\}
    \right).
  \]
\end{theorem}

\medskip

Formal versions of these statements are given in
\cref{th:prefix-select-tradeoffs,th:optimal-prefix-special-rank}.  The two
prefix-select structures expose the central space--query-time tradeoff:
for the same choice of $B$, the second uses a factor $B$ more space than the
first but improves the query time by the same factor.  Specializing to
$\ell=\lfloor\log_{\AlphabetSize}m\rfloor$ and an input-length budget
$m\ell\leq N$ yields the versions used with
\cref{th:overview-prefix-query-equivalences}.

\subsection{Data Structures and Query Algorithms}
\label{sec:overview-data-structures}

Our prefix-select and prefix-special-rank data structures follow the same
two-step blueprint described below.
\begin{itemize}[leftmargin=1.8em,labelsep=0.55em,itemsep=0pt,topsep=0.7ex,
    parsep=0pt,partopsep=0pt]
  \item \emph{Baseline structures.}
    For each type of query, we begin by designing two baseline data structures.
    Specifically, for prefix select, we give an
    $\Oh(m\ell^2\log\AlphabetSize)$-bit structure
    with $\Oh(1)$ query time and an $\Oh(m\ell\log\AlphabetSize)$-bit
    structure with $\Oh(\ell)$ query time
    (\cref{pr:large-space-prefix-select-baseline,pr:small-space-prefix-select-baseline});
    their tradeoffs correspond to $B\coloneqq \ell$ in \cref{th:overview-prefix-select-tradeoffs}.
    For prefix special rank, the structure of
    \cref{pr:prefix-special-rank-large-alphabets} uses
    $\Oh(m\ell(\log\AlphabetSize+\log\log m))$ bits and answers prefix special
    rank queries in $\Oh(1)$ time; these bounds are optimal when
    $\log\AlphabetSize=\Omega(\log\log m)$.
    Moreover, the structure of
    \cref{pr:short-string-general-prefix-rank} uses
    $\Oh(m\ell\log\AlphabetSize)$ bits and answers even the more general prefix
    rank queries in $\Oh(1)$ time, but it requires a stronger assumption
    $\AlphabetSize^\ell\leq\sqrt{\log m}$.
    The two prefix-select baselines and the large-alphabet
    prefix-special-rank baseline are complete solutions under the general
    assumption $\AlphabetSize^\ell\leq m$ and can already
    be used with \cref{th:overview-prefix-query-equivalences} to obtain
    suffix-array and inverse-suffix-array structures.  However, these
    structures do not, in general, yield the tradeoffs in \cref{tab:intro-tradeoffs}.

  \item \emph{Combining baselines using layering.}
    We next use layering to combine the baseline structures into the final data
    structures.  The key idea is to apply the baseline structures to parts of
    the input represented over different alphabets, with blocks of original characters
    treated as single characters over larger alphabets (\cref{lem:prefix-select-final-layer-composition,%
lem:reduce-prefix-select-to-prefix-select-large-space,lem:prefix-special-rank-final-layer-composition}).
    In other words, layering combines baseline structures on short strings over
    progressively larger alphabets into a structure for the original strings
    (\cref{%
th:prefix-select-tradeoffs,th:optimal-prefix-special-rank}).

    Although similar layering techniques have appeared
    before~\cite{GrossiV05,Rao02,breaking,ISA26}, the resulting constructions do not
    provide sublinear-time preprocessing throughout the space--query-time tradeoffs
    considered here.  For example, the prefix-select structure
    of~\cite[Proposition~4.6]{breaking} reduces the work within each layer to
    ordinary select queries and constructs select data structures separately
    across layers.  Our layering lemmas instead reduce each problem to an instance of the same
    prefix-query problem with shorter strings and typically larger
    alphabets.  Their construction algorithms can therefore exploit the
    overlap among components for different prefix lengths ($|X|$ for prefix
    select and $p$ for prefix special rank).  This enables sublinear
    preprocessing for small alphabets while retaining small space and fast
    queries.
\end{itemize}

\subsubsection{Prefix Select}

\paragraph{A Constant-Time Structure}

For the empty prefix, a query returns $r$ if $r\leq m$ and $\infty$
otherwise.  For every prefix length
$p\in[1\dd\ell]$, let
$A_p\in[0\dd\AlphabetSize^p)^m$ denote the string satisfying
$A_p[j]=\Val{\AlphabetSize}{W[j][1\dd p]}$.  If
$X\in[0\dd\AlphabetSize)^p$, then
$\PrefixSelect{W}{r}{X}=\Select{A_p}{r}{\Val{\AlphabetSize}{X}}$.
Equipping each $A_p$ with a constant-time select structure therefore gives
constant-time prefix select.  Since the alphabet of $A_p$ has size
$\AlphabetSize^p$, these structures occupy
$\sum_{p=1}^{\ell}\Oh(m\log(\AlphabetSize^p))
=\Oh(m\ell^2\log\AlphabetSize)$ bits.  To obtain an efficient packed-input
construction, however, we open the Elias--Fano select black box of
Belazzougui and Puglisi~\cite{BelazzouguiP16}, in the packed form stated by Gao,
He, and Nekrich~\cite{Gao0N20}, and construct its components jointly for all
prefix lengths.  The resulting structure answers
prefix-select queries in $\Oh(1)$ time within the same space bound.

Denote $M\coloneqq \AlphabetSize^\ell$, and let
$W_1,\ldots,W_{n_b}$ be a partition of $W$ into $n_b\coloneqq \lceil m/M\rceil$
consecutive blocks, each of length $M$ except possibly the last.  For every prefix $X$ and block
$W_t$, let $f_{X,t}$ be the number of strings in $W_t$ having prefix $X$.
Denote $F_X\coloneqq \sum_{t=1}^{n_b}f_{X,t}$, and let
$s_X\coloneqq \AlphabetSize^{\ell-|X|}$ be the number of distinct
length-$\ell$ strings having prefix $X$.

\smallskip

\emph{The Data Structure.}
  The key components for every prefix $X\in \IntegerAlphabet^{\leq \ell}$ form a variant
  of the Elias--Fano encoding of the sequence of positions with
  prefix $X$: a bitvector $B_X$ encodes their block numbers (the high parts) in unary,
  whereas a string $P_X$ stores their positions within the blocks (the low parts).
  More precisely,
  \[
    B_X\coloneqq \one^{f_{X,1}}\zero^{s_X}
         \one^{f_{X,2}}\zero^{s_X}\cdots
         \one^{f_{X,n_b}}\zero^{s_X}
  \]
  and $B_X$ is augmented with constant-time select support. Its $\one$-runs encode the frequencies of $X$ in the blocks.
  The string
  $P_X\in[0\dd M)^{F_X}$ stores the zero-based position within the block of
  each string having prefix $X$, that is,
  $P_X[r]=(\PrefixSelect{W}{r}{X}-1)\bmod M$.  Direct-access arrays store
  $F_X$ and pointers to the packed $P_X$, the bitvector $B_X$, and its select
  structure.

  Every string in $W$ contributes to $F_X$ once for each prefix length, so the
  strings $P_X$ contain $m(\ell+1)$ entries in total.  Each entry uses
  $\Oh(\ell\log\AlphabetSize)$ bits, so the strings $P_X$ use
  $\Oh(m\ell^2\log\AlphabetSize)$ bits.  For every
  $p\in[0\dd\ell]$, the zero-runs for one block have total length
  $\AlphabetSize^p\AlphabetSize^{\ell-p}=M$.  Since $n_bM<2m$, the
  bitvectors $B_X$ use $\Oh(m\ell)$ bits.
  The direct-access arrays have $\Oh(M)=\Oh(\AlphabetSize^\ell)$ word-sized
  entries;  since $\AlphabetSize^\ell\leq m$, they occupy
  $\Oh(\AlphabetSize^\ell\log m)
   =\Oh(m\ell\log\AlphabetSize)$ bits, so the index size is
  $\Oh(m\ell^2\log\AlphabetSize)$ bits, dominated by the $P_X$ strings.

\smallskip

\emph{Query Algorithm.}
  Consider a query $(X,r)$ whose requested occurrence exists, and let
  $z\coloneqq \Select{B_X}{r}{\one}$.  The value $z-r$ is the number of zeros preceding
  the $r$th one.  Every preceding block contributes $s_X$ zeros, so
  $(z-r)/s_X$ is the zero-based index of the block containing the answer.
  The entry $P_X[r]$ gives the zero-based position within that block.  Hence,
  $\PrefixSelect{W}{r}{X}=M(z-r)/s_X+P_X[r]+1$.
  All accesses and operations take constant time.  The formal statement,
  including the preprocessing bounds, is
  \cref{pr:large-space-prefix-select-baseline}.

\paragraph{A Linear-Space Structure}

The second structure uses $\Oh(m\ell\log\AlphabetSize)$ bits and answers a
query in $\Oh(\ell)$ time.  It is the $\AlphabetSize$-ary wavelet tree of
$W$~\cite{wt}; see also the survey~\cite{Navarro14}.
For every prefix $P\in \IntegerAlphabet^{<\ell}$, the node corresponding
to $P$ stores the characters immediately following $P$ in all strings
with prefix $P$, preserving their relative order in $W$.
Formally, for $P\in[0\dd\AlphabetSize)^{p}$, where $p\in [0\dd \ell)$, let $F_P$ be its
frequency in~$W$, and let $D_P\in[0\dd\AlphabetSize)^{F_P}$ be this node
string, defined so that $D_P[r]\coloneqq W[\PrefixSelect{W}{r}{P}][p+1]$ for $r\in[1\dd F_P]$.

\smallskip

\emph{The Data Structure.}
  The key components are one string $D_p$ for every depth
  $p\in[0\dd\ell)$, defined as the concatenation of the strings
  $D_P$ for all length-$p$ strings $P$, a constant-time select structure for each $D_p$, and
  offsets for the node strings $D_P$ within $D_p$. The classes
  of strings having each such prefix partition~$W$, and hence $|D_p|=m$.
  Each $D_p$ is equipped with the constant-time select support of
  \cite{BelazzouguiP16,Gao0N20}.  For a prefix $P$, let $b_P$ be the
  number of characters preceding $D_P$ in $D_p$.  We store
  $b_P$, $b_P+F_P$, and $\Rank{D_p}{b_P}{c}$ for every character $c$.
  The first two values delimit $D_P$, while the third translates a select
  query on $D_P$ into one on $D_p$.  The select structures for the $\ell$
  strings $D_p$ use $\Oh(m\ell\log\AlphabetSize)$ bits.  The arrays
  have $\Oh(\AlphabetSize^\ell)$ word-sized entries, and
  $\AlphabetSize^\ell\log m=\Oh(m\ell\log\AlphabetSize)$, so the data structure uses
  $\Oh(m\ell\log\AlphabetSize)$ bits in total.

\smallskip

\emph{Query Algorithm.}
  Consider a query string $X=Pc$, where $p\coloneqq |P|$ and $c$ is its last character,
  and suppose
  that the requested occurrence exists.  The $r$th string having prefix $Pc$
  corresponds to the $r$th occurrence of $c$ in $D_P$.  Its position in
  $D_P$ can be retrieved using the precomputed $\Rank{D_p}{b_P}{c}$ and a select query on $D_p$:
  \[
    q\coloneqq \Select{D_p}{\Rank{D_p}{b_P}{c}+r}{c}-b_P.
  \]
  Thus, $\PrefixSelect{W}{r}{Pc}=\PrefixSelect{W}{q}{P}$.  The query repeatedly
  replaces $(Pc,r)$ by $(P,q)$, removing one character while preserving the
  desired position in $W$.  At the empty prefix, the current rank is the
  answer.  Each iteration takes constant time and removes one character, so
  the query time is $\Oh(|X|)=\Oh(\ell)$.  The formal statement, including the
  preprocessing bounds, is described in \cref{pr:small-space-prefix-select-baseline}.

\paragraph{The Layering Lemma}

The layering lemma composes multiple data structures for short strings over
various alphabets into one for the original strings.  Its three forms
introduce the composition step by step.

\begin{itemize}[leftmargin=1.8em,labelsep=0.55em,itemsep=0.5ex,topsep=0.7ex,
    parsep=0pt,partopsep=0pt]
  \item \emph{A two-layer composition.}
    Let $\lambda\in[1\dd\ell)$ be a block length.  The coarse sequence
    $U[1\dd m]$ consists of strings of length $\lfloor\ell/\lambda\rfloor$
    over $[0\dd\AlphabetSize^\lambda)$ obtained by encoding every complete
    length-$\lambda$ block as one character.  More precisely,
    \[
      U[j][t]\coloneqq \Val{\AlphabetSize}{
        W[j][(t-1)\lambda+1\dd t\lambda]}
      \qquad\text{for }j\in[1\dd m]
      \text{ and }t\in[1\dd\lfloor\ell/\lambda\rfloor].
    \]
    For every prefix
    $P\in[0\dd\AlphabetSize)^{\leq \ell}$ whose length is a multiple of
    $\lambda$, let $V_P$ denote the local sequence satisfying
    \[
      V_P[r]\coloneqq
        W[\PrefixSelect{W}{r}{P}][|P|+1\dd \min(\ell,|P|+\lambda)]
      \qquad\text{for }
      r\in[1\dd\PrefixRank{W}{m}{P}].
    \]
    Thus, $V_P$ lists the blocks of length at most $\lambda$ immediately
    following the occurrences of $P$, preserving their order in $W$.  Suppose
    that the query is $(PY,r)$, where $P$ consists of complete blocks and $Y$
    is its final piece.  The query $\PrefixSelect{V_P}{r}{Y}$ returns the rank
    $r'$ of the desired string among the strings beginning with $P$.
    A prefix-select query on $U$, with the block encoding of $P$ and rank $r'$,
    then returns its position in $W$.  Thus, the reduction makes
    one local query on $V_P$ and one coarse query on $U$.

    To share a single local structure among all prefixes of the same length,
    for every multiple $p$ of $\lambda$, the sequences $V_P$ for
    $P\in[0\dd\AlphabetSize)^p$ are concatenated in lexicographic order into a
    sequence $V_p$; cumulative offsets translate a query to and from the
    segment corresponding to $P$
    (\cref{lem:prefix-select-final-layer-composition}).

  \item \emph{Divisible layering.}
    Consider now block lengths
    $1=\lambda_0<\lambda_1<\cdots<\lambda_q=\ell$ such that every length $\lambda_{i-1}$
    divides the next length $\lambda_{i}$.  At level $i$, a block of $\lambda_{i-1}$ original
    characters is one character, and each local structure handles strings of
    $\lambda_i/\lambda_{i-1}$ such characters.  The two-layer translation is
    repeated across the levels while maintaining a simple invariant: the
    current prefix and rank identify the same position in $W$ as the original
    query.  At level $i$, one local query removes the final piece of the current
    prefix and converts its rank into the rank of the desired string among
    those sharing the remaining prefix.  The updated pair is then passed to
    level $i+1$; once the prefix is empty, the current rank is the answer.
    Every query visits each level at most once.  The
    data structure stores $\ell/\lambda_i$ local structures at level $i$, so
    the index sizes add over all stored structures and the query times add over
    the levels (\cref{lem:prefix-select-divisible-layering}).

  \item \emph{General layering.}
    The general lemma removes the divisibility requirement for the final
    transition to $\ell$: it applies divisible layering through
    $\lambda_{q-1}$ and then uses the two-layer composition for the final
    transition to $\ell$.  This preserves the query invariant above.  At level
    $i$, the number of local structures becomes
    $\Oh(\lceil\ell/\lambda_i\rceil)$, so the same asymptotic sums bound the
    space and query time
    (\cref{lem:reduce-prefix-select-to-prefix-select-large-space}).
\end{itemize}

Every subquery is itself a prefix-select query on
block-encoded strings.  The same baseline structure can therefore be applied at
all levels, which is the sense in which layering amplifies baseline
structures.

\paragraph{The Resulting Tradeoffs}

For any $B\in[2\dd\ell]$, the level sequence can be chosen with
$q=\Oh(\log_B\ell)$ so that every local string has length at most $B$.
Applying \cref{lem:reduce-prefix-select-to-prefix-select-large-space} with
the constant-time structure of
\cref{pr:large-space-prefix-select-baseline} gives space
$\Oh(Bm\ell\log\AlphabetSize\log_B\ell)$ bits and query time
$\Oh(\log_B\ell)$.  Applying it instead with the linear-space structure of
\cref{pr:small-space-prefix-select-baseline} gives space
$\Oh(m\ell\log\AlphabetSize\log_B\ell)$ bits and query time
$\Oh(B\log_B\ell)$.  The choice of levels in
\cref{th:prefix-select-tradeoffs} also gives the stated preprocessing-time and
peak-space bounds.

\subsubsection{Prefix Special Rank}

\paragraph{Prefix Special Rank over Large Alphabets}

As in the constant-time prefix-select structure, we answer queries for the
empty prefix directly and treat all prefixes of each positive length as
characters.  For every $p\in[1\dd\ell]$, let
$A_p\in[0\dd\AlphabetSize^p)^m$ denote the string satisfying
$A_p[j]=\Val{\AlphabetSize}{W[j][1\dd p]}$.  Then,
$\PrefixSpecialRank{W}{j}{p}=\SpecialRank{A_p}{j}$.  Consequently, one
constant-time special-rank structure of Munro, Navarro, and Nekrich
\cite[Theorem~A.4.1]{MunroNN17} for each prefix length gives constant-time
prefix special rank.  The character $A_p[j]$ required by the special-rank
query is extracted directly from the packed representation of $W[j]$.
For each $p\in [1\dd \ell]$, the special-rank structure occupies
$\Oh(m(1+\log\log(\AlphabetSize^p)))$ auxiliary bits.  Summing over all
prefix lengths gives
$\Oh(m\ell(1+\log\log(\AlphabetSize^\ell)))$ bits.
Incorporating the packed representation of $W$ and using
$\AlphabetSize^\ell\leq m$, we get a total size bound of
$\Oh(m\ell(\log\AlphabetSize+\log\log m))$ bits.

To enable a fast construction algorithm, we modify the design
of~\cite{MunroNN17} and construct its components jointly for all positive
prefix lengths $p$.  Denote $b\coloneqq \lceil\log m\rceil$ and $B\coloneqq b^2$.  For each
  nonempty prefix $X$ of length at most $\ell$, the positions of the strings
  with prefix $X$ in $W$ are listed in increasing order and divided into
  consecutive groups of $B$ positions, except possibly for the last group.  We call these groups the
  $X$-buckets, and we say that $X$ is \emph{frequent} if it occurs more than $B$ times
  as a prefix in $W$, and \emph{rare} otherwise.  For a query $(j,p)$, the
  answer is
determined by the index of the $W[j][1\dd p]$-bucket containing $j$ and the
position of $j$ within the bucket.

\smallskip

\emph{The Data Structure.}
  The key components are two packed sequences $R$ and $L$, a prefix-frequency
  array, and a static deterministic dictionary $\mathcal D_X$ for every
  frequent string $X$~\cite{Ruzic08}.  Their roles mirror
  the two parts of the answer: $R$ stores the position within a bucket, whereas
  $L$ and $\mathcal D_X$ together identify the bucket.  Specifically, $L$ tells
  us how many leading bits of the queried position form the bucket \emph{code}, and
  $\mathcal D_X$ maps that code to the bucket index.  For every position $j$
  and prefix length $p\in[1\dd\ell]$, the
  entry $R[j][p]$ is
  $(\PrefixSpecialRank{W}{j}{p}-1)\bmod B$, i.e., the zero-based position in
  the corresponding bucket.  To identify the bucket, let $C_{X,k}$ be the
  longest common prefix of the length-$b$ binary encodings of the first and
  last positions (zero-based) in the $k$th
  $X$-bucket.  The length-$b$ binary encoding of every position
  in the bucket has prefix $C_{X,k}$, and different buckets of the same prefix
  $X$ have different such prefixes.  The entry $L[j][p]$ stores $|C_{X,k}|$
  for the bucket containing $j$, or zero if $X=W[j][1\dd p]$ occurs at most
  $B$ times.
  For every frequent string $X$, the dictionary
  $\mathcal D_X$ maps the common prefix $C_{X,k}$ to the $X$-bucket index $k$.
  The sequences $L$ and $R$ use $\Oh(m\ell\log\log m)$ bits.  At every positive
  prefix length, the prefix frequencies sum to $m$, so the dictionaries contain
  $\Oh(m\ell/B)$ pairs in total and use $\Oh(m\ell/\log m)$ bits.  The packed
  input and other auxiliary information use
  $\Oh(m\ell\log\AlphabetSize)$ further bits.  Thus, the total space is
  $\Oh(m\ell(\log\AlphabetSize+\log\log m))$ bits.

\smallskip

\emph{Query Algorithm.}
  A query with $p=0$ returns $j$ directly.  For $p\geq1$, denote
  $X\coloneqq W[j][1\dd p]$,
  $d\coloneqq L[j][p]$, and $\delta\coloneqq R[j][p]$.  The prefix-frequency array determines
  whether $X$ is frequent.  If it is not, there is only
  one $X$-bucket, and the answer is $\delta+1$.  Otherwise, compute the
  bitstring $C\in\{0,1\}^d$ consisting of the first $d$ bits of the length-$b$
  binary encoding of $j-1$.  The definition of $L$ implies that $C$ is the code
  of the $X$-bucket containing $j$.  The query computes
  $k\coloneqq \mathcal D_X[C]$ and returns $(k-1)B+\delta+1$.
  The prefix $X$ and the entries of $L$ and $R$ are extracted in constant time,
  and the arithmetic and dictionary lookup also take constant time.  The formal
  statement, including the preprocessing bounds, is given in
  \cref{pr:prefix-special-rank-large-alphabets}.

\paragraph{Prefix Rank on Small Universes}

The second structure applies if
$M\coloneqq \AlphabetSize^\ell\leq\sqrt{\log m}$.  It uses
$\Oh(m\ell\log\AlphabetSize)$ bits and answers the more general prefix-rank
queries in $\Oh(1)$ time.  Its design resembles both constant-time rank
structures for small alphabets~\cite{WaveletSuffixTree,Gao0N20} and our
constant-time prefix-select structure.

Let $W_1,\ldots,W_{n_b}$ be the partition of $W$ into
$n_b\coloneqq \lceil m/M\rceil$ consecutive blocks, each of length $M$ except possibly
the last.  For every prefix $X\in \IntegerAlphabet^{\le \ell}$, denote $s_X\coloneqq \AlphabetSize^{\ell-|X|}$, and let $f_{X,t}$ be the
number of strings in $W_t$ having prefix $X$.

\smallskip

\emph{The Data Structure.}
  The key components are, for every prefix
  $X\in\IntegerAlphabet^{\leq\ell}$, a bitvector $B_X$ with constant-time
  select support, and a lookup table $L_{\rm srank}$.
  As in the constant-time prefix-select structure, the bitvector $B_X\coloneqq \bigodot_{t=1}^{n_b}(\one^{f_{X,t}}\zero^{s_X})$ is stored together with a constant-time select structure
  accessible through a direct-access array of pointers, and all these components use
  $\Oh(m\ell\log\AlphabetSize)$ bits in total.

  Given any possible block $C\in (\IntegerAlphabet^{\ell})^{\leq M}$, a prefix
  $X\in\IntegerAlphabet^{\leq\ell}$, and $t\in[0\dd |C|]$, the table
  $L_{\rm srank}$ stores the number of strings among $C[1\dd t]$ having prefix
  $X$.  Since
  $\ell\lceil\log\AlphabetSize\rceil\leq2M/3$, encoding every character in
  $\lceil\log\AlphabetSize\rceil$ bits makes the entire block $C$ occupy at
  most $M\ell\lceil\log\AlphabetSize\rceil\leq2M^2/3\leq2\log m/3$ bits.
  Thus, it fits in one word and can be used as part of a lookup-table key;
  the table over all possible blocks and prefixes uses $o(m/\log m)$ bits.
  With the packed representation of $W$, the data structure uses
  $\Oh(m\ell\log\AlphabetSize)$ bits overall.

\smallskip

\emph{Query Algorithm.}
  A query with $j=0$ returns zero.  Otherwise, consider a prefix-rank query
  $(X,j)$ and denote
  $q\coloneqq \lfloor(j-1)/M\rfloor+1$ and $t\coloneqq j-(q-1)M$.  Thus, $W_q$ is the block
  containing position $j$, and $t$ is the position of $j$ within that block.
  We split the answer into two parts: $r_0$ counts strings with prefix $X$ in
  the complete blocks preceding $W_q$, while $r_1$ counts them among the first
  $t$ strings of $W_q$.

  The table $L_{\rm srank}$ supplies $r_1$ directly.  Extract $W_q$ from the
  packed representation of $W$; a lookup using $W_q$, $X$, and $t$ gives
  $r_1$.  A second lookup, with $t=|W_q|$, gives the frequency
  $f_{X,q}$ of $X$ in $W_q$.

  The bitvector $B_X$ supplies $r_0$ as follows.  Its $(qs_X)$th occurrence of
  $\zero$ is the last zero in the factor associated with $W_q$.  Its position
  therefore accounts for $qs_X$ zeros and all occurrences of $X$ through
  blocks $W_1,\ldots,W_q$.  Subtracting the zeros and the $f_{X,q}$ occurrences in $W_q$
  leaves exactly
  \[
    r_0\coloneqq \Select{B_X}{qs_X}{\zero}-qs_X-f_{X,q}
       =\sum_{h=1}^{q-1}f_{X,h}.
  \]
  Hence, the query returns $r_0+r_1=\PrefixRank{W}{j}{X}$.  The table lookups,
  select query, and all remaining operations take constant time.  The formal
  statement and preprocessing bounds are given in
  \cref{pr:short-string-general-prefix-rank}.

\paragraph{The Layering Lemma}

The prefix-special-rank layering lemma needs only a two-layer composition.
Let $\lambda\in[1\dd\ell)$ be a block length.  As for prefix select, let the
coarse sequence $U[1\dd m]$ be obtained by encoding every complete length-$\lambda$ block
as one character:
\[
  U[j][t]\coloneqq \Val{\AlphabetSize}{
    W[j][(t-1)\lambda+1\dd t\lambda]}
  \qquad\text{for }j\in[1\dd m]
  \text{ and }t\in[1\dd\lfloor\ell/\lambda\rfloor].
\]
For every prefix
$P\in[0\dd\AlphabetSize)^{\leq\ell}$ whose length is a multiple of $\lambda$,
let $V_P$ denote the local sequence satisfying
\[
  V_P[r]\coloneqq
    W[\PrefixSelect{W}{r}{P}]
      [|P|+1\dd\min(\ell,|P|+\lambda)]
  \qquad\text{for }r\in[1\dd\PrefixRank{W}{m}{P}].
\]
Thus, $V_P$ lists the blocks immediately following the occurrences of $P$, preserving
their order in $W$.

For $p=0$, the answer is $j$.  For a query $(j,p)$ with $p\geq1$, denote
$q\coloneqq \lfloor p/\lambda\rfloor$ and
$d\coloneqq p-q\lambda$.  Let $P\coloneqq W[j][1\dd q\lambda]$ be the part consisting of
complete blocks and $Y\coloneqq W[j][q\lambda+1\dd p]$ the remaining part.  The coarse
query
\[
  t\coloneqq \PrefixSpecialRank{U}{j}{q}
\]
gives the rank of $W[j]$ among the strings beginning with $P$.  Since $V_P$ preserves their order, its first
$t$ strings correspond exactly to the strings $W[i]$ with $i\leq j$ that begin
with $P$.  Consequently,
\[
  \PrefixSpecialRank{W}{j}{p}=\PrefixSpecialRank{V_P}{t}{d}.
\]
Indeed, $V_P[t]$ is the next block of $W[j]$, whose length-$d$ prefix is
$Y$.  Every nonempty query thus makes one coarse prefix-special-rank query on
$U$ and one local prefix-special-rank query on $V_P$.

As in the prefix-select composition, the sequences $V_P$ for prefixes of the
same length are concatenated in lexicographic order.  A cumulative position
offset translates $t$ to the corresponding position in the concatenation, and
subtracting the precomputed number of matching blocks preceding the segment
translates the answer back.  If the local and coarse prefix-special-rank
structures use $S_1$ and $S_2$ bits and have query times $Q_1$ and $Q_2$,
respectively, the resulting structure uses
$\Oh(\lceil\ell/\lambda\rceil S_1+S_2)$ bits and answers a query in
$\Oh(Q_1+Q_2)$ time.  The exact statement, including the preprocessing bounds,
is given in \cref{lem:prefix-special-rank-final-layer-composition}.

\paragraph{The Final Data Structure}

Denote
$\lambda\coloneqq \max\{1,\lfloor\log\log m/(2\log\AlphabetSize)\rfloor\}$.  First suppose
that $2\leq\lambda<\ell$.  Since $\AlphabetSize^\lambda\leq\sqrt{\log m}$, the
small-universe prefix-rank structure of
\cref{pr:short-string-general-prefix-rank} applies to the local strings; we use
it as a prefix-special-rank structure.
The coarse alphabet has size $\AlphabetSize^\lambda$, and
$\lambda\log\AlphabetSize=\Theta(\log\log m)$.  Hence,
\cref{pr:prefix-special-rank-large-alphabets} uses
$\Oh(m\ell\log\AlphabetSize)$ bits on the coarse strings.  Applying
\cref{lem:prefix-special-rank-final-layer-composition} yields the same overall
space bound and constant query time.  If $\lambda=1$, the large-alphabet
structure already satisfies the desired space bound.  Finally, if $\lambda>1$
and $\lambda\geq\ell$, the small-universe structure applies directly to $W$.
Together, the three cases attain all four bounds in
\cref{th:optimal-prefix-special-rank}.

\subsection{Construction Algorithms}
\label{sec:overview-construction-algorithms}

We next explain how the structures described above are constructed within
their packed preprocessing bounds.

\subsubsection{Auxiliary Tools}

\paragraph{Basic Tools for Packed Strings}

\Cref{sec:tools} supplies word-parallel primitives for prefix frequencies
(\cref{pr:packed-prefix-frequencies}), character-block encoding
(\cref{pr:packed-character-block-encoding}), and packed-string partitioning
(\cref{pr:packed-pointwise-concatenation-partition}).  Each runs in time linear
in the number of machine words involved and is used throughout the overview.

\paragraph{Routing and Prefix-Path Gathering}

The next two tools move short payloads between the input order and the groups
of strings sharing a prefix.
\begin{itemize}[leftmargin=1.8em,labelsep=0.55em,itemsep=0.5ex,topsep=0.7ex,
    parsep=0pt,partopsep=0pt]
  \item \emph{Grouping and routing.}
    Stable grouping rearranges payloads according to one-character keys while
    preserving their relative order within every group, whereas stable ungrouping
    reverses this rearrangement
    (\cref{pr:packed-binary-stable-grouping,pr:packed-stable-grouping}).
    Repeating stable grouping along the characters of the input yields
    multilevel routing.  Given one payload for each input string, multilevel
    routing returns, at every depth and for each prefix $X$ of that length, the
    payloads of the strings having prefix $X$ in their original relative order
    (\cref{pr:packed-stable-multilevel-routing,%
      pr:packed-stable-multilevel-routing-comparable}).

  \item \emph{Prefix-path gathering.}
    Suppose that every prefix $X\ne \emptystring$ has a list $P_X$ containing one
    payload for every occurrence of $X$, in prefix-occurrence order.
    Prefix-path gathering constructs,
    for every input string $W[j]$, the payload associated with its occurrence
    of each prefix on the path
    $W[j][1\dd 1],W[j][1\dd2],\ldots,W[j][1\dd\ell]$.  More precisely, its output
    satisfies
    $Q[j][p]=P_{W[j][1\dd p]}[\PrefixSpecialRank{W}{j}{p}]$
    (\cref{pr:packed-prefix-path-payload-gathering}).
\end{itemize}
Both tools are stated entirely for packed string sequences and aligned
payloads, independently of the prefix-select and prefix-special-rank
structures, and are of independent interest.  Here, they substantially
simplify the constructions by expressing the required transformations as
grouping and moving short payloads.

\paragraph{Wavelet Trees}

For every proper prefix $P$ of a string in $W$, the corresponding wavelet-tree
node string is precisely the string $D_P$ defined for the linear-space
prefix-select structure: it lists the characters following the occurrences of
$P$ in their original order.  Thus, constructing the wavelet tree constructs
all the strings $D_P$ simultaneously.  Fast construction algorithms for
binary wavelet trees were
    given by Babenko et al.~\cite{WaveletSuffixTree} and Munro, Nekrich, and
    Vitter~\cite{MunroNV16}.  For general alphabets, we give a new packed
    construction of all the wavelet-tree strings that refines the earlier construction
of~\cite[Lemma~6.4]{sss}.  Its running time is
\[
  \Oh\!\left(
    m\min\!\left\{
      \ell,
      \frac{\ell\log\AlphabetSize}{\sqrt{\log m}},
      \frac{(\ell\log\AlphabetSize)^2}{\log m}
    \right\}
  \right).
\]
The algorithm divides the wavelet-tree levels into strips and uses stable
grouping (\cref{pr:packed-stable-grouping}) to process several short fields
within one word.  Its peak space is
$\Oh(m\ell\log\AlphabetSize)$ bits
(\cref{pr:packed-wavelet-tree-construction}).  We also augment the wavelet
  tree with a very useful new operation (applied in constructions below).  Given $p\in [0\dd \ell)$ and $d\in[1\dd\ell-p]$, it returns, for
every $X\in \IntegerAlphabet^p$, the packed sequence of length-$d$ strings
immediately following the occurrences of $X$ as prefixes of the input strings.  The ordinary wavelet-tree
node strings $D_X$ are the special case $d=1$.  The resulting generalized
wavelet tree is constructed within the same time and space bounds and supports this
operation in
$\Oh(\AlphabetSize^p+\min\{m,m(d\log\AlphabetSize)^2/\log m\})$ time
(\cref{pr:generalized-wavelet-tree-construction}).

\subsubsection{Prefix Select}

\paragraph{The Constant-Time Structure}

Prefix frequencies (computed using \cref{pr:packed-prefix-frequencies})
provide the sizes of all stored objects.  To
construct the bitvectors $B_X$, we interleave every block of $W$ with a copy
of all length-$\ell$ strings.  We attach to each input string a payload $\one$, and to each appended string a payload $\zero$.  Binary stable ungrouping
(\cref{cor:packed-binary-string-ungrouping-comparable}) creates this
interleaving, and stable multilevel routing
(\cref{pr:packed-stable-multilevel-routing-comparable}) groups the payloads by
every prefix.  The resulting strings are exactly the bitvectors $B_X$.  To
construct $P_X$, we
instead attach the position of every input string within its
length-$\AlphabetSize^\ell$ block.  A second application of multilevel routing
(\cref{pr:packed-stable-multilevel-routing}) sends these positions to all
prefixes of the corresponding input string.  Taking the faster of these
routing procedures and direct scans constructs these components in preprocessing
time
$\Oh(m\min\{\ell,(\ell\log\AlphabetSize)^2/\log m\})$ and peak space
$\Oh(m\ell^2\log\AlphabetSize)$ bits, as stated in
\cref{pr:large-space-prefix-select-baseline}.

\paragraph{The Linear-Space Structure}

Prefix frequencies (computed using \cref{pr:packed-prefix-frequencies}) give
the group boundaries and cumulative occurrence offsets stored by this
structure.
The strings $D_X$ are precisely the strings returned
by the wavelet-tree construction
(\cref{pr:packed-wavelet-tree-construction}).  For every depth $p$, their
packed representations are concatenated in lexicographic order to form the
length-$m$ string $D_p$, after which a constant-time select structure~\cite{BelazzouguiP16,Gao0N20} is constructed for $D_p$.  Processing
the depths one at a time gives construction time
$\Oh(m\min\{\ell,(\ell\log\AlphabetSize)^2/\log m\})$ and peak space
$\Oh(m\ell\log\AlphabetSize)$ bits, as stated in \cref{pr:small-space-prefix-select-baseline}.

\paragraph{Layering}

The divisible and two-layer constructions each build one generalized wavelet
tree and use its new operation at the required block boundaries
(\cref{pr:generalized-wavelet-tree-construction}).  For a required prefix
length $p$, this operation returns, for every length-$p$ prefix $P$, the
sequence of blocks immediately following the occurrences of $P$.  These are
precisely the local sequences $V_P$ from the two-layer reduction.
Concatenating them in lexicographic order produces $V_p$, and interpreting
blocks of characters as larger meta-characters
(\cref{pr:packed-character-block-encoding}) produces the instance on which
the selected baseline structure is constructed.

The coarse sequence $U$ is obtained separately.  Pointwise partitioning
(\cref{pr:packed-pointwise-concatenation-partition}) isolates the complete
blocks of every input string, and block encoding
(\cref{pr:packed-character-block-encoding}) turns those blocks into the
characters of $U$.  Prefix frequencies (computed using
\cref{pr:packed-prefix-frequencies}) provide the boundaries of the $V_P$
segments within $V_p$ and the associated occurrence offsets.  The local
structures are processed one at a time.  The general layering lemma composes
the divisible and two-layer constructions.  These auxiliary structures and
operations, the packed transformations, and the baseline preprocessing calls
together determine the preprocessing time, with the baseline times adding over
the stored structures.  The levels in \cref{th:prefix-select-tradeoffs} grow
geometrically by factors at most $B$, with one level chosen as
$\Theta(\sqrt{\log m}/\log\AlphabetSize)$ whenever this scale lies between
$1$ and $\ell$; this aligns the hierarchy with the crossover between the two
baseline preprocessing bounds and keeps the full sum within the claimed bound.
The exact generic accounting is given in
\cref{lem:prefix-select-final-layer-composition,%
  lem:prefix-select-divisible-layering,%
  lem:reduce-prefix-select-to-prefix-select-large-space}.

\subsubsection{Prefix Special Rank}

\paragraph{Prefix Special Rank over Large Alphabets}

Prefix frequencies (computed using \cref{pr:packed-prefix-frequencies})
first identify the frequent prefixes.
For every bucket of a prefix $X$ whose frequency $F_X$ exceeds
$B=\ceil{\log m}^2$, a temporary prefix-select structure from
\cref{th:prefix-select-tradeoffs} locates the bucket's first and last positions
with two queries; comparing their binary encodings determines the bucket code.

Once these codes are known, we form two payload strings for every prefix $X$,
both ordered by the occurrences of $X$.  In the first string, the code length
of each bucket is repeated once for every occurrence in that bucket; for a rare
prefix, every value is zero.  These are precisely the values that must become
the entries of $L$.  The second string assigns the successive values
$0,1,\ldots,B-1$ within each bucket, producing the local ranks stored in $R$.
One application of prefix-path gathering
(\cref{pr:packed-prefix-path-payload-gathering}) scatters the first payloads to
their entries indexed by the input position and prefix length, thereby
constructing $L$; a second application constructs $R$ in the same way.  The
bucket codes also supply the key--value pairs for the static dictionaries
of~\cite{Ruzic08}.  There are
$\Oh(m\ell/(\log m)^2)$ buckets of frequent prefixes, so the prefix-select
queries, code comparisons, and dictionary constructions fit the following
bound.  Altogether, the construction takes
\[
  \Oh\!\left(
    m\min\!\left\{
      \ell,
      \frac{\ell(\log\AlphabetSize+\log\log m)}{\sqrt{\log m}},
      \frac{\ell^2(\log\AlphabetSize+\log\log m)^2}{\log m}
    \right\}
  \right)
\]
time and uses $\Oh(m\ell(\log\AlphabetSize+\log\log m))$ bits of peak space,
as stated in \cref{pr:prefix-special-rank-large-alphabets}.

\paragraph{Prefix Rank on Small Universes}

The bitvectors $B_X$ are exactly the same as in the constant-time
prefix-select structure.  The only other nontrivial component is the
universal table $L_{\rm srank}$.
The small-universe assumption $M\coloneqq \AlphabetSize^\ell \le \sqrt{\log m}$ makes
the fixed-width encoding of any block of $M$ input strings fit in one word, so
this table can be constructed by enumerating every possible block, prefix, and
within-block endpoint.
The construction takes $\Oh(m(\ell\log\AlphabetSize)^2/\log m)$ time and uses
$\Oh(m\ell\log\AlphabetSize)$ bits of peak space
(\cref{pr:short-string-general-prefix-rank}).

\paragraph{Layering}

The layering construction follows the two-layer prefix-select construction,
using analogous local sequences $V_p$ and a coarse sequence $U$, but
replacing its prefix-select components with prefix-special-rank components
(\cref{lem:prefix-special-rank-final-layer-composition}).
Using a baseline
directly when it gives the desired bounds, and the composition
otherwise, with every $V_p$ using
\cref{pr:short-string-general-prefix-rank} and $U$ using
\cref{pr:prefix-special-rank-large-alphabets},
yields the bounds of \cref{th:optimal-prefix-special-rank}.

\subsection{Preprocessing Lower Bounds}\label{sec:overview-lower}

\Cref{sec:conditional-lower-bounds} gives conditional lower bounds from the
following packed Dictionary Matching problem.  The input is a binary text $T$ of
length $n$ and $k=\Theta(n/\log n)$ binary patterns $P_i$, all of a common length
$m=\Theta(\log n)$.  Thus, the whole instance occupies $\Theta(n/\log n)$
words.  The premise is that no algorithm decides whether any
pattern $P_i$ occurs in the text $T$ in $o(n/\sqrt{\log n})$ time.  This is a
$\sqrt{\log n}$ factor above the time needed to read the packed input, and it
matches the preprocessing time of our constant-alphabet constructions.

For inverse suffix arrays, a reduction from~\cite{hierarchy} constructs in
$\bigO(n/\log n)$ time a binary string $S$ of length $N=\Theta(n)$ and two
positions $a_i,b_i$ per pattern such that
$P_i$ occurs in $T$ if and only if $\ISA{S}[a_i]+1<\ISA{S}[b_i]$.
Only $2k=\Theta(n/\log n)$ inverse suffix array queries are needed to verify this condition.
If these queries are answered in $Q(N)$ time after $P(N)$-time preprocessing,
Dictionary Matching therefore takes
$\bigO(n/\log n+P(N)+(n/\log n)Q(N))$ time.
Under the premise, $P(N)=o(N/\sqrt{\log N})$ rules out $Q(N)=o(\sqrt{\log N})$,
establishing the claimed conditional optimality
of our inverse-suffix-array preprocessing; see
\cref{cor:inverse-suffix-array-conditional-preprocessing-lower-bound}.

The suffix-array reduction works directly on the Dictionary Matching text $T$.
After sorting the packed patterns, the reduction samples every
$\lceil\log n\rceil$th suffix-array entry and merges the samples with the
patterns.  This brackets
each insertion rank within $\lceil\log n\rceil$ suffix-array positions, so
parallel binary search needs only $\bigO(\log\log n)$ query batches; final
comparisons determine which patterns occur.
If the suffix-array index has preprocessing and query times $P(n)$ and
$Q(n)$, the total Dictionary Matching time is
\[
  \bigO\left(
    P(n)+
    \frac{n\log\log n}{\log n}\bigl(Q(n)+1\bigr)
  \right).
\]
Consequently, the premise excludes
$P(n)=o(n/\sqrt{\log n})$ whenever
$Q(n)=o(\sqrt{\log n}/\log\log n)$.  This covers the first suffix-array
family when
$B=o(\sqrt{\log n}/\log\log n)$ and the second family throughout its stated
parameter range.  These are exactly the conditional-optimality ranges given
in the introduction; see
\cref{cor:suffix-array-tradeoffs-conditional-preprocessing-lower-bound}.

\section{Preliminaries}\label{sec:prelim}

\subsection{Basic Definitions}\label{sec:prelim-basic}

A \emph{string} is a finite sequence of characters from a given
\emph{alphabet} $\Sigma$.  The length of a string $S$ is denoted $|S|$. For
any $i \in [1 \dd |S|]$, the $i$th character of $S$ is denoted $S[i]$.
A~\emph{substring} or a \emph{factor} of $S$ is a string of the form
$S[i \dd j]$, where $i\in[1\dd |S|+1]$ and $j\in[0\dd |S|]$. If $i>j$,
$S[i \dd j]$ is the \emph{empty string}, also denoted by $\emptystring$.
For $i,j\in \mathbb{Z}$, we let
  $[i \dd j] = \{k \in \Z : i \leq k \leq j\}$,
  $[i \dd j) = \{k \in \Z : i \leq k < j\}$, and
  $(i \dd j] = \{k \in \Z: i < k \leq j\}$, and often write
  $S[i \dd j) = S[i \dd j-1]$ or $S(i \dd j] = S[i+1 \dd j]$,
  where $i \leq j$.
Substrings of the form $S[1 \dd j)$ and $S[i \dd |S|{+}1)$ are called
\emph{prefixes} and \emph{suffixes}, respectively. We use
$\revstr{S}$ to denote the \emph{reverse} of $S$, i.e.,
$S[|S|]\cdots S[2]S[1]$.
We denote the \emph{concatenation} of two strings $U$ and
$V$, that is, the string $U[1]\cdots U[|U|]V[1]\cdots V[|V|]$, by $UV$
or $U\cdot V$. Furthermore, $S^k = \bigodot_{i=1}^k S$ is the
concatenation of $k \in \Zz$ copies of $S$; note that $S^0 =
\emptystring$ is the empty string. An integer $p \in [1\dd |S|]$ is
a \emph{period} of $S$ if $S[i] = S[i + p]$ holds for every
$i \in [1 \dd |S|-p]$. We denote the smallest period of $S$ as
$\per{S}$.  For every $S \in \Sigma^{+}$, we define the infinite power
$S^{\infty}$ so that $S^{\infty}[i] = S[1 + (i-1) \bmod |S|]$ for
$i \in \Z$.  In particular, $S = S^{\infty}[1 \dd |S|]$. By $\lcp{U}{V}$
we denote the length of the longest common prefix of $U$ and $V$. For
any string $S \in \Sigma^{*}$ and any $j_1, j_2 \in [1 \dd |S|]$, we
denote $\LCE{S}{j_1}{j_2} = \lcp{S[j_1 \dd |S|]}{S[j_2 \dd |S|]}$. We
use $\preceq$ to denote the order on $\Sigma$, extended to the
\emph{lexicographic} order on $\Sigma^*$, so that $U,V\in \Sigma^*$
satisfy $U \preceq V$ if and only if either
\begin{enumerate*}[label=(\alph*)]
  \item $U$ is a prefix of $V$, or
  \item $U[1 \dd i) = V[1 \dd i)$ and
    $U[i]\prec V[i]$ holds for some $i\in [1\dd \min(|U|,|V|)]$.
\end{enumerate*}

\begin{definition}[Pattern occurrences and SA-interval]\label{def:occ}
  For any pattern $\Pat \in \Sigma^{*}$ and any text $\Text \in \Sigma^*$,
  we define
  \begin{align*}
    \OccTwo{\Pat}{\Text}
      &= \{j \in [1 \dd |\Text|] : j + |\Pat| \leq |\Text| + 1\text{ and }\Text[j \dd j + |\Pat|) = \Pat\},\\
    \RangeBegTwo{\Pat}{\Text}
      &= |\{j \in [1 \dd |\Text|] : \Text[j \dd |\Text|] \prec \Pat\}|,\\
    \RangeEndTwo{\Pat}{\Text}
      &= \RangeBegTwo{\Pat}{\Text} + |\OccTwo{\Pat}{\Text}|.
  \end{align*}
\end{definition}

\subsection{Model of Computation}\label{sec:prelim-model}

Throughout the paper, we use the standard word RAM model of computation~\cite{Hagerup98}
with $w$-bit \emph{machine words}, where $w =\Omega(\log N)$ (where $N$ is the size
of the input), and all standard bitwise and arithmetic operations take $\bigO(1)$ time.
If the unit of space is not specified (e.g., it is stated as \emph{space usage} or
\emph{space complexity}), then
the space is measured in $w$-bit words. In many situations, however, we explicitly
specify the unit; most commonly in this paper we measure the space \emph{in bits}.

\begin{remark}\label{rm:space}
  In this paper, we adopt the standard notion that the space usage/complexity of an algorithm is its peak memory usage.
  Since all our algorithms operate entirely in memory, this quantity includes the input and output size of an algorithm.
  Throughout the paper, the algorithms and data structures we construct and use have explicit representations: auxiliary
  memory and output data structures are materialized word by word during the computation. Thus, an algorithm running in
  $t$ time creates and retains at most $\bigO(t)$ words of auxiliary or constructed space. In particular, our data
  structures constructed in $t$ time use $\bigO(t)$ words of space.
\end{remark}

\subsection{Static Dictionaries}\label{sec:prelim-static-dictionaries}

\begin{theorem}[Static dictionaries~\cite{Ruzic08}]
  \label{th:static-dictionaries}
  Let $n\in\Z_{\geq2}$, and consider the word RAM model with word size
  $w\geq\log n$. Let $A \subseteq [0 \dd 2^w) \times [0 \dd 2^w)$
  be a set of $|A| = n$ key--value pairs such that $x_1 \neq x_2$ for all
  distinct $(x_1, y_1), (x_2, y_2) \in A$. Given the set $A$, we can
  deterministically construct a data structure occupying $\bigO(nw)$ bits
  such that, given any $x \in [0 \dd 2^w)$, it determines in $\bigO(1)$ time
  whether there exists a $y \in [0 \dd 2^w)$ satisfying $(x,y) \in A$ and, if
  so, returns the unique such $y$. The construction takes
  $\bigO(n (1 + \log \log n)^2)$ time and has a peak space usage of
  $\bigO(n (1 + \log \log n)^2)$ words.
\end{theorem}

\subsection{Packed Representation of Strings and String Sequences}\label{sec:prelim-packed}

\begin{definition}[Packed representation of a string]\label{def:packed-representation}
  In the word RAM model with word size $w$, let $\AlphabetSize \in \Z_{\geq 2}$
  be an alphabet size satisfying $\log \AlphabetSize \leq w$.
  The \emph{packed representation} of a string $S \in \IntegerAlphabet^{*}$,
  denoted $\PackedRepresentation{w}{\AlphabetSize}{S}$,
  is the unique sequence of integers $(x_1, x_2, \ldots, x_k)$ encoding $S$,
  satisfying the following properties:
  \begin{enumerate}
  \item Each $x_i$ is a $w$-bit word, i.e., $x_i \in [0 \dd 2^{w})$.
  \item Each symbol of $S$ is encoded using $b := \lceil \log \AlphabetSize \rceil$ bits.
  \item Each word stores $s := \lfloor w / b \rfloor$ symbols.
  \item The sequence consists of $k := \lceil |S| / s \rceil$ words.
  \item Any bits in a word not used to store a symbol from $S$ are set to
    $0$.
  \end{enumerate}
  Equivalently, for every $i \in [1 \dd k]$,
  \[
    x_i = \sum_{\delta=1}^{\min(s, |S|-(i-1)s)} S[(i-1)s+\delta] \cdot 2^{(\delta-1)b}.
  \]
  In other words, for every $j \in [1 \dd |S|]$, the symbol $S[j]$ is stored
  in word $x_i$ where $i = \lceil j/s \rceil$, at the relative position
  $\delta = j - (i-1)s$. The retrieval formula is thus given by:
  \[
    S[j] = \left\lfloor \frac{x_i}{2^{(\delta-1)b}} \right\rfloor \bmod 2^b.
  \]
  Using the inequalities $\lceil x \rceil < x+1$ and
  $\lfloor x \rfloor \geq x/2$ (valid for $x \geq 1$), the number of words
  $k$ can be bounded as follows:
  \[
    k =
    \left\lceil \frac{|S|}{\lfloor w / b \rfloor} \right\rceil
    \leq 1 + \frac{|S|}{\lfloor w / b \rfloor}
    \leq 1 + \frac{2 |S| b}{w}
    = \bigO\left(1 + \frac{|S| \log \AlphabetSize}{w} \right).
  \]
\end{definition}
\vspace{1ex}

\begin{definition}[Packed representation of a sequence of strings]\label{def:packed-sequence-representation}
  In the word RAM model with word size $w$, let
  $\AlphabetSize \in \Z_{\geq 2}$ satisfy $\log \AlphabetSize \leq w$, and
  let $m,\ell \in \Z_{\geq 0}$. For any sequence $W[1 \dd m]$ of $m$
  strings, each of length $\ell$ over alphabet $[0\dd\AlphabetSize)$, we
  define the \emph{packed sequence representation} of $W$ as the packed
  representation of the concatenation of its strings:
  \[
    \PackedSeqRepresentation{w}{\AlphabetSize}{W}
      :=\PackedRepresentation{w}{\AlphabetSize}{\textstyle\bigodot_{i=1,\ldots,m}W[i]}.
  \]
  In particular, word boundaries need not coincide with string boundaries.
  The symbol $W[i][j]$ occupies position $(i-1)\ell+j$ in the represented
  concatenation. By \cref{def:packed-representation}, the number of words in this sequence is
  \[
    k
      = \left\lceil
          \frac{m\ell}{\lfloor w/\lceil\log\AlphabetSize\rceil\rfloor}
        \right\rceil
      = \bigO\left(1 + \frac{m\ell \log \AlphabetSize}{w} \right).
  \]
\end{definition}
\vspace{1ex}

\begin{example}\label{ex:packed-representation}
  Let $w = 8$ and $\AlphabetSize = 3$. Then
  $b = \lceil \log \AlphabetSize \rceil = 2$ and
  $s = \lfloor w/b \rfloor = 4$, so each word stores $4$ symbols,
  each using $2$ bits. Consider the sequence
  \[
    W = \big((2,0,1),(0,2,1)\big) \in \big([0\dd 3)^3\big)^2,
  \]
  corresponding to $(\texttt{cab},\texttt{acb})$ under the mapping
  $0 = \texttt{a}$, $1 = \texttt{b}$, and $2 = \texttt{c}$.
  Its concatenation is $(2,0,1,0,2,1)$, so its packed representation consists
  of $k = \lceil (2\cdot 3)/4\rceil = 2$ words, namely
  \[
    \PackedSeqRepresentation{8}{3}{W}
      = (x_1, x_2)
      = \big((00010010)_2, (00000110)_2\big)
      = \big(18, 6\big).
  \]
  The first word stores all of $W[1]$ followed by $W[2][1]$, illustrating
  that word boundaries need not coincide with string boundaries. The second
  word stores $W[2][2]$ and $W[2][3]$ and pads its remaining bits with $0$.
  For example, $W[2][2]$ occupies position $j=(2-1)\cdot3+2=5$ in the
  concatenation, so $i = \lceil 5/4 \rceil = 2$ and $\delta = 1$. Thus
  \[
    W[2][2]
      = \left\lfloor \frac{x_2}{2^{(\delta-1)b}} \right\rfloor \bmod 2^b
      = \left\lfloor \frac{6}{2^0} \right\rfloor \bmod 4
      = 2.
  \]
\end{example}
\vspace{1ex}

\begin{remark}\label{rm:packed-representation}
  To avoid mixing low-level bit manipulations with higher-level algorithmic
  concepts, we treat packed strings as an abstract layer and use the
  proposition below as its interface. The stated operations can be
  implemented straightforwardly using standard word RAM operations.
\end{remark}

\begin{proposition}\label{pr:packed-representation}
  Let $\AlphabetSize$ and $\Textlen$ be integers such that $2 \leq \AlphabetSize \leq \Textlen$.
  Assume that we have computed the value $b = \lceil \log \AlphabetSize \rceil$.
  In the word RAM model with word size $w > \log \Textlen$, the following
  operations on packed representations (\cref{def:packed-representation}) are
  supported:
  \begin{enumerate}[itemsep=1ex]

  \item\label{pr:packed-representation-initialize}
    \ul{Initialize:}
    Given any $m \in [0 \dd \Textlen]$, the packed representation
    $\PackedRepresentation{w}{\AlphabetSize}{\zero^{m}}$
    of the string $\zero^{m}$ can be constructed in
    $\bigO(1 + (m \log \AlphabetSize) / w) \subseteq
    \bigO(1 + m / \log_{\AlphabetSize} \Textlen)$ time.

  \item\label{pr:packed-representation-access}
    \ul{Access:}
    Given the packed representation $\PackedRepresentation{w}{\AlphabetSize}{S}$ and the length $|S|$
    of any string $S \in \IntegerAlphabet^{\leq \Textlen}$
    and any position $i \in [1 \dd |S|]$, the symbol $S[i]$ can be computed
    in $\bigO(1)$ time.

  \item\label{pr:packed-representation-update}
    \ul{Update:}
    Let $S \in \IntegerAlphabet^{\leq \Textlen}$, $i \in [1 \dd |S|]$, and $a \in \IntegerAlphabet$.
    Given position $i$ and symbol $a$, the packed representation
    $\PackedRepresentation{w}{\AlphabetSize}{S}$ of $S$ can be updated in $\bigO(1)$
    time to a packed representation of the string obtained by setting $S[i] := a$.

  \item\label{pr:packed-representation-substring}
    \ul{Substring:}
    Given the packed representation $\PackedRepresentation{w}{\AlphabetSize}{S}$ and the length $|S|$
    of any string $S \in \IntegerAlphabet^{\leq \Textlen}$,
    for every $\ell \in [0 \dd |S|]$ and every $p \in [1 \dd |S|-\ell+1]$,
    the packed representation of the length-$\ell$ substring $S[p \dd p + \ell)$
    can be computed in
    $\bigO(1 + (\ell \log \AlphabetSize) / w) \subseteq
    \bigO(1 + \ell / \log_{\AlphabetSize} \Textlen)$ time.

  \item\label{pr:packed-representation-concat}
    \ul{Concatenate:}
    Given any $m \in [1 \dd \Textlen]$, a sequence
    $(\ell_i)_{i \in [1 \dd m]}$ of nonnegative integers satisfying
    $\sum_{i=1}^{m} \ell_i \leq \Textlen$, and, for every
    $i \in [1 \dd m]$, the packed representation
    $\PackedRepresentation{w}{\AlphabetSize}{S_i}$ of a string
    $S_i \in \IntegerAlphabet^{\ell_i}$,
    the packed representation $\PackedRepresentation{w}{\AlphabetSize}{S}$ of their concatenation
    $S = S_1 S_2 \cdots S_m$ can be computed in
    $\bigO(m + (|S| \log \AlphabetSize) / w) \subseteq
    \bigO(m + |S| / \log_{\AlphabetSize} \Textlen)$ time.

  \item\label{pr:packed-representation-split}
    \ul{Split:}
    Given any $m \in [1 \dd \Textlen]$, a sequence
    $(\ell_i)_{i \in [1 \dd m]}$ of nonnegative integers satisfying
    $\sum_{i=1}^{m} \ell_i \leq \Textlen$,
    together with the length $|S|$ and the packed representation
    $\PackedRepresentation{w}{\AlphabetSize}{S}$ of a string
    $S \in \IntegerAlphabet^{*}$ satisfying
    $|S| = \sum_{i=1}^{m} \ell_i$, the packed representation
    $\PackedRepresentation{w}{\AlphabetSize}{S_i}$
    of every string in the unique sequence $(S_i)_{i \in [1 \dd m]}$
    satisfying $S = S_1 S_2 \cdots S_m$ and $|S_i| = \ell_i$ can be
    computed in $\bigO(m + (|S| \log \AlphabetSize) / w) \subseteq
    \bigO(m + |S| / \log_{\AlphabetSize} \Textlen)$ time.

  \end{enumerate}
\end{proposition}

\subsection{Suffix Arrays}\label{sec:prelim-suffix-array}

\begin{figure}
  \centering
  \begin{tikzpicture}[yscale=0.35]
    \foreach \x [count=\i] in {a, aabba, abaabba, abba, abbabaabba,
        ba, baabba, babaabba, bba, bbabaabba}
      \draw (1.9, -1.05*\i) node[right]
        {$\texttt{\x}$};
    \draw(1.9,0) node[right] {\scriptsize $\Text[\SA{\Text}[i]\dd \Textlen]$};
    \foreach \x [count=\i] in {a, a, a, a, a, a, a, a, a, a}
      \draw (0.9, -1.05*\i) node {\footnotesize $\i$};
    \draw(0.9,0) node{\scriptsize $i$};
    \foreach \x [count=\i] in {10,6,4,7,1,9,5,3,8,2}
      \draw (1.5, -1.05*\i) node {$\x\vphantom{\textbf{\underline{7}}}$};
    \draw(1.5,0) node{\scriptsize $\SA{\Text}[i]$};
  \end{tikzpicture}
  \caption{\small A list of sorted suffixes of $\Text = \texttt{abbabaabba}$ along with the suffix array.}\label{fig:sa-example}
\end{figure}

\begin{definition}[Suffix array]\label{def:suffix-array}
  For any string $\Text \in \Sigma^{\Textlen}$ of length
  $\Textlen \geq 1$, the \emph{suffix array} $\SA{\Text}[1 \dd \Textlen]$
  of $\Text$ is a permutation of $[1 \dd \Textlen]$ such that
  $\Text[\SA{\Text}[1] \dd \Textlen] \prec
  \Text[\SA{\Text}[2] \dd \Textlen] \prec \cdots \prec
  \Text[\SA{\Text}[\Textlen] \dd \Textlen]$, i.e., $\SA{\Text}[i]$ is
  the starting position of the lexicographically $i$th suffix of
  $\Text$.
\end{definition}

\begin{definition}[Inverse suffix array]\label{def:inverse-suffix-array}
  The \emph{inverse suffix array} of a text $\Text \in \Sigma^{\Textlen}$
  of length $\Textlen \geq 1$
  is an array $\ISA{\Text}[1 \dd \Textlen]$ containing the inverse
  permutation of $\SA{\Text}$ (\cref{def:suffix-array}), i.e., $\ISA{\Text}[j] = i$ holds if and
  only if $\SA{\Text}[i] = j$. Equivalently, $\ISA{\Text}[j]$ stores
  the lexicographic \emph{rank} of $\Text[j \dd \Textlen]$ among the
  suffixes of $\Text$, that is,
  $\ISA{\Text}[j] = 1 + \RangeBegTwo{\Text[j \dd \Textlen]}{\Text}$
  (\cref{def:occ}).
\end{definition}

\begin{example}\label{ex:sa-and-isa}
  The suffix array for the example string
  $\Text = \texttt{abbabaabba}$
  is shown in \cref{fig:sa-example}. The inverse suffix array for the same
  example string is
  \[
    \ISA{\Text} = [5, 10, 8, 3, 7, 2, 4, 9, 6, 1].
  \]
\end{example}

\begin{remark}\label{rm:occ}
  Note that the two values $\RangeBegTwo{\Pat}{\Text}$ and
  $\RangeEndTwo{\Pat}{\Text}$ (\cref{def:occ}) are the endpoints of
  the so-called \emph{SA-interval} representing the occurrences of
  $\Pat$ in $\Text$, i.e.,
  \[
    \OccTwo{\Pat}{\Text} =
      \{\SA{\Text}[i] : i \in (\RangeBegTwo{\Pat}{\Text} \dd \RangeEndTwo{\Pat}{\Text}]\}
  \]
  holds for every $\Text \in \Sigma^{+}$ and $\Pat \in \Sigma^{*}$,
  including when $\Pat = \emptystring$ and when
  $\OccTwo{\Pat}{\Text} = \emptyset$.
\end{remark}

\begin{example}\label{ex:occ-and-ranges}
  For the example text $\Text$ in \cref{fig:sa-example} and pattern
  $\Pat = \texttt{abb}$,
  it holds
  $\OccTwo{\Pat}{\Text} = \{7,1\} = \{\SA{\Text}[i] : i \in (3 \dd 5]\}$,
  so $\RangeBegTwo{\Pat}{\Text} = 3$ and $\RangeEndTwo{\Pat}{\Text} = 5$.
\end{example}

\begin{theorem}[{\cite{Farach-ColtonFM00,KoA03,KarkkainenSB06}}]\label{th:sa-and-isa-construction}
  Let $\AlphabetSize, \Textlen \in \Z_{\geq 2}$ be such that $\AlphabetSize \leq \Textlen$.
  In the word RAM model
  with word size $w > \log \Textlen$, the suffix array $\SA{\Text}$ (\cref{def:suffix-array}) and
  the inverse suffix array $\ISA{\Text}$ (\cref{def:inverse-suffix-array}) of
  any text $\Text \in [0 \dd \AlphabetSize)^{\Textlen}$ can
  be computed in $\bigO(\Textlen)$ time, given the string $\Text$
  represented as an array of $\Textlen$ integers (i.e., using $\bigO(\Textlen)$
  words of space).
\end{theorem}

\begin{theorem}[{\cite{SaPerfectEquiv}}]
  \label{th:sa-isa-space-lower-bound}
  Let $\AlphabetSize,\Textlen\in\Z_{\geq2}$ satisfy
  $\AlphabetSize\leq\Textlen$.  Any data structure supporting suffix array
  queries for every text in $[0\dd\AlphabetSize)^{\Textlen}$ has worst-case
  space usage $\Omega(\Textlen\log\AlphabetSize)$ bits.  The same lower bound
  holds for inverse suffix array queries.
\end{theorem}

\begin{theorem}[{\cite{SaPerfectEquiv}}]
  \label{th:suffix-array-cell-probe-lower-bound}
  Let $N$ be a sufficiently large positive integer.  In the cell-probe model
  with $\Theta(\log N)$-bit cells, every deterministic data structure
  supporting suffix array queries for all binary texts of length at most $N$
  and using at most $S(N)$ bits, where $S(N)=\Omega(N)$, has worst-case query time
  \[
    \Omega\left(
      \frac{\log\log N}{\log((S(N)/N)\log\log N)}
    \right).
  \]
\end{theorem}

\subsection{Rank and Selection Queries}\label{sec:prelim-rank-select}

\begin{definition}[Rank and selection queries]\label{def:rank-select}
  Let $S \in \Sigma^{m}$ be a string.
  \begin{description}[style=sameline,itemsep=1ex]
  \item[Rank query:]
    For every $a \in \Sigma$ and $j \in [0 \dd m]$, we define
    $\Rank{S}{j}{a} := |\{i \in [1\dd j]: S[i] = a\}|$.
  \item[Special rank query:]
    For every $j \in [1 \dd m]$, we define
    $\SpecialRank{S}{j} := \Rank{S}{j}{S[j]}$.
  \item[Select query:]
    For every $a \in \Sigma$ and $r \in \Z_{\geq 1}$,
    we define $\Select{S}{r}{a}$ as the $r$th smallest element of
    the set $\{i \in [1 \dd m] : S[i] = a\}$ (if $r \leq \Rank{S}{m}{a}$),
    and $\Select{S}{r}{a} := \infty$ (otherwise).
  \end{description}
\end{definition}

\begin{example}\label{ex:rank-and-select-queries}
  Let $S = \texttt{abbaabaababbabbaabb}$. Then
  \[
    \Rank{S}{11}{\texttt{a}} = 6,\qquad
    \SpecialRank{S}{14} = 7,\qquad
    \Select{S}{5}{\texttt{a}} = 8,\qquad
    \Select{S}{11}{\texttt{b}} = \infty.
  \]
  Indeed, among the first $11$ symbols of $S$, the letter $\texttt{a}$
  occurs $6$ times. Moreover, $S[14] = \texttt{b}$, and $\texttt{b}$
  occurs $7$ times in the prefix of length $14$. Finally, the
  occurrences of $\texttt{a}$ in $S$ are at positions
  $1,4,5,7,8,10,13,16,17$. Hence, the fifth occurrence is at position $8$,
  while $\texttt{b}$ occurs only $10$ times in total.
\end{example}

\begin{theorem}[Rank and selection queries in
    bitvectors~\cite{WaveletSuffixTree,Clark98,Jac89,MunroNV16}]\label{th:bin-rank-select}
  Let $n \in \Z_{\geq 2}$. Consider a word RAM model with word size $w > \log n$.
  For every binary string $S \in \BinaryAlphabet^{\leq n}$, there exists a data structure of
  $\bigO(|S|)$ bits answering rank and selection queries (\cref{def:rank-select}) in
  $\bigO(1)$ time. Moreover, given the packed representation (\cref{def:packed-representation})
  of any $m$ binary strings of total length at most $n$, the data structures for all
  these strings can be constructed in $\bigO(m + n / \log n)$ time.
\end{theorem}

\begin{theorem}[{\cite{SaPerfectEquiv}}]\label{th:select-space-lower-bound-exact}
  Let $\AlphabetSize, \Textlen \in \Z_{\geq 2}$ satisfy
  $\AlphabetSize \leq \Textlen$.
  Any data structure supporting select queries (\cref{def:rank-select})
  for every string $\Text \in \IntegerAlphabet^{\Textlen}$ has worst-case
  space usage $\Omega(\Textlen \log \AlphabetSize)$ bits.
\end{theorem}

\begin{corollary}[{\cite{SaPerfectEquiv}}]\label{cor:select-space-lower-bound}
  Let $\AlphabetSize, \Textlen \in \Z_{\geq 2}$ satisfy
  $\AlphabetSize \leq \Textlen$.
  Any data structure supporting select queries (\cref{def:rank-select})
  for every string $\Text \in \IntegerAlphabet^{*}$ satisfying
  $\AlphabetSize \leq |\Text| \leq \Textlen$ has worst-case space usage
  $\Omega(\Textlen \log \AlphabetSize)$ bits.
\end{corollary}

\begin{theorem}[{\cite[Theorem~2.1]{SaPerfectEquiv}}]
  \label{th:special-rank-space-lower-bound-exact}
  Let $\AlphabetSize, \Textlen \in \Z_{\geq 2}$ satisfy
  $\AlphabetSize \leq \Textlen$.
  Any data structure supporting special rank queries
  (\cref{def:rank-select}) for every string
  $\Text \in \IntegerAlphabet^{\Textlen}$ has worst-case space usage
  $\Omega(\Textlen \log \AlphabetSize)$ bits.
\end{theorem}

\begin{corollary}[{\cite[Corollary~5.20]{SaPerfectEquiv}}]
  \label{cor:special-rank-space-lower-bound}
  Let $\AlphabetSize, \Textlen \in \Z_{\geq 2}$ satisfy
  $\AlphabetSize \leq \Textlen$.
  Any data structure supporting special rank queries
  (\cref{def:rank-select}) for every string
  $\Text \in \IntegerAlphabet^{*}$ satisfying
  $\AlphabetSize \leq |\Text| \leq \Textlen$ has worst-case space usage
  $\Omega(\Textlen \log \AlphabetSize)$ bits.
\end{corollary}

\subsection{Prefix Range Queries}\label{sec:prelim-prefix-range-queries}

\begin{definition}[Prefix range queries]\label{def:prefix-range-queries}
  Let $S \in (\Sigma^{*})^m$ be a sequence of strings over alphabet $\Sigma$.
  \begin{description}[style=sameline,itemsep=1ex]
  \item[Prefix rank query:]
    For every $X \in \Sigma^{*}$ and $j \in [0 \dd m]$, we define
    $\PrefixRank{S}{j}{X} := |\{i \in [1 \dd j]: X\text{ is a prefix of }S[i]\}|$.
  \item[Prefix special rank query:]
    For every $j \in [1 \dd m]$ and $p \in [0 \dd |S[j]|]$, we let
    $\PrefixSpecialRank{S}{j}{p} :=\allowbreak \PrefixRank{S}{j}{S[j][1 \dd p]}$.
  \item[Prefix select query:]
    For every $X \in \Sigma^{*}$ and $r \in \Z_{\geq 1}$,
    we define $\PrefixSelect{S}{r}{X}$ as the $r$th smallest element of
    the set $\{i \in [1 \dd m] : X\text{ is a prefix of }S[i]\}$
    (if $r \leq \PrefixRank{S}{m}{X}$),
    and $\PrefixSelect{S}{r}{X} := \infty$ (otherwise).
  \end{description}
\end{definition}

\noindent
\begin{minipage}[t]{\dimexpr\linewidth-0.18\linewidth\relax}
  \begin{example}\label{ex:prefix-range-queries}
    Let $S \in (\Sigma^*)^9$ be the sequence of strings shown on the right, and let
    $X = \texttt{bb}$. Then
    \begin{itemize}
    \item $\PrefixRank{S}{8}{X} = 2$,
    \item $\PrefixSpecialRank{S}{7}{2} = 2$,
    \item $\PrefixSelect{S}{3}{X} = 9$, and
    \item $\PrefixSelect{S}{4}{X} = \infty$.
    \end{itemize}
    Indeed, exactly the strings $S[4]$, $S[7]$, and $S[9]$ have prefix $X$.
    Moreover, $S[7][1 \dd 2] = \texttt{bb}$, so among the first $7$ strings,
    exactly two have prefix $S[7][1 \dd 2]$, namely $S[4]$ and $S[7]$.
  \end{example}
\end{minipage}\hfill%
\begin{minipage}[t]{0.18\linewidth}
  \vspace{0pt}
  \makebox[\linewidth][r]{%
    \begin{tabular}[t]{@{}r@{\ }c@{\ }l@{}}
      $S[1]$ & = & \texttt{caba}\\
      $S[2]$ & = & \texttt{baba}\\
      $S[3]$ & = & \texttt{abba}\\
      $S[4]$ & = & \texttt{bbab}\\
      $S[5]$ & = & \texttt{baaa}\\
      $S[6]$ & = & \texttt{aabb}\\
      $S[7]$ & = & \texttt{bbaa}\\
      $S[8]$ & = & \texttt{abab}\\
      $S[9]$ & = & \texttt{bbba}
    \end{tabular}%
  }
\end{minipage}%

\begin{theorem}[{\cite[Lemmas~5.12 and~5.21; Theorems~2.1
    and~5.10]{SaPerfectEquiv}}]
  \label{th:prefix-query-space-lower-bounds-exact}
  Let $\AlphabetSize,m\in\Z_{\geq2}$ and $\ell\in\Z_{\geq1}$ satisfy
  $\AlphabetSize^\ell\leq m$.  Any data structure supporting prefix rank,
  prefix special rank, or prefix select queries
  (\cref{def:prefix-range-queries}) for every sequence
  $W\in([0\dd\AlphabetSize)^\ell)^m$ has worst-case space usage
  $\Omega(m\ell\log\AlphabetSize)$ bits.  The bound holds separately for
  each of the three query types.
\end{theorem}
\begin{proof}
  Denote $k:=\AlphabetSize^\ell$. We identify the characters of $[0\dd k)$
  with the strings in $[0\dd\AlphabetSize)^\ell$.  Under this
  identification, rank, special rank, and select on a string in
  $[0\dd k)^m$ reduce, respectively, to prefix rank, prefix special rank,
  and prefix select on a sequence in $([0\dd\AlphabetSize)^\ell)^m$.
  By $k\leq m$,
  \cref{th:select-space-lower-bound-exact,th:special-rank-space-lower-bound-exact}
  yield
  $\Omega(m\log k)=\Omega(m\ell\log\AlphabetSize)$ bits for prefix select
  and prefix special rank.  The same bound holds for rank, since select can
  be implemented using rank by binary search without increasing the space
  usage, and hence also for prefix rank.
\end{proof}

\section{Basic Tools for Packed Strings}\label{sec:tools}

\subsection{Preliminaries}\label{sec:tools-preliminaries}

\begin{definition}[Integer value of a string]\label{def:val}
  Let $\AlphabetSize \in \Z_{\geq 2}$. For every $X \in [0 \dd \AlphabetSize)^{+}$,
  letting $m = |X|$, we define $\Val{\AlphabetSize}{X} := \sum_{i=0}^{m-1} X[m-i] \cdot \AlphabetSize^{i}$
  as the unique integer such that
  viewing its representation as a base-$\AlphabetSize$ number (padded on the
  left side, if needed, with zeros to length $m$) we obtain string $X$.
  We also define $\Val{\AlphabetSize}{\emptystring} := 0$.
\end{definition}

\begin{definition}[Basic integer coding of variable-length strings]\label{def:basic-int}
  Let $\AlphabetSize \in \Z_{\geq 2}$. For every $X \in \IntegerAlphabet^{*}$,
  letting $m = |X|$, we define
  $\BasicInt{\AlphabetSize}{X} := \AlphabetSize^{m} + \Val{\AlphabetSize}{X} \in [\AlphabetSize^{m} \dd 2\AlphabetSize^{m})$
  (see \cref{def:val}).
\end{definition}

\begin{example}\label{ex:val-and-basic-int}
  Let $\AlphabetSize = 4$ and let $X = (0, 2, 1)$. Then $|X| = 3$, and the
  unique integer whose base-$4$ representation padded to length $3$ is
  $021$ equals $x = 9$. Hence
  \[
    \Val{4}{X} = 9\qquad\text{ and }\qquad
    \BasicInt{4}{X} = 4^3 + 9 = 73 = (1021)_4.
  \]
\end{example}

\begin{observation}\label{ob:basic-int}
  Let $\AlphabetSize \in \Z_{\geq 2}$. For any $X, X' \in [0 \dd \AlphabetSize)^{*}$,
  $\BasicInt{\AlphabetSize}{X} < \BasicInt{\AlphabetSize}{X'}$ (\cref{def:basic-int}) holds if and only
  if $|X| < |X'|$, or $|X| = |X'|$ and $X \prec X'$.
\end{observation}

\begin{definition}[Alphabet mapping]\label{def:alphabet-map}
  Let $\AlphabetSize_1, \AlphabetSize_2 \in \Z_{\geq 2}$ be such that $\AlphabetSize_1 \geq \AlphabetSize_2$.
  Denote $k = \lceil \log_{\AlphabetSize_2} \AlphabetSize_1 \rceil$. For any $x \in [0 \dd \AlphabetSize_1)$,
  by $\AlphabetMap{\AlphabetSize_1}{\AlphabetSize_2}{x} \in [0 \dd \AlphabetSize_2)^k$ we denote the
  string of length $k$ containing the base-$\AlphabetSize_2$ representation of $x$
  (padded on the left side with leading zeros, if needed).
  For any string $S \in [0 \dd \AlphabetSize_1)^{*}$, we then let
  \[
    \AlphabetMap{\AlphabetSize_1}{\AlphabetSize_2}{S} :=
      \textstyle\bigodot_{i=1,\ldots,|S|} \AlphabetMap{\AlphabetSize_1}{\AlphabetSize_2}{S[i]}
      \in [0 \dd \AlphabetSize_2)^{|S|\cdot k}.
  \]
\end{definition}

\begin{remark}\label{rm:alphabet-map}
  Observe that the integer
  $k=\lceil\log_{\AlphabetSize_2}\AlphabetSize_1\rceil$ in
  \cref{def:alphabet-map} satisfies
  $k\geq\log_{\AlphabetSize_2}\AlphabetSize_1$, and hence
  $\AlphabetSize_2^k\geq\AlphabetSize_1$. Consequently, the
  base-$\AlphabetSize_2$ representation of every
  $x\in[0\dd\AlphabetSize_1)$ can be padded on the left with zeros to length
  $k$, as specified in \cref{def:alphabet-map}.
\end{remark}

\begin{proposition}[{\cite{SaPerfectEquiv}}]\label{pr:alphabet-map}
  Let $\Seqlen,\AlphabetSize_1,\AlphabetSize_2\in\Z_{\geq2}$ satisfy
  $\AlphabetSize_2\leq\AlphabetSize_1\leq\Seqlen$. In the word RAM model
  with word size $w\geq2\log\Seqlen$, we can in $\bigO(\sqrt\Seqlen)$ time
  construct a data structure that, given $|S|$ and
  $\PackedRepresentation{w}{\AlphabetSize_1}{S}$
  (\cref{def:packed-representation}) for any
  $S\in[0\dd\AlphabetSize_1)^{\leq\Seqlen}$, returns
  $\PackedRepresentation{w}{\AlphabetSize_2}
    {\AlphabetMap{\AlphabetSize_1}{\AlphabetSize_2}{S}}$
  (\cref{def:alphabet-map}) in
  $\bigO(1+|S|/\log_{\AlphabetSize_1}\Seqlen)$ time.
\end{proposition}

\begin{proposition}[{\cite{SaPerfectEquiv}}]\label{pr:val-encoding}
  Let $\Seqlen, \AlphabetSize \in \Z_{\geq 2}$ be such that $\AlphabetSize \leq \Seqlen$.
  Denote $\ell = \lfloor \log_{\AlphabetSize} \Seqlen \rfloor$.
  In the word RAM model with word size $w \geq 1 + \log\Seqlen$, we can in
  $\bigO(\sqrt{\Seqlen})$ time construct a data structure
  that, given $|X|$ and the packed representation $\PackedRepresentation{w}{\AlphabetSize}{X}$ (\cref{def:packed-representation})
  of any string $X \in \IntegerAlphabet^{\leq \ell}$,
  returns the integer $\Val{\AlphabetSize}{X}$ (\cref{def:val}) in $\bigO(1)$ time.
\end{proposition}

\begin{proposition}[{\cite{SaPerfectEquiv}}]\label{pr:val-decoding}
  Let $\Seqlen, \AlphabetSize \in \Z_{\geq 2}$ be such that $\AlphabetSize \leq \Seqlen$.
  Denote $\ell = \lfloor \log_{\AlphabetSize} \Seqlen \rfloor$.
  In the word RAM model with word size $w \geq 1 + \log\Seqlen$, we can in
  $\bigO(\sqrt{\Seqlen})$ time construct a data structure
  that, given $|X|$ and the integer $\Val{\AlphabetSize}{X}$ (\cref{def:val})
  for any string $X \in \IntegerAlphabet^{\leq \ell}$,
  returns the packed representation
  $\PackedRepresentation{w}{\AlphabetSize}{X}$ (\cref{def:packed-representation})
  in $\bigO(1)$ time.
\end{proposition}

\begin{proposition}[{\cite{SaPerfectEquiv}}]\label{pr:basic-int-encoding}
  Let $\Seqlen, \AlphabetSize \in \Z_{\geq 2}$ be such that $\AlphabetSize \leq \Seqlen$.
  Denote $\ell = \lfloor \log_{\AlphabetSize} \Seqlen \rfloor$.
  In the word RAM model with word size $w \geq 1 + \log\Seqlen$, we can in
  $\bigO(\sqrt{\Seqlen})$ time construct a data structure
  that, given $|X|$ and the packed representation $\PackedRepresentation{w}{\AlphabetSize}{X}$ (\cref{def:packed-representation})
  of any string $X \in \IntegerAlphabet^{\leq \ell}$,
  returns the integer $\BasicInt{\AlphabetSize}{X}$ (\cref{def:basic-int}) in $\bigO(1)$ time.
\end{proposition}

\begin{proposition}[{\cite{SaPerfectEquiv}}]\label{pr:packed-string-reversal}
  Let $\Seqlen,\AlphabetSize\in\Z_{\geq2}$ satisfy
  $\AlphabetSize\leq\Seqlen$. In the word RAM model with word size
  $w\geq1+\log\Seqlen$, we can in $\bigO(\sqrt\Seqlen)$ time construct a data
  structure occupying $\bigO(\sqrt\Seqlen)$ words that, given $|S|$ and
  $\PackedRepresentation{w}{\AlphabetSize}{S}$
  (\cref{def:packed-representation}) for any
  $S\in[0\dd\AlphabetSize)^{\leq\Seqlen}$, returns
  $\PackedRepresentation{w}{\AlphabetSize}{\revstr{S}}$ in
  $\bigO(1+|S|/\log_{\AlphabetSize}\Seqlen)$ time.
\end{proposition}

\subsection{Packed Prefix Frequencies}
  \label{sec:tools-packed-prefix-frequencies}

\begin{proposition}[Packed prefix frequencies]
  \label{pr:packed-prefix-frequencies}
  Let $\AlphabetSize,m\in\Z_{\geq2}$ and $\ell\in\Z_{\geq1}$ satisfy
  $\AlphabetSize^\ell\leq m$, and consider the word RAM model with word size
  $w=c\log m$, where $c\geq2$ is a constant.
  There is an algorithm that, given the packed sequence representation
  $\PackedSeqRepresentation{w}{\AlphabetSize}{W}$
  (\cref{def:packed-sequence-representation}) of a sequence
  $W\in([0\dd\AlphabetSize)^\ell)^m$, constructs an array
  $F[0\dd2\AlphabetSize^\ell)$ that
  satisfies (see \cref{def:basic-int,def:prefix-range-queries}):
  \[
    F[\BasicInt{\AlphabetSize}{X}]
      =\PrefixRank{W}{m}{X}
    \qquad
    \text{for every }
      X\in[0\dd\AlphabetSize)^{\leq\ell}
  \]
  All remaining entries of $F$ are set to zero. The construction takes
  $\bigO(m\ell\log\AlphabetSize/\log m)$ time and has a peak space usage of
  $\bigO(m\ell\log\AlphabetSize)$ bits.
\end{proposition}
\begin{proof}
  If $m<16$, scan every input string character by character, count the
  occurrences of each length-$\ell$ string, and sum these counts bottom-up
  over their prefixes. All arrays then have constant size, so the claimed
  bounds follow. Thus, assume that $m\geq16$.

  Compute $b:=\ceil{\log\AlphabetSize}$ by repeated doubling and
  $\AlphabetSize^\ell$ by repeated multiplication. This takes
  $\bigO(\log m)$ time because
  $\AlphabetSize\leq\AlphabetSize^\ell\leq m$ and $\ell\leq\log m$.
  Moreover, $b\leq2\log\AlphabetSize$. Using
  $\log(m/\AlphabetSize^\ell)\leq m/\AlphabetSize^\ell$, we obtain
  \begin{equation}\label{eq:packed-prefix-frequencies-universe}
    \AlphabetSize^\ell\log m
      =\AlphabetSize^\ell\left(
        \ell\log\AlphabetSize
        +\log\frac{m}{\AlphabetSize^\ell}
      \right)
      \leq m\ell\log\AlphabetSize+m
      =\bigO(m\ell\log\AlphabetSize),
    \qquad
    \AlphabetSize^\ell
      =\bigO\left(\frac{m\ell\log\AlphabetSize}{\log m}\right).
  \end{equation}
  Computing $b$ and $\AlphabetSize^\ell$ is also bounded by
  $\bigO(m\ell\log\AlphabetSize/\log m)$ because
  $m/\log m\geq\log m$ and $\ell\log\AlphabetSize\geq1$.

  Every application of
  \cref{pr:packed-representation}\eqref{pr:packed-representation-substring}
  below uses length parameter $m\ell$. Its assumptions hold because
  $\AlphabetSize\leq m\leq m\ell$ and
  $w\geq2\log m>\log(m\ell)$, where we use
  $\ell\leq\log m<m$. The supplied packed sequence representation is the
  packed representation of $W[1]\cdots W[m]$, whose length is $m\ell$.
  We construct $F$ in two steps.
  \begin{enumerate}

  \item \emph{Compute the frequencies of the length-$\ell$ strings.}
    Allocate $F[0\dd2\AlphabetSize^\ell)$ as a plain array of counters and
    initialize every entry to zero. Each counter fits in one word. By
    \cref{eq:packed-prefix-frequencies-universe}, the initialization takes
    $\bigO(m\ell\log\AlphabetSize/\log m)$ time, and $F$ uses
    $\bigO(m\ell\log\AlphabetSize)$ bits. We compute the frequencies using
    one of the following two methods.
    \begin{enumerate}

      \item \emph{Suppose that $\ell b\geq(\log m)/8$.} Construct the
        string-encoding structure from \cref{pr:basic-int-encoding} with
        parameters $m$ and $\AlphabetSize$. Its assumptions hold because
        $\AlphabetSize\leq m$,
        $\ell\leq\floor{\log_{\AlphabetSize}m}$, and
        $w\geq1+\log m$. For every $j\in[1\dd m]$, apply
        \cref{pr:packed-representation}\eqref{pr:packed-representation-substring}
        to the supplied representation, with starting position
        $(j-1)\ell+1$ and substring length $\ell$. Query this structure with
        $\ell$ and the returned representation to compute
        $y:=\BasicInt{\AlphabetSize}{W[j]}$, and increment $F[y]$. The
        substring and encoding queries take constant time. Since
        $b\leq2\log\AlphabetSize$, the assumption in this case gives
        $\ell\log\AlphabetSize\geq(\log m)/16$. Thus, the scan and the
        $\bigO(\sqrt m)$-time construction from
        \cref{pr:basic-int-encoding} take
        $\bigO(m\ell\log\AlphabetSize/\log m)$ time. By
        \cref{pr:basic-int-encoding,rm:space}, the string-encoding structure uses
        $\bigO(\sqrt m\log m)=\bigO(m\ell\log\AlphabetSize)$ bits.

      \item \emph{Suppose that $\ell b<(\log m)/8$.} Compute
        $t:=\floor{\log m/(4\ell b)}
        =\Theta(\log m/(\ell b))$. In particular, $t\geq2$ and
        $t\ell b\leq(\log m)/4$. Compute the frequencies in two substeps.
        \begin{enumerate}

          \item \emph{Construct $H$.} Process the first
            $t\floor{m/t}$ strings of $W$ in consecutive groups of $t$
            strings. The strings in one group contain
            $t\ell b\leq(\log m)/4$ symbol bits and therefore fit in one
            packed word. Allocate a plain counter array
            $H[0\dd2^{t\ell b})$ and initialize it to zero. For every
            $i\in[0\dd\floor{m/t})$, apply
            \cref{pr:packed-representation}\eqref{pr:packed-representation-substring} with starting position
            $it\ell+1$ and substring length $t\ell$. Let
            $Z\in[0\dd2^{t\ell b})$ be the sole word of the returned
            representation, and increment $H[Z]$.

          \item \emph{Update $F$.} Scan $Z\in[0\dd2^{t\ell b})$. Whenever
            $H[Z]>0$, every $b$-bit symbol field of $Z$ represents a character
            in $[0\dd\AlphabetSize)$ because $Z$ was obtained from an input
            block. Split its low-order $t\ell b$ bits into $t$ groups of
            $\ell$ symbol fields. For every decoded group
            $c_1,\ldots,c_\ell$, initialize $y:=1$, successively set
            $y:=\AlphabetSize y+c_p$ for $p=1,2,\ldots,\ell$, and add $H[Z]$
            to $F[y]$. By \cref{def:basic-int}, the final value of $y$ is the
            basic-integer code of the decoded group. This counts every
            occurrence in every block represented by $Z$, including repeated
            strings within one block. For every
            $j\in[t\floor{m/t}+1\dd m]$, read the $\ell$ characters of $W[j]$,
            compute $y:=\BasicInt{\AlphabetSize}{W[j]}$ by the same recurrence,
            and increment $F[y]$.
        \end{enumerate}

        The two substeps take
        \[
          \bigO\left(m/t+2^{t\ell b}(1+t\ell)+t\ell\right)
            =\bigO\left(
              \frac{m\ell b}{\log m}+m^{1/4}\log m
            \right)
            =\bigO\left(
              \frac{m\ell\log\AlphabetSize}{\log m}
            \right)
        \]
        time. Here, we use $1/t=\bigO(\ell b/\log m)$,
        $2^{t\ell b}\leq m^{1/4}$, $t\ell=\bigO(\log m)$,
        $b\leq2\log\AlphabetSize$, and
        $m^{1/4}\log m=\bigO(m/\log m)$. The array $H$ uses
        $\bigO(m^{1/4}\log m)=\bigO(m\ell\log\AlphabetSize)$ bits.
    \end{enumerate}

    In either case, $F[\BasicInt{\AlphabetSize}{X}]$ now equals the number
    of occurrences of $X$ in $W$ for every
    $X\in[0\dd\AlphabetSize)^\ell$.

  \item \emph{Compute the frequencies of the shorter prefixes.} Process
    $p=\ell-1,\ell-2,\ldots,0$ in decreasing order. For every
    $y\in[\AlphabetSize^p\dd2\AlphabetSize^p)$, set
    \[
      F[y]:=\sum_{c=0}^{\AlphabetSize-1}F[\AlphabetSize y+c].
    \]
    There is a unique $X\in[0\dd\AlphabetSize)^p$ satisfying
    $y=\BasicInt{\AlphabetSize}{X}$, and
    $\AlphabetSize y+c=\BasicInt{\AlphabetSize}{Xc}$ for every
    $c\in[0\dd\AlphabetSize)$. When length $p$ is processed, the entries
    corresponding to the extensions $Xc$ already contain their correct prefix
    frequencies. Their sum is the number of strings in $W$ having prefix $X$.
    The total number of additions is
    $\sum_{p=0}^{\ell-1}\AlphabetSize^{p+1}
      =\bigO(\AlphabetSize^\ell)$, so by
    \cref{eq:packed-prefix-frequencies-universe}, this step takes
    $\bigO(m\ell\log\AlphabetSize/\log m)$ time. The intervals
    $[\AlphabetSize^p\dd2\AlphabetSize^p)$, for
    $p\in[0\dd\ell]$, contain exactly the basic-integer codes of the strings
    in $[0\dd\AlphabetSize)^{\leq\ell}$. Since all other entries of $F$
    were initialized to zero and are never changed, they remain zero.
  \end{enumerate}

  At every point, the input representation and $F$ coexist with either the
  string-encoding structure from \cref{pr:basic-int-encoding} or $H$. Together,
  they occupy $\bigO(m\ell\log\AlphabetSize)$ bits. The total running time is
  $\bigO(m\ell\log\AlphabetSize/\log m)$.
\end{proof}

\subsection{Packed Encoding of Character Blocks}
  \label{sec:tools-packed-character-block-encoding}

\begin{proposition}[Packed encoding of character blocks]
  \label{pr:packed-character-block-encoding}
  Let $\AlphabetSize,m\in\Z_{\geq2}$ and $a,g,h\in\Z_{\geq1}$ satisfy
  $g\leq h$ and $\AlphabetSize^{ah}\leq m$. Denote
  $k:=\AlphabetSize^a$, and consider the word RAM model with word size
  $w=c\log m$, where $c\geq2$ is a constant. Let
  $U\in([0\dd\AlphabetSize)^{ag})^m$. Using the notation from
  \cref{def:val}, let $V\in([0\dd k)^h)^m$ denote the sequence satisfying
  \[
    V[j][t]:=
    \begin{cases}
      \Val{\AlphabetSize}{U[j][(t-1)a+1\dd ta]}, & t\leq g,\\
      0, & t>g,
    \end{cases}
    \qquad
    \text{for every }j\in[1\dd m]\text{ and }t\in[1\dd h].
  \]
  There is an algorithm that, given
  $\PackedSeqRepresentation{w}{\AlphabetSize}{U}$
  (\cref{def:packed-sequence-representation}), constructs
  $\PackedSeqRepresentation{w}{k}{V}$. The algorithm takes
  $\bigO(1+mh\log k/\log m)$ time and has a peak space usage of
  $\bigO(mh\log k)$ bits.
\end{proposition}
\begin{proof}
  If $m<16$, initialize every word of
  $\PackedSeqRepresentation{w}{k}{V}$ to zero. For every
  $j\in[1\dd m]$ and $t\in[1\dd g]$, scan
  $U[j][(t-1)a+1\dd ta]$ and write its base-$\AlphabetSize$ value to field
  $(j-1)h+t$ of the output. The conditions
  $\AlphabetSize^{ah}\leq m<16$ and $g\leq h$ imply that the input and output
  contain $\bigO(1)$ symbols, so the claimed bounds follow. Thus, assume that
  $m\geq16$.

  Compute $k$, $b:=\lceil\log\AlphabetSize\rceil$, and
  $v:=h\lceil\log k\rceil$ by repeated multiplication and doubling. Thus,
  $v$ is the number of bits in one output string. Since
  $b\leq2\log\AlphabetSize$, $\lceil\log k\rceil\leq2\log k$, and
  $\log k=a\log\AlphabetSize$, the number $ag b$ of bits in one input string
  is at most $2v$. Moreover, both $ag b$ and $v$ are at most
  $2ah\log\AlphabetSize\leq2\log m\leq w$. Thus, every $U[j]$ and $V[j]$
  fits in one word. Computing $k,b,v$ takes $\bigO(\log m)$ time and bits,
  which is bounded by the claimed time and space because
  $v=\Theta(h\log k)$, $m/\log m\geq\log m$, and $m\geq16$.

  Every application of \cref{pr:packed-representation} below uses length
  parameter $mah$. The common assumptions hold because
  $\AlphabetSize,k\leq m\leq mah$ and
  $mah\leq m\log m<m^2\leq2^w$, and hence $w>\log(mah)$. The represented
  concatenation of the input strings has length $mag\leq mah$.

  We use one of the following two constructions.
  \begin{enumerate}

  \item \emph{Construct the output when $v\leq(\log m)/8$.} Compute
    $r:=\floor{\log m/(8v)}$ and $s:=\ceil{m/r}$. Thus, $r\geq1$.
    For every $i\in[1\dd s]$, let
    $r_i:=\min\{r,m-(i-1)r\}$. Thus, batch $i$ consists of the $r_i$ strings
    beginning at position $(i-1)r+1$. We construct the output in two
    substeps.
    \begin{enumerate}

      \item \emph{Construct $T$.} Allocate a word array
        $T[0\dd2^{rag b})$ and initialize it to zero. Scan
        $x\in[0\dd2^{rag b})$. Whenever all $rag$ consecutive low-order
        $b$-bit fields of $x$ belong to $[0\dd\AlphabetSize)$, perform the
        following computation for every $i\in[1\dd r]$ and $t\in[1\dd g]$.
        Initialize $z:=0$, and process
        $p=(t-1)a+1,(t-1)a+2,\ldots,ta$ in increasing order by letting
        $\gamma$ be
        $b$-bit field $(i-1)ag+p$ of $x$ and setting
        $z:=\AlphabetSize z+\gamma$. Store the final value of $z$ in
        $\lceil\log k\rceil$-bit field $(i-1)h+t$ of $T[x]$. By
        \cref{def:val}, this is the required output field for the corresponding
        length-$a$ block.
        Every field with $t>g$ remains zero. If some input field of $x$ is at
        least $\AlphabetSize$, the entry $T[x]$ remains zero. Since
        $rag b\leq2rv\leq(\log m)/4$ and $rv\leq(\log m)/8$, both $x$ and
        $T[x]$ fit in one word. Exhaustive enumeration constructs $T$ in
        $m^{1/4}(\log m)^{\bigO(1)}=o(m/\log m)$ time and bits.

      \item \emph{Construct the packed representation of $V$.} Allocate an
        array $A_{\rm batch}[1\dd s]$ of words. For every
        $i\in[1\dd s]$, apply
        \cref{pr:packed-representation}\eqref{pr:packed-representation-substring}, with alphabet size
        $\AlphabetSize$, source length $mag$, starting position
        $((i-1)r)ag+1$, and substring length $r_iag$, to obtain the packed
        representation of the concatenation of the strings in batch $i$. Let
        $x$ be its sole word. By \cref{def:packed-representation}, the zero
        high fields of $x$ encode $r-r_i$ additional all-zero input strings,
        so $T[x]$ represents the $r_i$ required output strings followed by
        $r-r_i$ all-zero output strings.
        Store $T[x]$ in $A_{\rm batch}[i]$ as the packed representation of its
        first $r_i h$ fields. The substring interval is contained in
        $[1\dd mag]$. Finally, apply
        \cref{pr:packed-representation}\eqref{pr:packed-representation-concat}, with alphabet size $k$, to the
        $s$ representations whose words are stored in $A_{\rm batch}$ and
        whose lengths are $r_1h,\ldots,r_sh$. The number of representations
        satisfies $s\leq m$, their total length is $mh$, and their concatenation is
        $\PackedSeqRepresentation{w}{k}{V}$.
    \end{enumerate}

    Since $r\geq\log m/(16v)$, we have
    $s=\bigO(1+mv/\log m)
      =\bigO(1+mh\log k/\log m)$. The applications of
    \cref{pr:packed-representation}\eqref{pr:packed-representation-substring} and
    \cref{pr:packed-representation}\eqref{pr:packed-representation-concat}
    take $\bigO(1+mh\log k/\log m)$ total time because
    $ag\log\AlphabetSize\leq h\log k$. The
    $m^{1/4}(\log m)^{\bigO(1)}$ construction time of $T$ is also
    $\bigO(1+mh\log k/\log m)$. By \cref{rm:space}, the input and output
    representations, $T$, $A_{\rm batch}$, and the working space of the two
    packed operations use $\bigO(mv)=\bigO(mh\log k)$ bits in total.

  \item \emph{Construct the output when $v>(\log m)/8$.} We use two
    substeps.
    \begin{enumerate}

      \item \emph{Construct the packed representations of the output
        strings.} Allocate an array $A_V[1\dd m]$ of words. Construct the
        value-encoding structure from \cref{pr:val-encoding} with parameters
        $(m,\AlphabetSize)$ and the value-decoding structure from
        \cref{pr:val-decoding} with parameters $(m,k)$. Their word-size
        requirements follow from $w\geq1+\log m$. The value-encoding
        structure is applicable because $\AlphabetSize\leq m$ and
        $ag\leq ah\leq\floor{\log_{\AlphabetSize}m}$. The value-decoding
        structure is applicable because
        $k\leq k^h=\AlphabetSize^{ah}\leq m$ and
        $g\leq h\leq\floor{\log_k m}$. For every $j\in[1\dd m]$, apply
        \cref{pr:packed-representation}\eqref{pr:packed-representation-substring}, with alphabet size
        $\AlphabetSize$, source length $mag$, starting position $(j-1)ag+1$,
        and substring length $ag$, to obtain the packed representation of
        $U[j]$. The requested interval is contained in $[1\dd mag]$. Apply
        \cref{pr:val-encoding} with length $ag$ and the returned representation
        to compute $y:=\Val{\AlphabetSize}{U[j]}\in[0\dd k^g)$. Grouping the
        base-$\AlphabetSize$ digits into blocks of $a$ and using
        $k=\AlphabetSize^a$ gives
        \[
          y
            =\sum_{t=1}^g V[j][t]k^{g-t}
            =\Val{k}{V[j][1\dd g]}.
        \]
        Apply \cref{pr:val-decoding} with length $g$ and integer $y$ to
        obtain $\PackedRepresentation{w}{k}{V[j][1\dd g]}$. Because
        $v\leq w$, both this representation and the packed representation of
        $V[j]$ consist of one word. Every bit outside its first $g$ symbol
        fields is zero (\cref{def:packed-representation}), so the returned
        word also represents $V[j][1\dd g]\zero^{h-g}=V[j]$. Store it in
        $A_V[j]$.

      \item \emph{Construct the packed representation of $V$.} Apply
        \cref{pr:packed-representation}\eqref{pr:packed-representation-concat}, with alphabet size $k$, to
        the $m$ one-word representations stored in $A_V$, each of length $h$.
        Their total length is $mh$, and the resulting concatenation is
        $\PackedSeqRepresentation{w}{k}{V}$.
    \end{enumerate}

    The assumption $v>(\log m)/8$ implies that the $\bigO(m+\sqrt m)$ time
    used above is $\bigO(mv/\log m)$ and hence
    $\bigO(mh\log k/\log m)$. By \cref{rm:space}, the input and output
    representations, the two value structures, $A_V$, and the working space
    of the packed operations use
    $\bigO(mv)=\bigO(mh\log k)$ bits in total.
  \end{enumerate}

  Both constructions satisfy the claimed bounds.
\end{proof}

\subsection{Packed Reversal of Fixed-Length Strings}
  \label{sec:tools-packed-fixed-length-string-reversal}

\begin{proposition}[Packed reversal of fixed-length strings]
  \label{pr:packed-fixed-length-string-reversal}
  Let $\AlphabetSize,m\in\Z_{\geq2}$ and $\ell\in[1\dd m)$ satisfy
  $\AlphabetSize\leq m$. Consider the word RAM model with word size
  $w=c\log m$, where $c\geq2$ is a constant. Let
  $W\in([0\dd\AlphabetSize)^\ell)^m$, and let
  $W^{\rm rev}\in([0\dd\AlphabetSize)^\ell)^m$ denote the sequence satisfying
  $W^{\rm rev}[j]=\revstr{W[j]}$ for every $j\in[1\dd m]$. There is an
  algorithm that, given
  $\PackedSeqRepresentation{w}{\AlphabetSize}{W}$
  (\cref{def:packed-sequence-representation}), constructs
  $\PackedSeqRepresentation{w}{\AlphabetSize}{W^{\rm rev}}$. The algorithm
  takes $\bigO(1+m\ell\log\AlphabetSize/\log m)$ time and has a peak space
  usage of $\bigO(m\ell\log\AlphabetSize)$ bits.
\end{proposition}
\begin{proof}
  If $m<16$, scan every $W[j]$ and write its characters to the corresponding
  output string in reverse order. Since $\AlphabetSize\leq m$ and $\ell<m$,
  the claimed bounds follow. Thus, assume that $m\geq16$, and compute
  $b:=\lceil\log\AlphabetSize\rceil$ by repeated doubling. This takes
  $\bigO(\log m)$ time, which is bounded by the claim. Every application of
  \cref{pr:packed-representation} below uses length parameter $m\ell$. Its
  common requirements hold because
  $\AlphabetSize\leq m\leq m\ell<m^2\leq2^w$, and hence
  $w>\log(m\ell)$. The supplied packed sequence representation is also the
  packed representation of $W[1]\cdots W[m]$.

  We distinguish two cases.
  \begin{enumerate}

    \item \emph{Construct the output when $\ell b\leq(\log m)/8$.} Compute
      $r:=\floor{\log m/(8\ell b)}$ and $s:=\ceil{m/r}$. For every
      $i\in[1\dd s]$, let $r_i:=\min\{r,m-(i-1)r\}$.
      We use four substeps.
      \begin{enumerate}

        \item \emph{Construct the table $T$.} Allocate
          $T[0\dd2^{r\ell b})$ as a word array. For every
          $x\in[0\dd2^{r\ell b})$, $i\in[1\dd r]$, and
          $p\in[1\dd\ell]$, let
          $x_{i,p}:=\floor{x/2^{((i-1)\ell+p-1)b}}\bmod2^b$. The value
          $x_{i,p}$ is field $p$ in group $i$ of $x$, where groups and fields
          are numbered from least to most significant. Set
          \[
            T[x]
              :=\sum_{i=1}^r\sum_{p=1}^{\ell}
                x_{i,p}\,2^{((i-1)\ell+\ell-p)b}.
          \]
          Thus, field $p$ of group $i$ in $x$ becomes field $\ell-p+1$ of
          the same group in $T[x]$. Both $x$ and $T[x]$ occupy at most
          $(\log m)/8$ bits. Scanning every $x$ and evaluating the displayed
          sum takes $\bigO(2^{r\ell b}r\ell)
          =\bigO(m^{1/8}\log m)=o(m/\log m)$ time. The table uses
          $\bigO(2^{r\ell b}w)=\bigO(m^{1/8}\log m)$ bits.

        \item \emph{Construct the packed representations of the batches.}
          Apply \cref{pr:packed-representation}\eqref{pr:packed-representation-split}, with alphabet size
          $\AlphabetSize$, length parameter $m\ell$, and length sequence
          $(r_1\ell,\ldots,r_s\ell)$, to split the supplied representation
          into its $s$ consecutive batches.

        \item \emph{Construct the packed representations of the reversed
          batches.} Allocate a word array $A_{\rm batch}[1\dd s]$. For every
          $i\in[1\dd s]$, let $x_i$ denote the sole word of the packed
          representation of batch $i$. Its unused high fields are zero, so
          they represent $r-r_i$ additional all-zero strings of length
          $\ell$. By the definition of $T$, the first $r_i\ell$ fields of
          $T[x_i]$ represent the reversals of the strings in batch $i$, in
          their original order. Store $T[x_i]$ in $A_{\rm batch}[i]$ as the
          packed representation of these fields.

        \item \emph{Construct the packed representation of $W^{\rm rev}$.}
          Apply \cref{pr:packed-representation}\eqref{pr:packed-representation-concat}, with alphabet size
          $\AlphabetSize$, to the $s$ representations stored in
          $A_{\rm batch}$, whose lengths are
          $r_1\ell,\ldots,r_s\ell$. Their total length is $m\ell$, and their
          concatenation is
          $\PackedSeqRepresentation{w}{\AlphabetSize}{W^{\rm rev}}$.
      \end{enumerate}

      Since $r\geq\log m/(16\ell b)$ and
      $b\leq2\log\AlphabetSize$, it holds
      $s=\bigO(1+m\ell b/\log m)
        =\bigO(1+m\ell\log\AlphabetSize/\log m)$. The split and
      concatenation operations therefore take
      $\bigO(1+m\ell\log\AlphabetSize/\log m)$ total time, which also
      dominates the construction of $T$. The input and output
      representations, the representations of the $s$ batches, $T$,
      $A_{\rm batch}$, and the working space used by one split or concatenation
      operation together occupy $\bigO(m\ell\log\AlphabetSize)$ bits.

    \item \emph{Construct the output when $\ell b>(\log m)/8$.} We use four
      substeps.
      \begin{enumerate}

        \item \emph{Construct the reversal data structure.} Construct the
          packed-string reversal data structure from
          \cref{pr:packed-string-reversal} with parameters $m$ and
          $\AlphabetSize$. Its hypotheses hold because
          $\AlphabetSize\leq m$ and $w\geq1+\log m$.

        \item \emph{Construct the packed representations of the input
          strings.} Apply
          \cref{pr:packed-representation}\eqref{pr:packed-representation-split}, with alphabet size
          $\AlphabetSize$, length parameter $m\ell$, and a length sequence
          consisting of $m$ copies of $\ell$, to construct the packed
          representation of every $W[j]$.

        \item \emph{Construct the packed representations of the reversed
          strings.} For every $j\in[1\dd m]$, query the reversal data structure
          with length $\ell$ and the packed representation of $W[j]$, and
          retain the returned
          $\PackedRepresentation{w}{\AlphabetSize}{\revstr{W[j]}}$.

        \item \emph{Construct the packed representation of $W^{\rm rev}$.}
          Apply \cref{pr:packed-representation}\eqref{pr:packed-representation-concat}, with alphabet size
          $\AlphabetSize$, to the $m$ retained representations, each of length
          $\ell$, to construct
          $\PackedSeqRepresentation{w}{\AlphabetSize}{W^{\rm rev}}$.
      \end{enumerate}

      Since $b\leq2\log\AlphabetSize$, the case assumption implies
      $\ell\log\AlphabetSize>(\log m)/16$. Consequently, the
      $\bigO(\sqrt m)$ construction time, the split, the $m$ reversal
      queries, and the final concatenation take
      $\bigO(m\ell\log\AlphabetSize/\log m)$ total time. The input and output
      representations, the representations of $W[1],\ldots,W[m]$ and
      $\revstr{W[1]},\ldots,\revstr{W[m]}$, the reversal data structure, and
      the working space used by one reversal query, split, or concatenation
      operation together occupy
      $\bigO(m\ell\log\AlphabetSize)$ bits.\qedhere
  \end{enumerate}
\end{proof}

\subsection{Packed Pointwise Concatenation and Partition}
  \label{sec:tools-packed-pointwise-concatenation-partition}

\begin{proposition}[Packed pointwise concatenation and partition]
  \label{pr:packed-pointwise-concatenation-partition}
  Let $\AlphabetSize,m\in\Z_{\geq2}$ satisfy $\AlphabetSize\leq m$, let
  $\ell_1,\ell_2\in\Z_{\geq0}$ satisfy
  $1\leq\ell_1+\ell_2<m$, and consider the word RAM model with word size
  $w=c\log m$, where $c\geq2$ is a constant. For
  $U\in([0\dd\AlphabetSize)^{\ell_1})^m$ and
  $V\in([0\dd\AlphabetSize)^{\ell_2})^m$, let
  $W\in([0\dd\AlphabetSize)^{\ell_1+\ell_2})^m$ denote the sequence
  satisfying $W[j]=U[j]V[j]$ for every $j\in[1\dd m]$. Given $\ell_1$ and
  $\ell_2$, there are algorithms for the following two tasks on packed
  sequence representations (\cref{def:packed-sequence-representation}):
  \begin{enumerate}
    \item Given $\PackedSeqRepresentation{w}{\AlphabetSize}{U}$ and
      $\PackedSeqRepresentation{w}{\AlphabetSize}{V}$, construct
      $\PackedSeqRepresentation{w}{\AlphabetSize}{W}$.
    \item Given $\PackedSeqRepresentation{w}{\AlphabetSize}{W}$, construct
      $\PackedSeqRepresentation{w}{\AlphabetSize}{U}$ and
      $\PackedSeqRepresentation{w}{\AlphabetSize}{V}$.
  \end{enumerate}
  Each algorithm takes
  $\bigO(1+m(\ell_1+\ell_2)\log\AlphabetSize/\log m)$ time and has a peak
  space usage of
  $\bigO(m(\ell_1+\ell_2)\log\AlphabetSize)$ bits.
\end{proposition}
\begin{proof}
  If $\ell_1=0$, the concatenation algorithm returns the supplied
  representation of $V$ as the representation of $W$, and the partition
  algorithm returns the empty representation as the representation of $U$
  and the supplied representation of $W$ as the representation of $V$. The
  case $\ell_2=0$ is symmetric. Hence, assume that
  $\ell_1,\ell_2\geq1$, and denote $\ell:=\ell_1+\ell_2$. If
  $m<2^{16}$, scan the strings in order. For the first task, write $U[j]$
  followed by $V[j]$ for every $j$; for the second task, write the first
  $\ell_1$ and the final $\ell_2$ characters of every $W[j]$ to the two
  outputs. Since $m\ell=\bigO(1)$ in this case, the claim follows.
  Thus, assume that $m\geq2^{16}$, and compute
  $b:=\lceil\log\AlphabetSize\rceil$ by repeated doubling. This takes
  $\bigO(\log m)$ time and bits, which are bounded by the claimed time and
  peak space because $m/\log m\geq\log m$.

  Every application of \cref{pr:packed-representation} below uses alphabet
  size $\AlphabetSize$ and length parameter $m\ell$. Its common hypotheses
  hold because
  $\AlphabetSize\leq m\leq m\ell<m^2\leq2^w$, and hence
  $w>\log(m\ell)$. The number of strings supplied to any split or
  concatenation operation is at most $2m\leq m\ell$.

  We distinguish two cases.
  \begin{enumerate}

    \item \emph{Process strings satisfying $\ell b\leq(\log m)/8$.}
      Compute $r:=\floor{\log m/(8\ell b)}$ and $s:=\ceil{m/r}$. For every
      $i\in[1\dd s]$, let $r_i:=\min\{r,m-(i-1)r\}$. We use the following
      three substeps, executing only the second or the third substep according
      to the requested task.
      \begin{enumerate}

        \item \emph{Construct the lookup tables.} Allocate the word arrays
          $T_{\rm con}[0\dd2^{r\ell_1b})[0\dd2^{r\ell_2b})$,
          $T_1[0\dd2^{r\ell b})$, and $T_2[0\dd2^{r\ell b})$. For
          $x\in[0\dd2^{r\ell_1b})$, $y\in[0\dd2^{r\ell_2b})$, and
          $i\in[1\dd r]$, let
          $x_{i,p}:=\floor{x/2^{((i-1)\ell_1+p-1)b}}\bmod2^b$ for
          $p\in[1\dd\ell_1]$, and let
          $y_{i,p}:=\floor{y/2^{((i-1)\ell_2+p-1)b}}\bmod2^b$ for
          $p\in[1\dd\ell_2]$. Let $T_{\rm con}$ denote the table satisfying
          \[
            T_{\rm con}[x,y]
              :=\sum_{i=1}^r\left(
                  \sum_{p=1}^{\ell_1}
                    x_{i,p}2^{((i-1)\ell+p-1)b}
                  +\sum_{p=1}^{\ell_2}
                    y_{i,p}2^{((i-1)\ell+\ell_1+p-1)b}
                \right).
          \]
          Thus, $T_{\rm con}$ preserves every field and changes the order
          from the $r$ length-$\ell_1$ strings encoded by $x$ followed by the
          $r$ length-$\ell_2$ strings encoded by $y$ to their $r$ pointwise
          concatenations. This field permutation is bijective. For every
          $z\in[0\dd2^{r\ell b})$, let $T_1[z]$ and $T_2[z]$ be the unique
          values satisfying
          $T_{\rm con}[T_1[z],T_2[z]]=z$. Construct the tables by enumerating
          every pair $(x,y)$, evaluating $z:=T_{\rm con}[x,y]$, and storing
          $T_1[z]:=x$ and $T_2[z]:=y$. Every table index and entry fits in one
          word because $r\ell b\leq(\log m)/8$. The enumeration takes
          $\bigO(2^{r\ell b}r\ell)
            =\bigO(m^{1/8}\log m)=o(m/\log m)$ time. The tables use
          $\bigO(2^{r\ell b}w)=\bigO(m^{1/8}\log m)$ bits.

        \item \emph{Perform pointwise concatenation.} Apply
          \cref{pr:packed-representation}\eqref{pr:packed-representation-split} to the supplied
          representations of $U$ and $V$, using the length sequences
          $(r_1\ell_1,\ldots,r_s\ell_1)$ and
          $(r_1\ell_2,\ldots,r_s\ell_2)$, respectively. Each returned batch
          consists of one word. For every $i\in[1\dd s]$, let $x_i$ and
          $y_i$ denote the words representing batch $i$ of $U$ and $V$.
          The unused high fields of the final pair of words are zero, so they
          represent additional pairs of all-zero strings up to a batch of
          size $r$. The first $r_i\ell$ fields of
          $T_{\rm con}[x_i,y_i]$ therefore represent the required strings
          $W[(i-1)r+1],\ldots,W[(i-1)r+r_i]$. Apply
          \cref{pr:packed-representation}\eqref{pr:packed-representation-concat} to these $s$
          representations, with lengths $r_1\ell,\ldots,r_s\ell$, to
          construct $\PackedSeqRepresentation{w}{\AlphabetSize}{W}$.

        \item \emph{Perform pointwise partition.} Apply
          \cref{pr:packed-representation}\eqref{pr:packed-representation-split} to the supplied
          representation of $W$, using the length sequence
          $(r_1\ell,\ldots,r_s\ell)$. For every $i\in[1\dd s]$, let $z_i$
          denote the sole word representing the returned batch. The unused
          high fields of the final word are zero, so $T_1[z_i]$ and
          $T_2[z_i]$ represent the corresponding batches of $U$ and $V$,
          padded by all-zero strings to size $r$. Apply
          \cref{pr:packed-representation}\eqref{pr:packed-representation-concat} to the first
          $r_i\ell_1$ fields of $T_1[z_i]$, for $i=1,\ldots,s$, and likewise
          to the first $r_i\ell_2$ fields of $T_2[z_i]$. The two resulting
          representations are
          $\PackedSeqRepresentation{w}{\AlphabetSize}{U}$ and
          $\PackedSeqRepresentation{w}{\AlphabetSize}{V}$.
      \end{enumerate}

      Since $r\geq\log m/(16\ell b)$ and
      $b\leq2\log\AlphabetSize$, we have
      $s=\bigO(1+m\ell\log\AlphabetSize/\log m)$. For either task, the
      split and concatenation operations, the table queries, and the table
      construction take
      $\bigO(1+m(\ell_1+\ell_2)\log\AlphabetSize/\log m)$ total time. The
      representations of $U$, $V$, and $W$ needed for the requested task,
      all batch representations, the tables, and the working space used by
      one split or concatenation operation together occupy
      $\bigO(m(\ell_1+\ell_2)\log\AlphabetSize)$ bits.

    \item \emph{Process strings satisfying $\ell b>(\log m)/8$.} Execute
      only the substep corresponding to the requested task.
      \begin{enumerate}

        \item \emph{Perform pointwise concatenation.} Apply
          \cref{pr:packed-representation}\eqref{pr:packed-representation-split} to the supplied
          representations of $U$ and $V$, using $m$ copies of $\ell_1$ and
          $m$ copies of $\ell_2$, respectively. Apply
          \cref{pr:packed-representation}\eqref{pr:packed-representation-concat} to the returned
          representations in the order
          $U[1],V[1],U[2],V[2],\ldots,U[m],V[m]$. The result is
          $\PackedSeqRepresentation{w}{\AlphabetSize}{W}$.

        \item \emph{Perform pointwise partition.} Apply
          \cref{pr:packed-representation}\eqref{pr:packed-representation-split} to the supplied
          representation of $W$, using the alternating length sequence
          $(\ell_1,\ell_2,\ell_1,\ell_2,\ldots,\ell_1,\ell_2)$. Apply
          \cref{pr:packed-representation}\eqref{pr:packed-representation-concat} to the odd-indexed returned
          representations and separately to the even-indexed returned
          representations. The results are
          $\PackedSeqRepresentation{w}{\AlphabetSize}{U}$ and
          $\PackedSeqRepresentation{w}{\AlphabetSize}{V}$.
      \end{enumerate}

      Each task takes
      $\bigO(m+m\ell\log\AlphabetSize/\log m)$ time in split and
      concatenation operations. Since $b\leq2\log\AlphabetSize$, the case
      assumption gives $m=\bigO(m\ell\log\AlphabetSize/\log m)$. Thus, each
      task takes
      $\bigO(1+m(\ell_1+\ell_2)\log\AlphabetSize/\log m)$ time. The supplied
      and constructed representations of $U$, $V$, and $W$, the separately
      stored representations of their individual strings, and the working
      space used by one split or concatenation operation occupy
      $\bigO(m(\ell_1+\ell_2)\log\AlphabetSize)$ bits.\qedhere
  \end{enumerate}
\end{proof}

\subsection{Packed String Repetition}
  \label{sec:tools-packed-string-repetition}

\begin{proposition}[Packed string repetition]
  \label{pr:packed-string-repetition}
  Let $\AlphabetSize,\Textlen\in\Z_{\geq2}$ satisfy
  $\AlphabetSize\leq\Textlen$, assume that
  $\lceil\log\AlphabetSize\rceil$ has been computed, and consider the word
  RAM model with word size $w>\log\Textlen$. Let
  $S\in[0\dd\AlphabetSize)^+$ satisfy $|S|\leq\Textlen$, let
  $t\in[0\dd\Textlen]$, and let $R:=S^\infty[1\dd t+1)$, where $S^\infty$
  denotes the infinite repetition of $S$. There is an algorithm that, given
  $t$, $|S|$, and $\PackedRepresentation{w}{\AlphabetSize}{S}$
  (\cref{def:packed-representation}), constructs
  $\PackedRepresentation{w}{\AlphabetSize}{R}$. The algorithm takes
  $\bigO(1+\log(1+t/|S|)+t\log\AlphabetSize/w)$ time and has a peak space
  usage of $\bigO(w+\max\{|S|,t\}\log\AlphabetSize)$ bits.
\end{proposition}
\begin{proof}
  If $t\leq|S|$, apply
  \cref{pr:packed-representation}\eqref{pr:packed-representation-substring}, with length parameter
  $\Textlen$, string length $|S|$, starting position $1$, and substring
  length $t$, to construct $\PackedRepresentation{w}{\AlphabetSize}{R}$.
  The invocation is valid by the hypotheses of the claim and because
  $0\leq t\leq|S|\leq\Textlen$. It takes
  $\bigO(1+t\log\AlphabetSize/w)$ time. The supplied representation of $S$,
  the constructed representation of $R$, and the working space of the
  substring invocation use
  $\bigO(w+|S|\log\AlphabetSize)$ bits. The claim follows in this case.
  Hence, assume that $t>|S|$.

  Let $R^{(0)}:=S$ and $\mu_0:=|S|$. Starting with the supplied packed
  representation of $R^{(0)}$, for $i=0,1,\ldots$, while $\mu_i<t$, perform
  the following two steps.
  \begin{enumerate}

    \item \emph{Construct the packed representation of $H_i$.} Compute
      $h_i:=\min\{\mu_i,t-\mu_i\}$, and let
      $H_i:=R^{(i)}[1\dd h_i+1)$. Apply
      \cref{pr:packed-representation}\eqref{pr:packed-representation-substring}, with length parameter
      $\Textlen$, string length $\mu_i$, starting position $1$, and substring
      length $h_i$, to construct
      $\PackedRepresentation{w}{\AlphabetSize}{H_i}$. The invocation is valid
      because $0<h_i\leq\mu_i\leq t\leq\Textlen$.

    \item \emph{Construct the packed representation of $R^{(i+1)}$.} Let
      $R^{(i+1)}:=R^{(i)}H_i$. Apply
      \cref{pr:packed-representation}\eqref{pr:packed-representation-concat}, with length parameter
      $\Textlen$, $m=2$, length sequence $(\mu_i,h_i)$, and the packed
      representations of $R^{(i)}$ and $H_i$, to construct
      $\PackedRepresentation{w}{\AlphabetSize}{R^{(i+1)}}$. The invocation is
      valid because $2\leq\Textlen$ and
      $\mu_i+h_i\leq t\leq\Textlen$. Compute
      $\mu_{i+1}:=\mu_i+h_i$, and stop if $\mu_{i+1}=t$.
  \end{enumerate}

  Let $q$ denote the number of iterations. For every $i\in[0\dd q]$, we have
  $\mu_i=|R^{(i)}|$ and $R^{(i)}=S^\infty[1\dd\mu_i+1)$. The equalities hold
  for $i=0$. Every iteration preceding the last one doubles $\mu_i$, so
  $|S|$ divides $\mu_i$ at the beginning of every iteration. For any
  $i\in[0\dd q)$, assuming the equalities for $i$, the periodicity of
  $S^\infty$ gives
  \[
    H_i
      =S^\infty[1\dd h_i+1)
      =S^\infty[\mu_i+1\dd\mu_i+h_i+1).
  \]
  Hence, $R^{(i+1)}=R^{(i)}H_i=S^\infty[1\dd\mu_{i+1}+1)$, proving the
  equalities by induction. Since $\mu_q=t$, the constructed packed
  representation of $R^{(q)}$ is the required representation of $R$.

  Every iteration except possibly the last doubles the length. Thus,
  $q=\bigO(1+\log(1+t/|S|))$, and the values
  $\mu_0,\ldots,\mu_{q-1}$ form a geometric sequence with sum less than $2t$.
  Moreover, since $\mu_{i+1}=\mu_i+h_i$ for every $i\in[0\dd q)$, we have
  $\sum_{i=0}^{q-1}h_i=\mu_q-\mu_0<t$. The two applications of
  \cref{pr:packed-representation} in iteration $i$ take
  $\bigO(1+(\mu_i+h_i)\log\AlphabetSize/w)$ time. Summing over all iterations
  gives $\bigO(1+\log(1+t/|S|)+t\log\AlphabetSize/w)$ time.

  During iteration $i$, the supplied representation of $S$ and the packed
  representations of $R^{(i)}$, $H_i$, and $R^{(i+1)}$ coexist. Only the
  representation of $R^{(i+1)}$ is retained for the next iteration. Together
  with the working space of one application of
  \cref{pr:packed-representation}, these objects use
  $\bigO(w+\max\{|S|,t\}\log\AlphabetSize)$ bits.
\end{proof}

\subsection{Word-Size Conversion of Packed Representations}
  \label{sec:tools-packed-representation-word-size-conversion}

\begin{proposition}[Packed-representation word-size conversion]
  \label{pr:packed-representation-word-size-conversion}
  Let $\AlphabetSize,\Textlen\in\Z_{\geq2}$ satisfy
  $\AlphabetSize\leq\Textlen$, assume that
  $b:=\lceil\log\AlphabetSize\rceil$ has been computed, and let
  $w,\widehat w>\log\Textlen$ satisfy $\widehat w=\Theta(w)$. Denote
  $W:=\max\{w,\widehat w\}$. In the word RAM model with word size $W$, there
  is an algorithm that, given the length $|S|$ and
  $\PackedRepresentation{w}{\AlphabetSize}{S}$
  (\cref{def:packed-representation}) of any string
  $S\in\IntegerAlphabet^{\leq\Textlen}$, constructs
  $\PackedRepresentation{\widehat w}{\AlphabetSize}{S}$. The algorithm takes
  $\bigO(1+|S|\log\AlphabetSize/w)$ time and has a peak space usage of
  $\bigO(W+|S|\log\AlphabetSize)$ bits.
\end{proposition}
\begin{proof}
  Compute $s:=\floor{w/b}$, $\widehat s:=\floor{\widehat w/b}$,
  $k:=\ceil{|S|/s}$, and $\widehat k:=\ceil{|S|/\widehat s}$.
  Let $(x_1,\ldots,x_k)$ denote the given packed representation, and allocate
  the output sequence $(y_1,\ldots,y_{\widehat k})$. Store an integer buffer
  $Z$ in two $W$-bit words. Its low-order fields are the symbols that have been
  read from $S$ but not yet written, and a counter $r$ stores their number. We
  construct the output in two steps.
  \begin{enumerate}

    \item \emph{Construct the complete output words.} Initialize
      $Z:=0$, $r:=0$, and $p:=1$. For every $j=1,2,\ldots,k$, perform the
      following two substeps.
      \begin{enumerate}

        \item Let $a_j:=\min\{s,|S|-(j-1)s\}$ denote the number of symbols
          encoded by $x_j$. By \cref{def:packed-representation},
          $x_j<2^{a_jb}$. Append these symbols to the fields already in $Z$ by
          setting
          \[
            Z:=Z+x_j2^{rb}
            \qquad\text{and}\qquad
            r:=r+a_j.
          \]

        \item While $r\geq\widehat s$, set
          $y_p:=Z\bmod2^{\widehat s b}$, set
          $Z:=\floor{Z/2^{\widehat s b}}$ and
          $r:=r-\widehat s$, and increment $p$.
      \end{enumerate}

    \item \emph{Construct the final output word.} If $r>0$, set $y_p:=Z$.
  \end{enumerate}

  For every $j\in[0\dd k]$, after processing $x_1,\ldots,x_j$, the symbol
  fields in $y_1,\ldots,y_{p-1}$ followed by the $r$ fields in $Z$ from low
  to high order form exactly $S[1\dd\min\{js,|S|\}+1)$, and
  $r<\widehat s$ and $0\leq Z<2^{rb}$. This invariant holds for $j=0$.
  Appending $x_j$ extends the represented prefix by the next $a_j$ symbols
  and preserves $0\leq Z<2^{rb}$. Constructing $y_p$ from the low-order
  $\widehat s$ fields of $Z$ preserves their order, and the division and
  decrement restore $0\leq Z<2^{rb}$. Thus, the invariant holds after every
  iteration. The while loops construct $\floor{|S|/\widehat s}$ complete
  output words, and the assignment $y_p:=Z$ constructs one more word precisely when
  $\widehat s$ does not divide $|S|$. Hence, exactly
  $\ceil{|S|/\widehat s}=\widehat k$ words are constructed, including zero
  words when $S$ is empty. Every complete output word is smaller than
  $2^{\widehat s b}$, and, if $r>0$, the word $y_p$ is
  smaller than $2^{rb}$. Thus, all unused high bits are zero, and
  $(y_1,\ldots,y_{\widehat k})$ is
  $\PackedRepresentation{\widehat w}{\AlphabetSize}{S}$.

  Immediately before each input-word read and after the subsequent while
  loop, $Z$ contains fewer than $\widehat s$ fields. Immediately after
  appending the fields of $x_j$, it contains fewer than
  $s+\widehat s$ fields and therefore occupies fewer than
  $(s+\widehat s)b\leq w+\widehat w\leq2W$ bits. Thus, two $W$-bit words
  suffice for $Z$. Since $rb<\widehat s b\leq W$ before the fields of $x_j$
  are appended, multiplication by $2^{rb}$ is a shift by fewer than $W$ bits.
  If $\widehat s b=W$, reduction modulo $2^{\widehat s b}$ and division by
  $2^{\widehat s b}$ select the low-order and high-order word of $Z$,
  respectively. If $\widehat s b<W$, ordinary masks and shifts implement both
  operations.
  Therefore, appending fields, extracting an output word, and shifting $Z$
  use a constant number of standard operations on the two words storing $Z$.

  By \cref{def:packed-representation},
  $k=\bigO(1+|S|b/w)$ and
  $\widehat k=\bigO(1+|S|b/\widehat w)$. Since
  $\widehat w=\Theta(w)$ and
  $b=\Theta(\log\AlphabetSize)$, it follows that
  $k+\widehat k=\bigO(1+|S|\log\AlphabetSize/w)$. Every $x_j$ is read once
  and every $y_p$ is written once, proving the time bound. By
  \cref{def:packed-representation}, the input and output representations
  together occupy $\bigO(W+|S|\log\AlphabetSize)$ bits. The two-word
  representation of $Z$ and the counters and indices use $\bigO(W)$ bits.
  This proves the claimed bound on peak space usage.
\end{proof}

\section{Packed Routing and Wavelet-Tree Tools}\label{sec:packed-permuting}

\subsection{Stable Grouping and Ungrouping}\label{sec:packed-stable-grouping-and-ungrouping}

\subsubsection{Binary Alphabet}\label{sec:packed-permuting-binary-alphabet}

\begin{proposition}[Packed binary stable grouping and ungrouping]
  \label{pr:packed-binary-stable-grouping}
  Let $m\in\Z_{\geq2}$, and consider the word RAM model with word size
  $w=c\log m$ for a constant $c\geq2$. Let $b\in[1\dd w]$. Using the rank
  and select notation from \cref{def:rank-select}, for
  $B\in\{\zero,\one\}^m$ and $P\in[0\dd2^b)^m$, let
  $\srted{P}{B}\in[0\dd2^b)^m$ denote
  \begin{align*}
    \srted{P}{B}
      =\left(
        \bigodot_{r=1}^{\Rank{B}{m}{\zero}}
          P[\Select{B}{r}{\zero}]
      \right)
      \odot
      \left(
        \bigodot_{r=1}^{\Rank{B}{m}{\one}}
          P[\Select{B}{r}{\one}]
      \right).
  \end{align*}
  For each of the following tasks on packed representations
  (\cref{def:packed-representation}), there is an algorithm that takes
  \begin{align*}
    \bigO\left(\frac{mb}{\log m}\right)
  \end{align*}
  time and has a peak space usage of $\bigO(mb)$ bits.
  \begin{enumerate}
    \item Given $\PackedRepresentation{w}{2}{B}$ and
      $\PackedRepresentation{w}{2^b}{P}$, construct
      $\PackedRepresentation{w}{2^b}{\srted{P}{B}}$.
    \item Given $\PackedRepresentation{w}{2}{B}$ and
      $\PackedRepresentation{w}{2^b}{P'}$ for some
      $P'\in[0\dd2^b)^m$, construct
      $\PackedRepresentation{w}{2^b}{P}$ for the unique
      $P\in[0\dd2^b)^m$ satisfying $P'=\srted{P}{B}$.
  \end{enumerate}
\end{proposition}

\begin{proof}
  If $m<4$, perform either task directly, and hence the claim follows
  immediately. Thus, assume that $m\geq4$.
  We distinguish two cases.
  \begin{enumerate}

    \item \emph{Process wide payload fields.} Suppose that
      $b+1>(\log m)/16$. Process $B$ once to compute
      $z:=\Rank{B}{m}{\zero}$, and allocate and zero the
      $\lceil m/\floor{w/b}\rceil$ output words. For the first task, initialize
      $d_0:=1$ and $d_1:=z+1$. Process $j\in[1\dd m]$ in increasing order,
      write $P[j]$ at position $d_{B[j]}$ of the output, and increment
      $d_{B[j]}$. For the second task, initialize $s_0:=1$ and $s_1:=z+1$.
      Process $j\in[1\dd m]$ in increasing order, write $P'[s_{B[j]}]$ at
      position $j$ of the output, and increment $s_{B[j]}$. By
      \cref{def:packed-representation}, reading or writing one payload field
      uses shifts and masks in its containing word. Thus, either task takes
      $\bigO(m)$ time and has a peak space usage of $\bigO(mb)$ bits. Since
      $b+1\leq2b$, the case hypothesis implies $b=\Omega(\log m)$, so these
      bounds satisfy the claim.

    \item \emph{Process narrow payload fields.} Suppose that
      $b+1\leq(\log m)/16$. Set
      \[
        h:=\floor{\frac{\log m}{8(b+1)}},
        \qquad
        q:=\ceil{\frac{m}{h}},
        \qquad\text{and}\qquad
        h_i:=\min\{h,m-(i-1)h\}
      \]
      for every $i\in[1\dd q]$. Thus, $h\geq2$ and
      $\sum_{i=1}^q h_i=m$. An entry of the forward lookup table is specified
      by a binary string of length at most $h$ and an equally long string of
      $b$-bit payloads. It stores the packed subsequences $Z$ and $O$ of the
      payloads paired with $\zero$ and $\one$, respectively, together with
      their lengths. An entry of the inverse table is specified by the binary
      string and the two subsequences and stores their interleaving. A count
      table stores the number of occurrences of each bit in a binary string of
      length at most $h$. The tables occupy
      $m^{1/8}\polylog m=o(m/\log m)$ bits and can be
      constructed in the same time by exhaustive enumeration.
      Denote $T:=2m$. The case hypothesis implies $2^b<m<T$, and
      $T<m^2\leq2^w$. Apply
      \cref{pr:packed-representation}\eqref{pr:packed-representation-split}, with alphabet size $2$, length
      parameter $m$, and lengths $h_1,\ldots,h_q$, to split the supplied
      representation of $B$ into representations of strings
      $B_1,\ldots,B_q$. We perform the two tasks as follows.
      \begin{enumerate}

        \item To construct $\srted{P}{B}$, apply the same split operation to
          the supplied representation of $P$, with alphabet size $2^b$ and
          length parameter $T$, to obtain representations of
          $P_1,\ldots,P_q$. For every $i\in[1\dd q]$, query the forward table
          on $B_i,P_i$. Denote its returned representations by $Z_i,O_i$ and
          their lengths by $z_i,o_i$. Apply
          \cref{pr:packed-representation}\eqref{pr:packed-representation-concat}, with alphabet size $2^b$
          and length parameter $T$, to
          $Z_1,\ldots,Z_q,O_1,\ldots,O_q$, with lengths
          $z_1,\ldots,z_q,o_1,\ldots,o_q$. Its output is $\srted{P}{B}$.

        \item To reconstruct $P$, query the count table on every $B_i$, denote
          the returned counts by $z_i,o_i$, and apply
          \cref{pr:packed-representation}\eqref{pr:packed-representation-split}, with alphabet size $2^b$
          and length parameter $T$, to the supplied representation of $P'$,
          with source length $m$ and lengths
          $z_1,\ldots,z_q,o_1,\ldots,o_q$. Denote the returned representations
          by $Z_1,\ldots,Z_q,O_1,\ldots,O_q$, respectively. For every
          $i\in[1\dd q]$, query the inverse table on $B_i,Z_i,O_i$ and denote
          its output by $P_i$. Apply
          \cref{pr:packed-representation}\eqref{pr:packed-representation-concat} to
          $P_1,\ldots,P_q$, with lengths $h_1,\ldots,h_q$, alphabet size
          $2^b$, and length parameter $T$. Its output is $P$.
      \end{enumerate}
      All split and concatenation operations are valid because
      $q\leq m$, $2q\leq T$, $w>\log T$, and every supplied length is
      nonnegative and at most $m$. The lengths in each application sum to the
      length of its input or output.
      Since
      $h\geq\log m/(16(b+1))$, we have
      $q=\bigO(mb/\log m)$. Every table query takes constant time because one
      block occupies at most $(\log m)/8\leq w$ bits. The table construction,
      queries, and packed operations take
      $\bigO(q+mb/w)=\bigO(mb/\log m)$ time. The block representations and
      length arrays use $\bigO(qw)=\bigO(mb)$ bits, including word-alignment
      overhead. By \cref{rm:space}, one packed operation has a peak space usage
      of $\bigO(mb)$ bits. Together with the input and output representations
      and lookup tables, this proves the claimed space bound in the second
      case.
      \qedhere
  \end{enumerate}
\end{proof}

\begin{corollary}[Packed binary stable grouping and ungrouping of strings]
  \label{cor:packed-binary-string-grouping}
  Let $\AlphabetSize,m\in\Z_{\geq2}$ and $\ell\in\Z_{\geq1}$ satisfy
  $\AlphabetSize^\ell\leq m$, and consider the word RAM model with word size
  $w=c\log m$ for a constant $c\geq2$.
  Using the rank and select notation from \cref{def:rank-select}, for
  $B\in\{\zero,\one\}^m$ and
  $Q\in([0\dd\AlphabetSize)^\ell)^m$, let
  $\srted{Q}{B}\in([0\dd\AlphabetSize)^\ell)^m$ denote
  \begin{align*}
    \srted{Q}{B}
      =\left(
        \bigodot_{r=1}^{\Rank{B}{m}{\zero}}
          Q[\Select{B}{r}{\zero}]
      \right)
      \odot
      \left(
        \bigodot_{r=1}^{\Rank{B}{m}{\one}}
          Q[\Select{B}{r}{\one}]
      \right).
  \end{align*}
  There are algorithms for the following two tasks on packed representations
  of strings and sequences of strings
  (\cref{def:packed-representation,def:packed-sequence-representation}):
  \begin{itemize}
    \item given $\PackedRepresentation{w}{2}{B}$ and
      $\PackedSeqRepresentation{w}{\AlphabetSize}{Q}$, construct
      $\PackedSeqRepresentation{w}{\AlphabetSize}{\srted{Q}{B}}$, and
    \item given $\PackedRepresentation{w}{2}{B}$ and
      $\PackedSeqRepresentation{w}{\AlphabetSize}{Q'}$ for some
      $Q'\in([0\dd\AlphabetSize)^\ell)^m$, construct
      $\PackedSeqRepresentation{w}{\AlphabetSize}{Q}$ for the unique $Q$
      satisfying $Q'=\srted{Q}{B}$.
  \end{itemize}
  Each algorithm takes
  $\bigO(m\ell\log\AlphabetSize/\log m)$ time and has a peak space usage of
  $\bigO(m\ell\log\AlphabetSize)$ bits.
\end{corollary}
\begin{proof}
  Let $k:=\AlphabetSize^\ell$ and $b:=\lceil\log k\rceil$. For every
  $R\in([0\dd\AlphabetSize)^\ell)^m$, let
  $\widehat R\in[0\dd k)^m$ denote the string satisfying
  $\widehat R[j]=\Val{\AlphabetSize}{R[j]}$ for every $j\in[1\dd m]$.
  Apply \cref{pr:packed-character-block-encoding} to
  $\PackedSeqRepresentation{w}{\AlphabetSize}{R}$, with alphabet size
  $\AlphabetSize$, input length $m$, block length $\ell$, and $g=h=1$.
  Its hypotheses hold because $k\leq m$. Its output strings have length one,
  and their sole characters are $\widehat R[1],\ldots,\widehat R[m]$. By
  \cref{def:packed-sequence-representation}, its output is therefore
  $\PackedRepresentation{w}{k}{\widehat R}$. Since alphabets $k$ and $2^b$
  use the same number of bits per character, the output is also
  $\PackedRepresentation{w}{2^b}{\widehat R}$.

  For the first task, apply the preceding encoding to $Q$ and then apply the
  forward algorithm from
  \cref{pr:packed-binary-stable-grouping}, with payload width $b$, to
  $\PackedRepresentation{w}{2}{B}$ and
  $\PackedRepresentation{w}{2^b}{\widehat Q}$. Its output is
  $\PackedRepresentation{w}{2^b}{\widehat{\srted{Q}{B}}}$. For the second
  task, apply the preceding encoding to $Q'$ and then apply the inverse
  algorithm with the same payload width to
  $\PackedRepresentation{w}{2}{B}$ and
  $\PackedRepresentation{w}{2^b}{\widehat{Q'}}$. Its output is
  $\PackedRepresentation{w}{2^b}{\widehat Q}$ for the unique $Q$ satisfying
  $Q'=\srted{Q}{B}$. These applications are valid because
  $b\leq\lceil\log m\rceil\leq w$. The stated outputs follow because
  positionwise encoding by $\Val{\AlphabetSize}{\cdot}$ commutes with the
  stable permutation.

  Construct the alphabet-mapping data structure from \cref{pr:alphabet-map}
  with parameters $m$, $k$, and $\AlphabetSize$. Its hypotheses hold because
  $\AlphabetSize\leq k\leq m$ and $w\geq2\log m$. The representation returned
  in either task is also a representation over alphabet $k$. Apply the data
  structure to this length-$m$ string. For every
  $X\in[0\dd\AlphabetSize)^\ell$, we have
  \[
    \AlphabetMap{k}{\AlphabetSize}{\Val{\AlphabetSize}{X}}=X,
  \]
  including the leading zeros of $X$. The returned packed string is therefore
  the required packed sequence representation.

  Character-block encoding, binary grouping or ungrouping, and applying the
  alphabet-mapping data structure take $\bigO(1+m\log k/\log m)$ time.
  Constructing the alphabet-mapping data structure takes $\bigO(\sqrt m)$
  time. Since
  $\log k=\ell\log\AlphabetSize$ and $\sqrt m=\bigO(m/\log m)$, their total
  time is $\bigO(m\ell\log\AlphabetSize/\log m)$. By \cref{rm:space}, the
  alphabet-mapping data structure occupies $\bigO(\sqrt m\log m)$ bits. The
  peak space usage of any one application, together with the memory occupied
  by the supplied representations, the intermediate packed strings, the
  output, and the data structure, is
  $\bigO(m\log k+\sqrt m\log m)
    =\bigO(m\ell\log\AlphabetSize)$ bits.
\end{proof}

\begin{corollary}[Packed binary stable ungrouping of strings for comparable
    lengths]
  \label{cor:packed-binary-string-ungrouping-comparable}
  Let $\AlphabetSize,m,n\in\Z_{\geq2}$ and $\ell\in\Z_{\geq1}$ satisfy
  $\AlphabetSize^\ell\leq m<n=\bigO(m)$, and consider the word RAM model with
  word size $w=c\log m$ for a constant $c\geq2$. Given
  $B\in\{\zero,\one\}^n$, $Q'\in([0\dd\AlphabetSize)^\ell)^n$,
  $\PackedRepresentation{w}{2}{B}$, and
  $\PackedSeqRepresentation{w}{\AlphabetSize}{Q'}$
  (\cref{def:packed-representation,def:packed-sequence-representation}), there
  is an algorithm that constructs
  $\PackedSeqRepresentation{w}{\AlphabetSize}{Q}$ for the unique
  $Q\in([0\dd\AlphabetSize)^\ell)^n$ satisfying
  $Q'=\srted{Q}{B}$, where $\srted{Q}{B}$ is defined as in
  \cref{cor:packed-binary-string-grouping}, with $n$ in place of $m$. The
  algorithm takes
  $\bigO(m\ell\log\AlphabetSize/\log m)$ time and has a peak space usage of
  $\bigO(m\ell\log\AlphabetSize)$ bits.
\end{corollary}
\begin{proof}
  If $m<16$ or $n\ell\geq m^2$, perform the task directly, and hence the
  claim follows immediately. Since $n=\bigO(m)$ and $\ell=\bigO(\log m)$, the
  inequality $n\ell\geq m^2$ holds for only boundedly many values of $m$.
  Thus, assume that $m\geq16$ and $n\ell<m^2\leq2^w$, and denote $N:=n\ell$.
  In particular, $w>\log N$.

  Let $q:=\ceil{n/m}=\bigO(1)$. For every $i\in[1\dd q]$, let
  \[
    I_i:=[(i-1)m+1\dd\min\{im,n\}],
    \qquad m_i:=|I_i|,
    \qquad\text{and}\qquad t_i:=m-m_i.
  \]
  Let $B_i$ denote the restriction of $B$ to $I_i$, and let
  $\widehat B_i:=B_i\odot\zero^{t_i}$. Every application of
  \cref{pr:packed-representation} below uses length parameter $N$. Its common
  hypotheses hold because
  $2\leq\AlphabetSize\leq\AlphabetSize^\ell\leq m<n\leq N$ and $w>\log N$.
  Compute $\lceil\log\AlphabetSize\rceil$ by repeated doubling. This takes
  $\bigO(\log m)$ time and bits, which are bounded by the claim since
  $m/\log m\geq\log m$. Since $n>m\geq16$, we have $q\geq2$, and the supplied
  length sequences contain at most $2q\leq n\leq N$ values. Their sums are one
  of $m,n,m\ell,N$, and all initialization and substring lengths are at most
  $N$. Perform the following two steps.
  \begin{enumerate}

    \item \emph{Construct $\widehat B_i$, $z_i$, and $o_i$ for every $i$.}
      Apply \cref{pr:packed-representation}\eqref{pr:packed-representation-split} to the representation of $B$, with
      alphabet size $2$, source length $n$, and lengths $m_1,\ldots,m_q$.
      Its outputs represent $B_1,\ldots,B_q$. For every $i\in[1\dd q]$, use
      \cref{pr:packed-representation}\eqref{pr:packed-representation-initialize} for $\zero^{t_i}$ and
      \cref{pr:packed-representation}\eqref{pr:packed-representation-concat} with lengths $m_i,t_i$ to
      construct $\PackedRepresentation{w}{2}{\widehat B_i}$.
      For every $i\in[1\dd q]$, let
      $\widehat{\mathcal B}_i\in(\{\zero,\one\}^1)^m$ denote the sequence
      satisfying $\widehat{\mathcal B}_i[j][1]=\widehat B_i[j]$ for every
      $j\in[1\dd m]$. For every $i\in[1\dd q]$, its packed sequence
      representation is identical to
      $\PackedRepresentation{w}{2}{\widehat B_i}$. For every
      $i\in[1\dd q]$, apply \cref{pr:packed-prefix-frequencies}, with parameters
      $2,m,1$, to
      $\PackedSeqRepresentation{w}{2}{\widehat{\mathcal B}_i}$, denote the
      returned array by
      $F_i[0\dd4)$, and compute
      \[
        z_i:=F_i[2]-t_i
        \qquad\text{and}\qquad
        o_i:=F_i[3].
      \]
      These are the numbers of zeros and ones in $B_i$, respectively, because
      the final $t_i$ bits of $\widehat B_i$ are zero. Every application is
      valid because $2^1\leq m$, its input contains exactly $m$ strings, and
      the word size is $w=c\log m$. Execute the applications sequentially.

    \item \emph{Construct the output.} For every $i\in[1\dd q]$, let $D_i$
      denote the sequence consisting of $t_i$ copies of $\zero^\ell$. Compute
      $z:=\sum_{i\in[1\dd q]}z_i$. Let $Z_1,\ldots,Z_q$ denote the consecutive
      blocks of lengths $z_1,\ldots,z_q$ in the first $z$ strings of $Q'$, and
      let $O_1,\ldots,O_q$ denote the consecutive blocks of lengths
      $o_1,\ldots,o_q$ in the remaining strings. The supplied packed sequence
      representation is also the packed representation of
      $\bigodot_{j=1}^n Q'[j]$. Apply
      \cref{pr:packed-representation}\eqref{pr:packed-representation-split}, with alphabet size
      $\AlphabetSize$, source length $N$, and lengths
      $\ell z_1,\ldots,\ell z_q,\ell o_1,\ldots,\ell o_q$, to obtain the packed
      sequence representations of $Z_1,\ldots,Z_q,O_1,\ldots,O_q$.
      For every $i\in[1\dd q]$, use
      \cref{pr:packed-representation}\eqref{pr:packed-representation-initialize} for $\zero^{t_i\ell}$ and
      \cref{pr:packed-representation}\eqref{pr:packed-representation-concat} with lengths
      $\ell z_i,t_i\ell,\ell o_i$ to construct
      $\PackedSeqRepresentation{w}{\AlphabetSize}{Z_i\odot D_i\odot O_i}$,
      and apply the inverse algorithm from
      \cref{cor:packed-binary-string-grouping}, with parameters
      $\AlphabetSize,m,\ell$, to
      $\PackedRepresentation{w}{2}{\widehat B_i}$ and
      $\PackedSeqRepresentation{w}{\AlphabetSize}{Z_i\odot D_i\odot O_i}$.
      Both inputs have length exactly $m$, and
      $\AlphabetSize^\ell\leq m$, so all its hypotheses hold.
      Denote the sequence represented by its output by $R_i$, and let
      $Y_i:=R_i[1\dd m_i]$. Apply
      \cref{pr:packed-representation}\eqref{pr:packed-representation-substring} to the representation returned
      by the inverse algorithm, with alphabet size $\AlphabetSize$, source
      length $m\ell$, starting position $1$, and substring length $m_i\ell$, to
      construct
      $\PackedSeqRepresentation{w}{\AlphabetSize}{Y_i}$.
      Finally, apply \cref{pr:packed-representation}\eqref{pr:packed-representation-concat}, with alphabet size
      $\AlphabetSize$ and lengths $m_1\ell,\ldots,m_q\ell$, to these outputs.
      By \cref{def:packed-sequence-representation}, return its output as the
      packed sequence representation of $Y_1\odot\cdots\odot Y_q$.
  \end{enumerate}

  To prove correctness, let $Q_i$ be the restriction of $Q$ to $I_i$. The
  strings $Z_i$ and $O_i$ are exactly the subsequences of $Q_i$ selected by
  the zero and one positions of $B_i$, respectively. Therefore,
  $Z_i\odot D_i\odot O_i$ is the stable grouping of $Q_i\odot D_i$ according
  to $\widehat B_i$. The inverse algorithm from
  \cref{cor:packed-binary-string-grouping} therefore returns a representation
  of $R_i=Q_i\odot D_i$. Consequently, $Y_i=Q_i$ for every $i\in[1\dd q]$,
  and the final concatenation returns the required representation of $Q$.

  The $q=\bigO(1)$ applications of \cref{pr:packed-prefix-frequencies} take
  $\bigO(m/\log m)$ time and use $\bigO(m)$ bits of working space. The
  $q=\bigO(1)$ applications of \cref{cor:packed-binary-string-grouping} take
  $\bigO(m\ell\log\AlphabetSize/\log m)$ time and are executed sequentially.
  Since $q=\bigO(1)$ and $n=\bigO(m)$, all applications of
  \cref{pr:packed-representation} take
  $\bigO(m\ell\log\AlphabetSize/\log m)$ time in total. The supplied
  representations, the output, the arrays $F_i$, and the representations of
  $B_i,\widehat B_i,D_i,Z_i,O_i,Z_i\odot D_i\odot O_i,R_i,Y_i$, for every
  $i\in[1\dd q]$, occupy $\bigO(m\ell\log\AlphabetSize)$ bits, including
  word-alignment overhead. By \cref{rm:space}, one packed operation uses
  $\bigO(m\ell\log\AlphabetSize)$ bits. The packed operations and inverse
  algorithms execute sequentially and reuse their working space. Adding the
  space of one prefix-frequencies or inverse-grouping application proves the
  claimed bound.
\end{proof}

\subsubsection{Arbitrary Alphabet}\label{sec:packed-permuting-arbitrary-alphabet}

\begin{proposition}[Packed stable grouping and ungrouping]
  \label{pr:packed-stable-grouping}
  Let $\AlphabetSize,m\in\Z_{\geq2}$ satisfy $\AlphabetSize\leq m$, and
  consider the word RAM model with word size $w=c\log m$ for a constant
  $c\geq2$. Let $b\in[1\dd w]$, $W\in[0\dd\AlphabetSize)^m$, and
  $P\in[0\dd2^b)^m$. Using the rank and select notation from
  \cref{def:rank-select}, let
  $\srted{P}{W}\in[0\dd2^b)^m$ denote
  \begin{align*}
    \srted{P}{W}
      =\bigodot_{\alpha=0}^{\AlphabetSize-1}
        \bigodot_{j=1}^{\Rank{W}{m}{\alpha}}
          P[\Select{W}{j}{\alpha}].
  \end{align*}
  For each of the following tasks on packed representations
  (\cref{def:packed-representation}), there is an algorithm that takes
  \begin{align*}
    \bigO\!\left(
      \min\left\{
        m,
        \frac{m\log\AlphabetSize\,(b+\log\AlphabetSize)}{\log m}
      \right\}
    \right)
  \end{align*}
  time and has a peak space usage of
  $\bigO(m(b+\log\AlphabetSize))$ bits.
  \begin{enumerate}
    \item Given $\PackedRepresentation{w}{\AlphabetSize}{W}$ and
      $\PackedRepresentation{w}{2^b}{P}$, construct
      $\PackedRepresentation{w}{2^b}{\srted{P}{W}}$.
    \item Given $\PackedRepresentation{w}{\AlphabetSize}{W}$ and
      $\PackedRepresentation{w}{2^b}{P'}$ for some
      $P'\in[0\dd2^b)^m$, construct
      $\PackedRepresentation{w}{2^b}{P}$ for the unique
      $P\in[0\dd2^b)^m$ satisfying $P'=\srted{P}{W}$.
  \end{enumerate}
\end{proposition}
\begin{proof}
  If $m<2^{16}$, perform either task directly, and hence the claim follows
  immediately. Thus, assume that $m\geq2^{16}$ and compute
  $d:=\lceil\log\AlphabetSize\rceil$ by repeated doubling. Since
  $\AlphabetSize\leq m$, we have $d<1+\log m<m$.

  For the forward algorithm, let $W\in[0\dd\AlphabetSize)^m$ and
  $P\in[0\dd2^b)^m$. Given the packed representation
  $\PackedRepresentation{w}{\AlphabetSize}{W}$ of the string $W$ and the
  packed representation $\PackedRepresentation{w}{2^b}{P}$ of the string
  $P$, we construct
  $\PackedRepresentation{w}{2^b}{\srted{P}{W}}$. For the inverse algorithm,
  let $W\in[0\dd\AlphabetSize)^m$ and $P'\in[0\dd2^b)^m$. Given the packed
  representation $\PackedRepresentation{w}{\AlphabetSize}{W}$ of the string
  $W$ and the packed representation $\PackedRepresentation{w}{2^b}{P'}$ of
  the string $P'$, we construct
  $\PackedRepresentation{w}{2^b}{P}$ for the unique
  $P\in[0\dd2^b)^m$ satisfying $P'=\srted{P}{W}$. For either task, we execute
  only the faster of the following two algorithms.
  \begin{enumerate}

  \item \emph{Packed algorithm.} We process multiple fields per word. The
    algorithm proceeds in three substeps.
    \begin{enumerate}

      \item \emph{Encode and reverse the keys.} Let
        $D\in(\{\zero,\one\}^d)^m$ denote the sequence satisfying
        $D[j]=\AlphabetMap{\AlphabetSize}{2}{W[j]}$ for every
        $j\in[1\dd m]$, and let
        $R[1\dd m]\in(\{\zero,\one\}^d)^m$ denote the sequence satisfying
        $R[j]=\revstr{D[j]}$ for every $j\in[1\dd m]$. Construct the
        alphabet-mapping data structure from \cref{pr:alphabet-map} with
        sequence-length parameter $m$, source alphabet size $\AlphabetSize$,
        and target alphabet size $2$. Its hypotheses hold because
        $2\leq\AlphabetSize\leq m$ and $w\geq2\log m$. Apply it to the
        representation of $W$, with input length $m$, to
        construct $\PackedSeqRepresentation{w}{2}{D}$. Apply
        \cref{pr:packed-fixed-length-string-reversal}, with alphabet size $2$,
        sequence length $m$, and string length $d$, to construct
        $\PackedSeqRepresentation{w}{2}{R}$. This application is valid because
        $2\leq m$ and $1\leq d<m$. Discard the representation of $D$ and the
        alphabet-mapping data structure after constructing $R$.

      \item \emph{Construct the bitvectors for radix sorting.} For the
        correctness argument, for every $h\in[0\dd d]$, let
        $\pi_h[1\dd m]$ denote the permutation obtained by listing the groups
        of indices with the same length-$h$ prefix of $R$ in colexicographic
        order and listing the indices in increasing order within each group.
        For every $h\in[0\dd d)$, let
        $S_h\in(\{\zero,\one\}^{d-h})^m$ denote the sequence satisfying
        \[
          S_h[r]=R[\pi_h[r]][h+1\dd d]
          \qquad\text{for every }r\in[1\dd m].
        \]
        In particular, $\pi_0$ is the identity permutation and $S_0=R$.
        Starting with the representation of $S_0$, process
        $h=0,1,\ldots,d-1$ in increasing order. Let
        $B_h[1\dd m]\in\{\zero,\one\}^m$ and
        $T_h\in(\{\zero,\one\}^{d-h-1})^m$ denote the string and sequence
        satisfying
        \[
          B_h[r]=S_h[r][1]
          \qquad\text{and}\qquad
          T_h[r]=S_h[r][2\dd d-h]
        \]
        for every $r\in[1\dd m]$. Apply the partition algorithm from
        \cref{pr:packed-pointwise-concatenation-partition} to
        $\PackedSeqRepresentation{w}{2}{S_h}$, with alphabet size $2$,
        sequence length $m$, and component lengths $1$ and $d-h-1$. By
        \cref{def:packed-sequence-representation}, its first output is
        $\PackedRepresentation{w}{2}{B_h}$, and its second output is
        $\PackedSeqRepresentation{w}{2}{T_h}$. Retain the representation of
        $B_h$ and discard the representation of $S_h$.

        If $h<d-1$, apply the forward algorithm from
        \cref{cor:packed-binary-string-grouping} to the representations of
        $B_h$ and $T_h$, with alphabet size $2$, sequence length $m$, and
        string length $d-h-1$. Its hypotheses hold because the key string has
        length exactly $m$, the payload sequence contains exactly $m$ strings,
        $d-h-1\geq1$, and
        \[
          2^{d-h-1}\leq2^{d-1}<\AlphabetSize\leq m.
        \]
        For every $X\in\{\zero,\one\}^h$ and
        $\beta\in\{\zero,\one\}$, the identity
        $\revstr{X\beta}=\beta\revstr X$ shows that stable grouping by $B_h$
        orders the length-$(h+1)$ prefix classes in colexicographic order.
        Stability preserves increasing index order within every class. Thus,
        the output is $\PackedSeqRepresentation{w}{2}{S_{h+1}}$. Retain this
        representation for the next iteration and discard the representation
        of $T_h$. If $h=d-1$, discard the representation of the sequence
        $T_h$ of empty strings. Every pointwise-partition application is valid
        because its component lengths are nonnegative and their sum $d-h$
        belongs to $[1\dd d]$, where $d<m$.

      \item \emph{Group or ungroup the payloads.} For the first task, set
        $P^{(0)}:=P$. For every $h\in[0\dd d)$ in increasing order, apply the
        forward algorithm from \cref{pr:packed-binary-stable-grouping} to
        $\PackedRepresentation{w}{2}{B_h}$ and
        $\PackedRepresentation{w}{2^b}{P^{(h)}}$. Retain its output as
        $\PackedRepresentation{w}{2^b}{P^{(h+1)}}$. Once $P^{(h+1)}$ has been
        constructed, the packed representations of $P^{(h)}$ and $B_h$ are no
        longer retained by the algorithm. By the definition of $\pi_h$, the
        payloads in $P^{(h)}$ are aligned with the indices
        $\pi_h[1],\ldots,\pi_h[m]$.
        The stable grouping at iteration $h$ therefore constructs the payload
        sequence aligned with $\pi_{h+1}$. At $h=d$, the colexicographic order
        of the strings $R[j]$ coincides with the lexicographic order of the
        strings $D[j]$, which is the numerical order of the characters $W[j]$.
        Hence, $P^{(d)}=\srted{P}{W}$.
        For the second task, set $P^{(d)}:=P'$. Process
        $h=d-1,d-2,\ldots,0$ in decreasing order, applying the inverse
        algorithm from \cref{pr:packed-binary-stable-grouping} to
        $\PackedRepresentation{w}{2}{B_h}$ and
        $\PackedRepresentation{w}{2^b}{P^{(h+1)}}$. Its output is the unique
        sequence $P^{(h)}$ whose stable grouping according to $B_h$ is
        $P^{(h+1)}$. Once $P^{(h)}$ has been constructed, the packed
        representations of $P^{(h+1)}$ and $B_h$ are no longer retained by the
        algorithm. The final iteration recovers $P^{(0)}=P$.
    \end{enumerate}
    Computing $d$, constructing and applying the alphabet-mapping data
    structure, and reversing $D$ take $\bigO(md/\log m)$ time because
    $m\geq2^{16}$ implies $\log m\leq m/\log m$ and
    $\sqrt m\leq m/\log m$. At iteration $h$, pointwise partition takes
    \[
      \bigO\left(1+\frac{m(d-h)}{\log m}\right)
    \]
    time. If $h<d-1$, applying
    \cref{cor:packed-binary-string-grouping} takes
    $\bigO(m(d-h-1)/\log m)$ time. Summing over all iterations shows that
    constructing the bitvectors takes $\bigO(md^2/\log m)$ time. The $d$
    applications of \cref{pr:packed-binary-stable-grouping} take
    $\bigO(mdb/\log m)$ time. Thus, the packed algorithm takes
    $\bigO(md(d+b)/\log m)$ time.

    By \cref{rm:space}, the alphabet-mapping data structure occupies
    $\bigO(\sqrt m\log m)=\bigO(md)$ bits. The supplied key representation,
    the representation of $D$, the representations of $S_h$, $T_h$, and
    $S_{h+1}$ that coexist during one iteration, and the peak space usage of one application of
    alphabet mapping, fixed-length reversal, pointwise partition, or binary
    string grouping together occupy $\bigO(md)$ bits. The retained packed
    representations of $B_0,\ldots,B_{d-1}$ together occupy
    $\bigO(md+d\log m)=\bigO(md)$ bits, including all unused bits and
    word-alignment overhead. The representations of $P^{(h)}$ and
    $P^{(h+1)}$ used during one binary grouping or ungrouping application,
    together with its peak space usage, occupy
    $\bigO(mb)$ bits. The applications execute sequentially and reuse their
    temporary space. Hence, the packed algorithm has a peak space usage of
    $\bigO(m(d+b))$ bits.

  \item \emph{Direct algorithm.} We process one position at a time, i.e., use
  constant time per position. Compute the frequency of every
  $\alpha\in[0\dd\AlphabetSize)$ in $W$, and compute the starting position of
  its block in the stable order. Initialize the packed output with zeros. For
  the first task, initialize one counter at each starting position. Scan
  $j\in[1\dd m]$ in increasing order, store $P[j]$ at the next position of the
  block for $W[j]$, and increment its counter. For the second task, initialize
  the same counters, scan $j\in[1\dd m]$ in increasing order, retrieve the next
  payload from the block for $W[j]$ in $P'$, store it in $P[j]$, and increment
  the counter. Either task takes $\bigO(m+\AlphabetSize)=\bigO(m)$ time. The
  output,
  the frequency and counter arrays, and the packed input use
  $\bigO(mb+\AlphabetSize\log m)=\bigO(m(d+b))$ bits.
  The final equality uses
  $\AlphabetSize\log m=\bigO(m\log\AlphabetSize)=\bigO(md)$.
  \end{enumerate}

  Choosing the faster of these two algorithms and using
  $d=\Theta(\log\AlphabetSize)$ proves the claim.
\end{proof}

\begin{corollary}[Packed stable grouping and ungrouping of string payloads]
  \label{cor:packed-stable-string-grouping}
  Let $\AlphabetSize,k,m\in\Z_{\geq2}$ and $r\in\Z_{\geq1}$ satisfy
  $\AlphabetSize\leq m$ and $k^r\leq m$, and consider the word RAM model
  with word size $w=c\log m$ for a constant $c\geq2$. For
  $W\in[0\dd\AlphabetSize)^m$ and
  $Q\in([0\dd k)^r)^m$, let
  $\srted{Q}{W}\in([0\dd k)^r)^m$ denote
  \begin{align*}
    \bigodot_{i=1}^m \srted{Q}{W}[i]
      =\bigodot_{\alpha=0}^{\AlphabetSize-1}
        \bigodot_{j=1}^{\Rank{W}{m}{\alpha}}
          Q[\Select{W}{j}{\alpha}].
  \end{align*}
  Each of the following tasks on packed representations
  (\cref{def:packed-representation,def:packed-sequence-representation}) can
  be accomplished by an algorithm that takes
  \begin{align*}
    \bigO\!\left(
      \min\left\{
        m,
        \frac{m\log\AlphabetSize
          (r\log k+\log\AlphabetSize)}{\log m}
      \right\}
    \right)
  \end{align*}
  time and has a peak space usage of
  $\bigO(m(r\log k+\log\AlphabetSize))$ bits.
  \begin{enumerate}
    \item Given $\PackedRepresentation{w}{\AlphabetSize}{W}$ and
      $\PackedSeqRepresentation{w}{k}{Q}$, construct
      $\PackedSeqRepresentation{w}{k}{\srted{Q}{W}}$.
    \item Given $\PackedRepresentation{w}{\AlphabetSize}{W}$ and
      $\PackedSeqRepresentation{w}{k}{Q'}$ for some
      $Q'\in([0\dd k)^r)^m$, construct
      $\PackedSeqRepresentation{w}{k}{Q}$ for the unique
      $Q\in([0\dd k)^r)^m$ satisfying $Q'=\srted{Q}{W}$.
  \end{enumerate}
\end{corollary}
\begin{proof}
  Let $b:=\lceil\log(k^r)\rceil$. For every
  $R\in([0\dd k)^r)^m$, let $\widehat R\in[0\dd k^r)^m$ denote the string
  satisfying $\widehat R[j]=\Val{k}{R[j]}$ for every $j\in[1\dd m]$.
  Apply \cref{pr:packed-character-block-encoding} to
  $\PackedSeqRepresentation{w}{k}{R}$, with alphabet size $k$, input length
  $m$, block length $r$, and $g=h=1$. Its hypotheses hold because $k^r\leq m$,
  and its output is
  $\PackedRepresentation{w}{k^r}{\widehat R}$. Since both alphabets $k^r$
  and $2^b$ use $b$ bits per character, this is also
  $\PackedRepresentation{w}{2^b}{\widehat R}$.

  For the first task, apply the preceding encoding to $Q$, and then apply the
  forward algorithm from \cref{pr:packed-stable-grouping} to
  $\PackedRepresentation{w}{\AlphabetSize}{W}$ and
  $\PackedRepresentation{w}{2^b}{\widehat Q}$, with key alphabet size
  $\AlphabetSize$, input length $m$, and payload width $b$. Encoding the
  payload strings positionwise commutes with their stable permutation, so the
  output is
  $\PackedRepresentation{w}{2^b}{\widehat{\srted{Q}{W}}}$. For the second
  task, apply the encoding to $Q'$ and apply the inverse algorithm with the
  same parameters to $\PackedRepresentation{w}{\AlphabetSize}{W}$ and
  $\PackedRepresentation{w}{2^b}{\widehat{Q'}}$. Its output is
  $\PackedRepresentation{w}{2^b}{\widehat Q}$ for the unique $Q$ satisfying
  $Q'=\srted{Q}{W}$. These applications are valid because
  $\AlphabetSize\leq m$ and $b\leq2\log m\leq w$.

  Construct the alphabet-mapping data structure from \cref{pr:alphabet-map}
  with parameters $m$, $k^r$, and $k$. Its hypotheses hold because
  $k\leq k^r\leq m$ and $w\geq2\log m$. Every character returned by stable
  grouping or ungrouping is smaller than $k^r$. Since $k^r$ and $2^b$ use the
  same field width, the representations returned in the first and second tasks
  are also
  $\PackedRepresentation{w}{k^r}{\widehat{\srted{Q}{W}}}$ and
  $\PackedRepresentation{w}{k^r}{\widehat Q}$, respectively. In each task,
  apply the alphabet-mapping data structure to the corresponding
  representation, with length $m$. For every $X\in[0\dd k)^r$, the equality
  $\AlphabetMap{k^r}{k}{\Val{k}{X}}=X$ includes any leading zeros of $X$.
  Hence, the returned packed string is the required packed sequence
  representation.

  Applying \cref{pr:packed-character-block-encoding} and constructing and
  applying the data structure from \cref{pr:alphabet-map} take
  $\bigO(1+\sqrt m+mr\log k/\log m)$ time. Since
  $r\log k\leq\log m$ and $\log\AlphabetSize\geq1$, the term
  $mr\log k/\log m$ is bounded by both $m$ and
  $m\log\AlphabetSize(r\log k+\log\AlphabetSize)/\log m$.
  The same holds for $1+\sqrt m$ because
  $\sqrt m=\bigO(m/\log m)$. Thus, these applications are bounded by the
  claimed time, and \cref{pr:packed-stable-grouping}, with
  $b=\Theta(r\log k)$, satisfies the same bound.

  The supplied representations, the encoded payload, the grouped or ungrouped
  code string, the alphabet-mapping data structure, the output, and the
  working space used by one application together occupy
  $\bigO(m(r\log k+\log\AlphabetSize)+\sqrt m\log m)$ bits. Since
  $\sqrt m\log m=\bigO(m)$, this proves the claimed peak space usage.
\end{proof}

\subsection{Stable Multilevel Routing}\label{sec:packed-stable-multilevel-routing}

\begin{proposition}[Packed stable multilevel routing]
  \label{pr:packed-stable-multilevel-routing}
  Let $\AlphabetSize,m\in\Z_{\geq2}$ and $\ell\in\Z_{\geq1}$ satisfy
  $\AlphabetSize^\ell\leq m$, and consider the word RAM model with word size
  $w=c\log m$ for a constant $c\geq2$. Let $b\in[1\dd w]$,
  $W\in([0\dd\AlphabetSize)^\ell)^m$, and $P\in[0\dd2^b)^m$. Using the
  notation from \cref{def:prefix-range-queries}, for every
  $X\in[0\dd\AlphabetSize)^{\leq\ell}$, let $f_X$ denote
  $\PrefixRank{W}{m}{X}$, and let
  $P_X[1\dd f_X]\in[0\dd2^b)^{f_X}$ denote the string satisfying
  \begin{align*}
    P_X[r]=P[\PrefixSelect{W}{r}{X}]
    \qquad\text{for every }r\in[1\dd f_X].
  \end{align*}
  There is an algorithm that, given
  $\PackedSeqRepresentation{w}{\AlphabetSize}{W}$ and
  $\PackedRepresentation{w}{2^b}{P}$
  (\cref{def:packed-representation,def:packed-sequence-representation}),
  constructs, for every $p\in[0\dd\ell]$, an array
  $\mathcal P_p[0\dd\AlphabetSize^p)$ indexed according to \cref{def:val} and
  satisfying
  \begin{align*}
    \mathcal P_p[\Val{\AlphabetSize}{X}]
      =\PackedRepresentation{w}{2^b}{P_X}
    \qquad\text{for every }X\in[0\dd\AlphabetSize)^p.
  \end{align*}
  The arrays are returned one by one in increasing order of $p$. The
  algorithm takes
  \begin{align*}
    \bigO\!\left(
      \frac{m\ell\log\AlphabetSize
        (\ell\log\AlphabetSize+b)}{\log m}
    \right)
  \end{align*}
  time and has a peak space usage of
  $\bigO(m(\ell\log\AlphabetSize+b))$ bits.
\end{proposition}
\begin{proof}
  If $m<2^{16}$, construct the arrays directly, and hence the claim follows
  immediately. Thus, assume that $m\geq2^{16}$.

  For every $X\in[0\dd\AlphabetSize)^{\leq\ell}$, let
  $W_X\in([0\dd\AlphabetSize)^\ell)^{f_X}$ denote the subsequence of $W$
  consisting of the strings having prefix $X$, in their original relative
  order. For every $p\in[0\dd\ell]$, let
  $P_p[1\dd m]\in[0\dd2^b)^m$ denote the payload string satisfying
  \begin{equation}\label{eq:packed-routing-output}
    P_p
      =
      \bigodot_{X\in[0\dd\AlphabetSize)^p\text{ in colexicographic order}}P_X.
  \end{equation}
  For every $p\in[0\dd\ell)$, let
  $W_p[1\dd m]\in([0\dd\AlphabetSize)^{\ell-p})^m$ denote the sequence
  satisfying
  \begin{equation}\label{eq:packed-routing-colex-levels}
    \bigodot_{j=1}^m W_p[j]
      =
      \bigodot_{X\in[0\dd\AlphabetSize)^p\text{ in colexicographic order}}
        \bigodot_{j=1}^{f_X}W_X[j][p+1\dd\ell].
  \end{equation}
  Since the empty string is the unique length-zero prefix, $P_0=P$ and
  $W_0=W$. The supplied representations
  $\PackedRepresentation{w}{2^b}{P}$ and
  $\PackedSeqRepresentation{w}{\AlphabetSize}{W}$ are the representations of
  $P_0$ and $W_0$, respectively.

  Construct the output arrays in two steps.
  \begin{enumerate}

    \item\label{step:packed-stable-multilevel-routing-frequencies}
      \emph{Construct $A_{\rm freq}$ and $A_{\rm pow}$.} Apply
      \cref{pr:packed-prefix-frequencies}, with alphabet size $\AlphabetSize$,
      sequence length $m$, and string length $\ell$, to
      $\PackedSeqRepresentation{w}{\AlphabetSize}{W}$. Its hypothesis holds
      because $\AlphabetSize^\ell\leq m$. Retain its output as
      $A_{\rm freq}[0\dd2\AlphabetSize^\ell)$. Thus,
      $A_{\rm freq}[\BasicInt{\AlphabetSize}{X}]=f_X$ for every
      $X\in[0\dd\AlphabetSize)^{\leq\ell}$. Construct
      $A_{\rm pow}[0\dd\ell]$ by setting $A_{\rm pow}[0]:=1$ and
      $A_{\rm pow}[p]:=\AlphabetSize\cdot A_{\rm pow}[p-1]$ for every
      $p\in[1\dd\ell]$. Constructing the two arrays takes
      $\bigO(m\ell\log\AlphabetSize/\log m)$ time, and the arrays occupy
      $\bigO(m\ell\log\AlphabetSize)$ bits.

    \item\label{step:packed-stable-multilevel-routing-output-arrays}
      \emph{Construct the output arrays $\mathcal P_p$.} For every
      $p\in[0\dd\ell]$ in increasing order, perform the following substeps.
      \begin{enumerate}

        \item \emph{Construct $\mathcal P_p$.} Initialize every entry of
          $\mathcal P_p[0\dd\AlphabetSize^p)$ to store the empty packed string
          and set $s:=1$, $X:=0^p$, and $v:=0$. Scan
          $u\in[0\dd\AlphabetSize^p)$ in increasing order while maintaining
          $X\in[0\dd\AlphabetSize)^p$ satisfying
          $u=\Val{\AlphabetSize}{\revstr X}$ and
          $v=\Val{\AlphabetSize}{X}$. Between consecutive
          iterations, increment $X[1]$ and propagate carries toward $X[p]$.
          Update $v$ using $A_{\rm pow}$ whenever a character changes. Compute
          $x:=A_{\rm pow}[p]+v$ and retrieve
          $f:=A_{\rm freq}[x]$. If $f>0$, distinguish two cases. If $b<w$,
          apply the substring operation from
          \cref{pr:packed-representation}\eqref{pr:packed-representation-substring}
          to the packed representation
          $\PackedRepresentation{w}{2^b}{P_p}$, with alphabet size $2^b$,
          length parameter $\max\{m,2^b\}$, starting position $s$, and
          substring length $f$. Its hypotheses hold because
          $2^b\leq\max\{m,2^b\}$ and
          $w>\log\max\{m,2^b\}$. Store the returned packed representation in
          $\mathcal P_p[v]$. If $b=w$, copy the $f$
          consecutive words starting at word $s$ and store this packed
          representation in the same entry. In both cases, set $s:=s+f$. The
          entries that are not overwritten represent the empty strings $P_X$.
          After the scan, return
          $\mathcal P_p$. It holds $s=m+1$ by
          \cref{eq:packed-routing-output}.
          Copying a positive-length piece takes $\bigO(1+fb/w)$ time and sets
          every unused bit in every output word to zero. The scan changes
          $\bigO(\AlphabetSize^p)$ characters. Therefore, constructing all
          arrays takes
          \[
            \bigO\left(
              \sum_{p=0}^{\ell}\AlphabetSize^p
              +\frac{m\ell b}{\log m}
            \right)
          \]
          time. The monotonicity of $x/\log x$ gives
          \begin{equation}\label{eq:packed-prefix-bucket-universe}
            \AlphabetSize^\ell\log m
              =\bigO(m\ell\log\AlphabetSize),
          \end{equation}
          so this time is bounded by the claimed time. By
          \cref{def:packed-representation}, the representations stored in
          $\mathcal P_p$, including all unused bits and word-alignment
          overhead, together with the entries of $\mathcal P_p$, use
          $\bigO(mb+\AlphabetSize^p\log m)$ bits.

        \item \emph{If $p<\ell$, construct $P_{p+1}$ and, if $p<\ell-1$,
          $W_{p+1}$.} If $p=\ell$, terminate the construction. Otherwise, let
          $C_p[1\dd m]\in[0\dd\AlphabetSize)^m$ denote the string satisfying
          $C_p[j]=W_p[j][1]$ for every $j\in[1\dd m]$. Let
          $R_p\in([0\dd\AlphabetSize)^{\ell-p-1})^m$ denote the sequence
          satisfying $W_p[j]=(C_p[j])R_p[j]$ for every $j\in[1\dd m]$.
          Apply the partition algorithm from
          \cref{pr:packed-pointwise-concatenation-partition} to
          $\PackedSeqRepresentation{w}{\AlphabetSize}{W_p}$, with alphabet size
          $\AlphabetSize$, sequence length $m$, and component lengths $1$ and
          $\ell-p-1$. Its outputs are
          $\PackedSeqRepresentation{w}{\AlphabetSize}{((C_p[j]))_{j=1}^m}$ and
          $\PackedSeqRepresentation{w}{\AlphabetSize}{R_p}$. By
          \cref{def:packed-sequence-representation},
          $\PackedSeqRepresentation{w}{\AlphabetSize}{((C_p[j]))_{j=1}^m}$ is
          $\PackedRepresentation{w}{\AlphabetSize}{C_p}$. The application is
          valid because $p<\ell$ and $\AlphabetSize^\ell\leq m$ implies
          $\AlphabetSize\leq m$ and
          $1\leq\ell-p\leq\ell\leq\log m<m$.
          Apply the forward algorithm from
          \cref{pr:packed-stable-grouping}, with alphabet size $\AlphabetSize$,
          sequence length $m$, packed key representation
          $\PackedRepresentation{w}{\AlphabetSize}{C_p}$, packed payload
          representation $\PackedRepresentation{w}{2^b}{P_p}$, and payload
          width $b$, to construct
          $\PackedRepresentation{w}{2^b}{P_{p+1}}$. This application is valid
          because $\AlphabetSize\leq m$ and $b\leq w$. If $p<\ell-1$, apply
          the forward algorithm from
          \cref{cor:packed-stable-string-grouping}, with key alphabet size
          $\AlphabetSize$, payload alphabet size $\AlphabetSize$, sequence
          length $m$, and payload string length $\ell-p-1$, to
          $\PackedRepresentation{w}{\AlphabetSize}{C_p}$ and
          $\PackedSeqRepresentation{w}{\AlphabetSize}{R_p}$. Retain its output
          as $\PackedSeqRepresentation{w}{\AlphabetSize}{W_{p+1}}$. The
          application is valid because $\AlphabetSize\leq m$,
          $\ell-p-1\geq1$, and
          $\AlphabetSize^{\ell-p-1}\leq\AlphabetSize^\ell\leq m$. If
          $p=\ell-1$, the strings in $R_p$ are empty, so discard their
          representation without applying the corollary.
          For $X\in[0\dd\AlphabetSize)^p$ and
          $\alpha\in[0\dd\AlphabetSize)$, the identity
          $\revstr{X\alpha}=\alpha\revstr X$ shows that stable grouping by
          $C_p$ orders the prefix classes $X\alpha$ colexicographically.
          Stability preserves the original relative order inside every class.
          Hence, applying \cref{pr:packed-stable-grouping} constructs $P_{p+1}$
          as specified in \cref{eq:packed-routing-output}. If $p<\ell-1$,
          applying \cref{cor:packed-stable-string-grouping} constructs the
          suffix sequence $W_{p+1}$ specified in
          \cref{eq:packed-routing-colex-levels}.
          The pointwise partition takes
          $\bigO(1+m(\ell-p)\log\AlphabetSize/\log m)$ time. By
          \cref{pr:packed-stable-grouping,cor:packed-stable-string-grouping},
          the two grouping applications, when both are present, take
          \[
            \bigO\left(
              \frac{m\log\AlphabetSize
                ((\ell-p)\log\AlphabetSize+b)}{\log m}
            \right)
          \]
          time. The same bound applies when $p=\ell-1$ and only
          \cref{pr:packed-stable-grouping} is applied.
      \end{enumerate}
  \end{enumerate}

  Summing the bounds in
  Step~\ref{step:packed-stable-multilevel-routing-output-arrays} over all depths gives
  $\bigO(m\ell\log\AlphabetSize
    (\ell\log\AlphabetSize+b)/\log m)$ time.
  Step~\ref{step:packed-stable-multilevel-routing-frequencies} satisfies the same
  bound. At every depth $p<\ell$, the pointwise partition and the grouping
  applications are executed sequentially. Retain $C_p$ until the grouping
  applications at depth $p$ are complete. Discard $W_p$, $R_p$, and $P_p$
  after their respective final uses. Retain $P_{p+1}$ and, if $p<\ell-1$,
  $W_{p+1}$ for the next depth. By \cref{def:packed-representation}, the stored
  representations among $W_p,R_p,W_{p+1},C_p,P_p,P_{p+1}$, the arrays
  $A_{\rm freq}$ and $A_{\rm pow}$, and the output array $\mathcal P_p$ for the
  current depth occupy
  \[
    \bigO(m(\ell\log\AlphabetSize+b)+\AlphabetSize^p\log m)
      =\bigO(m(\ell\log\AlphabetSize+b))
  \]
  bits by \cref{eq:packed-prefix-bucket-universe}. The peak space usage of every
  application of \cref{pr:packed-pointwise-concatenation-partition},
  \cref{pr:packed-stable-grouping}, and
  \cref{cor:packed-stable-string-grouping} is
  $\bigO(m(\ell\log\AlphabetSize+b))$ bits. Since the applications are executed
  sequentially and the output arrays are returned one at a time, this proves
  the claimed peak space bound.
\end{proof}

\begin{proposition}[Packed stable multilevel routing for comparable lengths]
  \label{pr:packed-stable-multilevel-routing-comparable}
  Let $\AlphabetSize,m,n\in\Z_{\geq2}$ and $\ell\in\Z_{\geq1}$ satisfy
  $\AlphabetSize^\ell\leq m<n=\bigO(m)$, and consider the word RAM model with
  word size $w=c\log m$ for a constant $c\geq2$. Let $b\in[1\dd w]$,
  $W\in([0\dd\AlphabetSize)^\ell)^n$, and $P\in[0\dd2^b)^n$. For every
  $X\in[0\dd\AlphabetSize)^{\leq\ell}$, let $f_X$ and $P_X$ be defined as in
  \cref{pr:packed-stable-multilevel-routing}, with $n$ in place of $m$.
  There is an algorithm that, given
  $\PackedSeqRepresentation{w}{\AlphabetSize}{W}$ and
  $\PackedRepresentation{w}{2^b}{P}$
  (\cref{def:packed-representation,def:packed-sequence-representation}),
  constructs, for every $p\in[0\dd\ell]$, an array
  $\mathcal P_p[0\dd\AlphabetSize^p)$ indexed according to \cref{def:val} and
  satisfying
  \[
    \mathcal P_p[\Val{\AlphabetSize}{X}]
      =\PackedRepresentation{w}{2^b}{P_X}
    \qquad\text{for every }X\in[0\dd\AlphabetSize)^p.
  \]
  The arrays are returned one by one in increasing order of $p$. The algorithm
  takes
  \[
    \bigO\!\left(
      \frac{m\ell\log\AlphabetSize
        (\ell\log\AlphabetSize+b)}{\log m}
    \right)
  \]
  time and has a peak space usage of
  $\bigO(m(\ell\log\AlphabetSize+b))$ bits.
\end{proposition}
\begin{proof}
  Let $k:=\ceil{n/m}=\bigO(1)$ and $r:=km-n\in[0\dd m)$. If $m<16$ or
  $km\ell\geq m^2$, scan the $n$ key-payload pairs at every depth and append
  each payload to its prefix bucket. Since $k=\bigO(1)$ and
  $\ell\leq\log m$, the inequality $km\ell\geq m^2$ holds for only boundedly
  many values of $m$, so this direct construction satisfies both bounds.
  Thus, assume that $m\geq16$ and $km\ell<m^2\leq2^w$. Let $\overline W$ denote
  $W$ followed by $r$ copies of $0^\ell$, and let $\overline P:=P\odot0^r$.
  For every $i\in[1\dd k]$, let
  $W_i$ and $P_i$ denote the $i$th consecutive blocks of $m$ entries in
  $\overline W$ and $\overline P$, respectively. Thus,
  \[
    \overline W=W_1\odot\cdots\odot W_k
    \qquad\text{and}\qquad
    \overline P=P_1\odot\cdots\odot P_k.
  \]
  Compute $\lceil\log\AlphabetSize\rceil$ by repeated doubling. This takes
  $\bigO(\log m)$ time and bits, which are bounded by the claimed bounds because
  $m/\log m\geq\log m$. Denote $N_W:=km\ell$ and, when $b<w$,
  $N_P:=\max\{km,2^b\}$. Every application of
  \cref{pr:packed-representation} to keys in
  Step~\ref{step:packed-stable-multilevel-routing-comparable-representations} uses alphabet size
  $\AlphabetSize$ and length parameter $N_W$. When $b<w$, every application to
  payloads in that step uses alphabet size $2^b$ and length parameter $N_P$.
  Their common hypotheses hold because
  $\AlphabetSize\leq m\leq N_W<2^w$ and, when $b<w$,
  $2^b\leq N_P<2^w$. All lengths and numbers of pieces supplied to these
  applications are at most the corresponding length parameter.
  Perform the following steps.
  \begin{enumerate}

    \item\label{step:packed-stable-multilevel-routing-comparable-representations}
      \emph{Construct the representations of $W_i,P_i$ and the arrays
      $F_i$.} Apply
      \cref{pr:packed-representation}\eqref{pr:packed-representation-initialize} to construct the
      representation of $\zero^{r\ell}$. Apply
      \cref{pr:packed-representation}\eqref{pr:packed-representation-concat} to the supplied representation
      and the representation of $\zero^{r\ell}$, with lengths $n\ell,r\ell$, to
      construct $\PackedSeqRepresentation{w}{\AlphabetSize}{\overline W}$. Apply
      \cref{pr:packed-representation}\eqref{pr:packed-representation-split}, with source length $N_W$ and $k$
      copies of length $m\ell$. By \cref{def:packed-sequence-representation},
      the outputs are the packed sequence representations of $W_1,\ldots,W_k$.
      If $b<w$, apply
      \cref{pr:packed-representation}\eqref{pr:packed-representation-initialize} to construct the packed
      representation of $\zero^r$, apply
      \cref{pr:packed-representation}\eqref{pr:packed-representation-concat} to the supplied representation of
      $P$ and the representation of $\zero^r$, with lengths $n,r$, to construct
      $\PackedRepresentation{w}{2^b}{\overline P}$. Apply
      \cref{pr:packed-representation}\eqref{pr:packed-representation-split}, with $k$ copies of length $m$ and
      source length $km$, to obtain the representations of $P_1,\ldots,P_k$.
      If $b=w$, every payload occupies exactly one word. Allocate a local word
      array $A_P[1\dd km]$, copy the $n$ words of the supplied
      representation of $P$ into its first $n$ entries, and set its final $r$
      entries to zero. The $i$th block of $m$ words is
      $\PackedRepresentation{w}{2^b}{P_i}$.
      After the key split, discard the representations of $\zero^{r\ell}$ and
      $\overline W$. When $b<w$, similarly discard the representations of
      $\zero^r$ and $\overline P$ after the payload split. Retain all split
      outputs and, when $b=w$, retain $A_P$ until the routing applications
      finish.
      For every $i\in[1\dd k]$, apply
      \cref{pr:packed-prefix-frequencies}, with parameters
      $\AlphabetSize,m,\ell$, to
      $\PackedSeqRepresentation{w}{\AlphabetSize}{W_i}$. Retain the returned
      array $F_i[0\dd2\AlphabetSize^\ell)$ until $\mathcal P_\ell$ has been
      returned. It satisfies
      \[
        F_i[\BasicInt{\AlphabetSize}{X}]
          =\PrefixRank{W_i}{m}{X}
        \qquad
        \text{for every }X\in[0\dd\AlphabetSize)^{\leq\ell}.
      \]
      Every application is valid because $W_i$ contains exactly $m$ strings,
      $\AlphabetSize^\ell\leq m$, and the word size is $w=c\log m$.
      The packed operations take
      $\bigO(1+m(\ell\log\AlphabetSize+b)/\log m)$ time. When $b=w$, filling
      $A_P$ takes $\bigO(m)=\bigO(mb/\log m)$ time and satisfies the same bound.
      The $k$ applications of \cref{pr:packed-prefix-frequencies} take
      $\bigO(m\ell\log\AlphabetSize/\log m)$ time. The packed representations
      of $W_1,\ldots,W_k,P_1,\ldots,P_k$, the arrays $F_i$, the initialized zero
      suffixes, and the temporary representations of $\overline W$ and
      $\overline P$ occupy $\bigO(m(\ell\log\AlphabetSize+b))$ bits. When
      $b=w$, the representation of $\overline P$ is stored in $A_P$. The packed
      operations execute sequentially and reuse their working space. By
      \cref{rm:space}, one operation uses
      $\bigO(m(\ell\log\AlphabetSize+b))$ bits.

    \item\label{step:packed-stable-multilevel-routing-comparable-block-arrays}
      \emph{Construct the arrays $\mathcal P_{i,p}$.} Apply
      \cref{pr:packed-stable-multilevel-routing}, with parameters
      $\AlphabetSize,m,\ell,b$, to every pair of packed representations of
      $W_i$ and $P_i$. Execute the $k$ applications in lockstep. Denote the
      array returned by application $i$ at depth $p$ by
      $\mathcal P_{i,p}[0\dd\AlphabetSize^p)$.
      Every pair $W_i,P_i$ has length exactly $m$, and the inequalities
      $\AlphabetSize^\ell\leq m$ and $b\leq w$ hold. Every application uses
      the physical word size $w=c\log m$, so all the hypotheses of
      \cref{pr:packed-stable-multilevel-routing} hold.

    \item\label{step:packed-stable-multilevel-routing-comparable-output-arrays}
      \emph{Construct the arrays $\mathcal P_p$.} When
      $\mathcal P_{1,p},\ldots,\mathcal P_{k,p}$ have been returned, proceed in
      two substeps.
      \begin{enumerate}
        \item \emph{Construct $\widehat{\mathcal P}_{k,p}[0]$.} Compute
          $g_p:=F_k[\AlphabetSize^p]-r$. The value $g_p$ lies in
          $[0\dd F_k[\AlphabetSize^p]]$ because the final $r$ strings of $W_k$
          equal $0^\ell$. If $r=0$, let $\widehat{\mathcal P}_{k,p}[0]$ denote
          $\mathcal P_{k,p}[0]$. Otherwise, construct the packed representation
          $\widehat{\mathcal P}_{k,p}[0]$ of the length-$g_p$ prefix of the
          string represented by $\mathcal P_{k,p}[0]$. If $b<w$, apply the
          substring operation from
          \cref{pr:packed-representation}\eqref{pr:packed-representation-substring},
          with alphabet size $2^b$, length parameter $\max\{m,2^b\}$, source length
          $F_k[\AlphabetSize^p]$, starting position $1$, and substring length
          $g_p$. Its source is nonempty because $r>0$, the requested interval
          is valid because $0\leq g_p\leq F_k[\AlphabetSize^p]\leq m$, and
          $2^b\leq\max\{m,2^b\}<2^w$. If $b=w$, copy the first $g_p$ words of
          $\mathcal P_{k,p}[0]$.

        \item \emph{Construct the entries $\mathcal P_p[x]$.} Allocate
          $\mathcal P_p[0\dd\AlphabetSize^p)$ and scan
          $x\in[0\dd\AlphabetSize^p)$ in increasing order. For every
          $i\in[1\dd k]$, compute
          \[
            g_i:=
            \begin{cases}
              g_p & \text{if $i=k$ and $x=0$},\\
              F_i[\AlphabetSize^p+x] & \text{otherwise}.
            \end{cases}
          \]
          Concatenate the represented strings in increasing order of $i$,
          using $\widehat{\mathcal P}_{k,p}[0]$ in place of
          $\mathcal P_{k,p}[0]$ when $i=k$ and $x=0$. If $b<w$, apply the
          concatenate operation from \cref{pr:packed-representation}, with the
          $k$ named representations, lengths $g_1,\ldots,g_k$, alphabet size
          $2^b$, and length parameter $\max\{km,2^b\}$. Its hypotheses hold
          because the lengths sum to at most $n$, and
          $k,km,2^b<2^w$. If $b=w$, copy, in increasing order of $i$, the first
          $g_i$ words of $\mathcal P_{i,p}[x]$, using the first $g_p$ words of
          $\mathcal P_{k,p}[0]$ when $i=k$ and $x=0$. Store the resulting
          representation in $\mathcal P_p[x]$.

          After the scan, return $\mathcal P_p$, discard
          $\mathcal P_{1,p},\ldots,\mathcal P_{k,p}$ and
          $\widehat{\mathcal P}_{k,p}[0]$, and, if $p<\ell$, resume the
          applications from
          Step~\ref{step:packed-stable-multilevel-routing-comparable-block-arrays}
          at depth $p+1$.
      \end{enumerate}

      Across all depths,
      Step~\ref{step:packed-stable-multilevel-routing-comparable-output-arrays} takes
      $\bigO(\AlphabetSize^\ell+n\ell b/\log m)$ time. During depth $p$, the
      arrays $\mathcal P_{1,p},\ldots,\mathcal P_{k,p}$, the temporary
      representation $\widehat{\mathcal P}_{k,p}[0]$, and $\mathcal P_p$ use
      $\bigO(m(\ell\log\AlphabetSize+b))$ bits by
      \cref{pr:packed-stable-multilevel-routing}\eqref{eq:packed-prefix-bucket-universe}.
  \end{enumerate}

  For correctness, let $p\in[0\dd\ell]$ and
  $X\in[0\dd\AlphabetSize)^p$, and denote $x:=\Val{\AlphabetSize}{X}$. By
  \cref{def:basic-int},
  $\BasicInt{\AlphabetSize}{X}=\AlphabetSize^p+x$.
  Step~\ref{step:packed-stable-multilevel-routing-comparable-representations} and
  \cref{pr:packed-stable-multilevel-routing} imply that, for every
  $i\in[1\dd k]$, $F_i[\AlphabetSize^p+x]$ is the length of the string
  represented by $\mathcal P_{i,p}[x]$. For every $i\in[1\dd k]$, let
  $P^\circ_{i,X}$ denote the subsequence of
  $P[(i-1)m+1\dd\min\{im,n\}]$ whose corresponding strings in $W$ have prefix
  $X$. By \cref{pr:packed-stable-multilevel-routing},
  $\mathcal P_{i,p}[x]$ represents $P^\circ_{i,X}$, except possibly for
  $i=k$ and $x=0$. Every dummy key equals $0^\ell$ and therefore matches
  exactly the prefix $0^p$, including $p=0$, where $0^0$ is the empty prefix.
  Stability therefore makes $\mathcal P_{k,p}[0]$ represent
  $P^\circ_{k,0^p}\odot0^r$, whose length is
  $F_k[\AlphabetSize^p]$.
  Step~\ref{step:packed-stable-multilevel-routing-comparable-output-arrays} constructs
  $\widehat{\mathcal P}_{k,p}[0]$ as the packed representation of
  $P^\circ_{k,0^p}$ and uses it in place of $\mathcal P_{k,p}[0]$ in the
  concatenation. Since the intervals
  $[(i-1)m+1\dd\min\{im,n\}]$ are consecutive,
  $P_X=P^\circ_{1,X}\odot\cdots\odot P^\circ_{k,X}$. Therefore,
  $\mathcal P_p[x]=\PackedRepresentation{w}{2^b}{P_X}$.

  The $k=\bigO(1)$ applications in
  Step~\ref{step:packed-stable-multilevel-routing-comparable-block-arrays} take
  \[
    \bigO\!\left(
      \frac{m\ell\log\AlphabetSize
        (\ell\log\AlphabetSize+b)}{\log m}
    \right)
  \]
  time. Using $n=\bigO(m)$,
  $\ell\leq\ell\log\AlphabetSize$, and
  \cref{pr:packed-stable-multilevel-routing}\eqref{eq:packed-prefix-bucket-universe},
  both
  $\bigO(\AlphabetSize^\ell+n\ell b/\log m)$ and the time in
  Step~\ref{step:packed-stable-multilevel-routing-comparable-representations} are
  \[
    \bigO\!\left(
      \frac{m\ell\log\AlphabetSize
        (\ell\log\AlphabetSize+b)}{\log m}
    \right).
  \]
  The $k$ applications are advanced in lockstep, so only
  $\mathcal P_{1,p},\ldots,\mathcal P_{k,p}$ coexist. The packed
  representations of $W_1,\ldots,W_k,P_1,\ldots,P_k$, the arrays $F_i$, the
  working spaces of the $k$ applications, the arrays
  $\mathcal P_{1,p},\ldots,\mathcal P_{k,p}$, the temporary representation
  $\widehat{\mathcal P}_{k,p}[0]$, and $\mathcal P_p$ use
  $\bigO(m(\ell\log\AlphabetSize+b))$ bits. This proves both bounds.
\end{proof}

\subsection{Packed Wavelet-Tree Construction}\label{sec:packed-wavelet-tree-construction}

\begin{proposition}[Packed wavelet-tree construction]
  \label{pr:packed-wavelet-tree-construction}
  Let $\AlphabetSize,m\in\Z_{\geq2}$ and $\ell\in\Z_{\geq1}$ satisfy
  $\AlphabetSize^\ell\leq m$. Consider the word RAM model with word size
  $w=c\log m$, where $c\geq2$ is a constant. Let
  $W\in([0\dd\AlphabetSize)^\ell)^m$. Using the notation from
  \cref{def:prefix-range-queries}, for every
  $X\in[0\dd\AlphabetSize)^{<\ell}$, let
  $f_X$ denote $\PrefixRank{W}{m}{X}$, and let
  $C_X[1\dd f_X]\in[0\dd\AlphabetSize)^{f_X}$ denote the string satisfying
  \begin{align*}
    C_X[r]
      =W[\PrefixSelect{W}{r}{X}][|X|+1]
    \qquad\text{for every }r\in[1\dd f_X].
  \end{align*}
  There is an algorithm that, given
  $\PackedSeqRepresentation{w}{\AlphabetSize}{W}$
  (\cref{def:packed-sequence-representation}), constructs an array
  $\mathcal C[0\dd2\AlphabetSize^\ell)$. For every
  $X\in[0\dd\AlphabetSize)^{<\ell}$, the entry
  $\mathcal C[\BasicInt{\AlphabetSize}{X}]$ equals
  $\PackedRepresentation{w}{\AlphabetSize}{C_X}$
  (\cref{def:basic-int,def:packed-representation}). All remaining entries
  equal the empty packed representation. The algorithm takes
  \begin{align*}
    \bigO\!\left(
      m\min\!\left\{
        \ell,
        \frac{\ell\log\AlphabetSize}{\sqrt{\log m}},
        \frac{(\ell\log\AlphabetSize)^2}{\log m}
      \right\}
    \right)
  \end{align*}
  time and has a peak space usage of
  $\bigO(m\ell\log\AlphabetSize)$ bits.
\end{proposition}
\begin{proof}
  If $m<2^{16}$, construct the array directly, and hence the claim follows
  immediately. Thus, assume that $m\geq2^{16}$. Then
  $\AlphabetSize\leq m$ and $\ell\leq\log m<m$. The monotonicity of
  $x/\log x$ also gives
  \begin{equation}\label{eq:packed-wavelet-universe}
    \AlphabetSize^\ell\log m
      =\bigO(m\ell\log\AlphabetSize).
  \end{equation}

  For $p\in[0\dd\ell)$ and $t\in[1\dd\ell-p]$, let
  $W_{p,t}[1\dd m]\in([0\dd\AlphabetSize)^t)^m$ denote the sequence
  satisfying
  \begin{equation}\label{eq:packed-wavelet-strip}
    W_{p,t}
      =\bigodot_{X\in[0\dd\AlphabetSize)^p\text{ in colexicographic order}}
        \bigodot_{r=1}^{f_X}
          W[\PrefixSelect{W}{r}{X}][p+1\dd p+t].
  \end{equation}
  Let $\tau\in[1\dd\ell]$ be arbitrary. The algorithm partitions the levels into strips
  whose starting depths are the multiples of $\tau$ in $[0\dd\ell)$ and whose
  heights are $t_s=\min\{\tau,\ell-s\}$. It constructs the array in four
  steps.
  \begin{enumerate}

    \item \emph{Construct the character width, prefix frequencies, and the
      power array.} Compute $a:=\lceil\log\AlphabetSize\rceil$ by repeated
      doubling. Apply \cref{pr:packed-prefix-frequencies}, with parameters
      $\AlphabetSize$, $m$, and $\ell$, to the packed sequence representation
      $\PackedSeqRepresentation{w}{\AlphabetSize}{W}$, and retain its output as
      $A_{\rm freq}[0\dd2\AlphabetSize^\ell)$. Thus,
      \[
        A_{\rm freq}[\BasicInt{\AlphabetSize}{X}]
          =\PrefixRank{W}{m}{X}
          \qquad
          \text{for every }X\in[0\dd\AlphabetSize)^{\leq\ell}.
      \]
      Construct $A_{\rm pow}[0\dd\ell]$ by setting
      $A_{\rm pow}[0]:=1$ and
      $A_{\rm pow}[p]:=\AlphabetSize\cdot A_{\rm pow}[p-1]$ for every
      $p\in[1\dd\ell]$.
      Computing $a$ takes $\bigO(\log\AlphabetSize)$ time. Thus, this step
      takes $\bigO(m\ell\log\AlphabetSize/\log m)$ time and uses
      $\bigO(m\ell\log\AlphabetSize)$ bits.

    \item\label{step:packed-wavelet-tree-initialize-output}
      \emph{Initialize the output array.} Initialize every entry of
      $\mathcal C[0\dd2\AlphabetSize^\ell)$ with the empty packed
      representation.
      This takes $\bigO(\AlphabetSize^\ell)$ time and
      $\bigO(\AlphabetSize^\ell\log m)$ bits. Consequently, the entry for
      every zero-frequency prefix already stores the required representation.

    \item \emph{Construct the packed sequences at the strip starts.} Use the
      following two substeps.
      \begin{enumerate}
        \item \emph{Construct the sequence for the first strip.} Apply the
          partition algorithm from
          \cref{pr:packed-pointwise-concatenation-partition} to
          $\PackedSeqRepresentation{w}{\AlphabetSize}{W}$, with alphabet size
          $\AlphabetSize$, sequence length $m$, and component lengths $t_0$
          and $\ell-t_0$. This application is valid because
          $\AlphabetSize\leq m$ and $1\leq\ell<m$. By
          \cref{eq:packed-wavelet-strip}, its first output is
          $\PackedSeqRepresentation{w}{\AlphabetSize}{W_{0,t_0}}$.
          Retain this representation until
          Step~\ref{step:packed-wavelet-tree-construct-node-strings} processes
          the first strip.
        \item \emph{Construct the sequences for the remaining strips.} For
          every $i\in[1\dd\lceil\ell/\tau\rceil)$ in increasing order, set
          $s:=i\tau$ and $t:=t_s$. Let
          $P_s\in([0\dd\AlphabetSize)^s)^m$,
          $S_s\in([0\dd\AlphabetSize)^{\ell-s})^m$, and
          $U_s\in([0\dd\AlphabetSize)^t)^m$ denote the sequences satisfying
          $P_s[j]=W[j][1\dd s]$, $S_s[j]=W[j][s+1\dd\ell]$, and
          $U_s[j]=W[j][s+1\dd s+t]$ for every $j\in[1\dd m]$. Apply the
          partition algorithm, with alphabet size $\AlphabetSize$ and sequence
          length $m$, to the packed sequence representation
          $\PackedSeqRepresentation{w}{\AlphabetSize}{W}$, with
          component lengths $s$ and $\ell-s$, to construct the packed sequence
          representations of $P_s$ and $S_s$. Apply it with the same alphabet
          size and sequence length to
          $\PackedSeqRepresentation{w}{\AlphabetSize}{S_s}$, with component
          lengths $t$ and $\ell-s-t$. Its first output is
          $\PackedSeqRepresentation{w}{\AlphabetSize}{U_s}$. Both applications
          are valid because $\AlphabetSize\leq m$, $1\leq s<\ell<m$, and
          $1\leq t\leq\ell-s$.
          Let $P_s^{\rm rev}\in([0\dd\AlphabetSize)^s)^m$ and
          $K_s\in[0\dd\AlphabetSize^s)^m$ satisfy
          $P_s^{\rm rev}[j]=\revstr{P_s[j]}$ and
          $K_s[j]=\Val{\AlphabetSize}{P_s^{\rm rev}[j]}$ for every
          $j\in[1\dd m]$. Apply
          \cref{pr:packed-fixed-length-string-reversal} to
          $\PackedSeqRepresentation{w}{\AlphabetSize}{P_s}$, with alphabet
          size $\AlphabetSize$, sequence length $m$, and string length $s$, to
          construct
          $\PackedSeqRepresentation{w}{\AlphabetSize}{P_s^{\rm rev}}$. This
          application is valid because $\AlphabetSize\leq m$ and
          $1\leq s<m$. Apply
          \cref{pr:packed-character-block-encoding} to
          $\PackedSeqRepresentation{w}{\AlphabetSize}{P_s^{\rm rev}}$, with
          alphabet size $\AlphabetSize$, input length $m$, block length $s$,
          and $g=h=1$. The application is valid because
          $\AlphabetSize^s\leq\AlphabetSize^\ell\leq m$. Its output strings
          have length one and their characters form $K_s$, so its output is
          also $\PackedRepresentation{w}{\AlphabetSize^s}{K_s}$.
          Apply the forward algorithm from
          \cref{cor:packed-stable-string-grouping} to
          $\PackedRepresentation{w}{\AlphabetSize^s}{K_s}$ and
          $\PackedSeqRepresentation{w}{\AlphabetSize}{U_s}$, with key alphabet
          size $\AlphabetSize^s$, payload alphabet size $\AlphabetSize$,
          payload length $t$, and exactly $m$ positions. Its hypotheses hold
          because $\AlphabetSize^s\leq\AlphabetSize^\ell\leq m$ and
          $\AlphabetSize^t\leq\AlphabetSize^\ell\leq m$. Numerical order of
          the values $K_s[j]$ agrees with colexicographic order of the prefixes
          $W[j][1\dd s]$, and stability preserves increasing order of $j$
          within every prefix class. Hence, the output is exactly
          $\PackedSeqRepresentation{w}{\AlphabetSize}{W_{s,t}}$.
          Since $\ell\log\AlphabetSize\leq\log m$, constructing this
          representation for one nonzero strip start takes $\bigO(m)$ time.
          Retain the representation of $W_{s,t_s}$ until
          Step~\ref{step:packed-wavelet-tree-construct-node-strings} processes
          the strip.
      \end{enumerate}
      The strip heights sum to $\ell$, so the total length of the stored
      strings $W_{s,t_s}[j]$ for each position is $\ell$.
      There are $\lceil\ell/\tau\rceil-1$ nonzero strip starts. Together with
      the initialization in
      Step~\ref{step:packed-wavelet-tree-initialize-output}, constructing all
      strip starts takes
      \begin{equation}\label{eq:packed-wavelet-strip-starts}
        \bigO\left(
          m\left(\left\lceil\frac{\ell}{\tau}\right\rceil-1\right)
          +\frac{m\ell\log\AlphabetSize}{\log m}
          +\AlphabetSize^\ell
        \right)
      \end{equation}
      time.

    \item\label{step:packed-wavelet-tree-construct-node-strings}
      \emph{Construct and store the node strings.} Process the strips
      separately. For every $i\in[0\dd\lceil\ell/\tau\rceil)$ in increasing
      order, set $s:=i\tau$, process $p\in[s\dd s+t_s)$ in increasing order,
      set $t:=s+t_s-p$, and execute the following substeps.
      \begin{enumerate}

        \item At $p=s$, retrieve the stored packed sequence representation of
          $W_{s,t_s}$. At every later depth, use the packed sequence
          representation of $W_{p,t}$ returned at the preceding depth. Let
          $D_p[1\dd m]\in[0\dd\AlphabetSize)^m$ denote the string satisfying
          $D_p[j]=W_{p,t}[j][1]$ for every $j\in[1\dd m]$. If $t=1$, then
          \cref{def:packed-sequence-representation} identifies
          $\PackedSeqRepresentation{w}{\AlphabetSize}{W_{p,1}}$ with
          $\PackedRepresentation{w}{\AlphabetSize}{D_p}$. If $t>1$, let
          $R_{p,t}[1\dd m]\in([0\dd\AlphabetSize)^{t-1})^m$ denote the
          sequence satisfying
          $R_{p,t}[j]=W_{p,t}[j][2\dd t]$ for every $j\in[1\dd m]$. Apply the
          partition algorithm from
          \cref{pr:packed-pointwise-concatenation-partition} to
          $\PackedSeqRepresentation{w}{\AlphabetSize}{W_{p,t}}$, with alphabet
          size $\AlphabetSize$, sequence length $m$, and component lengths $1$
          and $t-1$. The application is valid because
          $\AlphabetSize\leq m$ and $2\leq t\leq\ell<m$. Its outputs are
          $\PackedRepresentation{w}{\AlphabetSize}{D_p}$ and
          $\PackedSeqRepresentation{w}{\AlphabetSize}{R_{p,t}}$.

        \item Allocate $A_{\rm len}[1\dd\AlphabetSize^p]$ and
          $A_{\rm index}[1\dd\AlphabetSize^p]$, set $X:=0^p$ and $v:=0$, and
          scan $u\in[0\dd\AlphabetSize^p)$ in increasing order
          while maintaining $u=\Val{\AlphabetSize}{\revstr X}$ and
          $v=\Val{\AlphabetSize}{X}$. Between consecutive iterations,
          increment $X[1]$, propagate carries toward $X[p]$, and update $v$
          using $A_{\rm pow}$. Retrieve
          $f:=A_{\rm freq}[A_{\rm pow}[p]+v]$, set $A_{\rm len}[u+1]:=f$, and
          set
          $A_{\rm index}[u+1]:=\BasicInt{\AlphabetSize}{X}=A_{\rm pow}[p]+v$.
          The lengths are nonnegative and sum to $m$. Apply the split
          operation from
          \cref{pr:packed-representation}\eqref{pr:packed-representation-split},
          with alphabet size
          $\AlphabetSize$, length parameter $m$, and $A_{\rm len}$, to $D_p$.
          Store its $j$th returned packed string in
          $\mathcal C[A_{\rm index}[j]]$ for every
          $j\in[1\dd\AlphabetSize^p]$. The application is valid because
          $\AlphabetSize^p\leq\AlphabetSize^\ell\leq m$,
          $\AlphabetSize\leq m$, and $w>\log m$. By
          \cref{eq:packed-wavelet-strip}, these strings are precisely the
          strings $C_X$ at depth $p$. The split operation returns packed
          representations with the zero padding required by
          \cref{def:packed-representation}.
          The storage for $A_{\rm len}$ and $A_{\rm index}$ is reused from one
          depth to the next.

        \item If $t>1$, apply the forward algorithm from
          \cref{cor:packed-stable-string-grouping} to
          $\PackedRepresentation{w}{\AlphabetSize}{D_p}$ and
          $\PackedSeqRepresentation{w}{\AlphabetSize}{R_{p,t}}$, with key and
          payload alphabet sizes $\AlphabetSize$, payload length $t-1$, and
          exactly $m$ positions. Its hypotheses hold because
          $\AlphabetSize\leq m$, $t-1\geq1$, and
          $\AlphabetSize^{t-1}\leq\AlphabetSize^\ell\leq m$. For
          $Y\in[0\dd\AlphabetSize)^p$ and
          $\alpha\in[0\dd\AlphabetSize)$, the identity
          $\revstr{Y\alpha}=\alpha\revstr Y$ shows that stable grouping by
          $D_p$ places the prefix classes $Y\alpha$ in colexicographic order.
          Stability preserves the original relative order within every class.
          The returned packed sequence representation is therefore
          $\PackedSeqRepresentation{w}{\AlphabetSize}{W_{p+1,t-1}}$, which
          supplies the next depth of the strip.
      \end{enumerate}
      At a depth with remaining height $t>1$, pointwise partition takes
      $\bigO(mt\log\AlphabetSize/\log m)$ time by
      \cref{pr:packed-pointwise-concatenation-partition}. Applying
      \cref{cor:packed-stable-string-grouping} takes
      $\bigO(mt(\log\AlphabetSize)^2/\log m)$ time. For a strip of height
      $t_0$, the sum of $t$ over the depths at which they are applied is at
      most $t_0(t_0-1)$. Since all strip heights sum to $\ell$ and are at most
      $\tau$, all such applications inside the strips take
      \begin{equation}\label{eq:packed-wavelet-strip-routing}
        \bigO\left(
          \frac{m\ell(\tau-1)(\log\AlphabetSize)^2}{\log m}
        \right)
      \end{equation}
      time. The scan at depth $p$ performs $\bigO(\AlphabetSize^p)$ character
      changes. Across all depths, constructing the index and length arrays and
      applying the split operation takes
      $\bigO(\AlphabetSize^\ell+m\ell\log\AlphabetSize/\log m)$ time.
  \end{enumerate}

  Combining the four steps and using
  \cref{eq:packed-wavelet-universe,eq:packed-wavelet-strip-starts,%
  eq:packed-wavelet-strip-routing} gives, for every
  $\tau\in[1\dd\ell]$,
  \begin{equation}\label{eq:packed-wavelet-strip-total}
    \bigO\left(
      m\left(\left\lceil\frac{\ell}{\tau}\right\rceil-1\right)
      +\frac{m\ell\log\AlphabetSize}{\log m}
      +\frac{m\ell(\tau-1)(\log\AlphabetSize)^2}{\log m}
    \right)
  \end{equation}
  time. Setting $\tau=1$ yields $\bigO(m\ell)$ time, and setting
  $\tau=\ell$ yields
  $\bigO(m(\ell\log\AlphabetSize)^2/\log m)$ time. It remains to obtain the
  bound $\bigO(m\ell\log\AlphabetSize/\sqrt{\log m})$ when
  $\log\AlphabetSize\leq\sqrt{\log m}<\ell\log\AlphabetSize$, since outside
  this range one of the preceding two bounds is no larger. In this case, let
  $\tau$ denote the value
  \[
    \left\lceil
      \frac{\sqrt{\log m}}{\log\AlphabetSize}
    \right\rceil\in[1\dd\ell].
  \]
  It holds
  $\lceil\ell/\tau\rceil-1<\ell/\tau$ and
  $\tau-1<\sqrt{\log m}/\log\AlphabetSize$. Substitution into
  \cref{eq:packed-wavelet-strip-total} gives
  $\bigO(m\ell\log\AlphabetSize/\sqrt{\log m})$ time. Choosing the best of
  the three values of $\tau$ proves the claimed time bound.

  The total length of the node strings at each depth is $m$. By
  \cref{def:packed-representation}, their packed representations, including
  all unused bits and word-alignment overhead, use
  $\bigO(m\ell\log\AlphabetSize+\AlphabetSize^\ell\log m)$ bits. The entries
  of $\mathcal C$, $A_{\rm freq}$, and the arrays
  $A_{\rm len}$ and $A_{\rm index}$ for one depth $p$ use
  $\bigO(\AlphabetSize^\ell\log m)$ bits, which is
  $\bigO(m\ell\log\AlphabetSize)$ by
  \cref{eq:packed-wavelet-universe}. The input, $A_{\rm pow}$, and all stored
  packed sequence representations of $W_{s,t_s}$ use
  $\bigO(m\ell\log\AlphabetSize)$ bits because the strip heights sum to
  $\ell$. For one nonzero strip start or one depth $p$, the representation of
  $W_{p,t}$ and, when $t>1$, the representation of $W_{p+1,t-1}$, the sequences
  $P_s,P_s^{\rm rev},S_s,U_s$, and $R_{p,t}$, the strings $K_s$ and $D_p$,
  and the working space used by one application of
  \cref{pr:packed-pointwise-concatenation-partition,%
  pr:packed-fixed-length-string-reversal,%
  pr:packed-character-block-encoding,cor:packed-stable-string-grouping}
  together use $\bigO(m\ell\log\AlphabetSize)$ bits. Processing the strip
  starts and depths sequentially proves the claimed bound on peak space usage.
\end{proof}

\subsection{Generalized Wavelet-Tree Construction}\label{sec:generalized-wavelet-tree-construction}

\begin{proposition}[Generalized wavelet-tree construction]\label{pr:generalized-wavelet-tree-construction}
  Let $\AlphabetSize,m\in\Z_{\geq2}$ and $\ell\in\Z_{\geq1}$ satisfy
  $\AlphabetSize^\ell\leq m$. Consider the word RAM model with word size
  $w=c\log m$, where $c\geq2$ is a constant. Let
  $W[1\dd m]\in([0\dd\AlphabetSize)^\ell)^m$ be given by
  $\PackedSeqRepresentation{w}{\AlphabetSize}{W}$
  (\cref{def:packed-sequence-representation}). For every
  $X\in[0\dd\AlphabetSize)^{\leq\ell}$, let $f_X$ denote
  $\PrefixRank{W}{m}{X}$. For every
  $p\in[0\dd\ell)$, $d\in[1\dd\ell-p]$, and
  $X\in[0\dd\AlphabetSize)^p$, let
  $W_{X,d}[1\dd f_X]\in([0\dd\AlphabetSize)^d)^{f_X}$ denote the sequence
  satisfying
  \[
    W_{X,d}[r]
      =W[\PrefixSelect{W}{r}{X}][p+1\dd p+d]
      \qquad\text{for every }r\in[1\dd f_X],
  \]
  using the notation from \cref{def:prefix-range-queries}.

  There exists a data structure that, given $p$ and $d$ as above, returns an
  array $\mathcal W_{p,d}[0\dd\AlphabetSize^p)$ of packed sequence
  representations such that
  \[
    \mathcal W_{p,d}[\Val{\AlphabetSize}{X}]
      =\PackedSeqRepresentation{w}{\AlphabetSize}{W_{X,d}}
      \qquad\text{for every }X\in[0\dd\AlphabetSize)^p,
  \]
  where the array is indexed according to \cref{def:val}. The data structure
  has the following complexities:
  \begin{itemize}
  \item space usage $\bigO(m\ell\log\AlphabetSize)$ bits,
  \item deterministic construction time
    \[
      \bigO\left(
        m\min\left\{
          \ell,
          \frac{\ell\log\AlphabetSize}{\sqrt{\log m}},
          \frac{(\ell\log\AlphabetSize)^2}{\log m}
        \right\}
      \right),
    \]
  \item peak space usage during construction
    $\bigO(m\ell\log\AlphabetSize)$ bits,
  \item query time
    \[
      \bigO\left(
        \AlphabetSize^p+
        \min\left\{
          m,
          \frac{m(d\log\AlphabetSize)^2}{\log m}
        \right\}
      \right),
    \]
  \item additional query space usage
    $\bigO(\AlphabetSize^p\log m+md\log\AlphabetSize)$ bits, including the
    returned array and packed sequence representations.
  \end{itemize}
\end{proposition}
\begin{proof}
  Let $a$ denote $\lceil\log\AlphabetSize\rceil$. It holds
  $\ell a\leq2\ell\log\AlphabetSize\leq2\log m\leq w$. The monotonicity of
  $x/\log x$ on $[4\dd\infty)$ implies
  $\AlphabetSize^\ell\log m
    =\bigO(m\ell\log\AlphabetSize)$ when $\AlphabetSize^\ell\geq4$. When
  $\AlphabetSize^\ell<4$, the same bound follows from $\log m=\bigO(m)$.
  Consequently,
  \begin{equation}\label{eq:generalized-wavelet-tree-universe}
    \AlphabetSize^\ell\log m
      =\bigO(m\ell\log\AlphabetSize)
    \qquad\text{and}\qquad
    \AlphabetSize^\ell
      =\bigO\left(\frac{m\ell\log\AlphabetSize}{\log m}\right).
  \end{equation}

  For every $p\in[0\dd\ell)$ and $d\in[1\dd\ell-p]$, let
  $W_{p,d}[1\dd m]\in([0\dd\AlphabetSize)^d)^m$ denote the concatenation
  of the sequences $W_{X,d}$ in colexicographic order of $X$, that is, in
  increasing order of
  $\Val{\AlphabetSize}{\revstr X}$. Thus,
  \[
    W_{p,d}
      =\bigodot_{X\in[0\dd\AlphabetSize)^p\text{ in increasing order of }
        \Val{\AlphabetSize}{\revstr X}}W_{X,d}.
  \]
  For every $p\in[0\dd\ell)$, let
  $D_p[1\dd m]\in[0\dd\AlphabetSize)^m$ denote the string satisfying
  $D_p[j]=W_{p,1}[j][1]$ for every $j\in[1\dd m]$.

  \DSComponents
  The data structure consists of the following components:
  \begin{enumerate}

  \item The packed sequence representation
    $\PackedSeqRepresentation{w}{\AlphabetSize}{W}$, the value $a$, and the
    array $A_{\rm pow}[0\dd\ell]$, where
    $A_{\rm pow}[t]=\AlphabetSize^t$ for every $t\in[0\dd\ell]$. These objects use
    $\bigO(m\ell\log\AlphabetSize)$ bits.

  \item We also store the data structure from
    \cref{pr:val-encoding} with parameters $m$ and $\AlphabetSize$. It uses
    $\bigO(\sqrt m\log m)$ bits, which is
    $\bigO(m\ell\log\AlphabetSize)$ bits.

  \item The prefix-frequency array
    $A_{\rm pfreq}[0\dd2\AlphabetSize^\ell)$. It satisfies
    $A_{\rm pfreq}[\BasicInt{\AlphabetSize}{X}]=f_X$ for every
    $X\in[0\dd\AlphabetSize)^{\leq\ell}$. All remaining entries are zero.
    By \cref{eq:generalized-wavelet-tree-universe}, it uses
    $\bigO(\AlphabetSize^\ell\log m)
      =\bigO(m\ell\log\AlphabetSize)$ bits.

  \item An array $A_D[0\dd\ell)$ and the packed representations
    $\PackedRepresentation{w}{\AlphabetSize}{D_p}$ for every
    $p\in[0\dd\ell)$. The entry $A_D[p]$ stores a pointer to the
    corresponding packed representation. The
    $\ell$ packed strings contain $m\ell$ characters altogether. Including
    their word-alignment overhead and the pointer array, this component uses
    $\bigO(m\ell\log\AlphabetSize)$ bits.
  \end{enumerate}

  In total, the data structure uses
  $\bigO(m\ell\log\AlphabetSize)$ bits.

  \DSQueries
  Given any $p\in[0\dd\ell)$ and any $d\in[1\dd\ell-p]$, we construct and
  return an array
  $\mathcal W_{p,d}[0\dd\AlphabetSize^p)$ satisfying
  \[
    \mathcal W_{p,d}[\Val{\AlphabetSize}{X}]
      =\PackedSeqRepresentation{w}{\AlphabetSize}{W_{X,d}}
    \qquad
    \text{for every }X\in[0\dd\AlphabetSize)^p
  \]
  as follows:
  \begin{enumerate}

  \item Initialize $\mathcal W_{p,d}[0\dd\AlphabetSize^p)$ with empty packed
    sequence representations. Compute whether
    $d^2a^2\geq\log m$. This step takes $\bigO(\AlphabetSize^p)$ time and
    uses $\bigO(\AlphabetSize^p\log m)$ bits.

  \item\label{step:generalized-wavelet-tree-query-direct}
    Suppose first that $d^2a^2\geq\log m$. For every
    $x\in[0\dd\AlphabetSize^p)$, retrieve
    $f:=A_{\rm pfreq}[A_{\rm pow}[p]+x]$. If $f>0$, allocate space for the
    packed sequence representation of $f$ strings of length $d$ in
    $\mathcal W_{p,d}[x]$. For every such entry, maintain the next output
    position and a zero-initialized word containing the symbols not yet
    written to an output word. Each output word has
    $\lfloor w/a\rfloor$ symbol fields in its lowest bits, and the remaining
    high bits stay zero. For every
    $j\in[1\dd m]$, let $X$ and $Y$ denote
    $W[j][1\dd p]$ and $W[j][p+1\dd p+d]$, respectively. Apply the substring
    operation from
    \cref{pr:packed-representation}\eqref{pr:packed-representation-substring}
    to the character concatenation
    represented by the stored $\PackedSeqRepresentation{w}{\AlphabetSize}{W}$,
    whose length is $m\ell$. Use length parameter $m\ell$, starting position
    $(j-1)\ell+1$, and substring length $p$ to extract the packed
    representation of $X$. Apply the same operation with starting position
    $(j-1)\ell+p+1$ and substring length $d$ to extract the packed
    representation of $Y$. If $p=0$,
    set $x:=0$. Otherwise, use \cref{pr:val-encoding} to compute
    $x:=\Val{\AlphabetSize}{X}$. This invocation is valid because
    $p<\ell\leq\lfloor\log_{\AlphabetSize}m\rfloor$. Append $Y$ to the
    representation stored in $\mathcal W_{p,d}[x]$ using shifts and masks.
    Whenever all $\lfloor w/a\rfloor$ symbol fields of a word have been
    filled, write it and initialize the next word to zero. After the scan, for
    every entry whose pending word contains at least one filled symbol field,
    write that word. Since each word is initialized with zeros and only its
    symbol fields are updated, all bits not used by a symbol remain zero.
    Return $\mathcal W_{p,d}$.
    The allocation takes
    $\bigO(\AlphabetSize^p+md\log\AlphabetSize/\log m)$ time. The substring
    invocations are valid because
    $\AlphabetSize\leq m\leq m\ell<m^2$ and
    $w\geq2\log m>\log(m\ell)$. Every string $W[j]$ occupies at most
    $\ell a\leq w$ bits, so the two extractions and appending $Y$ to its
    partially filled output word take constant time per value of $j$. Since
    $d\log\AlphabetSize\leq\log m$, this step takes
    $\bigO(\AlphabetSize^p+m)$ time. For every $X$, the
    strings appended to $\mathcal W_{p,d}[\Val{\AlphabetSize}{X}]$ occur in
    increasing order of their positions in $W$. This is precisely the order
    in $W_{X,d}$.

  \item\label{step:generalized-wavelet-tree-query-reconstruct}
    It remains to consider $d^2a^2<\log m$. We reconstruct the packed
    sequence representation of $W_{p,d}$ as follows.
    \begin{enumerate}
      \item Retrieve $\PackedRepresentation{w}{\AlphabetSize}{D_{p+d-1}}$
        using $A_D[p+d-1]$. This entry exists because $p+d-1<\ell$. By
        the definition of $D_{p+d-1}$ and
        \cref{def:packed-sequence-representation}, it is also
        $\PackedSeqRepresentation{w}{\AlphabetSize}{W_{p+d-1,1}}$.
      \item If $d>1$, process $q=p+d-2,p+d-3,\ldots,p$ in decreasing order
        and let $t:=p+d-q$. When $q=p+d-2$, use the representation of
        $W_{p+d-1,1}$ retrieved above as the representation of
        $W_{q+1,t-1}$. Every later iteration uses the representation of
        $W_{q+1,t-1}$ constructed in the iteration for $q+1$. Let
        $R_{q,t}\in([0\dd\AlphabetSize)^{t-1})^m$ denote the sequence satisfying
        $R_{q,t}[j]=W_{q,t}[j][2\dd t]$ for every $j\in[1\dd m]$.
        Retrieve $\PackedRepresentation{w}{\AlphabetSize}{D_q}$ using
        $A_D[q]$. Apply the inverse algorithm from
        \cref{cor:packed-stable-string-grouping} to
        $\PackedRepresentation{w}{\AlphabetSize}{D_q}$ and
        $\PackedSeqRepresentation{w}{\AlphabetSize}{W_{q+1,t-1}}$, with key
        and payload alphabet sizes $\AlphabetSize$, payload length $t-1$, and
        exactly $m$ positions. Its hypotheses hold because
        $\AlphabetSize\leq m$, $t-1\geq1$, and
        $\AlphabetSize^{t-1}\leq\AlphabetSize^\ell\leq m$.
        To identify the output, consider a length-$q$ prefix $X$ and a
        character $\gamma$. The definitions give
        $D_q[j]=W_{q,t}[j][1]$ for every $j\in[1\dd m]$. Thus, stable
        grouping of $R_{q,t}$ according to $D_q$ orders the rows first by
        $\gamma$ and then by colexicographic order of $X$, while preserving
        original order within every prefix class. The identity
        $\revstr{X\gamma}=\gamma\revstr X$ shows that this is precisely the
        order of the length-$(q+1)$ prefix classes. Hence, the stable grouping
        of $R_{q,t}$ according to $D_q$ is $W_{q+1,t-1}$, and the inverse
        algorithm returns
        $\PackedSeqRepresentation{w}{\AlphabetSize}{R_{q,t}}$.
        The representation of $D_q$ is also
        $\PackedSeqRepresentation{w}{\AlphabetSize}{W_{q,1}}$. Apply the
        concatenation algorithm from
        \cref{pr:packed-pointwise-concatenation-partition} to
        $\PackedSeqRepresentation{w}{\AlphabetSize}{W_{q,1}}$ and
        $\PackedSeqRepresentation{w}{\AlphabetSize}{R_{q,t}}$, with alphabet
        size $\AlphabetSize$, sequence length $m$, and component lengths $1$
        and $t-1$. The application is valid because $\AlphabetSize\leq m$ and
        $2\leq t\leq d\leq\ell<m$, and its output is
        $\PackedSeqRepresentation{w}{\AlphabetSize}{W_{q,t}}$. For $q>p$,
        this is the representation required by the iteration for $q-1$.
    \end{enumerate}
    When $d>1$, the iteration for $q=p$ constructs the representation of
    $W_{p,d}$. When $d=1$, the loop is empty and the representation retrieved
    in the first substep is already that of $W_{p,1}$. At remaining length
    $t$, inverse grouping takes
    $\bigO(mt(\log\AlphabetSize)^2/\log m)$ time, while pointwise concatenation
    takes $\bigO(1+mt\log\AlphabetSize/\log m)$ time. The time for pointwise
    concatenation is bounded by the time for inverse grouping. Summing over
    $t\in[2\dd d]$ gives
    $\bigO(md^2(\log\AlphabetSize)^2/\log m)$ time.

  \item\label{step:generalized-wavelet-tree-query-split}
    Allocate temporary arrays $A_{\rm len}[1\dd\AlphabetSize^p]$ and
    $A_{\rm index}[1\dd\AlphabetSize^p]$. Scan
    $u\in[0\dd\AlphabetSize^p)$ in increasing order while maintaining the
    packed representation of the length-$p$ string $X$ satisfying
    $u=\Val{\AlphabetSize}{\revstr X}$. At the start of the scan, set
    $X:=0^p$, which is the empty string when $p=0$. Between consecutive
    iterations when $p>0$, increment $X[1]$ and propagate carries toward
    $X[p]$. Use \cref{pr:val-encoding} to compute
    $x:=\Val{\AlphabetSize}{X}$. This invocation is valid because
    $p<\ell\leq\lfloor\log_{\AlphabetSize}m\rfloor$. Retrieve
    $f:=A_{\rm pfreq}[A_{\rm pow}[p]+x]$, set
    $A_{\rm len}[u+1]:=fd$, and set $A_{\rm index}[u+1]:=x$. Apply the
    split operation from
    \cref{pr:packed-representation}\eqref{pr:packed-representation-split},
    with length parameter $m\ell$ and supplied string length $md$, to the
    character concatenation represented by
    $\PackedSeqRepresentation{w}{\AlphabetSize}{W_{p,d}}$ and the lengths
    $A_{\rm len}[1\dd\AlphabetSize^p]$. The $i$th returned packed string
    contains $A_{\rm len}[i]$ characters, comprising $A_{\rm len}[i]/d$
    strings of length $d$ for one prefix class, and is therefore its packed
    sequence representation. For every $i\in[1\dd\AlphabetSize^p]$, store the
    $i$th returned representation in
    $\mathcal W_{p,d}[A_{\rm index}[i]]$. Return $\mathcal W_{p,d}$.
    Maintaining $X$ throughout the scan requires
    $\bigO(\AlphabetSize^p)$ character updates. The lengths supplied to the
    split operation are nonnegative and sum to $md$. The invocation is valid
    because $\AlphabetSize^p\leq\AlphabetSize^\ell\leq m\leq m\ell$,
    $md\leq m\ell<m^2$, and $w\geq2\log m>\log(m\ell)$. Thus, this step takes
    $\bigO(\AlphabetSize^p+md\log\AlphabetSize/\log m)$ time. By the
    definition of $W_{p,d}$, its pieces occur in colexicographic order. For
    the $i$th prefix class $X$ in this order,
    $A_{\rm index}[i]=\Val{\AlphabetSize}{X}$, so its piece is stored in the
    entry stated in the claim. Since $d\log\AlphabetSize\geq1$, the second
    term in the time bound is bounded by
    $\bigO(md^2(\log\AlphabetSize)^2/\log m)$.
  \end{enumerate}

  We use Step~\ref{step:generalized-wavelet-tree-query-direct} when
  $m\leq md^2a^2/\log m$ and
  Steps~\ref{step:generalized-wavelet-tree-query-reconstruct}
  and~\ref{step:generalized-wavelet-tree-query-split} otherwise. Since
  $a=\Theta(\log\AlphabetSize)$, their bounds give the claimed query time.
  In Step~\ref{step:generalized-wavelet-tree-query-reconstruct}, the
  representations of $W_{q+1,t-1}$, $R_{q,t}$, $W_{q,1}$, and
  $W_{q,t}$ and the peak space used by one inverse grouping or pointwise
  concatenation application together occupy
  $\bigO(md\log\AlphabetSize)$ bits. In
  Step~\ref{step:generalized-wavelet-tree-query-direct}, the returned representations
  contain $md$ characters, and their word-alignment overhead and partially
  filled output words occupy
  $\bigO(\AlphabetSize^p\log m+md\log\AlphabetSize)$ bits. In
  Step~\ref{step:generalized-wavelet-tree-query-split}, the
  returned representations, the temporary arrays, and the peak space used by
  the split application satisfy the same bound. Thus, the additional space
  used by every query is
  $\bigO(\AlphabetSize^p\log m+md\log\AlphabetSize)$ bits, including its output.

  \DSConstruction
  We construct the four components in four steps.
  \begin{enumerate}

  \item \emph{Store the input and construct the power array.} Retain
    $\PackedSeqRepresentation{w}{\AlphabetSize}{W}$. Compute $a$ by repeated
    doubling. For every $t\in[0\dd\ell]$, compute
    $A_{\rm pow}[t]=\AlphabetSize^t$ by repeated multiplication. This step
    takes $\bigO(\ell\log m)$ time. The stored input and the power array occupy
    $\bigO(m\ell\log\AlphabetSize)$ bits.

  \item \emph{Construct the data structure for computing
    $\Val{\AlphabetSize}{X}$.} Apply
    \cref{pr:val-encoding} with parameters $m$ and $\AlphabetSize$. The
    invocation is valid because $\AlphabetSize\leq m$ and
    $w\geq1+\log m$. Moreover,
    $\ell\leq\lfloor\log_{\AlphabetSize}m\rfloor$, so the structure supports
    every string used below. It takes $\bigO(\sqrt m)$ time. The resulting
    data structure occupies
    $\bigO(\sqrt m\log m)$ bits.

    \item \emph{Construct the prefix-frequency array.} Apply
    \cref{pr:packed-prefix-frequencies}, with parameters $\AlphabetSize$, $m$,
    and $\ell$, to the packed sequence representation
    $\PackedSeqRepresentation{w}{\AlphabetSize}{W}$ and retain
    its output as $A_{\rm pfreq}$. The invocation is valid because
    $\AlphabetSize^\ell\leq m$ and $w=c\log m$ with $c\geq2$.
    This takes $\bigO(m\ell\log\AlphabetSize/\log m)$ time using
    $\bigO(m\ell\log\AlphabetSize)$ bits of working space.

  \item\label{step:generalized-wavelet-tree-construction-node-strings}
    \emph{Construct the packed strings $D_p$.} Apply
    \cref{pr:packed-wavelet-tree-construction}, with parameters
    $\AlphabetSize$, $m$, and $\ell$, to the packed sequence representation
    $\PackedSeqRepresentation{w}{\AlphabetSize}{W}$ and store its returned
    array as $A_{\rm node}$ until this step is complete. The invocation is
    valid because
    $\AlphabetSize^\ell\leq m$ and $w=c\log m$ with $c\geq2$. Allocate
    $A_D[0\dd\ell)$ and proceed as follows. For every
    $p\in[0\dd\ell)$, allocate temporary arrays
    $A_{\rm ptr}[1\dd A_{\rm pow}[p]]$ and
    $A_{\rm len}[1\dd A_{\rm pow}[p]]$, and scan
    $u\in[0\dd A_{\rm pow}[p])$ in increasing order while maintaining the
    packed representation of the length-$p$ string $X$ satisfying
    $u=\Val{\AlphabetSize}{\revstr X}$. At the start of the scan, set
    $X:=0^p$, which is the empty string when $p=0$. Between consecutive
    iterations when $p>0$, increment $X[1]$ and propagate carries toward
    $X[p]$. Use \cref{pr:val-encoding} to compute
    $\Val{\AlphabetSize}{X}$, set
    $x:=A_{\rm pow}[p]+\Val{\AlphabetSize}{X}
      =\BasicInt{\AlphabetSize}{X}$, and retrieve
    $f:=A_{\rm pfreq}[x]$. Store a pointer to the packed string in
    $A_{\rm node}[x]$ at $A_{\rm ptr}[u+1]$ and set $A_{\rm len}[u+1]:=f$.
    After the scan, apply the concatenation operation from
    \cref{pr:packed-representation}\eqref{pr:packed-representation-concat},
    with length parameter $m$, to
    the packed strings addressed by $A_{\rm ptr}[1\dd A_{\rm pow}[p]]$ and the
    lengths in $A_{\rm len}[1\dd A_{\rm pow}[p]]$. Store a pointer to the
    resulting $\PackedRepresentation{w}{\AlphabetSize}{D_p}$ in $A_D[p]$.
    The enumeration lists the length-$p$ prefixes in colexicographic order.
    The packed string in $A_{\rm node}[x]$ contains the next characters of
    the strings having prefix $X$, in their original order. Hence, the
    resulting packed string is $D_p$. The supplied lengths are
    nonnegative and sum to $m$. Moreover,
    $A_{\rm pow}[p]=\AlphabetSize^p\leq\AlphabetSize^\ell\leq m$ and
    $\AlphabetSize\leq m$, while
    $w\geq2\log m>\log m$, so every concatenation invocation is valid.
    Across all depths, maintaining the counters requires
    $\bigO(\AlphabetSize^\ell)$ character updates in packed strings. The
    concatenations and scans take
    $\bigO(\AlphabetSize^\ell+m\ell\log\AlphabetSize/\log m)$ time, which is
    $\bigO(m\ell\log\AlphabetSize/\log m)$ by
    \cref{eq:generalized-wavelet-tree-universe}. Thus, this step takes the
    time of applying \cref{pr:packed-wavelet-tree-construction} plus this
    quantity. It has a peak space usage of
    $\bigO(m\ell\log\AlphabetSize)$ bits and retains the array $A_D$ and its
    packed strings.
  \end{enumerate}

  Applying \cref{pr:packed-wavelet-tree-construction} takes the construction
  time in the claim. Every other step takes
  \[
    \bigO\left(
      \frac{m\ell\log\AlphabetSize}{\log m}
      +\sqrt m+\ell\log m
    \right)
  \]
  time. The quantity $m\ell\log\AlphabetSize/\log m$ is bounded by each of the
  three expressions in the
  minimum in the claim because
  $\log\AlphabetSize\leq\log m$ and
  $\ell\log\AlphabetSize\geq1$. The quantity
  $\sqrt m+\ell\log m$ satisfies the same bounds
  because $\ell\geq1$, $\log\AlphabetSize\geq1$, and
  $(\log m)^2=\bigO(m)$. Thus, the complete construction has the claimed time
  bound.
  The four stored components together use
  $\bigO(m\ell\log\AlphabetSize)$ bits, as shown above. The array
  $A_{\rm node}$ used in
  Step~\ref{step:generalized-wavelet-tree-construction-node-strings} satisfies the same bound by
  \cref{pr:packed-wavelet-tree-construction}. In any one construction step,
  all temporary arrays and packed strings also use this many bits in total.
  Since the steps are executed sequentially, this proves the bound on peak
  space usage during construction.
\end{proof}

\subsection{Prefix-Path Payload Gathering}\label{sec:prefix-path-payload-gathering}

\begin{proposition}[Packed prefix-path payload gathering]
  \label{pr:packed-prefix-path-payload-gathering}
  Let $\AlphabetSize,m\in\Z_{\geq2}$ and $\ell\in\Z_{\geq1}$ satisfy
  $\AlphabetSize^\ell\leq m$. Consider the word RAM model with word size
  $w=c\log m$, where $c\geq2$ is a constant. Let $k\in\Z_{\geq2}$ satisfy
  $\lceil\log k\rceil\leq\sqrt{\log m}$, and let
  $W[1\dd m]\in([0\dd\AlphabetSize)^\ell)^m$.
  Using the notation from \cref{def:prefix-range-queries}, for every nonempty
  $X\in[0\dd\AlphabetSize)^{\leq\ell}$, let $f_X$ denote
  $\PrefixRank{W}{m}{X}$, and suppose that
  $P_X[1\dd f_X]\in[0\dd k)^{f_X}$. Suppose that we are given an array
  $A_{\rm pay}[0\dd2\AlphabetSize^\ell)$ whose entry at
  $\BasicInt{\AlphabetSize}{X}$ stores a pointer to
  $\PackedRepresentation{w}{k}{P_X}$ whenever $f_X>0$
  (\cref{def:basic-int,def:packed-representation}).
  All remaining entries are null.
  Let $Q[1\dd m]\in([0\dd k)^\ell)^m$ denote the sequence satisfying
  \begin{align*}
    Q[j][p]
      =P_{W[j][1\dd p]}[\PrefixSpecialRank{W}{j}{p}]
      \qquad
      \text{for every }j\in[1\dd m]\text{ and }p\in[1\dd\ell].
  \end{align*}
  There is a deterministic algorithm that constructs
  $\PackedSeqRepresentation{w}{k}{Q}$ from
  $\PackedSeqRepresentation{w}{\AlphabetSize}{W}$
  (\cref{def:packed-sequence-representation}) and $A_{\rm pay}$. The
  algorithm takes
  \begin{align*}
    \bigO\left(
      m\min\left\{
        \ell,
        \frac{\ell(\log\AlphabetSize+\log k)}{\sqrt{\log m}},
        \frac{\ell^2(\log\AlphabetSize+\log k)^2}{\log m}
      \right\}
    \right)
  \end{align*}
  time and has a peak space usage of
  $\bigO(m\ell(\log\AlphabetSize+\log k))$ bits.
\end{proposition}
\begin{proof}
  If $m<2^{16}$, construct $\PackedSeqRepresentation{w}{k}{Q}$ directly, and
  hence the claim follows immediately. Thus, assume that $m\geq2^{16}$. Compute
  \[
    d:=\lceil\log k\rceil,
    \qquad
    z:=\lceil\log\AlphabetSize\rceil,
    \qquad
    h:=\left\lfloor\frac{\sqrt{\log m}}{d}\right\rfloor,
    \qquad
    q:=\left\lceil\frac{\ell}{h}\right\rceil.
  \]
  The assumption on $k$ gives $h\geq1$. For every $s\in[0\dd q)$, let
  \[
    a_s=sh,
    \qquad
    b_s=\min\{(s+1)h,\ell\},
    \qquad
    h_s=b_s-a_s.
  \]
  Let $Q_s[1\dd m]\in([0\dd k)^{h_s})^m$ denote the sequence satisfying
  $Q_s[j]=Q[j][a_s+1\dd b_s]$ for every $j\in[1\dd m]$.

  For every $p\in[0\dd\ell]$, let $\pi_p$ denote the permutation of
  $[1\dd m]$ that orders $j$ by increasing
  $\Val{\AlphabetSize}{\revstr{W[j][1\dd p]}}$, breaking ties by increasing
  $j$. For every $p\in[1\dd\ell]$, let
  $P_p[1\dd m]\in[0\dd k)^m$ denote the concatenation of the strings $P_X$
  for $X\in[0\dd\AlphabetSize)^p$ in colexicographic order. For every
  $r\in[1\dd m]$, let $j$ denote $\pi_p[r]$. Then
  \begin{equation}\label{eq:packed-path-level-payload}
    P_p[r]
      =P_{W[j][1\dd p]}[\PrefixSpecialRank{W}{j}{p}]
      =Q[j][p].
  \end{equation}
  The algorithm constructs $\PackedSeqRepresentation{w}{k}{Q}$ in three steps.
  \begin{enumerate}

    \item\label{step:packed-prefix-path-level-payload-strings} \emph{Construct
      the level and payload strings.} Proceed in three substeps.
      \begin{enumerate}

        \item Apply \cref{pr:packed-wavelet-tree-construction}, with parameters
          $\AlphabetSize$, $m$, and $\ell$, to the packed sequence representation
          $\PackedSeqRepresentation{w}{\AlphabetSize}{W}$ and retain its output
          as $A_{\rm node}$. Apply \cref{pr:packed-prefix-frequencies} with the
          same parameters and packed input, and retain its output as
          $A_{\rm freq}$. Construct $A_{\rm pow}[0\dd\ell]$ by setting
          $A_{\rm pow}[0]:=1$ and
          $A_{\rm pow}[p]:=\AlphabetSize\cdot A_{\rm pow}[p-1]$ for every
          $p\in[1\dd\ell]$.

        \item For every $p\in[0\dd\ell)$, let
          $D_p[1\dd m]\in[0\dd\AlphabetSize)^m$ denote the string satisfying
          \[
            D_p[r]=W[\pi_p[r]][p+1]
            \qquad
            \text{for every }r\in[1\dd m].
          \]
          To construct its packed representation, allocate
          $A_{\rm ptr}[1\dd\AlphabetSize^p]$ and
          $A_{\rm len}[1\dd\AlphabetSize^p]$, set $X:=0^p$ and $v:=0$, and
          scan $u\in[0\dd\AlphabetSize^p)$ in increasing order
          while maintaining $u=\Val{\AlphabetSize}{\revstr X}$ and
          $v=\Val{\AlphabetSize}{X}$. Between consecutive iterations,
          increment $X[1]$, propagate carries toward $X[p]$, and update $v$
          using $A_{\rm pow}$. Retrieve
          $f:=A_{\rm freq}[A_{\rm pow}[p]+v]$. Store in $A_{\rm ptr}[u+1]$ a
          pointer to the packed string stored in
          $A_{\rm node}[A_{\rm pow}[p]+v]$ and set $A_{\rm len}[u+1]:=f$.
          Apply the concatenation
          operation from
          \cref{pr:packed-representation}\eqref{pr:packed-representation-concat},
          with alphabet size
          $\AlphabetSize$, length parameter $m$, and the two arrays, to construct
          $\PackedRepresentation{w}{\AlphabetSize}{D_p}$.

        \item For every $p\in[1\dd\ell]$, allocate
          $A_{\rm ptr}[1\dd\AlphabetSize^p]$ and
          $A_{\rm len}[1\dd\AlphabetSize^p]$, and set $t:=0$, $X:=0^p$, and
          $v:=0$. Scan $u\in[0\dd\AlphabetSize^p)$ in increasing order while
          maintaining $u=\Val{\AlphabetSize}{\revstr X}$ and
          $v=\Val{\AlphabetSize}{X}$. Between consecutive iterations,
          increment $X[1]$, propagate carries toward $X[p]$, and update $v$
          using $A_{\rm pow}$. Retrieve
          $f:=A_{\rm freq}[A_{\rm pow}[p]+v]$. Whenever $f>0$, set $t:=t+1$,
          store in $A_{\rm ptr}[t]$ the pointer from
          $A_{\rm pay}[A_{\rm pow}[p]+v]$, and set $A_{\rm len}[t]:=f$. The
          first $t$ lengths are positive and sum to $m$. Apply the concatenation
          operation from
          \cref{pr:packed-representation}\eqref{pr:packed-representation-concat},
          with alphabet size
          $k$, length parameter $m$, and the first $t$ entries of the two
          arrays, to construct $\PackedRepresentation{w}{k}{P_p}$.
      \end{enumerate}
      All these concatenation applications are valid because
      $\AlphabetSize\leq m$, $k\leq2^d\leq2^{\sqrt{\log m}}\leq m$,
      and $w>\log m$. In every application, the supplied lengths are
      nonnegative and sum to $m$, and their number is at most
      $\AlphabetSize^p\leq\AlphabetSize^\ell\leq m$.
      Retain all packed representations of $D_p$ and $P_p$ until
      Step~\ref{step:packed-prefix-path-strip-sequences}.
      The arrays $A_{\rm ptr}$ and $A_{\rm len}$ are reused from one value of
      $p$ to the next.
      The scans list the prefixes in colexicographic order. Hence, the strings
      $D_p$ and $P_p$ have their stated meanings. Moreover, for
      $X\in[0\dd\AlphabetSize)^p$ and
      $\alpha\in[0\dd\AlphabetSize)$, the identity
      $\revstr{X\alpha}=\alpha\revstr X$ shows that stable grouping according
      to $D_p$ transforms the order $\pi_p$ into the order $\pi_{p+1}$.
      At depth $p$, each scan performs $\bigO(\AlphabetSize^p)$ character
      changes. Thus, all such scans perform $\bigO(\AlphabetSize^\ell)$
      character changes.
      By the monotonicity of $x/\log x$,
      \begin{equation}\label{eq:packed-path-universe}
        \AlphabetSize^\ell\log m
          =\bigO(m\ell\log\AlphabetSize).
      \end{equation}
      Thus, applying
      \cref{pr:packed-wavelet-tree-construction,pr:packed-prefix-frequencies},
      constructing the temporary arrays, and concatenating all strings takes
      the time from \cref{pr:packed-wavelet-tree-construction} plus
      \[
        \bigO\left(
          \frac{m\ell(\log\AlphabetSize+\log k)}{\log m}
        \right).
      \]
      This is bounded by each of the three bounds in the claim.

    \item\label{step:packed-prefix-path-strip-sequences} \emph{Construct $Q_s$
      for every strip.} Initialize $A_{\rm strip}[0\dd q)$ with null pointers.
      For every $s\in[0\dd q)$,
      let $a$ and $b$ denote $a_s$ and $b_s$, respectively, and execute the
      following substeps.
      \begin{enumerate}

        \item \emph{Route the payload strings to the strip boundary.} For every
          $p\in[a+1\dd b]$, let
          $R_p[1\dd m]\in([0\dd k)^{b-p+1})^m$ denote the sequence satisfying
          $R_p[r]=Q[j][p\dd b]$, where $j=\pi_p[r]$, for every
          $r\in[1\dd m]$. For the same values of $p$, let
          $S_{p-1}[1\dd m]\in([0\dd k)^{b-p+1})^m$ denote the sequence
          satisfying $S_{p-1}[r]=Q[j][p\dd b]$, where $j=\pi_{p-1}[r]$, for
          every $r\in[1\dd m]$. By
          \cref{eq:packed-path-level-payload,def:packed-sequence-representation},
          the stored $\PackedRepresentation{w}{k}{P_b}$ is also
          $\PackedSeqRepresentation{w}{k}{R_b}$.
          Process $p=b,b-1,\ldots,a+1$ in decreasing order.
          Step~\ref{step:packed-prefix-path-level-payload-strings} shows that
          $R_p$ is the stable grouping of $S_{p-1}$ according to $D_{p-1}$.
          Apply the inverse algorithm from
          \cref{cor:packed-stable-string-grouping} to
          $\PackedRepresentation{w}{\AlphabetSize}{D_{p-1}}$ and
          $\PackedSeqRepresentation{w}{k}{R_p}$, with key alphabet size
          $\AlphabetSize$, payload alphabet size $k$, payload length $b-p+1$,
          and exactly $m$ positions. Its hypotheses hold because
          $\AlphabetSize\leq m$, $b-p+1\geq1$, and
          $k^{b-p+1}\leq2^{h_sd}\leq2^{hd}
            \leq2^{\sqrt{\log m}}\leq m$.
          Its output is $\PackedSeqRepresentation{w}{k}{S_{p-1}}$. If
          $p>a+1$, then \cref{eq:packed-path-level-payload} identifies the
          stored $\PackedRepresentation{w}{k}{P_{p-1}}$ with the packed
          sequence representation of the length-one strings
          $Q[\pi_{p-1}[r]][p-1]$ for $r\in[1\dd m]$. Apply the concatenation
          algorithm from \cref{pr:packed-pointwise-concatenation-partition} to
          the stored $\PackedRepresentation{w}{k}{P_{p-1}}$, viewed as the
          packed sequence representation identified above, and
          $\PackedSeqRepresentation{w}{k}{S_{p-1}}$, with alphabet size $k$,
          sequence length $m$, and component lengths $1$ and $b-p+1$. Its
          output is $\PackedSeqRepresentation{w}{k}{R_{p-1}}$. The application
          is valid because $k\leq m$ and
          $2\leq b-p+2\leq h_s\leq\ell<m$. By induction, the final inverse
          grouping constructs $\PackedSeqRepresentation{w}{k}{S_a}$.

        \item \emph{Restore the original order.} If $a=0$, then $\pi_0$ is the
          identity permutation and $S_0=Q_s$. Store a pointer to its packed
          sequence representation in $A_{\rm strip}[s]$. If $a>0$, let
          $U_a[1\dd m]\in([0\dd\AlphabetSize)^a)^m$ denote the sequence
          satisfying $U_a[j]=W[j][1\dd a]$ for every $j\in[1\dd m]$, let
          $U_a^{\rm rev}[1\dd m]\in([0\dd\AlphabetSize)^a)^m$ denote the
          sequence satisfying $U_a^{\rm rev}[j]=\revstr{U_a[j]}$ for every
          $j\in[1\dd m]$, and let
          $K_a[1\dd m]\in[0\dd\AlphabetSize^a)^m$ denote the string satisfying
          \[
            K_a[j]
              =\Val{\AlphabetSize}{\revstr{W[j][1\dd a]}}
              \qquad
              \text{for every }j\in[1\dd m].
          \]
          Apply the partition operation from
          \cref{pr:packed-pointwise-concatenation-partition} to
          $\PackedSeqRepresentation{w}{\AlphabetSize}{W}$, with alphabet size
          $\AlphabetSize$, sequence length $m$, and component lengths $a$ and
          $\ell-a$. Retain its first output as
          $\PackedSeqRepresentation{w}{\AlphabetSize}{U_a}$ and discard the
          representation of the length-$(\ell-a)$ suffix sequence. The
          application is valid because $\AlphabetSize\leq m$ and
          $1\leq a<\ell<m$. Apply
          \cref{pr:packed-fixed-length-string-reversal} to the packed sequence
          representation
          $\PackedSeqRepresentation{w}{\AlphabetSize}{U_a}$, with alphabet size
          $\AlphabetSize$, sequence length $m$, and string length $a$, to construct
          $\PackedSeqRepresentation{w}{\AlphabetSize}{U_a^{\rm rev}}$. Next,
          apply \cref{pr:packed-character-block-encoding} to the packed sequence
          representation
          $\PackedSeqRepresentation{w}{\AlphabetSize}{U_a^{\rm rev}}$, with
          alphabet size $\AlphabetSize$, input length $m$, block length $a$,
          $g=1$, and output length $1$. The sole
          character of its $j$th output string is
          $\Val{\AlphabetSize}{U_a^{\rm rev}[j]}=K_a[j]$, so its returned packed
          sequence representation is also
          $\PackedRepresentation{w}{\AlphabetSize^a}{K_a}$. The reversal and
          character-block encoding applications are valid because
          $\AlphabetSize^a\leq\AlphabetSize^\ell\leq m$ and
          $1\leq a<\ell<m$. The pointwise partition takes
          $\bigO(1+m\ell\log\AlphabetSize/\log m)=\bigO(m)$ time, and the other
          two applications together take
          $\bigO(1+ma\log\AlphabetSize/\log m)=\bigO(m)$ time. Stable grouping
          $Q_s$ according to $K_a$ produces $S_a$. Apply the inverse algorithm
          from \cref{cor:packed-stable-string-grouping} to
          $\PackedRepresentation{w}{\AlphabetSize^a}{K_a}$ and
          $\PackedSeqRepresentation{w}{k}{S_a}$, with key alphabet size
          $\AlphabetSize^a$, payload alphabet size $k$, payload length $h_s$,
          and exactly $m$ positions. Its hypotheses hold because
          $\AlphabetSize^a\leq\AlphabetSize^\ell\leq m$ and
          $k^{h_s}\leq2^{h_sd}\leq2^{\sqrt{\log m}}\leq m$. Its output is
          $\PackedSeqRepresentation{w}{k}{Q_s}$. Store a pointer to this
          representation in $A_{\rm strip}[s]$.
      \end{enumerate}
      We next analyze this step. For a strip $s\in[0\dd q)$ and a relative
      height $r\in[1\dd h_s]$, one application of the inverse algorithm from
      \cref{cor:packed-stable-string-grouping} takes
      \[
        \bigO\left(
          \min\left\{m,
            \frac{m z(rd+z)}{\log m}
          \right\}
        \right)
      \]
      time. The pointwise concatenation following this application, when
      present, takes $\bigO(1+mrd/\log m)$ time. Using the inverse-grouping
      bound $\bigO(m)$ together with the pointwise-concatenation bound gives
      $\bigO(m\ell)$ time because $rd\leq h_sd\leq\sqrt{\log m}$.
      Using the inverse-grouping bound
      $\bigO(mz(rd+z)/\log m)$, the pointwise-concatenation bound, and
      $\sum_{s=0}^{q-1}h_s^2\leq\ell\min\{h,\ell\}$ gives
      \[
        \bigO\left(
          \frac{m\ell^2(z+d)^2}{\log m}
        \right)
      \]
      time.
      To prove the bound
      $\bigO(m\ell(z+d)/\sqrt{\log m})$, first suppose that
      $z+d\geq\sqrt{\log m}$. The $\bigO(m\ell)$ bound just obtained satisfies
      this bound. Suppose now that $z+d<\sqrt{\log m}$. Since
      $h\leq\sqrt{\log m}/d$, it holds
      $\sum_{s=0}^{q-1}h_s^2\leq\ell\sqrt{\log m}/d$. Summing the bounds for
      inverse grouping and pointwise concatenation gives
      \[
        \bigO\left(
          \frac{m}{\log m}\left(
            zd\sum_{s=0}^{q-1}h_s^2
            +z^2\ell
            +d\sum_{s=0}^{q-1}h_s^2
          \right)
        \right)
        =\bigO\left(
          \frac{m\ell(z+d)}{\sqrt{\log m}}
        \right).
      \]
      It also holds $h\geq\sqrt{\log m}/(2d)$. Constructing $U_a$,
      $U_a^{\rm rev}$, and $K_a$ and applying the inverse algorithm from
      \cref{cor:packed-stable-string-grouping}
      takes $\bigO(m)$ time at each nonzero strip
      boundary. There are fewer than $\ell/h$ such boundaries, so their total
      time is $\bigO(m\ell d/\sqrt{\log m})$, which is
      $\bigO(m\ell(z+d)/\sqrt{\log m})$.
      If $q=1$, there are no nonzero boundaries. If $q>1$, then
      $\ell d>\sqrt{\log m}$, and hence
      \begin{equation}\label{eq:packed-path-boundary-time}
        \frac{m\ell d}{\sqrt{\log m}}
          <\frac{m\ell^2d^2}{\log m}
          \leq\frac{m\ell^2(z+d)^2}{\log m}.
      \end{equation}
      Thus, the time for these operations at nonzero strip boundaries is also
      $\bigO(m\ell^2(z+d)^2/\log m)$. This proves that the step satisfies all
      three expressions in the minimum stated in the claim.

    \item\label{step:packed-prefix-path-concatenate-strips} \emph{Concatenate
      the strip sequences pointwise.} If $q=1$, then $Q_0=Q$, so return the
      packed sequence representation addressed by
      $A_{\rm strip}[0]$. Suppose that $q>1$. Place the representations of
      $Q_0,Q_1,\ldots,Q_{q-1}$, in this order, at the leaves of an ordered full
      binary tree with $q$ leaves and height $\lceil\log q\rceil$. Every node
      represents a consecutive interval of strips and stores their common row
      length. Process the tree bottom-up. Let $L$ and $R$ denote the sequences
      at two sibling nodes, and let
      $\ell_1$ and $\ell_2$ denote their row lengths. Apply the concatenation
      algorithm from \cref{pr:packed-pointwise-concatenation-partition} to
      $\PackedSeqRepresentation{w}{k}{L}$ and
      $\PackedSeqRepresentation{w}{k}{R}$, with alphabet size $k$, sequence
      length $m$, and component lengths $\ell_1$ and $\ell_2$. Store the
      returned representation at their parent. The application is valid
      because $k\leq m$ and
      $1\leq\ell_1+\ell_2\leq\ell<m$. Inductively, a node contains the
      pointwise concatenation of its strips in increasing order, so the
      representation stored at the root is $\PackedSeqRepresentation{w}{k}{Q}$.
      Return this representation. The tree has $q-1$ internal nodes and at most
      $\lceil\log q\rceil$ levels of applications at internal nodes. At every such
      level, the sum of the row lengths returned by its applications is at most
      $\ell$. Hence, all applications take
      \[
        \bigO\left(
          q+\frac{m\ell\log k\log q}{\log m}
        \right)
        =\bigO\left(\frac{m\ell d}{\sqrt{\log m}}\right)
      \]
      time, where the equality follows from
      $q\leq\ell$, $\log q\leq\log\ell\leq\log\log m\leq\sqrt{\log m}$,
      and $\sqrt{\log m}\leq m$. This is bounded by
      $\bigO(m\ell)$ and
      $\bigO(m\ell(z+d)/\sqrt{\log m})$ because
      $d\leq\sqrt{\log m}$ and $d\leq z+d$. Since $q>1$,
      \cref{eq:packed-path-boundary-time} also bounds it by
      $\bigO(m\ell^2(z+d)^2/\log m)$.
  \end{enumerate}

  All applications of the inverse algorithm are executed sequentially. At every
  depth, the node strings have total length $m$, as do the payload strings.
  Hence, $W$, the input packed strings addressed by $A_{\rm pay}$, the packed
  node strings stored in $A_{\rm node}$, all strings $P_p$ and $D_p$, all
  sequences $Q_s$, and the output contain
  $\bigO(m\ell(z+d))$ data bits. By
  \cref{def:packed-representation}, their packed representations, including
  all unused bits and word-alignment overhead, use
  $\bigO(m\ell(z+d)+\AlphabetSize^\ell\log m)$ bits. The entries of
  $A_{\rm pay}$, $A_{\rm node}$, and $A_{\rm freq}$, and the arrays
  $A_{\rm ptr}$ and $A_{\rm len}$ for the depth currently being processed use
  $\bigO(\AlphabetSize^\ell\log m)$ bits. This is
  $\bigO(m\ell\log\AlphabetSize)$ bits by
  \cref{eq:packed-path-universe}. During
  Step~\ref{step:packed-prefix-path-strip-sequences}, only a constant number of the
  sequences $R_p$, $S_{p-1}$, and $R_{p-1}$ for the current strip coexist.
  During Step~\ref{step:packed-prefix-path-concatenate-strips}, discard two
  child representations after constructing their
  parent. The representations and tree nodes present at any time occupy
  $\bigO(m\ell d+q\log m)=\bigO(m\ell d)$ bits, including word-alignment
  overhead, because $q\leq\ell$ and $\log m\leq m$. The arrays $A_{\rm pow}$
  and $A_{\rm strip}$ use $\bigO(m\ell d)$ bits. The sequences $U_a$ and
  $U_a^{\rm rev}$ and the string $K_a$ constructed for one nonzero strip
  boundary use $\bigO(m\ell z)$ bits. The peak space usage of every application
  of \cref{pr:packed-wavelet-tree-construction},
  \cref{pr:packed-prefix-frequencies},
  \cref{cor:packed-stable-string-grouping},
  \cref{pr:packed-pointwise-concatenation-partition},
  \cref{pr:packed-fixed-length-string-reversal}, and
  \cref{pr:packed-character-block-encoding} is
  $\bigO(m\ell(z+d))$ bits.
  Since $z=\Theta(\log\AlphabetSize)$ and $d=\Theta(\log k)$, the time and
  space bounds proved in the three steps complete the proof.
\end{proof}

\section{Prefix Select Queries}\label{sec:prefix-select}

\subsection{Problem Definition}\label{sec:prefix-select-problem-def}

\begin{framed}
  \noindent
  \probname{Indexing for Prefix Select Queries over Alphabet $[0 \dd \AlphabetSize)$}
  \begin{bfdescription}
  \item[Input:]
    Let $\AlphabetSize, m \in \Z_{\geq 2}$ and $\ell \in \Z_{\geq 1}$ be such that $\AlphabetSize^{\ell} \leq m$.
    Consider the word RAM model with word size $w$ such that $w \geq \log \AlphabetSize$.
    The input to the problem is the packed sequence representation
    $\PackedSeqRepresentation{w}{\AlphabetSize}{W}$
    (\cref{def:packed-sequence-representation}) of a sequence $W[1 \dd m]$ of $m$
    strings of length $\ell$ over alphabet $[0 \dd \AlphabetSize)$.
  \item[Output:]
    A data structure that, given length $|X|$ and the packed representation $\PackedRepresentation{w}{\AlphabetSize}{X}$
    of any string $X \in \IntegerAlphabet^{\leq \ell}$ (\cref{def:packed-representation}), and any $r \in [1 \dd m]$,
    returns $\PrefixSelect{W}{r}{X}$ (\cref{def:prefix-range-queries}),
    i.e., the $r$th smallest element of the set
    $\{j \in [1 \dd m] : X\text{ is a prefix of }W[j]\}$
    (if $r \leq \PrefixRank{W}{m}{X}$) or $\infty$ (otherwise).
  \end{bfdescription}
\end{framed}

\subsection{Basic Constant-Time Structure}\label{sec:basic-constant-time-prefix-select}

\begin{proposition}\label{pr:large-space-prefix-select-baseline}
  Let $\AlphabetSize,m\in\Z_{\geq2}$ and $\ell\in\Z_{\geq1}$ be such that
  $\AlphabetSize^\ell\leq m$. Consider the word RAM model with word size
  $w = c\log m$, where $c\geq2$ is a constant.
  There exists a data structure for the problem of indexing for prefix
  select queries over alphabet $[0 \dd \AlphabetSize)$ (see \cref{sec:prefix-select-problem-def})
  that achieves the following complexities:
  \begin{itemize}
  \item space usage $\bigO(m \ell^2 \log \AlphabetSize)$ bits,
  \item preprocessing time
    $\bigO(m \min(\ell, \tfrac{(\ell\log\AlphabetSize)^2}{\log m}))$,
  \item preprocessing space $\bigO(m \ell^2 \log \AlphabetSize)$ bits,
  \item query time $\bigO(1)$.
  \end{itemize}
\end{proposition}
\begin{proof}

  If $m<64$, construct and store all answers directly, and hence the claim
  follows immediately. Thus, assume that $m\geq64$.

  Denote $a:=\lceil\log\AlphabetSize\rceil$, $M:=\AlphabetSize^\ell$,
  $b:=\lceil\log M\rceil$, and $n_b:=\lceil m/M\rceil$. Let
  $A_{\rm pow}[0\dd\ell]$ denote the array
  satisfying $A_{\rm pow}[d]=\AlphabetSize^d$ for every $d\in[0\dd\ell]$.

  Let $W_1,W_2,\ldots,W_{n_b}$ denote the partition of $W$ into consecutive
  blocks, each of length $M$ except possibly the last one. For every
  $X\in[0\dd\AlphabetSize)^{\leq\ell}$ and $t\in[1\dd n_b]$, denote
  $f_{X,t}:=\PrefixRank{W_t}{|W_t|}{X}$
  (\cref{def:prefix-range-queries}). For every
  $X\in[0\dd\AlphabetSize)^{\leq\ell}$, denote
  $F_X:=\sum_{t=1}^{n_b}f_{X,t}$. Thus,
  $F_X=\PrefixRank{W}{m}{X}$. Let
  $A_{\rm freq}[0\dd2\AlphabetSize^{\ell})$ denote the array satisfying
  $A_{\rm freq}[\BasicInt{\AlphabetSize}{X}]=F_X$ (\cref{def:basic-int})
  for every such $X$. All remaining entries of $A_{\rm freq}$ are zero.

  For every $X\in[0\dd\AlphabetSize)^{\leq\ell}$, denote
  $s_X:=\AlphabetSize^{\ell-|X|}$, and let $B_X$ denote the bitvector
  satisfying
  \[
    B_X=
      \one^{f_{X,1}}\zero^{s_X}
      \one^{f_{X,2}}\zero^{s_X}\cdots
      \one^{f_{X,n_b}}\zero^{s_X}
  \]
  and let $P_X$ denote the string of length $F_X$ over alphabet
  $[0\dd M)$ satisfying
  $P_X[r]=(\PrefixSelect{W}{r}{X}-1)\bmod M$ for every
  $r\in[1\dd F_X]$.

  \DSComponents
  The data structure consists of the following components:
  \begin{enumerate}

    \item The value $a$ and the data structure from
      \cref{pr:basic-int-encoding} with parameters $m$ and $\AlphabetSize$.
      They use $\bigO(\sqrt m\log m)$ bits.

    \item The array $A_{\rm pow}[0\dd\ell]$. It uses
      $\bigO(\ell\log m)$ bits.

    \item The array $A_{\rm freq}[0\dd2\AlphabetSize^{\ell})$.
      Every entry uses $\bigO(\log m)$ bits, so this array uses
      $\bigO(\AlphabetSize^{\ell}\log m)$ bits. To bound this quantity, observe
      that the function $x/\log x$ is increasing for $x\geq3$. If
      $\AlphabetSize^\ell\geq3$, it follows from
      $\AlphabetSize^{\ell}\leq m$ that
      $\AlphabetSize^{\ell}/\log(\AlphabetSize^{\ell})=\bigO(m/\log m)$, or,
      equivalently,
      $\AlphabetSize^{\ell}\log m
      =\bigO(m\log(\AlphabetSize^{\ell}))
      =\bigO(m\ell\log\AlphabetSize)$. If $\AlphabetSize^\ell=2$, the same
      bound follows directly from $\AlphabetSize=2$ and $\ell=1$. Thus, the
      final space bound for
      $A_{\rm freq}$ is $\bigO(m\ell\log\AlphabetSize)$ bits.

    \item The data structures from \cref{th:bin-rank-select} for the bitvectors
      $B_X$, for every $X\in[0\dd\AlphabetSize)^{\leq\ell}$, together with an
      array $A_{\rm sel}[0\dd2\AlphabetSize^\ell)$. Its entry at
      $\BasicInt{\AlphabetSize}{X}$ stores a pointer to the data structure for
      $B_X$. All remaining entries of $A_{\rm sel}$ are zero. Let
      $N_B:=\sum_{X\in[0\dd\AlphabetSize)^{\leq\ell}}|B_X|$.
      Every string in $W$ has exactly one prefix of each length in
      $[0\dd\ell]$, and hence $\sum_{X \in [0 \dd \AlphabetSize)^{\leq \ell}} F_X = m(\ell+1)$.
      Moreover,
      \begin{align*}
        N_B
          &=  \sum_{X\in[0\dd\AlphabetSize)^{\leq\ell}} F_X
            + n_b \sum_{p=0}^{\ell}\AlphabetSize^{p} \AlphabetSize^{\ell-p}
          =  m(\ell+1) + n_b(\ell+1)M
          =  (m+n_bM)(\ell+1)
          =  \Theta(m\ell).
      \end{align*}
      The returned data structures use $\bigO(N_B)=\bigO(m\ell)$ bits in
      total. The array
      $A_{\rm sel}[0 \dd 2\AlphabetSize^{\ell})$
      needs $\bigO(\AlphabetSize^{\ell}\log m)
      =\bigO(m\ell\log\AlphabetSize)$ bits.
      Thus, this component uses
      $\bigO(m\ell \log\AlphabetSize)$ bits in total.

    \item The value $b$, the packed representations
      $\PackedRepresentation{w}{M}{P_X}$ for every
      $X\in[0\dd\AlphabetSize)^{\leq\ell}$, and an array
      $A_{\rm pos}[0\dd2\AlphabetSize^\ell)$. Its entry at
      $\BasicInt{\AlphabetSize}{X}$ stores a pointer to the packed
      representation of $P_X$. All remaining entries of $A_{\rm pos}$ are
      zero. Since $|P_{X}| = F_{X}$,
      the total length of $P_{X}$ over all $X \in [0 \dd \AlphabetSize)^{\leq \ell}$ is
      $\sum_{X \in [0 \dd \AlphabetSize)^{\leq \ell}} F_{X} = m(\ell+1)$. Thus, by
      \cref{def:packed-representation}, the total space of the packed representations is
      $\bigO(\sum_{X \in [0 \dd \AlphabetSize)^{\leq \ell}} (1+(|P_{X}|\ell \log \AlphabetSize)/\log m))
      = \bigO(\AlphabetSize^{\ell} + m\ell^2\log\AlphabetSize/\log m)$ words, i.e.,
      $\bigO(\AlphabetSize^{\ell} \log m + m\ell^2\log\AlphabetSize) = \bigO(m \ell^2 \log \AlphabetSize)$ bits.
      The array adds only
      $\bigO(\AlphabetSize^\ell\log m)
      \subseteq \bigO(m\ell\log\AlphabetSize)$ bits.
      Thus, this component uses
      $\bigO(m\ell^2\log\AlphabetSize)$ bits in total.
  \end{enumerate}

  In total, the data structure uses $\bigO(m\ell^2\log\AlphabetSize)$ bits.

  \DSQueries
  Let $X\in[0\dd\AlphabetSize)^{\leq\ell}$ and $r\in[1\dd m]$. Given the
  length $|X|$, the packed representation
  $\PackedRepresentation{w}{\AlphabetSize}{X}$ (\cref{def:packed-representation})
  of the string $X$, and the position $r$, we compute
  $\PrefixSelect{W}{r}{X}$ as follows:
  \begin{enumerate}

  \item Using \cref{pr:basic-int-encoding}, compute
    $x:=\BasicInt{\AlphabetSize}{X}$. This step takes $\bigO(1)$ time.

  \item If $r>A_{\rm freq}[x]$, in $\bigO(1)$ time return $\infty$ and terminate the query.
    Henceforth, let us assume that $r \leq A_{\rm freq}[x]$.

  \item Using the data structure from \cref{th:bin-rank-select} constructed for the bitvector $B_{X}$
    (and accessed via the pointer in $A_{\rm sel}[x]$),
    in $\bigO(1)$ time compute $z:=\Select{B_X}{r}{1}$.

  \item In $\bigO(1)$ time retrieve the pointer to the packed representation
    $\PackedRepresentation{w}{M}{P_X}$ using $A_{\rm pos}[x]$ and its length
    $F_X=A_{\rm freq}[x]$. Using
    \cref{pr:packed-representation}\eqref{pr:packed-representation-access},
    in $\bigO(1)$ time compute $j':=P_X[r]$.

  \item In $\bigO(1)$ time compute
    $s:=A_{\rm pow}[\ell-|X|]=\AlphabetSize^{\ell-|X|}$ and set
    $j:=M((z-r)/s)+j'+1$. Let $W_t$ denote the block containing the $r$th
    string in $W$ that has $X$ as a prefix. The $r$th one in $B_X$ belongs to the
    run $\one^{f_{X,t}}$. Exactly $t-1$ zero runs precede this run, and each
    has length $s$. Therefore, $z-r=(t-1)s$, so $(z-r)/s=t-1$ is an integer
    and is the zero-based index of $W_t$. By the definition of $P_X$, the
    value $j'$ is the zero-based position of the same string inside $W_t$.
    Consequently, $j=\PrefixSelect{W}{r}{X}$, and we return $j$ as the answer.
    This step takes $\bigO(1)$ time.
  \end{enumerate}

  In total, the query takes $\bigO(1)$ time.

  \DSConstruction
  Given the packed sequence representation
  $\PackedSeqRepresentation{w}{\AlphabetSize}{W}$ of the sequence $W$, we
  construct the data structure as follows.
  \begin{enumerate}

  \item\label{large-space-baseline-construction-step-1}
    \emph{Construct the character width and string-encoding structure}:
    Compute $a=\lceil\log\AlphabetSize\rceil$ by repeated
    doubling, and apply \cref{pr:basic-int-encoding} with parameters $m$ and
    $\AlphabetSize$. This step takes
    $\bigO(\log\AlphabetSize+\sqrt m)=\bigO(\sqrt m)$ time, has peak space usage
    $\bigO(\sqrt m\log m)$ bits, and the value $a$ and the constructed data
    structure occupy $\bigO(\sqrt m\log m)$ bits.

  \item\label{large-space-baseline-construction-step-2}
    \emph{Construct $A_{\rm pow}$}: Set $A_{\rm pow}[0]:=1$ and
    $A_{\rm pow}[d]:=\AlphabetSize\cdot A_{\rm pow}[d-1]$ for every
    $d\in[1\dd\ell]$. This step takes $\bigO(\ell)$ time, uses
    $\bigO(1)$ words of temporary space, has peak space usage
    $\bigO(\ell\log m)$ bits, and the constructed array $A_{\rm pow}$ occupies
    $\bigO(\ell\log m)$ bits.

  \item\label{large-space-baseline-construction-step-3}
    \emph{Construct $A_{\rm freq}$}: Apply
    \cref{pr:packed-prefix-frequencies}, with parameters $\AlphabetSize$, $m$,
    and $\ell$, to the packed sequence representation
    $\PackedSeqRepresentation{w}{\AlphabetSize}{W}$, using
    $A_{\rm freq}[0\dd2\AlphabetSize^{\ell})$ as its output array. This
    application of \cref{pr:packed-prefix-frequencies} initializes all entries
    of $A_{\rm freq}$ to zero and sets
    $A_{\rm freq}[\BasicInt{\AlphabetSize}{X}]=F_X$ for every
    $X\in[0\dd\AlphabetSize)^{\leq\ell}$. This step takes
    $\bigO(m\ell\log\AlphabetSize/\log m)$ time, has peak space usage
    $\bigO(m\ell\log\AlphabetSize)$ bits, and the constructed array
    $A_{\rm freq}$ occupies
    $\bigO(\AlphabetSize^\ell\log m)
      =\bigO(m\ell\log\AlphabetSize)$ bits.

  \item\label{large-space-baseline-construction-step-4}
    \emph{Construct the bitvectors $B_X$ and augment them with
      $\bigO(1)$-time select support}: Use the faster of the following two
    alternatives.
    \begin{enumerate}

    \item \emph{Naive construction}: The algorithm
      running in $\bigO(m\ell)$ time proceeds as follows.
      \begin{enumerate}

      \item \emph{Initialization}:
        Initialize $A_{\rm sel}[0\dd2\AlphabetSize^{\ell})$ to zero. Also
        initialize the temporary arrays
        $A_{\rm bit}[0\dd2\AlphabetSize^{\ell})$ and
        $A_{\rm count}[0\dd2\AlphabetSize^{\ell})$ to zero.
        For every
        $k \in [0\dd\ell]$ and every
        $x \in [A_{\rm pow}[k]\dd2A_{\rm pow}[k])$, use
        \cref{pr:packed-representation}\eqref{pr:packed-representation-initialize}
        to construct the packed representation of an all-zero bitvector of
        length $A_{\rm freq}[x]+n_bA_{\rm pow}[\ell-k]$, store its pointer in
        $A_{\rm bit}[x]$, and set $A_{\rm count}[x]:=1$.
        In this application of \cref{pr:packed-representation}, we use the
        parameter $\Textlen=6m(\ell+1)$, and we use the same value in every
        subsequent application of that proposition. This bounds both the
        length and the alphabet size of every packed string used in these
        applications. Since
        $\ell\leq\log m$, we have
        $w\geq2\log m>\log(6m(\ell+1))$ for all $m\geq64$.
        Initializing the three arrays and all packed bitvectors takes
        $\bigO(\AlphabetSize^\ell+N_B/\log m)=\bigO(m\ell)$ time. The packed
        bitvectors and the three arrays use
        $\bigO(N_B+\AlphabetSize^\ell\log m)
        =\bigO(m\ell\log\AlphabetSize)$ bits. This step has peak space
        usage $\bigO(m\ell\log\AlphabetSize)$ bits. The packed bitvectors and
        the arrays $A_{\rm bit}$, $A_{\rm count}$, and $A_{\rm sel}$, which
        are needed in the subsequent substeps, occupy
        $\bigO(m\ell\log\AlphabetSize)$ bits.

      \item \emph{Compute the bitvectors $B_X$}: For every
        $t\in[1\dd n_b]$, execute the following steps.
        \begin{enumerate}

        \item For every
          $j=(t-1)M+1,\ldots,\min\{tM,m\}$ and every
          $k\in[0\dd\ell]$, use
          \cref{pr:packed-representation}\eqref{pr:packed-representation-substring}
          to compute the packed representation
          $\PackedRepresentation{w}{\AlphabetSize}{X}$ of the length-$k$
          prefix $X=W[j][1\dd k+1)$, starting at position $(j-1)\ell+1$ in
          the concatenation that represents $W$ (see
          \cref{def:packed-sequence-representation}). Using
          \cref{pr:basic-int-encoding}, in $\bigO(1)$ time compute
          $x:=\BasicInt{\AlphabetSize}{X}$. Then, set the bit at position
          $A_{\rm count}[x]$ in the bitvector pointed to by $A_{\rm bit}[x]$ to
          one using
          \cref{pr:packed-representation}\eqref{pr:packed-representation-update},
          and increment $A_{\rm count}[x]$.

        \item For every
          $k\in[0\dd\ell]$ and
          $x\in[A_{\rm pow}[k]\dd2A_{\rm pow}[k])$, set
          $A_{\rm count}[x]:=A_{\rm count}[x]+A_{\rm pow}[\ell-k]$.
        \end{enumerate}

        In total, the above computation takes $\bigO(m\ell + \AlphabetSize^{\ell} n_b) = \bigO(m\ell)$ time.
        It has peak space usage $\bigO(m\ell\log\AlphabetSize)$ bits. The
        packed bitvectors and the arrays $A_{\rm bit}$ and $A_{\rm sel}$,
        which are used in the next substep, occupy
        $\bigO(m\ell\log\AlphabetSize)$ bits.

      \item\label{large-space-baseline-construction-rank-select}
        \emph{Augment the bitvectors $B_X$ with select support}:
        Iterate over every $k\in[0\dd\ell]$ and every
        $x\in[A_{\rm pow}[k]\dd2A_{\rm pow}[k])$. Apply
        \cref{th:bin-rank-select} jointly to all these bitvectors, using
        $n:=3m(\ell+1)$ as an upper bound on their total length. This is a
        valid upper bound because $N_B=(m+n_bM)(\ell+1)<3m(\ell+1)$, and
        $n<m^2$ follows from $\ell\leq\log m$ and $m\geq64$. For every pair
        $(k,x)$, supply the length
        $A_{\rm freq}[x]+n_bA_{\rm pow}[\ell-k]$ and the packed representation
        pointed to by $A_{\rm bit}[x]$. Store the returned pointer to the data structure
        for the bitvector addressed by $x$ in $A_{\rm sel}[x]$.
        In total, this step takes
        $\bigO(\AlphabetSize^\ell+m\ell/\log m)=\bigO(m\ell)$ time, and the
        returned data structures use $\bigO(N_B)=\bigO(m\ell)$ bits.
        This step has peak space usage
        $\bigO(m\ell\log\AlphabetSize)$ bits. The select structures and
        $A_{\rm sel}$ that remain stored occupy
        $\bigO(m\ell\log\AlphabetSize)$ bits.
      \end{enumerate}

      In total, the direct construction takes $\bigO(m\ell)$ time and has peak
      space usage $\bigO(m\ell\log\AlphabetSize)$ bits. The constructed select
      structures and $A_{\rm sel}$ occupy
      $\bigO(m\ell\log\AlphabetSize)$ bits.

    \item\label{large-space-baseline-packed-construction}
      \emph{Packed construction}: The algorithm
      running in $\bigO(m(\ell\log\AlphabetSize)^2/\log m)$ time proceeds as
      follows.
      \begin{enumerate}

      \item \emph{Construct an auxiliary bitvector $B$}: Let $B$ denote the
        auxiliary bitvector
        $\bigodot_{t=1}^{n_b}(\one^{|W_t|}\zero^M)$ of length $m+n_bM$.
        In this step, we construct its packed representation
        $\PackedRepresentation{w}{2}{B}$
        (\cref{def:packed-representation}). Set
        $n_{\rm full}:=\floor{m/M}$ and
        $d:=m-n_{\rm full}M$, and proceed as follows.
        \begin{enumerate}

        \item Let $D$ denote the bitvector $\one^M\zero^M$. In the word RAM,
          the packed representation of a bitvector of the form
          $\one^k\zero^p$ can be constructed in
          $\bigO(1+(k+p)/w)\subseteq\bigO(1+(k+p)/\log m)$ time. Use this
          observation with $k=p=M$ to construct the packed representation
          $\PackedRepresentation{w}{2}{D}$.

        \item Let
          $B_{\rm full}$ denote
          $D^\infty[1\dd2n_{\rm full}M+1)=D^{n_{\rm full}}$. Apply
          \cref{pr:packed-string-repetition}, with
          $\Textlen=6m(\ell+1)$ and target length $2n_{\rm full}M$, to
          construct the packed representation
          $\PackedRepresentation{w}{2}{B_{\rm full}}$.
          We use the same value of $\Textlen$ in every subsequent application
          of
          \cref{pr:packed-representation,pr:packed-string-repetition} in the
          packed construction. Since $\ell\leq\log m$, we have
          $w\geq2\log m>\log(6m(\ell+1))$ for $m\geq64$.

        \item Let $B_{\rm last}$ denote the bitvector
          $\one^d\zero^{M\lceil d/M\rceil}$. Since $d\in[0\dd M)$, this
          bitvector is empty when $d=0$ and equals $\one^d\zero^M$
          otherwise. Apply the observation from the first substep with $k=d$
          and $p=M\lceil d/M\rceil$ to construct the packed representation
          $\PackedRepresentation{w}{2}{B_{\rm last}}$.

        \item Use
          \cref{pr:packed-representation}\eqref{pr:packed-representation-concat}
          to concatenate $B_{\rm full}$ and $B_{\rm last}$, thereby
          constructing the bitvector $B$ specified above.
        \end{enumerate}

        Constructing the packed representations of $D$ and $B_{\rm last}$
        takes $\bigO(1+M/\log m)$ time. By
        \cref{pr:packed-string-repetition}, constructing the packed
        representation of $B_{\rm full}$ takes
        $\bigO(1+\log n_{\rm full}+n_{\rm full}M/\log m)$ time. The final
        concatenation takes
        $\bigO(1+m/\log m)$ time. Since
        $\log n_{\rm full}\leq\log m=\bigO(m/\log m)$, this step takes
        $\bigO(m/\log m)$ time and uses $\bigO(m)$ bits of additional space.
        Together with $a$, the string-encoding structure, $A_{\rm pow}$, and
        $A_{\rm freq}$, this step has
        peak space usage $\bigO(m\ell\log\AlphabetSize)$ bits. The packed
        representation of $B$, which is used in the subsequent substeps,
        occupies $\bigO(m)$ bits.

      \item \emph{Construct an auxiliary sequence $W'$}: Let
        $\mathcal U[1\dd M]$ denote the sequence of all strings in
        $[0\dd\AlphabetSize)^\ell$ in colexicographic order, i.e., in
        lexicographic order of their reversals. Let $W'$ denote the auxiliary
        sequence
        $\bigodot_{t=1}^{n_b}(W_t\mathbin\odot\mathcal U)$. In this step, we
        construct the packed sequence representation
        $\PackedSeqRepresentation{w}{\AlphabetSize}{W'}$
        (\cref{def:packed-sequence-representation}). Compute its length
        $m':=|W'|=m+n_bM$. Proceed as follows.
        \begin{enumerate}

        \item Use
          \cref{pr:packed-representation}\eqref{pr:packed-representation-initialize}
          to initialize
          $\PackedSeqRepresentation{w}{\AlphabetSize}{\mathcal U}$ as the
          packed representation of an all-zero sequence of $M$ strings of
          length $\ell$. Then, for every $u\in[0\dd M)$ and every
          $i\in[1\dd\ell]$, use
          \cref{pr:packed-representation}\eqref{pr:packed-representation-update}
          at character position $u\ell+i$ to set
          $\mathcal U[u+1][i]:=
          \floor{u/A_{\rm pow}[i-1]}\bmod\AlphabetSize$. The reverse of
          $\mathcal U[u+1]$ is the length-$\ell$ base-$\AlphabetSize$
          representation of $u$, so the reversals of the strings in
          $\mathcal U$ are in lexicographic order. The construction performs
          $M\ell$ character updates. The bound for $A_{\rm freq}$ above gives
          $M\log m=\bigO(m\log M)
          =\bigO(m\ell\log\AlphabetSize)$. Since
          $\log\AlphabetSize\geq1$, it follows that
          $M\ell=\bigO(m(\ell\log\AlphabetSize)^2/\log m)$. This substep uses
          $\bigO(1)$ words of temporary space in addition to $\mathcal U$ and
          has peak space usage $\bigO(m\ell\log\AlphabetSize)$ bits.

        \item Let $\mathcal V$ denote $\mathcal U^{n_b}$. Starting with
          $\PackedSeqRepresentation{w}{\AlphabetSize}{\mathcal U}$, use
          \cref{pr:packed-string-repetition} with target length $n_bM\ell$ to
          construct $\PackedSeqRepresentation{w}{\AlphabetSize}{\mathcal V}$.
          This substep takes
          $\bigO(1+\log n_b+n_bM\ell\log\AlphabetSize/\log m)
          =\bigO(m\ell\log\AlphabetSize/\log m)$ time, where we use
          $n_bM<2m$ and
          $\log n_b\leq\log m=\bigO(m/\log m)$. It has peak space usage
          $\bigO(m\ell\log\AlphabetSize)$ bits.

        \item Let $G$ denote $\mathcal V\mathbin\odot W$. We use
          \cref{pr:packed-representation}\eqref{pr:packed-representation-concat}
          to construct
          $\PackedSeqRepresentation{w}{\AlphabetSize}{G}$ by concatenating the
          packed sequence representations of $\mathcal V$ and $W$. The
          hypothesis $\AlphabetSize^\ell\leq m$ of
          \cref{cor:packed-binary-string-ungrouping-comparable} holds. Moreover,
          $m'\in[2m\dd3m)$. Apply
          \cref{cor:packed-binary-string-ungrouping-comparable}, with
          reference length $m$ and input length $n=m'$, to the packed
          representations of $B$ and $G$. The result is
          $\PackedSeqRepresentation{w}{\AlphabetSize}{W'}$: the zero positions
          of $B$ correspond, in order, to the strings in $\mathcal V$, while
          its one positions correspond, in order, to the strings in $W$.
          Consequently,
          $\srted{W'}{B}=\mathcal V\mathbin\odot W=G$. Constructing $G$ and
          applying
          \cref{cor:packed-binary-string-ungrouping-comparable} takes
          $\bigO(m\ell\log\AlphabetSize/\log m)$ time.
        \end{enumerate}

        In total, this step takes
        $\bigO(m(\ell\log\AlphabetSize)^2/\log m)$ time, has peak space usage
        $\bigO(m\ell\log\AlphabetSize)$ bits. The packed representations of
        $B$ and $W'$, which are used in the next substep, occupy
        $\bigO(m\ell\log\AlphabetSize)$ bits.

      \item \emph{Construct the bitvectors $B_X$}: Initialize
        $A_{\rm bit}[0\dd2\AlphabetSize^{\ell})$ to zero. Apply
        \cref{pr:packed-stable-multilevel-routing-comparable}, with reference
        length $m$, input length $n=m'$, and payload width one, to the packed
        representations of $W'$ and $B$. For every
        $k\in[0\dd\ell]$, let
        $A^{\rm route}_{B,k}[0\dd\AlphabetSize^k)$ denote the returned
        level-$k$ array. For every
        $y\in[0\dd A_{\rm pow}[k])$, set
        $x:=A_{\rm pow}[k]+y$, and let $X\in[0\dd\AlphabetSize)^k$ denote the
        unique string satisfying $y=\Val{\AlphabetSize}{X}$. By the output
        guarantee of
        \cref{pr:packed-stable-multilevel-routing-comparable}, for every
        $t\in[1\dd n_b]$, the portion of
        $A^{\rm route}_{B,k}[y]$ corresponding to the consecutive factor
        $W_t\mathbin\odot\mathcal U$ of $W'$ is
        $\one^{f_{X,t}}\zero^{\AlphabetSize^{\ell-k}}
        =\one^{f_{X,t}}\zero^{s_X}$. Thus,
        $A^{\rm route}_{B,k}[y]=\PackedRepresentation{w}{2}{B_X}$. Moreover,
        $x=\BasicInt{\AlphabetSize}{X}$ by \cref{def:basic-int}. Store in
        $A_{\rm bit}[x]$ a pointer to $A^{\rm route}_{B,k}[y]$. After
        storing its pointers, discard the level-$k$ array but retain the
        packed bitvectors to which its entries point.
        Applying \cref{pr:packed-stable-multilevel-routing-comparable} takes
        $\bigO(m(\ell\log\AlphabetSize)^2/\log m)$ time and uses
        $\bigO(m\ell\log\AlphabetSize)$ bits of working space. Initializing
        $A_{\rm bit}$ and storing all pointers takes
        $\bigO(\AlphabetSize^\ell)$ additional time, which is bounded by
        $\bigO(m(\ell\log\AlphabetSize)^2/\log m)$ because
        $\AlphabetSize^\ell\log m
        =\bigO(m\ell\log\AlphabetSize)$. The returned bitvectors, including
        the unused bits in the final word of each nonempty separately stored
        representation, and $A_{\rm bit}$ use
        $\bigO(N_B+\AlphabetSize^\ell\log m)
        =\bigO(m\ell\log\AlphabetSize)$ bits. Thus, this step has peak space
        usage $\bigO(m\ell\log\AlphabetSize)$ bits. The packed bitvectors and
        $A_{\rm bit}$, which are used in the next substep, occupy
        $\bigO(m\ell\log\AlphabetSize)$ bits.

      \item \emph{Augment the bitvectors $B_X$ with support for select queries}:
        Initialize $A_{\rm sel}[0\dd2\AlphabetSize^{\ell})$ to
        zero. Iterate over every $k\in[0\dd\ell]$ and every
        $x\in[A_{\rm pow}[k]\dd2A_{\rm pow}[k])$. For each such pair,
        $A_{\rm bit}[x]$ points to one bitvector. Apply
        \cref{th:bin-rank-select} jointly to all these bitvectors, using
        $n:=3m(\ell+1)$ as an upper bound on their total length. For every pair
        $(k,x)$, supply the length
        $A_{\rm freq}[x]+n_bA_{\rm pow}[\ell-k]$ and the packed representation
        pointed to by $A_{\rm bit}[x]$, and store the returned pointer in
        $A_{\rm sel}[x]$. The bound on the total length follows from
        $N_B=(m+n_bM)(\ell+1)<3m(\ell+1)$, and $n<m^2$ follows from
        $\ell\leq\log m$ and $m\geq64$. This step takes
        $\bigO(\AlphabetSize^\ell+m\ell/\log m)$ time. By the monotonicity of
        $x/\log x$, we have
        $\AlphabetSize^\ell\log m
        =\bigO(m\ell\log\AlphabetSize)$, which shows that this time is
        $\bigO(m\ell\log\AlphabetSize/\log m)$, and hence is dominated by
        $\bigO(m(\ell\log\AlphabetSize)^2/\log m)$. This step has peak space
        usage $\bigO(m\ell\log\AlphabetSize)$ bits. The select structures and
        $A_{\rm sel}$ that remain stored occupy
        $\bigO(m\ell\log\AlphabetSize)$ bits.
      \end{enumerate}

      In total, the packed construction takes
      $\bigO(m(\ell\log\AlphabetSize)^2/\log m)$ time and has peak space
      usage $\bigO(m\ell\log\AlphabetSize)$ bits. The constructed select
      structures and $A_{\rm sel}$ occupy
      $\bigO(m\ell\log\AlphabetSize)$ bits.
    \end{enumerate}

    Taking the faster alternative, this step takes the minimum of
    $\bigO(m\ell)$ and
    $\bigO(m(\ell\log\AlphabetSize)^2/\log m)$ time. Its peak space usage is
    $\bigO(m\ell\log\AlphabetSize)$ bits. The resulting select structures and
    $A_{\rm sel}$ occupy $\bigO(m\ell\log\AlphabetSize)$ bits.

  \item\label{large-space-baseline-construction-step-5}
    \emph{Construct $b$, $A_{\rm pos}$, and the strings $P_X$}: Compute
    $b=\lceil\log M\rceil$ by repeated doubling and retain it. It holds
    $b=\Theta(\ell\log\AlphabetSize)$. Since $M\leq m$
    and $m\geq64$, we have
    $b\leq\log M+1\leq\log m+1\leq2\log m\leq w$.
    Computing $b$ takes $\bigO(\log M)$ time and one word of space, which are
    bounded by both alternatives below.
    Use the faster of the following two alternatives.
    \begin{enumerate}

    \item \emph{Naive construction}: This alternative proceeds as follows.
      \begin{enumerate}

      \item \emph{Compute the sizes and allocate the strings $P_X$}:
        Initialize $A_{\rm pos}$ to zero. Allocate the temporary
        array $A_{\rm count}[0\dd2\AlphabetSize^{\ell})$ and initialize with zeros.
        For every $k\in[0\dd\ell]$ and every
        $x\in[A_{\rm pow}[k]\dd2A_{\rm pow}[k])$, use
        \cref{pr:packed-representation}\eqref{pr:packed-representation-initialize}
        to construct the packed representation of the
        length-$A_{\rm freq}[x]$ all-zero string over alphabet $[0\dd M)$,
        store its pointer in $A_{\rm pos}[x]$, and set
        $A_{\rm count}[x]:=1$.
        The initialization takes
        $\bigO(\AlphabetSize^\ell+m(\ell+1)b/\log m)$ time. The quantity
        $\AlphabetSize^\ell$ is at most $m$, while
        $b=\bigO(\ell\log\AlphabetSize)
        \subseteq\bigO(\log m)$. Hence, the time is $\bigO(m\ell)$. The packed
        strings, including all unused bits and word-alignment overhead, and the
        two arrays use
        $\bigO(m(\ell+1)b+\AlphabetSize^\ell\log m)
        =\bigO(m\ell^2\log\AlphabetSize)$ bits. Thus, this step has peak
        space usage $\bigO(m\ell^2\log\AlphabetSize)$ bits. The packed strings
        and the arrays $A_{\rm pos}$ and $A_{\rm count}$, which are used in the
        next substep, occupy $\bigO(m\ell^2\log\AlphabetSize)$ bits.

      \item \emph{Fill the strings $P_X$}: For every $j\in[1\dd m]$, set
        $q:=(j-1)\bmod M$. For every $k\in[0\dd\ell]$, use
        \cref{pr:packed-representation}\eqref{pr:packed-representation-substring}
        to extract the packed representation
        $\PackedRepresentation{w}{\AlphabetSize}{X}$ of the length-$k$ prefix
        $X$ of $W[j]$, starting at position $(j-1)\ell+1$ in the concatenation
        that represents $W$ (see \cref{def:packed-sequence-representation}).
        Using \cref{pr:basic-int-encoding}, in $\bigO(1)$ time compute
        $x:=\BasicInt{\AlphabetSize}{X}$. Then, use
        \cref{pr:packed-representation}\eqref{pr:packed-representation-update}
        to set the symbol at position $A_{\rm count}[x]$ in the string pointed
        to by $A_{\rm pos}[x]$ to $q$, and increment $A_{\rm count}[x]$. Every
        extracted prefix has length at most $\ell\leq\log_{\AlphabetSize}m$, so the
        $m(\ell+1)$ accesses into $W$ take $\bigO(m\ell)$ time by
        \cref{pr:packed-representation}\eqref{pr:packed-representation-substring}.
        The $m(\ell+1)$ calls to the data structure from
        \cref{pr:basic-int-encoding} and the $m(\ell+1)$ character updates in
        the packed strings $P_X$ also take $\bigO(m\ell)$ time. Thus, this step
        takes $\bigO(m\ell)$ time, uses
        $\bigO(\AlphabetSize^{\ell}\log m)
        =\bigO(m\ell\log\AlphabetSize)$ bits of temporary space, and has peak
        space usage $\bigO(m\ell^2\log\AlphabetSize)$ bits. The array
        $A_{\rm pos}$ and the packed strings $P_X$, for
        $X\in[0\dd\AlphabetSize)^{\leq\ell}$, occupy
        $\bigO(m\ell^2\log\AlphabetSize)$ bits.
      \end{enumerate}

      In total, the direct construction takes $\bigO(m\ell)$ time, has peak
      space usage $\bigO(m\ell^2\log\AlphabetSize)$ bits. The constructed
      packed strings $P_X$ and $A_{\rm pos}$ occupy
      $\bigO(m\ell^2\log\AlphabetSize)$ bits.

    \item \emph{Packed construction}: This alternative proceeds as follows.
      \begin{enumerate}

      \item \emph{Construct an auxiliary string $R$}: Let
        $R[1\dd m]\in[0\dd2^b)^m$ denote the string satisfying
        $R[j]=(j-1)\bmod M$ for every $j\in[1\dd m]$. To construct its packed
        representation, use
        \cref{pr:packed-representation}\eqref{pr:packed-representation-initialize}
        to initialize a string $Q[1\dd M]$ over alphabet $[0\dd2^b)$ to
        zero. For every $q\in[0\dd M)$, use
        \cref{pr:packed-representation}\eqref{pr:packed-representation-update}
        to set $Q[q+1]:=q$. Thus, $Q=(0,1,\ldots,M-1)$ and
        $R=Q^\infty[1\dd m+1)$. Apply
        \cref{pr:packed-string-repetition}, with target length $m$ and length
        upper bound $6m(\ell+1)$, to construct the packed representation
        $\PackedRepresentation{w}{2^b}{R}$. The invocation is valid because
        $\lceil\log(2^b)\rceil=b$ has been computed,
        $2^b<2M\leq2m\leq6m(\ell+1)$, the source and target lengths are at
        most $m$, and
        $w\geq2\log m>\log(6m(\ell+1))$. Initializing and updating $Q$ takes
        $\bigO(M)$ time. Since
        $n_b=\ceil{m/M}$, \cref{pr:packed-string-repetition} constructs $R$ in
        $\bigO(1+\log n_b+mb/\log m)$ time. By the monotonicity of
        $x/\log x$, we have
        $M=\bigO(m\log M/\log m)=\bigO(mb/\log m)$. Moreover,
        $\log n_b\leq\log m=\bigO(m/\log m)
        \subseteq\bigO(mb/\log m)$. Therefore, this step takes
        $\bigO(m\ell\log\AlphabetSize/\log m)$ time, uses
        $\bigO(m\ell\log\AlphabetSize)$ bits of temporary space, has peak
        space usage $\bigO(m\ell\log\AlphabetSize)$ bits. The packed
        representation of $R$, which is used in the next substep, occupies
        $\bigO(m\ell\log\AlphabetSize)$ bits.

      \item \emph{Construct strings $P_X$}: Proceed as follows.
        \begin{enumerate}

        \item Initialize
          $A_{\rm pos}[0\dd2\AlphabetSize^{\ell})$ to zero.

        \item Apply
          \cref{pr:packed-stable-multilevel-routing}, with input length $m$
          and payload width $b$, to the packed sequence
          representation of $W$ and the packed representation
          $\PackedRepresentation{w}{2^b}{R}$. For every
          $k\in[0\dd\ell]$, let
          $A^{\rm route}_{R,k}[0\dd\AlphabetSize^k)$ denote the returned
          level-$k$ array.

        \item For every
          $k\in[0\dd\ell]$ and $y\in[0\dd A_{\rm pow}[k])$, set
          $x:=A_{\rm pow}[k]+y$, and let $X\in[0\dd\AlphabetSize)^k$ denote the
          unique string satisfying $y=\Val{\AlphabetSize}{X}$.
          By the output guarantee of \cref{pr:packed-stable-multilevel-routing}, for
          every $r\in[1\dd F_X]$, the $r$th payload in
          $A^{\rm route}_{R,k}[y]$ is
          \[
            R[\PrefixSelect{W}{r}{X}]
              =(\PrefixSelect{W}{r}{X}-1)\bmod M=P_X[r].
          \]
          Thus,
          $A^{\rm route}_{R,k}[y]=\PackedRepresentation{w}{2^b}{P_X}$. Since
          $\lceil\log(2^b)\rceil=b=\lceil\log M\rceil$ and every payload is
          smaller than $M$, this representation is identical to
          $\PackedRepresentation{w}{M}{P_X}$. Moreover,
          $x=\BasicInt{\AlphabetSize}{X}$ by \cref{def:basic-int}. Store in
          $A_{\rm pos}[x]$ a pointer to $A^{\rm route}_{R,k}[y]$.
        \end{enumerate}

        Applying \cref{pr:packed-stable-multilevel-routing} takes
        $\bigO(m(\ell\log\AlphabetSize)^2/\log m)$ time and uses
        $\bigO(m\ell\log\AlphabetSize)$ bits of temporary space. Initializing
        $A_{\rm pos}$ and storing all pointers takes
        $\bigO(\AlphabetSize^\ell)$ additional time. Since
        $\AlphabetSize^\ell\log m
        =\bigO(m\ell\log\AlphabetSize)$ and
        $\ell\log\AlphabetSize\geq1$, this is upper bounded by
        $\bigO(m(\ell\log\AlphabetSize)^2/\log m)$. There are
        $\bigO(\AlphabetSize^\ell)$ returned packed strings, and their total
        length is
        $\sum_{X\in[0\dd\AlphabetSize)^{\leq\ell}}F_X=m(\ell+1)$. Hence,
        including all unused bits and word-alignment overhead, these strings
        and $A_{\rm pos}$ occupy
        $\bigO(\AlphabetSize^\ell\log m+m(\ell+1)b)
        =\bigO(m\ell^2\log\AlphabetSize)$ bits. Thus, this step takes
        $\bigO(m(\ell\log\AlphabetSize)^2/\log m)$ time, uses
        $\bigO(m\ell\log\AlphabetSize)$ bits of temporary space, has peak
        space usage $\bigO(m\ell^2\log\AlphabetSize)$ bits. The packed strings
        $P_X$ and $A_{\rm pos}$ occupy
        $\bigO(m\ell^2\log\AlphabetSize)$ bits.
      \end{enumerate}

      In total, the packed construction takes
      $\bigO(m(\ell\log\AlphabetSize)^2/\log m)$ time. The value $b$, the
      constructed packed strings $P_X$, and $A_{\rm pos}$ occupy
      $\bigO(m\ell^2\log\AlphabetSize)$ bits, and the construction uses
      $\bigO(m\ell\log\AlphabetSize)$ additional temporary bits. Its peak
      space usage is $\bigO(m\ell^2\log\AlphabetSize)$ bits.
    \end{enumerate}

    Taking the faster alternative, this step takes the minimum of
    $\bigO(m\ell)$ and
    $\bigO(m(\ell\log\AlphabetSize)^2/\log m)$ time. Its peak space usage is
    $\bigO(m\ell^2\log\AlphabetSize)$ bits. The value $b$, the resulting packed
    strings $P_X$, and $A_{\rm pos}$ occupy
    $\bigO(m\ell^2\log\AlphabetSize)$ bits.
  \end{enumerate}

  Steps~\ref{large-space-baseline-construction-step-4}
  and~\ref{large-space-baseline-construction-step-5} each take the minimum of
  $\bigO(m\ell)$ and
  $\bigO(m(\ell\log\AlphabetSize)^2/\log m)$ time. Constructing $a$, the
  string-encoding structure, $A_{\rm pow}$, and $A_{\rm freq}$ takes
  $\bigO(m\ell\log\AlphabetSize/\log m)$ time in total. The construction of
  $A_{\rm freq}$ has this bound, while the other objects are dominated because
  $\sqrt m+\ell=\bigO(m\ell\log\AlphabetSize/\log m)$. This quantity is bounded
  by both alternatives because $1\leq\ell\log\AlphabetSize\leq\log m$.
  Thus, the total preprocessing time is
  \[
    \bigO\!\left(
      m\min\!\left\{\ell,
        \frac{(\ell\log\AlphabetSize)^2}{\log m}
      \right\}
    \right).
  \]
  The peak preprocessing-space usage is
  $\bigO(m\ell^2\log\AlphabetSize)$ bits.
\end{proof}

\begin{theorem}\label{th:large-space-select-baseline}
  Let $\AlphabetSize,m\in\Z_{\geq2}$ satisfy $\AlphabetSize\leq m$.
  Consider the word RAM model with word size $w=c\log m$, where $c\geq2$ is a
  constant. There exists a data structure that, given the packed representation
  $\PackedRepresentation{w}{\AlphabetSize}{S}$
  (\cref{def:packed-representation}) of any string
  $S\in[0\dd\AlphabetSize)^m$, answers $\Select{S}{r}{a}$ queries
  (\cref{def:rank-select}) for every $a\in[0\dd\AlphabetSize)$ and
  $r\in[1\dd m]$ in $\bigO(1)$ time, uses
  $\bigO(m\log\AlphabetSize)$ bits, and can be constructed in
  \[
    \bigO\!\left(
      m\min\!\left\{1,
        \frac{(\log\AlphabetSize)^2}{\log m}
      \right\}
    \right)
  \]
  time using $\bigO(m\log\AlphabetSize)$ bits of preprocessing space.
\end{theorem}
\begin{proof}
  The claim follows immediately by applying
  \cref{pr:large-space-prefix-select-baseline} with $\ell=1$.
\end{proof}

\subsection{Basic Linear-Space Structure}\label{sec:basic-linear-space-prefix-select}

\begin{proposition}\label{pr:small-space-prefix-select-baseline}
  Let $\AlphabetSize,m\in\Z_{\geq2}$ and $\ell\in\Z_{\geq1}$ satisfy
  $\AlphabetSize^\ell\leq m$. Consider the word RAM model with word size
  $w=c\log m$, where $c\geq2$ is a constant. There exists a data structure
  for the problem of indexing for prefix select queries over alphabet
  $[0\dd\AlphabetSize)$ (see \cref{sec:prefix-select-problem-def}) that
  achieves the following complexities:
  \begin{itemize}
  \item space usage $\bigO(m\ell\log\AlphabetSize)$ bits,
  \item preprocessing time
    $\bigO(m\min(\ell,
      \tfrac{(\ell\log\AlphabetSize)^2}{\log m}))$,
  \item preprocessing space $\bigO(m\ell\log\AlphabetSize)$ bits,
  \item query time $\bigO(\ell)$.
  \end{itemize}
\end{proposition}
\begin{proof}

  If $m<64$, construct and store all answers directly, and hence the claim
  follows immediately. Thus, assume that $m\geq64$.

  For every $X\in[0\dd\AlphabetSize)^{\leq\ell}$, denote
  $F_X:=\PrefixRank{W}{m}{X}$ (see \cref{def:prefix-range-queries}).
  For every $p\in[0\dd\ell]$, let
  $(X_{p,j})_{j\in[1\dd\AlphabetSize^p]}$ denote the sequence of all strings
  in $[0\dd\AlphabetSize)^p$ in lexicographic order.

  For every $p\in[0\dd\ell)$ and
  $X\in[0\dd\AlphabetSize)^p$, let
  $D_X\in[0\dd\AlphabetSize)^{F_X}$ denote the string satisfying
  $D_X[r]=W[\PrefixSelect{W}{r}{X}][p+1]$ for every
  $r\in[1\dd F_X]$. Thus, $D_X$ lists the characters following $X$ in the
  strings of $W$ having $X$ as a prefix, in their order in $W$. For every
  $p\in[0\dd\ell)$, let $D_p$ denote the concatenation of these strings in
  lexicographic order.
  Formally, $D_p:=\bigodot_{j=1}^{\AlphabetSize^p}D_{X_{p,j}}$. The
  length-$p$ prefixes partition $W$, and hence $|D_p|=m$.

  For every $p\in[0\dd\ell]$ and $j\in[1\dd\AlphabetSize^p]$, denote by
  $b_{X_{p,j}}$ the total frequency of the strings preceding $X_{p,j}$ in
  lexicographic order. Formally,
  $b_{X_{p,j}}:=\sum_{i=1}^{j-1}F_{X_{p,i}}$. Denote
  $e_{X_{p,j}}:=b_{X_{p,j}}+F_{X_{p,j}}$. Thus, for every $p<\ell$ and
  $X\in[0\dd\AlphabetSize)^p$, the value $b_X$ is the number of characters
  preceding $D_X$ in $D_p$, and $D_X=D_p(b_X\dd e_X]$. For every
  $p\in[0\dd\ell]$, let
  $L_p^{\rm beg}[0\dd\AlphabetSize^p)$ and
  $L_p^{\rm end}[0\dd\AlphabetSize^p)$ denote the arrays satisfying
  $L_p^{\rm beg}[\Val{\AlphabetSize}{X}]=b_X$ and
  $L_p^{\rm end}[\Val{\AlphabetSize}{X}]=e_X$, respectively, for every
  $X\in[0\dd\AlphabetSize)^p$. For every
  $p\in[0\dd\ell)$, let
  $L_p^{\rm rank}[0\dd\AlphabetSize^{p+1})$ denote the array satisfying
  $L_p^{\rm rank}[\Val{\AlphabetSize}{Xc}]
  =\Rank{D_p}{b_X}{c}$ for every $X\in[0\dd\AlphabetSize)^p$ and
  $c\in[0\dd\AlphabetSize)$, using the rank notation from
  \cref{def:rank-select}.

  \DSComponents
  The data structure consists of the following components:
  \begin{enumerate}

    \item The data structure from \cref{pr:val-encoding} with parameters $m$
      and $\AlphabetSize$. It uses
      $\bigO(\sqrt m\log m)=\bigO(m\ell\log\AlphabetSize)$ bits.

    \item The arrays $L_i^{\rm beg}[0\dd\AlphabetSize^i)$ and
      $L_i^{\rm end}[0\dd\AlphabetSize^i)$ for every $i\in[0\dd\ell]$,
      together with $L_i^{\rm rank}[0\dd\AlphabetSize^{i+1})$ for every
      $i\in[0\dd\ell)$. All these arrays are stored in plain form. They
      have $\bigO(\AlphabetSize^\ell)$ entries in total, each using
      $\bigO(\log m)$ bits. The function $x/\log x$ is increasing for
      $x\geq3$. Thus, if $\AlphabetSize^\ell\geq3$, the inequality
      $\AlphabetSize^\ell\leq m$ implies
      $\AlphabetSize^\ell\log m
      =\bigO(m\ell\log\AlphabetSize)$. If $\AlphabetSize^\ell=2$, the same
      bound follows directly from $\AlphabetSize=2$ and $\ell=1$. Hence, this
      component uses
      $\bigO(m\ell\log\AlphabetSize)$ bits in total.

    \item The data structures from \cref{th:large-space-select-baseline}
      constructed for the strings $D_p$, for every $p\in[0\dd\ell)$,
      together with an array $A_{\rm sel}[0\dd\ell)$. Its entry
      $A_{\rm sel}[p]$ stores a pointer to the data structure for $D_p$.
      Every $D_p$ has length $m$, so the
      data structures use $\bigO(m\ell\log\AlphabetSize)$ bits in total.
      The pointer array adds $\bigO(\ell\log m)$ bits. Thus, this component
      uses $\bigO(m\ell\log\AlphabetSize)$ bits in total.
  \end{enumerate}

  In total, the data structure uses $\bigO(m\ell\log\AlphabetSize)$ bits.

  \DSQueries
  Let $X\in[0\dd\AlphabetSize)^{\leq\ell}$ and $r\in[1\dd m]$. Given the
  length $|X|$, the packed representation
  $\PackedRepresentation{w}{\AlphabetSize}{X}$
  (\cref{def:packed-representation}) of the string $X$, and the position
  $r$, we compute $\PrefixSelect{W}{r}{X}$ as follows:
  \begin{enumerate}

  \item Using \cref{pr:val-encoding}, in $\bigO(1)$ time compute
    $x_{\rm cur}:=\Val{\AlphabetSize}{X}$ and set $d_{\rm cur}:=|X|$.

  \item In $\bigO(1)$ time retrieve
    $b:=L_{d_{\rm cur}}^{\rm beg}[x_{\rm cur}]$ and
    $e:=L_{d_{\rm cur}}^{\rm end}[x_{\rm cur}]$. If $r>e-b$, return
    $\infty$ and terminate the query. Otherwise, set $r_{\rm cur}:=r$.
    Throughout the query algorithm, we decrease $d_{\rm cur}$ until it
    reaches zero, while updating $x_{\rm cur}$ and $r_{\rm cur}$ to
    maintain the following invariants:
    \begin{itemize}
    \item \emph{Invariant 1:}
      $x_{\rm cur}=\Val{\AlphabetSize}{X[1\dd d_{\rm cur}]}$,
    \item \emph{Invariant 2:}
      $r_{\rm cur}\in
      [1\dd F_{X[1\dd d_{\rm cur}]}]$,
    \item \emph{Invariant 3:}
      $\PrefixSelect{W}{r}{X}
      =\PrefixSelect{W}{r_{\rm cur}}{X[1\dd d_{\rm cur}]}$.
    \end{itemize}
    The invariants hold after the first two steps.

  \item While $d_{\rm cur}>0$, perform the following steps:
    \begin{enumerate}

    \item Set $p:=d_{\rm cur}-1$,
      $c:=x_{\rm cur}\bmod\AlphabetSize$, and
      $x_{\rm pref}:=\lfloor x_{\rm cur}/\AlphabetSize\rfloor$.

    \item Retrieve $b_{\rm pref}:=L_p^{\rm beg}[x_{\rm pref}]$ and
      $r_{\rm base}:=L_p^{\rm rank}[x_{\rm cur}]$.

    \item Using the data structure pointed to by $A_{\rm sel}[p]$,
      compute $r_{\rm global}:=
      \Select{D_p}{r_{\rm base}+r_{\rm cur}}{c}$
      (see \cref{th:large-space-select-baseline}).

    \item Set $r_{\rm cur}:=r_{\rm global}-b_{\rm pref}$,
      $x_{\rm cur}:=x_{\rm pref}$, and $d_{\rm cur}:=p$.
    \end{enumerate}
    Each of the above substeps takes $\bigO(1)$ time.
    To justify an iteration, let $\hat d$, $\hat x$, and $\hat r$ denote
    the values of $d_{\rm cur}$, $x_{\rm cur}$, and $r_{\rm cur}$ at its
    beginning. The first substep sets $p=\hat d-1$. Let $P$ denote
    $X[1\dd p]$ and let $\hat c$ denote $X[\hat d]$, so that
    $X[1\dd\hat d]=P\hat c$. By Invariant~1, we have
    $\hat x=\AlphabetSize\cdot\Val{\AlphabetSize}{P}+\hat c$. Hence, the first
    substep sets $c=\hat c$ and
    $x_{\rm pref}=\Val{\AlphabetSize}{P}$. By the definition of
    $L_p^{\rm rank}$, $r_{\rm base}=\Rank{D_p}{b_P}{c}$. Moreover, the
    segment $D_P=D_p(b_P\dd e_P]$ contains exactly $F_{Pc}$ occurrences of
    $c$. By Invariant~2, $\hat r\leq F_{Pc}$. Hence, the argument of
    the select query is in $[1\dd m]$, and the query returns a position
    inside this segment. Let $q$ denote $r_{\rm global}-b_P$. Then
    $q\in[1\dd F_P]$, and the $q$th occurrence of $P$ in $W$ is precisely
    the $\hat r$th occurrence of $Pc$ in $W$. Therefore,
    $\PrefixSelect{W}{\hat r}{Pc}
    =\PrefixSelect{W}{q}{P}$. The assignments in the fourth substep
    set $d_{\rm cur}=p$, $x_{\rm cur}=\Val{\AlphabetSize}{P}$, and
    $r_{\rm cur}=q$. Hence, they preserve Invariants~1 and~2. The equality
    above together with Invariant~3 before the iteration preserves
    Invariant~3.

  \item Once $d_{\rm cur}=0$, Invariant~3 gives
    $\PrefixSelect{W}{r}{X}
    =\PrefixSelect{W}{r_{\rm cur}}{\emptystring}=r_{\rm cur}$. Return
    $r_{\rm cur}$ as the answer.
  \end{enumerate}

  There are $|X|\leq\ell$ iterations, and hence the query takes
  $\bigO(\ell)$ time in total.

  \DSConstruction
  Given the packed sequence representation
  $\PackedSeqRepresentation{w}{\AlphabetSize}{W}$
  (\cref{def:packed-sequence-representation}) of the sequence $W$, we
  construct the data structure as follows:
  \begin{enumerate}

  \item \emph{Construct the value-encoding data structure}: Apply
    \cref{pr:val-encoding} with parameters $m$ and $\AlphabetSize$. This
    step takes $\bigO(\sqrt m)$ time, has peak space usage
    $\bigO(\sqrt m\log m)$ bits, and the constructed data structure occupies
    $\bigO(\sqrt m\log m)$ bits.

  \item \emph{Construct the boundary and rank arrays}: Proceed as follows:
    \begin{enumerate}

    \item Apply \cref{pr:packed-prefix-frequencies}, with parameters
      $\AlphabetSize$, $m$, and $\ell$, to
      $\PackedSeqRepresentation{w}{\AlphabetSize}{W}$, using a temporary output
      array $A_{\rm pfreq}[0\dd2\AlphabetSize^\ell)$.
      Thus,
      $A_{\rm pfreq}[\BasicInt{\AlphabetSize}{X}]=F_X$ for every
      $X\in[0\dd\AlphabetSize)^{\leq\ell}$
      (see \cref{def:basic-int}). Applying
      \cref{pr:packed-prefix-frequencies} takes
      $\bigO(m\ell\log\AlphabetSize/\log m)$ time and has peak space
      usage $\bigO(m\ell\log\AlphabetSize)$ bits.

    \item Set $u:=1$. For every $p\in[0\dd\ell]$, we have
      $u=\AlphabetSize^p$ at the beginning of the iteration. Allocate
      $L_p^{\rm beg}[0\dd u)$ and $L_p^{\rm end}[0\dd u)$ in plain
      form, and set a counter $s:=0$. Then, for every
      $x\in[0\dd u)$ in increasing order, set
      $L_p^{\rm beg}[x]:=s$, set
      $s:=s+A_{\rm pfreq}[u+x]$, and set $L_p^{\rm end}[x]:=s$. For the
      unique string
      $X\in[0\dd\AlphabetSize)^p$ satisfying
      $x=\Val{\AlphabetSize}{X}$, the index $u+x$ equals
      $\BasicInt{\AlphabetSize}{X}$. Increasing $x$ enumerates the
      length-$p$ strings lexicographically, so the two assignments have
      the values specified in the definitions of the arrays. If
      $p<\ell$, set $u:=\AlphabetSize\cdot u$ for the next iteration.

    \item Set $u:=1$. For every $p\in[0\dd\ell)$, we have
      $u=\AlphabetSize^p$ at the beginning of the iteration. Allocate
      $L_p^{\rm rank}[0\dd\AlphabetSize\cdot u)$ in plain form. For every
      $c\in[0\dd\AlphabetSize)$, set a counter $s:=0$ and iterate over
      all $x\in[0\dd u)$ in increasing order. In every
      iteration, set $y:=\AlphabetSize\cdot x+c$, set
      $L_p^{\rm rank}[y]:=s$, and set
      $s:=s+A_{\rm pfreq}[\AlphabetSize\cdot u+y]$. If
      $x=\Val{\AlphabetSize}{X}$, then
      $y=\Val{\AlphabetSize}{Xc}$, and the counter before the update is
      the number of occurrences of $c$ in the strings $D_Y$ over all
      length-$p$ strings $Y$ preceding $X$ lexicographically: every
      $D_Y$ contains exactly $F_{Yc}$ occurrences of $c$, and when $Y$
      was processed, the update added
      $A_{\rm pfreq}[\BasicInt{\AlphabetSize}{Yc}]=F_{Yc}$. This counter
      is precisely $\Rank{D_p}{b_X}{c}$, as required. After processing
      all $c$ and $x$, set $u:=\AlphabetSize\cdot u$ for the next iteration.
    \end{enumerate}

    The loops in the second and third substeps perform
    $\bigO(\AlphabetSize^\ell)$ operations in total. By the monotonicity
    of $x/\log x$,
    $\AlphabetSize^\ell
    =\bigO(m\ell\log\AlphabetSize/\log m)$. Therefore, this step takes
    $\bigO(m\ell\log\AlphabetSize/\log m)$ time, has peak space usage
    $\bigO(m\ell\log\AlphabetSize)$ bits. The constructed boundary and rank
    arrays occupy
    $\bigO(m\ell\log\AlphabetSize)$ bits.

  \item \emph{Construct the select structures for the wavelet-tree
    levels}: Allocate $A_{\rm sel}[0\dd\ell)$ and compute
    $\lceil\log\AlphabetSize\rceil$ by repeated doubling in
    $\bigO(\log m)$ time. Proceed as follows:
    \begin{enumerate}

    \item Apply \cref{pr:packed-wavelet-tree-construction}, with parameters
      $\AlphabetSize$, $m$, and $\ell$, to
      $\PackedSeqRepresentation{w}{\AlphabetSize}{W}$. Let
      $A_{\rm node}[0\dd2\AlphabetSize^\ell)$ denote its returned array. For
      every
      $X\in[0\dd\AlphabetSize)^{<\ell}$, the entry at index
      $\BasicInt{\AlphabetSize}{X}$ contains
      $\PackedRepresentation{w}{\AlphabetSize}{D_X}$. Applying
      \cref{pr:packed-wavelet-tree-construction} takes time bounded by
      $\bigO(m\min\{\ell,
      (\ell\log\AlphabetSize)^2/\log m\})$, has peak space usage
      $\bigO(m\ell\log\AlphabetSize)$ bits, and returns an array and
      packed strings using $\bigO(m\ell\log\AlphabetSize)$ bits.

    \item Set $u:=1$. For every $p\in[0\dd\ell)$, execute the following
      steps. At the beginning of the iteration, we have
      $u=\AlphabetSize^p$.
      \begin{enumerate}

      \item Allocate temporary arrays
        $A_{\rm len}[1\dd u]$ and $A_{\rm ptr}[1\dd u]$. For every
        $x\in[0\dd u)$ in increasing order, retrieve
        $f:=L_p^{\rm end}[x]-L_p^{\rm beg}[x]$, set
        $A_{\rm len}[x+1]:=f$, and store in $A_{\rm ptr}[x+1]$ a pointer to
        the packed representation
        in $A_{\rm node}[u+x]$. For the unique
        $X\in[0\dd\AlphabetSize)^p$ satisfying
        $x=\Val{\AlphabetSize}{X}$, we have
        $u+x=\BasicInt{\AlphabetSize}{X}$ and $f=F_X=|D_X|$.
        Therefore, $A_{\rm ptr}$ contains pointers to the packed
        representations of the strings $D_X$ in lexicographic order,
        $A_{\rm len}$ contains their lengths, and these lengths sum to $m$.

      \item Apply
        \cref{pr:packed-representation}\eqref{pr:packed-representation-concat},
        with length parameter $m$, to the $u$ packed strings and their lengths
        stored in $A_{\rm ptr}$ and $A_{\rm len}$. Its requirements hold because
        $u=\AlphabetSize^p\leq m$, $\AlphabetSize\leq m$,
        $w\geq2\log m>\log m$, and the lengths sum to $m$. Since the
        strings are supplied in lexicographic order, the result is
        $\PackedRepresentation{w}{\AlphabetSize}{D_p}$.

      \item Apply \cref{th:large-space-select-baseline}, with parameters
        $\AlphabetSize$ and $m$, to the packed representation
        $\PackedRepresentation{w}{\AlphabetSize}{D_p}$ and store the pointer
        to the returned data structure in $A_{\rm sel}[p]$. Set
        $u:=\AlphabetSize\cdot u$ for the next iteration.
      \end{enumerate}
    \end{enumerate}

    Across all levels, constructing the temporary arrays and applying
    packed concatenation takes
    $\bigO(\AlphabetSize^\ell+
    m\ell\log\AlphabetSize/\log m)
    =\bigO(m\ell\log\AlphabetSize/\log m)$ time. This is bounded by both
    terms in the claimed preprocessing time. Applying
    \cref{th:large-space-select-baseline} at every level takes
    $\bigO(m\ell\min\{1,
    (\log\AlphabetSize)^2/\log m\})$ time, which is also bounded by both
    terms in the claimed preprocessing time. Together with applying
    \cref{pr:packed-wavelet-tree-construction}, this step takes
    $\bigO(m\min\{\ell,
    (\ell\log\AlphabetSize)^2/\log m\})$ time. Its peak space usage is
    $\bigO(m\ell\log\AlphabetSize)$ bits. The constructed select structures
    and $A_{\rm sel}$ occupy $\bigO(m\ell\log\AlphabetSize)$ bits.
  \end{enumerate}

  Combining the three steps, the total preprocessing time is
  $\bigO(m\min\{\ell,
  (\ell\log\AlphabetSize)^2/\log m\})$. The peak preprocessing-space
  usage is $\bigO(m\ell\log\AlphabetSize)$ bits.
\end{proof}

\subsection{The Layering Lemma}\label{sec:prefix-select-self-reduction}

\begin{lemma}[A two-layer composition lemma]
  \label{lem:prefix-select-final-layer-composition}
  Let $\AlphabetSize,m\in\Z_{\geq2}$,
  $\ell\in\Z_{\geq2}$, and $\lambda\in[1\dd\ell)$ satisfy
  $\AlphabetSize^\ell\leq m$. Denote
  $k:=\AlphabetSize^\lambda$ and $h:=\lfloor\ell/\lambda\rfloor$, and consider the word
  RAM model with word size $w=c\log m$, where $c\geq2$ is a constant.
  Assume the existence of the following two data structures:
  \begin{itemize}
  \item a prefix-select data structure for $m$ strings of length $\lambda$ over
    $[0\dd\AlphabetSize)$, with space usage $S_1$ bits, preprocessing time
    $P_{t,1}$, preprocessing space $P_{s,1}$ bits, and query time $Q_1$,
  \item a prefix-select data structure for $m$ strings of length $h$ over
    $[0\dd k)$, with space usage $S_2$ bits, preprocessing time $P_{t,2}$,
    preprocessing space $P_{s,2}$ bits, and query time $Q_2$.
  \end{itemize}
  Then there exists a data structure that, given
  $\PackedSeqRepresentation{w}{\AlphabetSize}{W}$
  (\cref{def:packed-sequence-representation}) for a sequence
  $W[1\dd m]$ of $m$ strings of length $\ell$ over
  $[0\dd\AlphabetSize)$, answers prefix select queries
  (\cref{def:prefix-range-queries}) on $W$ and has the following
  complexities:
  \begin{itemize}
  \item space usage
    $\bigO(\lceil\ell/\lambda\rceil\cdot S_1+S_2)$ bits,
  \item preprocessing time
    \[
      \bigO\!\left(
        m\min\!\left\{
          \ell,
          \frac{\ell\log\AlphabetSize}{\sqrt{\log m}},
          \frac{(\ell\log\AlphabetSize)^2}{\log m}
        \right\}
        +\left\lceil\frac{\ell}{\lambda}\right\rceil P_{t,1}+P_{t,2}
      \right),
    \]
  \item preprocessing space
    $\bigO(\lceil\ell/\lambda\rceil\cdot S_1+S_2
    +\max\{P_{s,1},P_{s,2}\})$ bits,
  \item query time $\bigO(Q_1+Q_2)$.
  \end{itemize}
\end{lemma}
\begin{proof}

  If $m<64$, construct and store an array $A_{\rm ans}$. For every
  $p\in[0\dd\ell]$, $x\in[0\dd\AlphabetSize^p)$, and $r\in[1\dd m]$, its
  entry $A_{\rm ans}[p][x][r]$ stores $\PrefixSelect{W}{r}{X}$ for the unique
  $X\in[0\dd\AlphabetSize)^p$ satisfying
  $x=\Val{\AlphabetSize}{X}$ (see \cref{def:val}). All parameters and input
  sizes are then bounded by a constant, and hence the claim follows
  immediately. Thus, assume that $m\geq64$.
  The monotonicity of $x/\log x$ and
  $\AlphabetSize^\ell\leq m$ give
  $\AlphabetSize^\ell\log m
  =\bigO(m\ell\log\AlphabetSize)$.
  Denote
  \[
    T_0:=m\min\!\left\{
      \ell,
      \frac{\ell\log\AlphabetSize}{\sqrt{\log m}},
      \frac{(\ell\log\AlphabetSize)^2}{\log m}
    \right\}.
  \]

  For every $X\in[0\dd\AlphabetSize)^{\leq\ell}$, denote
  $F_X:=\PrefixRank{W}{m}{X}$ (see \cref{def:prefix-range-queries}). For every
  $p\in[0\dd\ell]$, let
  $(X_{p,j})_{j\in[1\dd\AlphabetSize^p]}$ denote the sequence of all strings
  in $[0\dd\AlphabetSize)^p$ in lexicographic order.

  For every $p\in[0\dd\ell)$, $X\in[0\dd\AlphabetSize)^p$, and
  $d\in[1\dd\ell-p]$, let $W_{X,d}$ denote the sequence of $F_X$ strings of
  length $d$ satisfying
  $W_{X,d}[j]=W[\PrefixSelect{W}{j}{X}][p+1\dd p+d]$ for every
  $j\in[1\dd F_X]$. For every $p\in[0\dd\ell)$ and
  $d\in[1\dd\ell-p]$, let $W_{p,d}$ denote their concatenation in
  lexicographic order. Formally,
  $W_{p,d}:=\bigodot_{j=1}^{\AlphabetSize^p}W_{X_{p,j},d}$. The length-$p$
  prefixes partition $W$, and hence $|W_{p,d}|=m$.

  For every $X\in[0\dd\AlphabetSize)^{\leq\ell}$, denote by $b_X$ the
  number of strings in $W$ whose length-$|X|$ prefix is lexicographically
  smaller than $X$. Formally,
  $b_X:=|\{i\in[1\dd m]:W[i][1\dd |X|]\prec X\}|$. Equivalently, for every
  $p\in[0\dd\ell]$ and $j\in[1\dd\AlphabetSize^p]$, it holds
  $b_{X_{p,j}}=\sum_{s=1}^{j-1}F_{X_{p,s}}$. Denote
  $e_X:=b_X+F_X$. Thus, whenever $p<\ell$ and
  $X\in[0\dd\AlphabetSize)^p$ and $d\in[1\dd\ell-p]$, the sequence
  $W_{X,d}$ occupies positions $(b_X\dd e_X]$ in $W_{p,d}$. For every
  $p\in[0\dd\ell]$, let
  $L_p^{\rm beg}[0\dd\AlphabetSize^p)$ and
  $L_p^{\rm end}[0\dd\AlphabetSize^p)$ denote the arrays satisfying
  $L_p^{\rm beg}[\Val{\AlphabetSize}{X}]=b_X$ and
  $L_p^{\rm end}[\Val{\AlphabetSize}{X}]=e_X$, respectively, for every
  $X\in[0\dd\AlphabetSize)^p$.

  For every $p\in[0\dd\ell)$ and $d\in[1\dd\ell-p]$, let
  $L_{p,d}^{\rm rank}[0\dd\AlphabetSize^{p+d})$ denote the array satisfying
  $L_{p,d}^{\rm rank}[\Val{\AlphabetSize}{XY}]
  =\PrefixRank{W_{p,d}}{b_X}{Y}$ for every
  $X\in[0\dd\AlphabetSize)^p$ and
  $Y\in[0\dd\AlphabetSize)^d$. Thus, the entry counts the occurrences of
  $Y$ preceding the segment $W_{X,d}$ in $W_{p,d}$.

  For every multiple $p<\ell$ of $\lambda$, let $V_p[1\dd m]$ denote the sequence
  of strings of length $\lambda$ over $[0\dd\AlphabetSize)$ satisfying
  $V_p[j]=W_{p,\min\{\lambda,\ell-p\}}[j]
  \zero^{\lambda-\min\{\lambda,\ell-p\}}$ for every $j\in[1\dd m]$. Let
  $U[1\dd m]$ denote the sequence of strings of length $h$ over $[0\dd k)$
  satisfying
  $U[j][t]=\Val{\AlphabetSize}{W[j][(t-1)\lambda+1\dd t\lambda]}$ for every
  $j\in[1\dd m]$ and $t\in[1\dd h]$.

  For every positive multiple $p=t\lambda$ of $\lambda$ satisfying
  $p\leq \lambda\lfloor(\ell-1)/\lambda\rfloor$, we define
  $L_p^{\rm pat}[0\dd\AlphabetSize^p)$ as follows: For every
  $P\in[0\dd\AlphabetSize)^p$, $L_p^{\rm pat}[\Val{\AlphabetSize}{P}]$ stores
  the packed representation over alphabet $[0\dd k)$ of the length-$t$
  string whose $s$th character is
  $\Val{\AlphabetSize}{P[(s-1)\lambda+1\dd s\lambda]}$.

  \DSComponents
  The data structure consists of the following components:
  \begin{enumerate}

  \item The values $\lambda,h,k$ and $\lceil\log\AlphabetSize\rceil$, the array
    $A_{\rm pow}[0\dd\ell]$, and the data structure from
    \cref{pr:val-encoding} with parameters $m$ and $\AlphabetSize$. The array
    satisfies $A_{\rm pow}[j]=\AlphabetSize^j$ for every
    $j\in[0\dd\ell]$. These objects use
    $\bigO((\ell+\sqrt m)\log m)$ bits.

  \item The arrays $L_p^{\rm beg}$ and $L_p^{\rm end}$ for every
    $p\in[0\dd\ell]$, the arrays $L_{p,d}^{\rm rank}$ for every multiple
    $p<\ell$ of $\lambda$ and every
    $d\in[1\dd\min\{\lambda,\ell-p\}]$, and the arrays
    $L_p^{\rm pat}$ for every positive multiple
    $p\leq \lambda\lfloor(\ell-1)/\lambda\rfloor$ of $\lambda$, all stored in plain form. The
    pairs $(p,d)$ used by the rank arrays contain every value
    $p+d\in[1\dd\ell]$ exactly once.
    Therefore, all rank arrays have $\bigO(\AlphabetSize^\ell)$ entries.
    The boundary and pattern arrays satisfy the same bound. Each entry uses
    one word. Indeed, boundary and rank entries belong to $[0\dd m]$, and
    a pattern entry for a prefix of length $p$ uses
    $(p/\lambda)\lceil\log k\rceil
    \leq2p\log\AlphabetSize\leq2\log m\leq w$ bits. Thus, this component uses
    $\bigO(\AlphabetSize^\ell\log m)$ bits.

  \item An array $A_{\rm sel}[0\dd\lceil\ell/\lambda\rceil)$, copies of the first
    prefix-select data structure from the claim constructed for the sequences
    $V_p$, and one copy of the second prefix-select data structure from the
    claim constructed for $U$. For every multiple $p<\ell$ of $\lambda$, the entry
    $A_{\rm sel}[p/\lambda]$ stores a pointer to the prefix-select structure for
    $V_p$. This component uses
    $\bigO(\lceil\ell/\lambda\rceil\cdot S_1+S_2)$ bits.
  \end{enumerate}

  By \cref{th:prefix-query-space-lower-bounds-exact}, exact prefix select
  on $m$ strings of length $\lambda$ over $[0\dd\AlphabetSize)$ requires
  $S_1=\Omega(m\lambda\log\AlphabetSize)$ bits.
  Since $\lambda\lceil\ell/\lambda\rceil\geq\ell$, the structures constructed for the
  sequences $V_p$ use $\Omega(m\ell\log\AlphabetSize)$ bits. This dominates all
  auxiliary arrays because
  $\AlphabetSize^\ell\log m
  =\bigO(m\ell\log\AlphabetSize)$. Hence, the stated space bound holds.

  \DSQueries
  Let $X\in[0\dd\AlphabetSize)^{\leq\ell}$ and $r\in[1\dd m]$. Given the
  length $|X|$, the packed representation
  $\PackedRepresentation{w}{\AlphabetSize}{X}$ of the string $X$, and the
  position $r$, we compute $\PrefixSelect{W}{r}{X}$ as follows:
  \begin{enumerate}

  \item Using \cref{pr:val-encoding}, in $\bigO(1)$ time compute
    $x:=\Val{\AlphabetSize}{X}$. Retrieve
    $b:=L_{|X|}^{\rm beg}[x]$ and $e_X:=L_{|X|}^{\rm end}[x]$. If
    $r>e_X-b$, return $\infty$.

  \item If $|X|=0$, return $r$. Otherwise, set
    $p:=\lambda\lfloor(|X|-1)/\lambda\rfloor$, $d:=|X|-p$,
    and $x_{\rm pref}:=\lfloor x/A_{\rm pow}[d]\rfloor$. Let $Y$ denote
    $X[p+1\dd p+d]$. Use
    \cref{pr:packed-representation}\eqref{pr:packed-representation-substring}
    on the packed representation $\PackedRepresentation{w}{\AlphabetSize}{X}$
    with alphabet size $\AlphabetSize$, length parameter $m$, source length $|X|$, starting
    position $p+1$, and substring length $d$. This returns
    $\PackedRepresentation{w}{\AlphabetSize}{Y}$. The invocation is valid
    because $\AlphabetSize\leq m$, $|X|\leq\ell<m$, and
    $w\geq2\log m>\log m$.

  \item Retrieve $b_{\rm pref}:=L_p^{\rm beg}[x_{\rm pref}]$ and
    $r_{\rm base}:=L_{p,d}^{\rm rank}[x]$. Query the structure pointed to by
    $A_{\rm sel}[p/\lambda]$ with the length-$d$ pattern $Y$ and occurrence
    number $r_{\rm base}+r$. Store its answer as $r_{\rm global}$ and set
    $r':=r_{\rm global}-b_{\rm pref}$.

  \item If $p=0$, return $r'$. Otherwise, retrieve
    $\gamma:=L_p^{\rm pat}[x_{\rm pref}]$. Query the structure constructed
    for $U$ with the length-$p/\lambda$ pattern represented by
    $\gamma$ and occurrence number $r'$. Return its answer.
  \end{enumerate}

  To prove correctness, let $P$ denote $X[1\dd p]$, so that $X=PY$. The first
  $b_P$ strings of $V_p$ precede its segment for $P$, and the prefix $Y$
  occurs in that segment exactly $F_{PY}$ times. Hence,
  $r_{\rm base}$ is the number of its occurrences before the segment. The
  validity check before the local prefix-select query ensures $r\leq F_{PY}$,
  so
  $r_{\rm base}+r\leq r_{\rm base}+F_{PY}
  \leq\PrefixRank{V_p}{m}{Y}\leq m$. Hence, the query to the prefix-select
  structure constructed for $V_p$ returns a position in $(b_P\dd e_P]$.
  Thus, $r'$ is the occurrence number of the same string among the strings
  of $W$ having prefix $P$. If $p=0$, this is its position in $W$.
  Otherwise, the pattern represented by $\gamma$ is precisely the string
  obtained by partitioning $P$ into consecutive length-$\lambda$ blocks and
  encoding each block by its base-$\AlphabetSize$ value, so the query to the
  prefix-select structure constructed for $U$ returns the position of that
  string in $W$.
  The query makes at most one query to the structure for $V_p$ and one query
  to the structure for $U$. It otherwise takes $\bigO(1)$ time, so its total
  time is
  $\bigO(Q_1+Q_2)$.

  \DSConstruction
  Given $\PackedSeqRepresentation{w}{\AlphabetSize}{W}$, construct the data
  structure as follows. Retain the packed input until its partition into
  length-$h\lambda$ prefixes and length-$(\ell-h\lambda)$ suffixes has been
  constructed.
  \begin{enumerate}

  \item\label{step:prefix-select-layering-parameters} \emph{Construct the
    parameters, power array, and string-encoding structure}: Compute $h$, and
    compute $\lceil\log\AlphabetSize\rceil$ by repeated doubling. Set
    $A_{\rm pow}[0]:=1$, and set
    $A_{\rm pow}[d]:=\AlphabetSize\cdot A_{\rm pow}[d-1]$ for every
    $d\in[1\dd\ell]$. Set $k:=A_{\rm pow}[\lambda]$, and apply
    \cref{pr:val-encoding} with parameters $m$ and $\AlphabetSize$.
    Its hypotheses hold because $\AlphabetSize\leq m$ and
    $w\geq1+\log m$. Retain the resulting string-encoding structure. This
    step takes $\bigO(\ell+\sqrt m)=\bigO(T_0)$ time and has peak space usage
    $\bigO(m\ell\log\AlphabetSize)$ bits. The values
    $\lambda,h,k,\lceil\log\AlphabetSize\rceil$, together with $A_{\rm pow}$ and
    the string-encoding structure, occupy
    $\bigO(m\ell\log\AlphabetSize)$ bits.

  \item\label{step:prefix-select-layering-lookup-arrays} \emph{Construct the
    boundary, rank, and pattern arrays}: Apply
    \cref{pr:packed-prefix-frequencies}, with parameters $\AlphabetSize$, $m$,
    and $\ell$, to
    $\PackedSeqRepresentation{w}{\AlphabetSize}{W}$ and retain its output as
    $A_{\rm pfreq}[0\dd2\AlphabetSize^\ell)$, indexed using the basic-integer
    encoding from \cref{def:basic-int}. Thus,
    $A_{\rm pfreq}[\BasicInt{\AlphabetSize}{X}]=F_X$ for every
    $X\in[0\dd\AlphabetSize)^{\leq\ell}$. Proceed in three substeps.
    \begin{enumerate}

    \item\label{step:prefix-select-layering-boundary-arrays} Set $u:=1$.
      For every $p\in[0\dd\ell]$ in increasing order,
      allocate $L_p^{\rm beg}[0\dd u)$ and $L_p^{\rm end}[0\dd u)$, and set
      a counter $s:=0$. For every $x\in[0\dd u)$ in increasing order, set
      $L_p^{\rm beg}[x]:=s$, increase $s$ by $A_{\rm pfreq}[u+x]$, and set
      $L_p^{\rm end}[x]:=s$. At the beginning of the iteration for $p$, it
      holds $u=\AlphabetSize^p$. Hence, these assignments construct the two
      arrays from their definitions. Indeed, if
      $x=\Val{\AlphabetSize}{X}$, then
      $u+x=\BasicInt{\AlphabetSize}{X}$ (see \cref{def:basic-int}). If
      $p<\ell$, set $u:=\AlphabetSize\cdot u$.

    \item\label{step:prefix-select-layering-rank-arrays} For every multiple
      $p<\ell$ of $\lambda$ and every
      $d\in[1\dd\min\{\lambda,\ell-p\}]$, allocate $L_{p,d}^{\rm rank}$. For every
      $y\in[0\dd A_{\rm pow}[d])$, set a counter $s:=0$. Then scan
      $x\in[0\dd A_{\rm pow}[p])$ in increasing order, set
      $z:=x\cdot A_{\rm pow}[d]+y$, set
      $L_{p,d}^{\rm rank}[z]:=s$, and increase $s$ by
      $A_{\rm pfreq}[A_{\rm pow}[p+d]+z]$. If
      $x=\Val{\AlphabetSize}{X}$ and $y=\Val{\AlphabetSize}{Y}$, then
      $z=\Val{\AlphabetSize}{XY}$ and
      $A_{\rm pow}[p+d]+z=\BasicInt{\AlphabetSize}{XY}$. Thus, the last
      assignment increases $s$ by $F_{XY}$. Before the iteration for $x$,
      the counter equals $\PrefixRank{W_{p,d}}{b_X}{Y}$. Thus, these
      assignments construct $L_{p,d}^{\rm rank}$ from its definition. Discard
      $A_{\rm pfreq}$ after all boundary and rank arrays have been
      constructed.

    \item Compute $\lceil\log k\rceil$ by repeated doubling. For every
      $t\in[1\dd\lfloor(\ell-1)/\lambda\rfloor]$, set $p:=t\lambda$ and allocate
      $L_p^{\rm pat}[0\dd k^t)$. Enumerate
      $y\in[0\dd k^t)$ in increasing order, starting with the packed string
      $\zero^t$. Store the current length-$t$ base-$k$ representation in
      $L_p^{\rm pat}[y]$. If $y<k^t-1$, obtain the representation for $y+1$
      by adding one in base $k$ and propagating carries. The number of
      character changes over the entire enumeration is $\bigO(k^t)$. For the
      unique $P\in[0\dd\AlphabetSize)^p$ satisfying
      $y=\Val{\AlphabetSize}{P}$, partitioning $P$ into consecutive
      length-$\lambda$ blocks and encoding each block by its base-$\AlphabetSize$
      value gives a string whose base-$k$ value is $y$. Thus, these
      assignments construct $L_p^{\rm pat}$ from its definition.
    \end{enumerate}

    Applying \cref{pr:packed-prefix-frequencies} takes
    $\bigO(m\ell\log\AlphabetSize/\log m)$ time. The boundary, rank, and
    pattern arrays have $\bigO(\AlphabetSize^\ell)$ entries altogether.
    Since $\AlphabetSize^\ell
    =\bigO(m\ell\log\AlphabetSize/\log m)$, this step takes
    $\bigO(m\ell\log\AlphabetSize/\log m)=\bigO(T_0)$ time and has peak
    space usage $\bigO(m\ell\log\AlphabetSize)$ bits. The arrays that remain
    stored occupy $\bigO(m\ell\log\AlphabetSize)$ bits.

  \item\label{step:prefix-select-layering-structures} \emph{Construct the
    local and coarse data structures}: Construct the data structure from
    \cref{pr:generalized-wavelet-tree-construction} for
    $\PackedSeqRepresentation{w}{\AlphabetSize}{W}$ with parameters
    $\AlphabetSize$, $m$, and $\ell$, initialize
    $A_{\rm sel}[0\dd\lceil\ell/\lambda\rceil)$ with null pointers, and proceed in
    two substeps.
    \begin{enumerate}

    \item\label{step:prefix-select-layering-local-structures} For every
      multiple $p<\ell$ of $\lambda$, set
      $d:=\min\{\lambda,\ell-p\}$ and perform the
      following three substeps.
      \begin{enumerate}

      \item\label{step:prefix-select-layering-local-concatenation} Query the
        data structure from
        \cref{pr:generalized-wavelet-tree-construction} with parameters $p$
        and $d$. Retain its returned array as
        $\mathcal W_{p,d}[0\dd\AlphabetSize^p)$. Allocate temporary arrays
        $A_{\rm ptr}[1\dd\AlphabetSize^p]$ and
        $A_{\rm len}[1\dd\AlphabetSize^p]$. For every
        $x\in[0\dd\AlphabetSize^p)$, store in $A_{\rm ptr}[x+1]$ a pointer to
        the packed sequence representation in $\mathcal W_{p,d}[x]$, and set
        $A_{\rm len}[x+1]
        :=d(L_p^{\rm end}[x]-L_p^{\rm beg}[x])$.
        Apply the concatenation operation from
        \cref{pr:packed-representation}\eqref{pr:packed-representation-concat}
        with alphabet size $\AlphabetSize$ to the representations pointed to
        by $A_{\rm ptr}$, using the lengths in $A_{\rm len}$ and length
        parameter $m\ell$. This constructs
        $\PackedSeqRepresentation{w}{\AlphabetSize}{W_{p,d}}$. The number of
        supplied sequences is $\AlphabetSize^p\leq m\leq m\ell$, and their
        lengths sum to $md\leq m\ell$. Moreover,
        $w\geq2\log m>\log(m\ell)$, so this invocation is valid.

      \item\label{step:prefix-select-layering-local-encoding} Apply
        \cref{pr:packed-character-block-encoding} to
        $\PackedSeqRepresentation{w}{\AlphabetSize}{W_{p,d}}$ with alphabet
        size $\AlphabetSize$, sequence length $m$, block length $1$, input
        length $g=d$, and output length $h=\lambda$. This constructs
        $\PackedSeqRepresentation{w}{\AlphabetSize}{V_p}$ by appending the
        required zero padding to every string. The application is valid
        because $d\leq\lambda$ and
        $\AlphabetSize^\lambda\leq\AlphabetSize^\ell\leq m$.

      \item The inequality $\AlphabetSize^\lambda\leq m$ verifies the input-size
        hypothesis for the first prefix-select data structure from the claim.
        Apply its preprocessing algorithm, with parameters
        $\AlphabetSize$, $m$, and $\lambda$, to the packed sequence representation
        $\PackedSeqRepresentation{w}{\AlphabetSize}{V_p}$, and store a pointer
        to the output in $A_{\rm sel}[p/\lambda]$.
      \end{enumerate}
      Discard $\mathcal W_{p,d}$, $A_{\rm ptr}$, $A_{\rm len}$, and the packed
      representation of $W_{p,d}$ before processing the next value of $p$.
      After processing all values of $p$, discard the generalized-wavelet-tree
      data structure.

    \item\label{step:prefix-select-layering-coarse-structure} Let
      $W^{\rm pref}[1\dd m]$ and $W^{\rm suf}[1\dd m]$ denote the sequences
      satisfying $W^{\rm pref}[j]=W[j][1\dd h\lambda]$ and
      $W^{\rm suf}[j]=W[j][h\lambda+1\dd\ell]$ for every $j\in[1\dd m]$.
      Construct the coarse structure in three substeps.
      \begin{enumerate}

      \item\label{step:prefix-select-layering-coarse-partition} Apply the
        partition operation of
        \cref{pr:packed-pointwise-concatenation-partition} to
        $\PackedSeqRepresentation{w}{\AlphabetSize}{W}$, with alphabet size
        $\AlphabetSize$, sequence length $m$, and component lengths $h\lambda$
        and $\ell-h\lambda$. Retain
        $\PackedSeqRepresentation{w}{\AlphabetSize}{W^{\rm pref}}$, and
        discard $\PackedSeqRepresentation{w}{\AlphabetSize}{W^{\rm suf}}$
        and the packed input. This application is valid because
        $1\leq h\lambda\leq\ell$ and
        $\AlphabetSize^\ell\leq m$ implies
        $\AlphabetSize\leq m$ and $\ell<m$.

      \item\label{step:prefix-select-layering-coarse-encoding} Apply
        \cref{pr:packed-character-block-encoding} to
        $\PackedSeqRepresentation{w}{\AlphabetSize}{W^{\rm pref}}$ with
        alphabet size $\AlphabetSize$, sequence length $m$, block length $\lambda$,
        and input-length and output-length parameters both equal to $h$. This
        constructs $\PackedSeqRepresentation{w}{k}{U}$. The application is
        valid because $\lambda h\leq\ell$, and hence
        $k^h=\AlphabetSize^{\lambda h}\leq\AlphabetSize^\ell\leq m$.

      \item The inequality $k^h\leq m$ verifies the input-size hypothesis for
        the second prefix-select data structure from the claim. Apply its
        preprocessing algorithm, with parameters $k$, $m$, and $h$, to the
        packed sequence representation
        $\PackedSeqRepresentation{w}{k}{U}$, and retain the resulting data
        structure.
      \end{enumerate}
      Discard the packed representation of $W^{\rm pref}$ after the
      prefix-select structure has been constructed.
    \end{enumerate}

    Constructing the data structure from
    \cref{pr:generalized-wavelet-tree-construction} takes $\bigO(T_0)$ time.
    Its queries used to construct the sequences $V_p$ take
    \[
      \bigO\!\left(
        \AlphabetSize^\ell
        +m\left\lceil\frac{\ell}{\lambda}\right\rceil
          \min\!\left\{1,
            \frac{(\lambda\log\AlphabetSize)^2}{\log m}
          \right\}
      \right)
    \]
    time. Replacing the minimum by $1$ gives $\bigO(m\ell)$. Replacing it by
    $\tfrac{(\lambda\log\AlphabetSize)^2}{\log m}$ gives
    $\bigO(\tfrac{m(\ell\log\AlphabetSize)^2}{\log m})$. For the bound
    $\bigO(\tfrac{m\ell\log\AlphabetSize}{\sqrt{\log m}})$, the minimum is
    bounded by $\tfrac{(\lambda\log\AlphabetSize)^2}{\log m}$ when
    $\lambda\log\AlphabetSize\leq\sqrt{\log m}$ and by $1$ otherwise. The
    additive term $\AlphabetSize^\ell$ satisfies all three bounds because
    $\AlphabetSize^\ell\leq m$ and
    $\AlphabetSize^\ell
    =\bigO(m\ell\log\AlphabetSize/\log m)$.
    Thus, these queries take $\bigO(T_0)$ time.
    Across all values of $p$, constructing $A_{\rm ptr}$ and $A_{\rm len}$
    and applying concatenation and character-block encoding takes
    $\bigO(\AlphabetSize^\ell+\ell
      +m\ell\log\AlphabetSize/\log m)=\bigO(T_0)$ time. The partition and
    character-block encoding used to construct $U$ take
    $\bigO(1+m\ell\log\AlphabetSize/\log m)=\bigO(T_0)$ time. The times used
    in Steps~\ref{step:prefix-select-layering-parameters}
    and~\ref{step:prefix-select-layering-lookup-arrays} are also
    $\bigO(T_0)$. Adding the invocations of the two prefix-select
    preprocessing algorithms gives total preprocessing time
    $\bigO(T_0+\lceil\ell/\lambda\rceil P_{t,1}+P_{t,2})$.
    During Step~\ref{step:prefix-select-layering-local-structures}, the
    packed input, generalized-wavelet-tree structure, $\mathcal W_{p,d}$,
    $A_{\rm ptr}$, $A_{\rm len}$, the packed representations of $W_{p,d}$ and
    $V_p$, and the working space of the concatenation and character-block
    encoding use $\bigO(m\ell\log\AlphabetSize)$ additional bits. The packed
    input and the packed representations of $W^{\rm pref}$ and
    $W^{\rm suf}$ coexist with the working space of the partition in
    Step~\ref{step:prefix-select-layering-coarse-structure}. After
    $W^{\rm suf}$ is discarded, the packed representations of
    $W^{\rm pref}$ and $U$ coexist with the working space of the
    character-block encoding. The space in either case is
    $\bigO(m\ell\log\AlphabetSize)$ bits. This additional space,
    $A_{\rm pow}$, the string-encoding structure, and the boundary, rank, and
    pattern arrays use $\bigO(m\ell\log\AlphabetSize)$ bits, which is
    dominated by $\bigO(\lceil\ell/\lambda\rceil S_1+S_2)$. The two prefix-select
    preprocessing algorithms are invoked sequentially, so their preprocessing
    spaces contribute at most $\max\{P_{s,1},P_{s,2}\}$ bits. Thus, the peak
    space usage is
    $\bigO(\lceil\ell/\lambda\rceil S_1+S_2
    +\max\{P_{s,1},P_{s,2}\})$ bits.
    \qedhere
  \end{enumerate}
\end{proof}

\begin{lemma}[Divisible layering lemma]
  \label{lem:prefix-select-divisible-layering}
  Let $\AlphabetSize,m\in\Z_{\geq2}$ and $\ell\in\Z_{\geq1}$ satisfy
  $\AlphabetSize^\ell\leq m$. Consider the word RAM model with word size
  $w=c\log m$, where $c\geq2$ is a constant. Assume that there exists a
  data structure answering prefix select queries
  (\cref{def:prefix-range-queries}) that, given
  $\PackedSeqRepresentation{w}{k}{V}$
  (\cref{def:packed-sequence-representation}) for a sequence $V[1\dd m]$
  of $m$ strings of length $h$ over $[0\dd k)$, for every
  $k\in\Z_{\geq2}$ and $h\in\Z_{\geq1}$ satisfying $k^h\leq m$, has the
  following complexities:
  \begin{itemize}
  \item space usage $S(k,h,m)$ bits,
  \item preprocessing time $P_t(k,h,m)$,
  \item preprocessing space $P_s(k,h,m)$ bits,
  \item query time $Q(k,h,m)$.
  \end{itemize}

  Consider a sequence $(\lambda_i)_{i\in[0\dd q]}$, where $q\geq1$, such
  that $1=\lambda_0\leq\lambda_1\leq\cdots\leq\lambda_q=\ell$,
  $\lambda_{i-1}$ is a proper divisor of $\lambda_i$ for every
  $i\in[1\dd q)$, and $\lambda_{q-1}\mid\lambda_q$. Then there exists a
  data structure that, given
  $\PackedSeqRepresentation{w}{\AlphabetSize}{W}$ for a sequence
  $W[1\dd m]$ of $m$ strings of length $\ell$ over
  $[0\dd\AlphabetSize)$, answers prefix select queries on $W$ and has the
  following complexities:
  \begin{itemize}
  \item space usage
    $\bigO(\sum_{i\in[1\dd q]}r_i\cdot S(k_i,h_i,m))$ bits,
  \item preprocessing time
    \[
      \bigO\!\left(
        m\min\!\left\{
          \ell,
          \frac{\ell\log\AlphabetSize}{\sqrt{\log m}},
          \frac{(\ell\log\AlphabetSize)^2}{\log m}
        \right\}
        +\sum_{i\in[1\dd q]}r_i\cdot P_t(k_i,h_i,m)
      \right),
    \]
  \item preprocessing space
    $\bigO(\sum_{i\in[1\dd q]}r_i\cdot S(k_i,h_i,m)
    +\max_{i\in[1\dd q]}P_s(k_i,h_i,m))$ bits,
  \item query time $\bigO(\sum_{i\in[1\dd q]}Q(k_i,h_i,m))$,
  \end{itemize}
  where, for every $i\in[1\dd q]$,
  $k_i:=\AlphabetSize^{\lambda_{i-1}}$,
  $h_i:=\lambda_i/\lambda_{i-1}$, and
  $r_i:=\ell/\lambda_i$.
\end{lemma}
\begin{proof}

  If $m<64$, perform the construction directly, and hence the claim follows
  immediately. Thus, assume that $m\geq64$. Every
  $\lambda_i$ divides $\ell$. Moreover,
  $\lambda_i\geq2\lambda_{i-1}$ for every $i\in[1\dd q)$, and hence
  $q=\bigO(1+\log\ell)
  =\bigO(\min\{\ell,\sqrt{\log m}\})$. The monotonicity of $x/\log x$ and
  $\AlphabetSize^\ell\leq m$ give
  $\AlphabetSize^\ell\log m
  =\bigO(m\ell\log\AlphabetSize)$.

  Let $F_X$,
  $(X_{p,j})_{j\in[1\dd\AlphabetSize^p]}$, $W_{X,d}$, $W_{p,d}$, $b_X$,
  $e_X$, $L_p^{\rm beg}$, $L_p^{\rm end}$, and $L_{p,d}^{\rm rank}$ be
  defined as in the proof of
  \cref{lem:prefix-select-final-layer-composition}.

  For every $i\in[1\dd q]$, every
  $p\in[0\dd\ell-\lambda_i]$, and every $j\in[1\dd m]$, let
  $V_{i,p}[j]\in[0\dd k_i)^{h_i}$ denote the string satisfying

  \[
    V_{i,p}[j][t]=
      \Val{\AlphabetSize}{W_{p,\lambda_i}[j]
      [(t-1)\lambda_{i-1}+1\dd t\lambda_{i-1}]}
      \qquad\text{for every }t\in[1\dd h_i].
  \]

  For every $i\in[1\dd q]$ and every
  $d\in[1\dd\lambda_i]$ divisible by $\lambda_{i-1}$, letting
  $t:=d/\lambda_{i-1}$, we define the array
  $L_{i,d}^{\rm pat}[0\dd\AlphabetSize^d)$ as follows: For every
  $Y\in[0\dd\AlphabetSize)^d$, $L_{i,d}^{\rm pat}[\Val{\AlphabetSize}{Y}]$
  stores the packed representation over alphabet
  $[0\dd k_i)$ of the length-$t$ string whose $s$th character is
  $\Val{\AlphabetSize}{Y[(s-1)\lambda_{i-1}+1\dd
  s\lambda_{i-1}]}$.

  \DSComponents
  The data structure consists of the following components:
  \begin{enumerate}

  \item The arrays $A_{\rm pow}[0\dd\ell]$ and $A_\lambda[0\dd q]$, and the
    data structure from \cref{pr:val-encoding} with parameters $m$ and
    $\AlphabetSize$. The arrays satisfy $A_{\rm pow}[j]=\AlphabetSize^j$ for
    every $j\in[0\dd\ell]$ and $A_\lambda[i]=\lambda_i$ for every
    $i\in[0\dd q]$. These objects use
    $\bigO((\ell+q+\sqrt m)\log m)$ bits.

  \item The arrays $L_p^{\rm beg}[0\dd\AlphabetSize^p)$ and
    $L_p^{\rm end}[0\dd\AlphabetSize^p)$ for every $p\in[0\dd\ell]$. They
    are stored in plain form. The arrays contain
    $\bigO(\AlphabetSize^\ell)$ entries and use
    $\bigO(\AlphabetSize^\ell\log m)$ bits.

  \item The arrays $L_{i,t\lambda_{i-1}}^{\rm pat}$ for every
    $i\in[1\dd q]$ and $t\in[1\dd h_i]$ and, for every
    $i\in[1\dd q]$, $a\in[0\dd r_i)$, and $t\in[1\dd h_i]$, a separate copy
    of $L_{a\lambda_i,t\lambda_{i-1}}^{\rm rank}$ associated with $i$. For fixed $i$,
    the exponents
    $a\lambda_i+t\lambda_{i-1}$ enumerate the positive multiples of
    $\lambda_{i-1}$ at most $\ell$. Hence, the rank arrays contain fewer
    than $2\AlphabetSize^\ell$ entries. The pattern arrays contain fewer
    than $2\AlphabetSize^{\lambda_i}\leq2\AlphabetSize^\ell$ entries.
    Every rank entry uses one word. Every pattern occupies one word because
    $h_i\lceil\log k_i\rceil\leq2\lambda_i\log\AlphabetSize
    \leq2\log m\leq w$. Thus, this component uses
    $\bigO(q\AlphabetSize^\ell\log m)$ bits.

  \item For every $i\in[1\dd q]$ and $a\in[0\dd r_i)$, the data structure
    from the claim that answers prefix select queries, constructed for
    $V_{i,a\lambda_i}$,
    together with the arrays $A_i^{\rm ds}[0\dd r_i)$ for every
    $i\in[1\dd q]$. The entry $A_i^{\rm ds}[a]$ stores a pointer to the
    structure for $V_{i,a\lambda_i}$. Each structure is applicable because
    $k_i^{h_i}=\AlphabetSize^{\lambda_i}\leq m$. These structures use
    $\bigO(\sum_{i\in[1\dd q]}r_i\cdot S(k_i,h_i,m))$ bits.
  \end{enumerate}

  An exact prefix-select structure for $m$ strings of length $h$ over
  $[0\dd k)$ needs $\Omega(mh\log k)$ bits because its answers to all
  length-$h$ patterns determine the input. Consequently, for every
  $i\in[1\dd q]$, we have
  $r_i\cdot S(k_i,h_i,m)=\Omega(m\ell\log\AlphabetSize)$. The sum of these
  bounds dominates $\bigO(q\AlphabetSize^\ell\log m)$ and the space used by
  $A_{\rm pow}$, $A_\lambda$, the string-encoding structure, and the boundary,
  rank, and pattern arrays. Therefore, the total space usage is
  $\bigO(\sum_{i\in[1\dd q]}r_i\cdot S(k_i,h_i,m))$ bits.

  \DSQueries
  Let $X\in[0\dd\AlphabetSize)^{\leq\ell}$ and $r\in[1\dd m]$. Given the
  length $|X|$, the packed representation
  $\PackedRepresentation{w}{\AlphabetSize}{X}$ of the string $X$, and the
  position $r$, we compute $\PrefixSelect{W}{r}{X}$ as follows:
  \begin{enumerate}

  \item Using \cref{pr:val-encoding}, in $\bigO(1)$ time compute
    $x_{\rm cur}:=\Val{\AlphabetSize}{X}$.

  \item Retrieve $b:=L_{|X|}^{\rm beg}[x_{\rm cur}]$ and
    $e_X:=L_{|X|}^{\rm end}[x_{\rm cur}]$. If $r>e_X-b$, return $\infty$.
    Otherwise, set $i:=1$, $e:=|X|$, and $r_{\rm cur}:=r$. Throughout the
    query algorithm, maintain the following invariants:
    \begin{itemize}
    \item \emph{Invariant 1:} If $e>0$, then $i\in[1\dd q]$ and
      $\lambda_{i-1}\mid e$.
    \item \emph{Invariant 2:}
      $x_{\rm cur}=\Val{\AlphabetSize}{X[1\dd e]}$.
    \item \emph{Invariant 3:}
      $r_{\rm cur}\in[1\dd F_{X[1\dd e]}]$.
    \item \emph{Invariant 4:}
      $\PrefixSelect{W}{r}{X}
      =\PrefixSelect{W}{r_{\rm cur}}{X[1\dd e]}$.
    \end{itemize}
    The invariants hold after the first two steps.

  \item While $e>0$, perform the following steps:
    \begin{enumerate}

    \item Set $p:=A_\lambda[i]\cdot
      \lfloor(e-1)/A_\lambda[i]\rfloor$, $d:=e-p$, and
      $t:=d/A_\lambda[i-1]$.

    \item Set $x_{\rm pref}:=\lfloor
      x_{\rm cur}/A_{\rm pow}[d]\rfloor$ and
      $x_{\rm suf}:=x_{\rm cur}\bmod A_{\rm pow}[d]$.

    \item Retrieve $b_{\rm pref}:=L_p^{\rm beg}[x_{\rm pref}]$, retrieve
      $r_{\rm base}:=L_{p,d}^{\rm rank}[x_{\rm cur}]$ from the copy associated
      with $i$, and retrieve
      $\gamma:=L_{i,d}^{\rm pat}[x_{\rm suf}]$.

    \item\label{step:divisible-layering-query-level} Query the prefix-select
      structure constructed for $V_{i,p}$ and pointed to by
      $A_i^{\rm ds}[p/A_\lambda[i]]$, using occurrence number
      $r_{\rm base}+r_{\rm cur}$ and the length-$t$ pattern represented by
      $\gamma$. Store the answer as $r_{\rm global}$.

    \item Set $r_{\rm cur}:=r_{\rm global}-b_{\rm pref}$,
      $x_{\rm cur}:=x_{\rm pref}$, $e:=p$, and $i:=i+1$.
    \end{enumerate}
    Every substep other than the prefix select query in
    Step~\ref{step:divisible-layering-query-level} takes
    $\bigO(1)$ time. To justify an iteration, let $\hat i$, $\hat e$,
    $\hat x$, and $\hat r$ denote the values of $i$, $e$, $x_{\rm cur}$,
    and $r_{\rm cur}$ at its beginning. Let $\hat p$ denote
    $\lambda_{\hat i}\lfloor(\hat e-1)/\lambda_{\hat i}\rfloor$, let
    $\hat d$ denote $\hat e-\hat p$, let $P$ denote $X[1\dd\hat p]$, and let
    $Y$ denote $X(\hat p\dd\hat e]$. Invariant~1 and the divisibility of the
    levels
    imply that $\hat p=a\lambda_{\hat i}$ for some
    $a\in[0\dd r_{\hat i})$ and
    $\hat d=t\lambda_{\hat i-1}$ for some
    $t\in[1\dd h_{\hat i}]$. Thus, every accessed array and structure is
    stored. Invariant~2 gives
    $\hat x=A_{\rm pow}[\hat d]\cdot
    \Val{\AlphabetSize}{P}+\Val{\AlphabetSize}{Y}$, so the second substep
    computes the values of $P$ and $Y$. The first $b_P$ strings in
    $V_{\hat i,\hat p}$ correspond to the first $b_P$ strings in
    $W_{\hat p,\hat d}$. A string in $V_{\hat i,\hat p}$ starts with the
    pattern represented by $\gamma$ exactly when the corresponding string in
    $W_{\hat p,\hat d}$ equals $Y$. Hence, $r_{\rm base}$ counts the relevant
    occurrences before the segment for $P$, and this segment contains exactly $F_{PY}$
    of them. The pattern represented by $\gamma$ therefore occurs at least
    $r_{\rm base}+F_{PY}$ times in $V_{\hat i,\hat p}$. Invariant~3 gives
    $\hat r\leq F_{PY}$. Hence, $r_{\rm base}+\hat r\leq m$, and the query to
    the prefix-select structure constructed for $V_{\hat i,\hat p}$ returns
    a position $r_{\rm global}\in(b_P\dd e_P]$. Let
    $r'$ denote $r_{\rm global}-b_P$. Then $r'\in[1\dd F_P]$, and the $r'$th
    occurrence of $P$ in $W$ is the $\hat r$th occurrence of $PY$.
    Consequently,
    $\PrefixSelect{W}{\hat r}{PY}=\PrefixSelect{W}{r'}{P}$. The assignments
    in the fifth substep preserve Invariants~2--4. If the new value of $e$
    is positive, then $\hat i<q$ and $\lambda_{\hat i}\mid e$, so
    Invariant~1 is also preserved. If $\hat i=q$, then $\hat p=0$ and the
    loop terminates.

  \item Once $e=0$, Invariant~4 gives
    $\PrefixSelect{W}{r}{X}
    =\PrefixSelect{W}{r_{\rm cur}}{\emptystring}=r_{\rm cur}$. Return
    $r_{\rm cur}$.
  \end{enumerate}

  The loop executes at most once for every $i\in[1\dd q]$. Every prefix select
  query in the loop takes at least constant time,
  so its cost dominates the constant work in the same iteration. The query takes
  $\bigO(\sum_{i\in[1\dd q]}Q(k_i,h_i,m))$ time.

  \DSConstruction
  Given $\PackedSeqRepresentation{w}{\AlphabetSize}{W}$, construct the data
  structure as follows:
  \begin{enumerate}

  \item \emph{Construct the parameter arrays and value-encoding data structure}:
    Set $A_{\rm pow}[0]:=1$ and, for every $j\in[1\dd\ell]$, set
    $A_{\rm pow}[j]:=\AlphabetSize\cdot A_{\rm pow}[j-1]$. For every
    $i\in[0\dd q]$, set $A_\lambda[i]:=\lambda_i$. Apply
    \cref{pr:val-encoding} with parameters $m$ and $\AlphabetSize$. For use
    during construction, compute and keep
    $\lceil\log\AlphabetSize\rceil$ and $\lceil\log k_i\rceil$ for every
    $i\in[1\dd q]$ by repeated doubling. This step takes
    $\bigO(\ell+q\log m+\sqrt m)$ time,
    has peak space usage $\bigO((\ell+q+\sqrt m)\log m)$ bits. The arrays,
    the value-encoding data structure, and the logarithms stored for later steps
    occupy
    $\bigO((\ell+q+\sqrt m)\log m)$ bits.

  \item \emph{Construct the boundary arrays}: Apply
    \cref{pr:packed-prefix-frequencies}, with parameters $\AlphabetSize$, $m$,
    and $\ell$, to the packed sequence representation
    $\PackedSeqRepresentation{w}{\AlphabetSize}{W}$, using a
    temporary output array $A_{\rm pfreq}[0\dd2\AlphabetSize^\ell)$. Set $u:=1$.
    For every $p\in[0\dd\ell]$, allocate
    $L_p^{\rm beg}[0\dd u)$ and $L_p^{\rm end}[0\dd u)$, and set a counter
    $s:=0$. For every $x\in[0\dd u)$ in increasing order, set
    $L_p^{\rm beg}[x]:=s$, set
    $s:=s+A_{\rm pfreq}[u+x]$, and set $L_p^{\rm end}[x]:=s$. At the
    beginning of the iteration for $p$, we have $u=\AlphabetSize^p$. Hence, for the unique
    length-$p$ string $X$ satisfying $x=\Val{\AlphabetSize}{X}$, the index
    $u+x$ equals $\BasicInt{\AlphabetSize}{X}$. Increasing $x$ enumerates the
    strings lexicographically, so these assignments construct the arrays from
    their definitions. If $p<\ell$, set $u:=\AlphabetSize\cdot u$. This step takes
    $\bigO(m\ell\log\AlphabetSize/\log m)$ time, has peak space usage
    $\bigO(m\ell\log\AlphabetSize)$ bits. The constructed boundary arrays
    occupy
    $\bigO(\AlphabetSize^\ell\log m)$ bits.

  \item \emph{Construct the rank and pattern arrays}: Proceed as follows:
    \begin{enumerate}

    \item For every $i\in[1\dd q]$ and $t\in[1\dd h_i]$, set
      $d:=t\lambda_{i-1}$ and allocate
      $L_{i,d}^{\rm pat}[0\dd k_i^t)$. Enumerate
      $y\in[0\dd k_i^t)$ in increasing order, starting with the packed string
      $\zero^t$. Store the current length-$t$ base-$k_i$ representation in
      $L_{i,d}^{\rm pat}[y]$. If $y<k_i^t-1$, obtain the representation for
      $y+1$ by adding one in base $k_i$ and propagating carries. The number
      of changes to successive characters decreases geometrically, so all
      $k_i^t$ representations require $\bigO(k_i^t)$ character updates in
      packed strings. For the unique $Y\in[0\dd\AlphabetSize)^d$ satisfying
      $y=\Val{\AlphabetSize}{Y}$, partitioning $Y$ into consecutive
      length-$\lambda_{i-1}$ blocks and encoding each block by its
      base-$\AlphabetSize$ value gives a string whose base-$k_i$ value is $y$.
      Hence,
      increasing $y$ produces precisely the packed strings specified in the
      definition.

    \item For every $i\in[1\dd q]$, $a\in[0\dd r_i)$, and
      $t\in[1\dd h_i]$, set $p:=a\lambda_i$,
      $d:=t\lambda_{i-1}$, and $B:=A_{\rm pow}[d]$, and allocate the
      separate copy of $L_{p,d}^{\rm rank}[0\dd A_{\rm pow}[p+d])$ associated
      with $i$. For every
      $y\in[0\dd B)$, set a counter $s:=0$. Then, for every
      $x\in[0\dd A_{\rm pow}[p])$, set
      $z:=x\cdot B+y$, set $L_{p,d}^{\rm rank}[z]:=s$, and set
      $s:=s+L_{p+d}^{\rm end}[z]-L_{p+d}^{\rm beg}[z]$. If
      $x=\Val{\AlphabetSize}{X}$ and
      $y=\Val{\AlphabetSize}{Y}$, then
      $z=\Val{\AlphabetSize}{XY}$. Before processing $X$, the counter is
      $\PrefixRank{W_{p,d}}{b_X}{Y}$, as required.
    \end{enumerate}

    For every fixed $i$, either substep writes fewer than
    $2\AlphabetSize^\ell$ entries. Thus, this step takes
    $\bigO(q\AlphabetSize^\ell)$ time. The constructed rank and pattern arrays
    occupy
    $\bigO(q\AlphabetSize^\ell\log m)$ bits. The time is
    $\bigO(m\ell)$ because $q=\bigO(\ell)$ and
    $\AlphabetSize^\ell\leq m$. It is
    $\bigO(m\ell\log\AlphabetSize/\sqrt{\log m})$ because
    $q=\bigO(\sqrt{\log m})$ and
    $\AlphabetSize^\ell
    =\bigO(m\ell\log\AlphabetSize/\log m)$. Finally, it is
    $\bigO(m(\ell\log\AlphabetSize)^2/\log m)$ because
    $q=\bigO(\ell\log\AlphabetSize)$ and the same bound on
    $\AlphabetSize^\ell$ applies.

  \item \emph{Construct the data structures from the claim}: Apply
    \cref{pr:generalized-wavelet-tree-construction}, with parameters
    $\AlphabetSize$, $m$, and $\ell$, to the packed sequence representation
    $\PackedSeqRepresentation{w}{\AlphabetSize}{W}$. For every
    $i\in[1\dd q]$, allocate $A_i^{\rm ds}[0\dd r_i)$ and, for every
    $a\in[0\dd r_i)$, set $p:=a\lambda_i$ and perform the following
    steps:
    \begin{enumerate}

    \item Query the structure from
      \cref{pr:generalized-wavelet-tree-construction} with parameters $p$ and
      $\lambda_i$. Let $A_{p,\lambda_i}[0\dd\AlphabetSize^p)$ denote its
      returned array. Allocate temporary arrays
      $A_{\rm ptr}[1\dd\AlphabetSize^p]$ and
      $A_{\rm len}[1\dd\AlphabetSize^p]$. For every
      $x\in[0\dd\AlphabetSize^p)$ in increasing order, set
      $f:=L_p^{\rm end}[x]-L_p^{\rm beg}[x]$, store in
      $A_{\rm ptr}[x+1]$ a pointer to $A_{p,\lambda_i}[x]$, and set
      $A_{\rm len}[x+1]:=\lambda_i\cdot f$. Apply
      \cref{pr:packed-representation}\eqref{pr:packed-representation-concat}
      to the $\AlphabetSize^p$ representations and lengths, with length
      parameter $m\ell$. The call is valid because
      $\AlphabetSize^p\leq m\leq m\ell$, the supplied lengths sum to
      $m\lambda_i$, and
      $w\geq2\log m>\log(m\ell)$. Since
      increasing $x$ enumerates the prefixes lexicographically, the result is
      $\PackedSeqRepresentation{w}{\AlphabetSize}{W_{p,\lambda_i}}$.

    \item Apply \cref{pr:packed-character-block-encoding} to the packed sequence
      representation
      $\PackedSeqRepresentation{w}{\AlphabetSize}{W_{p,\lambda_i}}$, with
      alphabet size $\AlphabetSize$, source length $m$, string length
      $\lambda_i$, block length $\lambda_{i-1}$, and $g=h=h_i$, to construct
      $\PackedSeqRepresentation{w}{k_i}{V_{i,p}}$. This application is valid
      because
      $\AlphabetSize^{\lambda_{i-1}h_i}
      =\AlphabetSize^{\lambda_i}\leq m$.

    \item Apply the preprocessing algorithm for the data structure from the
      claim, with parameters $k_i$, $m$, and $h_i$, to the packed sequence
      representation
      $\PackedSeqRepresentation{w}{k_i}{V_{i,p}}$, and store a pointer to the
      returned structure in $A_i^{\rm ds}[a]$.
    \end{enumerate}

    For every fixed $i$,
    $\sum_{a=0}^{r_i-1}\AlphabetSize^{a\lambda_i}
    =\bigO(\AlphabetSize^\ell)$. The queries to the data structure from
    \cref{pr:generalized-wavelet-tree-construction} and the packed concatenations
    therefore take
    \[
      \bigO\!\left(
        q\AlphabetSize^\ell+m
        \sum_{i\in[1\dd q]}r_i
        \min\!\left\{1,
          \frac{\lambda_i^2(\log\AlphabetSize)^2}{\log m}
        \right\}
      \right)
    \]
    time. This quantity is bounded by each of
    $\bigO(m\ell)$,
    $\bigO(m\ell\log\AlphabetSize/\sqrt{\log m})$, and
    $\bigO(m(\ell\log\AlphabetSize)^2/\log m)$. The bound
    $\bigO(m\ell)$ follows from
    $\sum_{i\in[1\dd q]}r_i=\bigO(\ell)$, and the bound
    $\bigO(m(\ell\log\AlphabetSize)^2/\log m)$ follows from
    $\sum_{i\in[1\dd q]}r_i\lambda_i^2=\bigO(\ell^2)$. The term
    $q\AlphabetSize^\ell$ satisfies these two bounds because
    $q=\bigO(\ell)$, $\AlphabetSize^\ell\leq m$,
    $q=\bigO(\ell\log\AlphabetSize)$, and
    $\AlphabetSize^\ell
    =\bigO(m\ell\log\AlphabetSize/\log m)$. To prove the bound
    $\bigO(m\ell\log\AlphabetSize/\sqrt{\log m})$, consider separately the
    indices $i\in[1\dd q]$ satisfying
    $\lambda_i\leq\sqrt{\log m}/\log\AlphabetSize$ and those satisfying
    $\lambda_i>\sqrt{\log m}/\log\AlphabetSize$. For the indices satisfying
    $\lambda_i\leq\sqrt{\log m}/\log\AlphabetSize$, the minimum is bounded by
    $\lambda_i^2(\log\AlphabetSize)^2/\log m$. Since
    $r_i=\ell/\lambda_i$ and the values $\lambda_i$ grow geometrically, the
    contribution of these indices to the sum is
    $\bigO(\ell\log\AlphabetSize/\sqrt{\log m})$. For the remaining indices,
    the minimum is bounded by $1$. The values $r_i=\ell/\lambda_i$ decrease
    geometrically, so the contribution of these indices satisfies the same
    bound. Finally,
    $q\AlphabetSize^\ell
    =\bigO(m\ell\log\AlphabetSize/\sqrt{\log m})$ follows from
    $q=\bigO(\sqrt{\log m})$ and
    $\AlphabetSize^\ell
    =\bigO(m\ell\log\AlphabetSize/\log m)$.
    Applying \cref{pr:packed-character-block-encoding} takes
    $\bigO(1+m\lambda_i\log\AlphabetSize/\log m)$ time for every
    $a\in[0\dd r_i)$. All applications therefore take
    $\bigO(\ell+qm\ell\log\AlphabetSize/\log m)$ time. This is bounded by
    $\bigO(m\ell)$ because $q\log\AlphabetSize\leq\log m$, by
    $\bigO(m\ell\log\AlphabetSize/\sqrt{\log m})$ because
    $q=\bigO(\sqrt{\log m})$, and by
    $\bigO(m(\ell\log\AlphabetSize)^2/\log m)$ because
    $q=\bigO(\ell\log\AlphabetSize)$. Calls to the preprocessing algorithm for
    the data structure from the claim take
    $\bigO(\sum_{i\in[1\dd q]}r_i\cdot P_t(k_i,h_i,m))$ time. Including
    the construction of \cref{pr:generalized-wavelet-tree-construction}, this
    proves the preprocessing-time bound in the claim.
    We process one pair $(i,a)$ at a time. While processing this pair, the data
    structure from \cref{pr:generalized-wavelet-tree-construction}, the
    returned array $A_{p,\lambda_i}$, the temporary arrays $A_{\rm ptr}$ and
    $A_{\rm len}$, the concatenated sequence $W_{p,\lambda_i}$, the resulting
    packed sequence representation of $V_{i,p}$, and the additional working
    space used by \cref{pr:packed-character-block-encoding} together use
    $\bigO(m\ell\log\AlphabetSize)$ bits. The data structures from the claim
    already constructed for the sequences $V_{i,p}$ use at most
    $\bigO(\sum_{i\in[1\dd q]}r_i\cdot S(k_i,h_i,m))$ bits. As shown above
    when bounding the space of the components, this sum also dominates the
    $\bigO(m\ell\log\AlphabetSize)$ temporary space. Applications of the
    preprocessing algorithm for the data structure from the claim are run one
    at a time, so at
    most $\max_{i\in[1\dd q]}P_s(k_i,h_i,m)$ additional bits are needed for
    one such algorithm. Thus, the peak space usage is
    $\bigO(\sum_{i\in[1\dd q]}r_i\cdot S(k_i,h_i,m)
    +\max_{i\in[1\dd q]}P_s(k_i,h_i,m))$ bits.
  \end{enumerate}

  The quantity
  $m\min\{\ell,\ell\log\AlphabetSize/\sqrt{\log m},
  (\ell\log\AlphabetSize)^2/\log m\}$ in the preprocessing-time bound is
  $\Omega(m/\log m)$ and dominates the
  $\bigO(\ell+q\log m+\sqrt m)$ time used to construct $A_{\rm pow}$,
  $A_\lambda$, the string-encoding structure, and the character widths.
  Together with the bounds proved for constructing the boundary, rank, and
  pattern arrays and the structures for $V_{i,p}$, this establishes the claimed
  preprocessing time. The space bounds for the components and the construction
  establish the claimed preprocessing space.
\end{proof}

\begin{lemma}[General layering lemma]
  \label{lem:reduce-prefix-select-to-prefix-select-large-space}
  Let $\AlphabetSize,m\in\Z_{\geq2}$ and $\ell\in\Z_{\geq1}$ satisfy
  $\AlphabetSize^\ell\leq m$. Consider the word RAM model with word size
  $w=c\log m$, where $c\geq2$ is a constant. Assume that there exists a
  data structure answering prefix select queries
  (\cref{def:prefix-range-queries}) that, given
  $\PackedSeqRepresentation{w}{k}{V}$ for a sequence $V[1\dd m]$ of
  $m$ strings of length $h$ over $[0\dd k)$, for every
  $k\in\Z_{\geq2}$ and $h\in\Z_{\geq1}$ satisfying $k^h\leq m$, has the
  following complexities:
  \begin{itemize}
  \item space usage $S(k,h,m)$ bits,
  \item preprocessing time $P_t(k,h,m)$,
  \item preprocessing space $P_s(k,h,m)$ bits,
  \item query time $Q(k,h,m)$.
  \end{itemize}

  Consider a sequence $(\lambda_i)_{i\in[0\dd q]}$, where $q\geq1$, such
  that $1=\lambda_0\leq\lambda_1\leq\cdots\leq\lambda_q=\ell$ and
  $\lambda_{i-1}$ is a proper divisor of $\lambda_i$ for every
  $i\in[1\dd q)$. Then there exists a data structure that, given
  $\PackedSeqRepresentation{w}{\AlphabetSize}{W}$
  (\cref{def:packed-sequence-representation}) for a sequence
  $W[1\dd m]$ of $m$ strings of length $\ell$ over
  $[0\dd\AlphabetSize)$, answers prefix select queries on $W$ and has the
  following complexities:
  \begin{itemize}
  \item space usage
    $\bigO(\sum_{i\in[1\dd q]}r_i\cdot S(k_i,h_i,m))$ bits,
  \item preprocessing time
    \[
      \bigO\!\left(
        m\min\!\left\{
          \ell,
          \frac{\ell\log\AlphabetSize}{\sqrt{\log m}},
          \frac{(\ell\log\AlphabetSize)^2}{\log m}
        \right\}
        +\sum_{i\in[1\dd q]}r_i\cdot P_t(k_i,h_i,m)
      \right),
    \]
  \item preprocessing space
    $\bigO(\sum_{i\in[1\dd q]}r_i\cdot S(k_i,h_i,m)
    +\max_{i\in[1\dd q]}P_s(k_i,h_i,m))$ bits,
  \item query time $\bigO(\sum_{i\in[1\dd q]}Q(k_i,h_i,m))$,
  \end{itemize}
  where, for every $i\in[1\dd q]$,
  $k_i:=\AlphabetSize^{\lambda_{i-1}}$,
  $h_i:=\lfloor\lambda_i/\lambda_{i-1}\rfloor$, and
  $r_i:=\lceil\ell/\lambda_i\rceil$.
\end{lemma}
\begin{proof}
  If $\lambda_{q-1}\mid\ell$, apply
  \cref{lem:prefix-select-divisible-layering}. Its values $h_i$ and $r_i$
  agree with those in the present claim, so all four bounds follow directly.

  Suppose that $\lambda_{q-1}\nmid\ell$, and denote
  $a:=\lambda_{q-1}$. Necessarily, $q\geq2$ and $a<\ell$. Apply
  \cref{lem:prefix-select-divisible-layering} to sequences of $m$ strings of
  length $a$ over $[0\dd\AlphabetSize)$, using the levels
  $\lambda_0,\ldots,\lambda_{q-1}$. This gives a prefix-select structure
  with space usage
  $\bigO(\sum_{i\in[1\dd q)}(a/\lambda_i)
  S(k_i,h_i,m))$ bits, preprocessing time
  \[
    \bigO\!\left(
      m\min\!\left\{
        a,
        \frac{a\log\AlphabetSize}{\sqrt{\log m}},
        \frac{(a\log\AlphabetSize)^2}{\log m}
      \right\}
      +\sum_{i\in[1\dd q)}\frac{a}{\lambda_i}P_t(k_i,h_i,m)
    \right).
  \]
  Its preprocessing space is
  $\bigO(\sum_{i\in[1\dd q)}(a/\lambda_i)S(k_i,h_i,m)
  +\max_{i\in[1\dd q)}P_s(k_i,h_i,m))$ bits. Its query time is
  $\bigO(\sum_{i\in[1\dd q)}Q(k_i,h_i,m))$.

  This structure serves as the structure for length-$a$ strings in
  \cref{lem:prefix-select-final-layer-composition}. The prefix-select structure
  on the block-encoded sequence in that lemma is the prefix-select data
  structure from the claim, with
  parameters
  $k_q=\AlphabetSize^a$ and $h_q=\lfloor\ell/a\rfloor$.
  Its requirements hold because
  $k_q^{h_q}\leq\AlphabetSize^\ell\leq m$. We have
  \[
    \left\lceil\frac{\ell}{a}\right\rceil
      \frac{a}{\lambda_i}
      <\frac{2\ell}{\lambda_i}
      \leq2r_i
    \qquad\text{for every }i\in[1\dd q).
  \]
  Moreover, multiplying each of
  $a$, $a\log\AlphabetSize/\sqrt{\log m}$, and
  $(a\log\AlphabetSize)^2/\log m$ by
  $\lceil\ell/a\rceil$ gives at most a constant multiple of the
  corresponding expression with $\ell$ in place of $a$. Substituting the
  four bounds above into
  \cref{lem:prefix-select-final-layer-composition} gives the claimed space,
  preprocessing-time, and query-time bounds. For preprocessing space, denote
  by $S_1$ and $P_{s,1}$ the space usage and preprocessing space of the
  structure for length-$a$ strings. Its construction gives
  $P_{s,1}=\bigO(S_1+\max_{i\in[1\dd q)}P_s(k_i,h_i,m))$. The term
  $\lceil\ell/a\rceil S_1$ in
  \cref{lem:prefix-select-final-layer-composition} dominates $S_1$, and
  combining the two preprocessing-space maxima gives
  $\max_{i\in[1\dd q]}P_s(k_i,h_i,m)$. This proves the claimed
  preprocessing-space bound as well.
\end{proof}

\subsection{Putting Everything Together}\label{sec:prefix-select-summary}

\begin{theorem}\label{th:prefix-select-tradeoffs}
  Let $\AlphabetSize,m,\ell\in\Z_{\geq2}$ satisfy
  $\AlphabetSize^\ell\leq m$. Consider the word RAM model with word size
  $w=c\log m$, where $c\geq2$ is a constant. For every
  $B\in[2\dd\ell]$, there exist two data structures for the problem of
  indexing for prefix select queries over alphabet
  $[0\dd\AlphabetSize)$ (see \cref{sec:prefix-select-problem-def}) with the
  following complexities:
  \begin{itemize}
  \item A small-space tradeoff:
    \begin{itemize}
    \item space usage
      $\bigO(m\ell\log\AlphabetSize\log_B\ell)$ bits,
    \item preprocessing time
      $\bigO(m\min(\ell,
        \tfrac{\ell\log\AlphabetSize}{\sqrt{\log m}},
        \tfrac{(\ell\log\AlphabetSize)^2}{\log m}))$,
    \item preprocessing space
      $\bigO(m\ell\log\AlphabetSize\log_B\ell)$ bits,
    \item query time $\bigO(B\log_B\ell)$.
    \end{itemize}
  \item A large-space tradeoff:
    \begin{itemize}
    \item space usage
      $\bigO(Bm\ell\log\AlphabetSize\log_B\ell)$ bits,
    \item preprocessing time
      $\bigO(m\min(\ell,
        \tfrac{\ell\log\AlphabetSize}{\sqrt{\log m}},
        \tfrac{(\ell\log\AlphabetSize)^2}{\log m}))$,
    \item preprocessing space
      $\bigO(Bm\ell\log\AlphabetSize\log_B\ell)$ bits,
    \item query time $\bigO(\log_B\ell)$.
    \end{itemize}
  \end{itemize}
\end{theorem}
\begin{proof}
  Denote $\tau:=\sqrt{\log m}/\log\AlphabetSize$ and
  $\rho:=\min(\ell,\max(1,\tau))$. We define the index $q$ and the sequence
  $(\lambda_i)_{i\in[0\dd q]}$ in two cases.
  \begin{itemize}

  \item If $\rho=1$, denote $q:=\lceil\log_B\ell\rceil$. Let
    $(\lambda_i)_{i\in[0\dd q]}$ denote the sequence satisfying
    \[
      \lambda_i=
      \begin{cases}
        B^i, & \text{if }i\in[0\dd q),\\
        \ell, & \text{if }i=q.
      \end{cases}
    \]

  \item If $\rho>1$, denote
    $a:=\lceil\log_B\rho\rceil-1$,
    $c_\rho:=\lceil\rho/B^a\rceil$,
    $\mu:=c_\rho B^a$,
    $b:=\max(0,\lceil\log_B(\ell/\mu)\rceil)$, and
    $q:=a+b+1$. Let $(\lambda_i)_{i\in[0\dd q]}$ denote the sequence
    satisfying
    \[
      \lambda_i=
      \begin{cases}
        B^i, & \text{if }i\in[0\dd a],\\
        \mu B^{i-a-1}, & \text{if }i\in[a+1\dd a+b],\\
        \ell, & \text{if }i=q.
      \end{cases}
    \]
    The middle range is empty when $b=0$.
  \end{itemize}
  In the second case, $B^a<\rho\leq B^{a+1}$, and hence
  $c_\rho\in[2\dd B]$ and $\rho\leq\mu<\rho+B^a<2\rho$.
  Both definitions satisfy $\lambda_0=1$ and $\lambda_q=\ell$.
  Moreover, $\lambda_{i-1}$ is a proper divisor of $\lambda_i$ and
  $\lambda_i/\lambda_{i-1}\in[2\dd B]$ for every $i\in[1\dd q)$,
  whereas $1<\lambda_q/\lambda_{q-1}\leq B$.

  It holds $q=\bigO(\log_B\ell)$. This is immediate
  when $\rho=1$. If $\rho>1$, then
  $a<\log_B\rho\leq\log_B\ell$ and
  $b\leq1+\max(0,\log_B(\ell/\mu))$, which gives
  $q\leq2+\log_B\ell$.

  For every $i\in[1\dd q]$, denote
  $k_i:=\AlphabetSize^{\lambda_{i-1}}$,
  $h_i:=\lfloor\lambda_i/\lambda_{i-1}\rfloor$, and
  $r_i:=\lceil\ell/\lambda_i\rceil$, as in
  \cref{lem:reduce-prefix-select-to-prefix-select-large-space}. The
  properties above imply $1\leq h_i\leq B$,
  $h_i\log k_i\leq\lambda_i\log\AlphabetSize$, and
  $r_i\leq2\ell/\lambda_i$. In particular,
  $k_i^{h_i}\leq\AlphabetSize^{\lambda_i}\leq m$, so
  \cref{pr:small-space-prefix-select-baseline,pr:large-space-prefix-select-baseline}
  can be used with these parameters.

  First apply \cref{lem:reduce-prefix-select-to-prefix-select-large-space}
  using \cref{pr:small-space-prefix-select-baseline}. At level $i$, the
  $r_i$ structures use
  $\bigO((\ell/\lambda_i)mh_i\log k_i)
  =\bigO(m\ell\log\AlphabetSize)$ bits. Summing over
  $q=\bigO(\log_B\ell)$ levels gives space usage
  $\bigO(m\ell\log\AlphabetSize\log_B\ell)$ bits.

  By \cref{pr:small-space-prefix-select-baseline},
  $P_t(k_i,h_i,m)=\bigO(mh_i)$. Consequently,
  $r_iP_t(k_i,h_i,m)=\bigO(m\ell/\lambda_{i-1})$.
  The alternative bound
  $P_t(k_i,h_i,m)=\bigO(m(h_i\log k_i)^2/\log m)$ gives
  $r_iP_t(k_i,h_i,m)
  =\bigO(m\ell\lambda_i(\log\AlphabetSize)^2/\log m)$.
  The additive term
  $m\min\{\ell,\ell\log\AlphabetSize/\sqrt{\log m},
  (\ell\log\AlphabetSize)^2/\log m\}$ in the preprocessing-time bound of
  \cref{lem:reduce-prefix-select-to-prefix-select-large-space} has the claimed
  bound, so it remains to bound the sum of
  $r_iP_t(k_i,h_i,m)$ over $i\in[1\dd q]$. We consider three cases.
  \begin{itemize}

  \item Suppose that $1<\tau<\ell$. In this case,
    $\rho=\tau$ and $\tau\leq\mu<2\tau$. For the indices
    $i\in[1\dd q]$ satisfying $\lambda_i\leq\mu$, the minimum is bounded by
    $(h_i\log k_i)^2/\log m$. By the definition of the
    levels, the sum of the corresponding values $\lambda_i$ is
    $\bigO(\mu)$, and hence
    \[
      \sum_{i\in[1\dd q]:\lambda_i\leq\mu}r_iP_t(k_i,h_i,m)
      \subseteq\bigO\!\left(\frac{m\ell(\log\AlphabetSize)^2}{\log m}
      \sum_{i\in[1\dd q]:\lambda_i\leq\mu}\lambda_i\right)
      \subseteq\bigO\!\left(\frac{m\ell\log\AlphabetSize}{\sqrt{\log m}}\right).
    \]
    For the indices $i\in[1\dd q]$ satisfying
    $\lambda_i>\mu$, the minimum is bounded by $h_i$. If this set is nonempty,
    the corresponding values $\lambda_{i-1}$ form a geometrically
    increasing sequence starting at $\mu$. The sum of their reciprocals is
    therefore $\bigO(1/\mu)$, and hence
    \[
      \sum_{i\in[1\dd q]:\lambda_i>\mu}r_iP_t(k_i,h_i,m)
      \subseteq\bigO\!\left(m\ell
      \sum_{i\in[1\dd q]:\lambda_i>\mu}\frac{1}{\lambda_{i-1}}\right)
      \subseteq\bigO\!\left(\frac{m\ell}{\mu}\right)
      \subseteq\bigO\!\left(\frac{m\ell\log\AlphabetSize}{\sqrt{\log m}}\right).
    \]
    By $\tau>1$ and $\tau<\ell$, 
    $\ell\log\AlphabetSize/\sqrt{\log m}$ is the minimum of the three
    expressions in the claim.

  \item Suppose that $\tau\geq\ell$. For every $i\in[1\dd q]$, the minimum
    is bounded by $(h_i\log k_i)^2/\log m$.
    Here, $\rho=\ell$ and $b=0$. Thus,
    $\sum_{i=1}^q\lambda_i=\bigO(\ell)$, and
    \[
      \sum_{i=1}^q r_iP_t(k_i,h_i,m)
      \subseteq\bigO\!\left(\frac{m\ell(\log\AlphabetSize)^2}{\log m}
      \sum_{i=1}^q\lambda_i\right)
      \subseteq\bigO\!\left(\frac{m(\ell\log\AlphabetSize)^2}{\log m}\right).
    \]
    The inequality
    $\ell\log\AlphabetSize\leq\sqrt{\log m}$ implies that this is the
    claimed minimum.

  \item Suppose that $\tau\leq1$. For every $i\in[1\dd q]$, the minimum is
    bounded by $h_i$. Since
    $\sum_{i=1}^q1/\lambda_{i-1}=\bigO(1)$,
    \[
      \sum_{i=1}^q r_iP_t(k_i,h_i,m)
      \subseteq\bigO\!\left(m\ell\sum_{i=1}^q\frac{1}{\lambda_{i-1}}\right)
      \subseteq\bigO(m\ell).
    \]
    The inequality
    $\sqrt{\log m}\leq\log\AlphabetSize$ implies that this is the claimed
    minimum.
  \end{itemize}
  This proves the preprocessing-time bound for the small-space tradeoff.

  By \cref{pr:small-space-prefix-select-baseline}, the maximum preprocessing
  space of one structure over all levels is
  $\bigO(m\ell\log\AlphabetSize)$ bits. Since $\log_B\ell\geq1$, this is
  $\bigO(m\ell\log\AlphabetSize\log_B\ell)$ bits. Together with the
  $\bigO(m\ell\log\AlphabetSize\log_B\ell)$-bit space bound,
  \cref{lem:reduce-prefix-select-to-prefix-select-large-space}
  gives the claimed preprocessing space. Finally, the query-time bound in
  \cref{lem:reduce-prefix-select-to-prefix-select-large-space} is
  $\bigO(\sum_{i=1}^q h_i)=\bigO(B\log_B\ell)$.

  Next apply \cref{lem:reduce-prefix-select-to-prefix-select-large-space}
  using \cref{pr:large-space-prefix-select-baseline}. Its preprocessing-time
  analysis is unchanged. Since $h_i\leq B$, the $r_i$ structures at level
  $i$ use $\bigO((\ell/\lambda_i)mh_i^2\log k_i)
  =\bigO(Bm\ell\log\AlphabetSize)$ bits. Summing over the levels gives space
  $\bigO(Bm\ell\log\AlphabetSize\log_B\ell)$ bits. By
  \cref{pr:large-space-prefix-select-baseline}, the
  maximum preprocessing space of one structure over all levels is
  $\bigO(Bm\ell\log\AlphabetSize)$ bits. Since $\log_B\ell\geq1$, this is
  $\bigO(Bm\ell\log\AlphabetSize\log_B\ell)$ bits. Together with the
  $\bigO(Bm\ell\log\AlphabetSize\log_B\ell)$-bit space bound, this proves the
  claimed preprocessing space. By
  \cref{pr:large-space-prefix-select-baseline,lem:reduce-prefix-select-to-prefix-select-large-space},
  each query takes
  $\bigO(\sum_{i=1}^q1)=\bigO(\log_B\ell)$ time.
\end{proof}

The following corollary gives a version of
\cref{th:prefix-select-tradeoffs} in which the word size is based on an upper
bound $N$ on $m\ell$. The range of $B$ and the bounds for both tradeoffs are
also stated in terms of $N$.

\begin{corollary}\label{cor:prefix-select-tradeoffs-input-length}
  Let $\AlphabetSize,N\in\Z_{\geq2}$ satisfy $\AlphabetSize\leq N$.
  Consider the word RAM model with word size $w=2\log N$. Denote
  $K_N:=\max(2,\lceil\log_{\AlphabetSize}N\rceil)$. For every
  $B\in[2\dd K_N]$, every input to the problem of indexing for prefix select
  queries from \cref{sec:prefix-select-problem-def} whose parameters satisfy
  \[
    m\geq\AlphabetSize,\qquad
    \ell=\lfloor\log_{\AlphabetSize}m\rfloor,\qquad
    m\ell\leq N
  \]
  admits two data structures with the following complexities:
  \begin{itemize}
  \item A data structure with:
    \begin{itemize}
    \item space usage
      $\bigO(N\log\AlphabetSize\log_B K_N)$ bits,
    \item preprocessing time
      $\bigO(N\min(1,\log\AlphabetSize/\sqrt{\log N}))$,
    \item preprocessing space
      $\bigO(N\log\AlphabetSize\log_B K_N)$ bits,
    \item query time $\bigO(B\log_B K_N)$.
    \end{itemize}
  \item A data structure with:
    \begin{itemize}
    \item space usage
      $\bigO(BN\log\AlphabetSize\log_B K_N)$ bits,
    \item preprocessing time
      $\bigO(N\min(1,\log\AlphabetSize/\sqrt{\log N}))$,
    \item preprocessing space
      $\bigO(BN\log\AlphabetSize\log_B K_N)$ bits,
    \item query time $\bigO(\log_B K_N)$.
    \end{itemize}
  \end{itemize}
\end{corollary}
\begin{proof}
  If $N<2^{16}$, construct and store all answers directly, and hence the claim
  follows immediately. Thus, assume that $N\geq2^{16}$. Let $W$ be the input
  sequence of an arbitrary prefix-select input satisfying the three conditions
  in the claim, and let $B\in[2\dd K_N]$ be arbitrary. Since
  $m\geq\AlphabetSize$, it holds $\ell\geq1$. The definition of $\ell$ also
  gives $\AlphabetSize^\ell\leq m$.

  The word size in \cref{th:prefix-select-tradeoffs} is based on $\log m$, and
  that theorem assumes $\ell\geq2$. If $m\geq\sqrt N$ and $\ell\geq2$, then
  the word size $2\log N$ has the form required by the theorem. We use an
  explicit data structure when $m<\sqrt N$ or $\ell=1$.

  We first describe an explicit data structure that will be used in the first
  two cases.
  \begin{itemize}

  \item \emph{Definitions.} For every
    $X\in[0\dd\AlphabetSize)^{\leq\ell}$, let $f_X$ denote
    $\PrefixRank{W}{m}{X}$, and let
    $P_X[1\dd f_X]\in[1\dd m]^{f_X}$ denote the sequence satisfying
    \[
      P_X[r]=\PrefixSelect{W}{r}{X}
      \qquad\text{for every }r\in[1\dd f_X].
    \]
    Let $P$ denote the concatenation of the sequences $P_X$ in increasing
    order of $\BasicInt{\AlphabetSize}{X}$
    (\cref{def:basic-int,ob:basic-int}). For every $X$, let $b_X$ denote the
    number of entries preceding $P_X$ in $P$. Let
    $A_{\rm len}[0\dd2\AlphabetSize^\ell)
      \in[0\dd m]^{2\AlphabetSize^\ell}$ denote the array satisfying
    \[
      A_{\rm len}[\BasicInt{\AlphabetSize}{X}]=f_X
      \qquad
      \text{for every }X\in[0\dd\AlphabetSize)^{\leq\ell},
    \]
    while all remaining entries are zero. Let
    $A_{\rm beg}[0\dd2\AlphabetSize^\ell)
      \in[0\dd m(\ell+1)]^{2\AlphabetSize^\ell}$ denote the array satisfying
    \[
      A_{\rm beg}[y]=\sum_{z\in[0\dd y)}A_{\rm len}[z]
      \qquad
      \text{for every }y\in[0\dd2\AlphabetSize^\ell).
    \]
    It holds $A_{\rm beg}[\BasicInt{\AlphabetSize}{X}]=b_X$ for every $X$.
    Denote $d_P:=\lceil\log(m+1)\rceil$.

  \item \emph{Components.} The explicit data structure consists of the
    following components:
    \begin{enumerate}

      \item The string-encoding structure from
        \cref{pr:basic-int-encoding} with parameters $m$ and
        $\AlphabetSize$. By
        \cref{pr:basic-int-encoding,rm:space}, it uses
        $\bigO(\sqrt m\log N)$ bits.

      \item The arrays $A_{\rm len}$ and $A_{\rm beg}$ in plain form. Each
        array has $2\AlphabetSize^\ell\leq2m$ entries. Since
        $m(\ell+1)\leq2N$, every entry fits in one $w$-bit word. Thus, this
        component uses $\bigO(m\log N)$ bits.

      \item The value $d_P$ and the packed representation
        $\PackedRepresentation{w}{m+1}{P}$. At each prefix length, every
        string $W[j]$ belongs to exactly one prefix class and therefore
        corresponds to one entry in exactly one sequence $P_X$. Hence,
        $|P|=m(\ell+1)$. By \cref{def:packed-representation}, this
        component uses
        $\bigO(\log N+m\ell\log m)$ bits.
    \end{enumerate}

    Therefore, the total space usage of the explicit data structure is
    $\bigO(m\ell\log m+m\log N)$ bits.

  \item \emph{Queries.} Given the length $|X|$, the packed representation
    $\PackedRepresentation{w}{\AlphabetSize}{X}$ of any string
    $X\in[0\dd\AlphabetSize)^{\leq\ell}$, and any occurrence number
    $r\in[1\dd m]$, we compute $\PrefixSelect{W}{r}{X}$ as follows:
    \begin{enumerate}

      \item Compute $y:=\BasicInt{\AlphabetSize}{X}$ using
        \cref{pr:basic-int-encoding}.

      \item If $r>A_{\rm len}[y]$, return $\infty$. Otherwise, compute
        $z:=A_{\rm beg}[y]+r$.

      \item Apply
        \cref{pr:packed-representation}\eqref{pr:packed-representation-access}
        to return $P[z]$.
    \end{enumerate}

    Every step takes constant time. If $r>A_{\rm len}[y]=f_X$, fewer than
    $r$ strings have prefix $X$, and hence
    $\PrefixSelect{W}{r}{X}=\infty$. Otherwise, the definitions of $P_X$ and
    $A_{\rm beg}$ imply that
    $P[z]=P_X[r]=\PrefixSelect{W}{r}{X}$.

  \item \emph{Construction.} When the explicit data structure is used in
    either of the first two cases, construct it from the supplied packed
    sequence representation
    $\PackedSeqRepresentation{w}{\AlphabetSize}{W}$ as follows:
    \begin{enumerate}

      \item \emph{Construct the character width and string-encoding
        structure.} Compute $d_W:=\lceil\log\AlphabetSize\rceil$ by repeated
        doubling, and apply \cref{pr:basic-int-encoding} with parameters $m$
        and $\AlphabetSize$. Its word-size assumption holds because $m\leq N$
        and $w=2\log N\geq1+\log m$. This step takes
        $\bigO(\log\AlphabetSize+\sqrt m)=\bigO(\sqrt m)$ time.

      \item\label{step:explicit-prefix-select-construct-boundaries} \emph{Construct
        $A_{\rm len}$ and $A_{\rm beg}$.} Initialize
        $A_{\rm len}$ with zeros. Scan $j\in[1\dd m]$ in increasing order.
        For every $j$, initialize $y:=1$, increase $A_{\rm len}[y]$ by one,
        and then scan $p\in[1\dd\ell]$ in increasing order. In the iteration
        for $p$, use
        \cref{pr:packed-representation}\eqref{pr:packed-representation-access}
        to retrieve $W[j][p]$ from position $(j-1)\ell+p$ of the represented
        concatenation. Set $y:=\AlphabetSize y+W[j][p]$, and increase
        $A_{\rm len}[y]$ by one. By \cref{def:basic-int}, the value of $y$
        after the update is
        $\BasicInt{\AlphabetSize}{W[j][1\dd p]}$. Construct $A_{\rm beg}$ by
        a prefix-sum scan over $A_{\rm len}$. Thus, the first scan constructs
        $A_{\rm len}$, and the prefix-sum scan constructs $A_{\rm beg}$.
        This step takes $\bigO(m\ell)$ time.

      \item \emph{Construct $d_P$ and the packed representation of $P$.}
        Compute $d_P=\lceil\log(m+1)\rceil$ by repeated doubling and retain it.
        Apply
        \cref{pr:packed-representation}\eqref{pr:packed-representation-initialize},
        with length parameter $2N$, to construct
        $\PackedRepresentation{w}{m+1}{\zero^{m(\ell+1)}}$. This packed
        representation will be updated to obtain
        $\PackedRepresentation{w}{m+1}{P}$. The invocation is valid because
        $m+1\leq2N$, $m(\ell+1)\leq2N$, and
        $w=2\log N>\log(2N)$. Allocate a temporary array
        $A_{\rm cur}[0\dd2\AlphabetSize^\ell)
          \in[0\dd m(\ell+1)]^{2\AlphabetSize^\ell}$, and set
        $A_{\rm cur}[y]:=A_{\rm beg}[y]$ for every $y$. Scan
        $j\in[1\dd m]$ in increasing order. For every $j$, initialize
        $y:=1$, write $j$ at position $A_{\rm cur}[y]+1$ of the packed
        representation using
        \cref{pr:packed-representation}\eqref{pr:packed-representation-update},
        and increase $A_{\rm cur}[y]$ by one. Next, scan
        $p\in[1\dd\ell]$ in increasing order. In the iteration for $p$,
        retrieve $W[j][p]$ as in
        Step~\ref{step:explicit-prefix-select-construct-boundaries}, set
        $y:=\AlphabetSize y+W[j][p]$, write $j$ at position
        $A_{\rm cur}[y]+1$ of the packed representation, and increase
        $A_{\rm cur}[y]$ by one. Since $A_{\rm cur}[y]$ starts at
        $A_{\rm beg}[y]$, the increasing scan over $j$ writes the entries of
        every $P_X$ in increasing order. Thus, the packed representation becomes
        $\PackedRepresentation{w}{m+1}{P}$. This step takes
        $\bigO(m\ell+\log m)=\bigO(m\ell)$ time, and the temporary array uses
        $\bigO(m\log N)$ bits.
    \end{enumerate}

    Since $\ell\geq1$, the three steps take $\bigO(m\ell)$ time altogether.
    Including the supplied input and the temporary array $A_{\rm cur}$, the
    preprocessing space is
    $\bigO(m\ell\log m+m\log N)$ bits.
  \end{itemize}

  We now distinguish three cases.
  \begin{enumerate}

    \item \emph{Suppose that $m<\sqrt N$.}
      The explicit data structure described above applies. Since
      $\ell\leq\log m\leq\log N$ and $N\geq2^{16}$, its space usage and
      preprocessing space are at most
      $\bigO(\sqrt N(\log N)^2)=\bigO(N)$ bits. The preprocessing time is
      \[
        \bigO(m\ell)
          =\bigO(\sqrt N\log N)
          =\bigO(N/\sqrt{\log N}).
      \]
      The $\bigO(N)$-bit space and $\bigO(N/\sqrt{\log N})$ preprocessing time
      satisfy both alternatives in the claim because
      $\log\AlphabetSize\geq1$ and $\log_B K_N\geq1$.

    \item \emph{Suppose that $m\geq\sqrt N$ and $\ell=1$.}
      The explicit data structure described above applies. Since $\ell=1$, it holds
      $\log m<2\log\AlphabetSize$. Moreover,
      $m\geq\sqrt N$ implies $\log N\leq2\log m$, and hence
      $\log\AlphabetSize=\Omega(\log N)$. Therefore, the space usage and
      preprocessing space are $\bigO(N\log\AlphabetSize)$ bits, and the
      preprocessing time is $\bigO(m)=\bigO(N)$. More precisely,
      $\log\AlphabetSize\geq\tfrac14\log N$, so
      $\log\AlphabetSize/\sqrt{\log N}\geq1$ because $N\geq2^{16}$.
      Thus, the required preprocessing-time bound is $\bigO(N)$. The constant
      query time and the space bounds satisfy both alternatives in the claim.

    \item \emph{Suppose that $m\geq\sqrt N$ and $\ell\geq2$.}
      Since $m\ell\leq N$, it holds $m\leq N$. Hence,
      $\log m\in[\tfrac12\log N,\log N]$ and
      $w=c_m\log m$ for a constant $c_m\in[2,4]$. Since $m\leq N$, it also
      holds $\ell\leq K_N$. Denote $B_m:=\min(B,\ell)$. Then
      $B_m\in[2\dd\ell]$. If $B\leq\ell$, then
      $\log_{B_m}\ell=\log_B\ell\leq\log_B K_N$. If $B>\ell$, then
      $\log_{B_m}\ell=1\leq\log_B K_N$. Consequently,
      $\log_{B_m}\ell\leq\log_B K_N$ and
      $B_m\log_{B_m}\ell\leq B\log_B K_N$.
      Apply \cref{th:prefix-select-tradeoffs} with $B_m$. Since
      $m\ell\log\AlphabetSize\leq N\log\AlphabetSize$, its small-space
      alternative has space usage and preprocessing space
      $\bigO(N\log\AlphabetSize\log_B K_N)$ bits and query time
      $\bigO(B\log_B K_N)$. Its large-space alternative has space usage and
      preprocessing space
      $\bigO(BN\log\AlphabetSize\log_B K_N)$ bits and query time
      $\bigO(\log_B K_N)$. The common preprocessing time is upper bounded
      both by $m\ell\leq N$ and by
      \[
        \frac{m\ell\log\AlphabetSize}{\sqrt{\log m}}
          =\bigO\left(
            \frac{N\log\AlphabetSize}{\sqrt{\log N}}
          \right).
      \]
      Thus, both data structures satisfy all the required bounds in this case.
  \end{enumerate}
  Since $W$ and $B$ were arbitrary, the three cases prove the corollary.
\end{proof}

\subsection{Consequences}\label{sec:prefix-select-consequences}

\begin{framed}
  \noindent
  \probname{Indexing for Suffix Array Queries over Alphabet $[0 \dd \AlphabetSize)$}
  \begin{bfdescription}
  \item[Input:]
    Let $\AlphabetSize,\Textlen\in\Z_{\geq2}$ satisfy
    $\AlphabetSize\leq\Textlen$. Consider the word RAM model with word size
    $w$ such that $w\geq\log\AlphabetSize$. The input is the packed representation
    $\PackedRepresentation{w}{\AlphabetSize}{\Text}$
    (\cref{def:packed-representation}) of a text
    $\Text\in[0\dd\AlphabetSize)^{\Textlen}$.
  \item[Output:]
    A data structure that, given $i\in[1\dd\Textlen]$, returns
    $\SA{\Text}[i]$ (\cref{def:suffix-array}), i.e., the starting position
    of the lexicographically $i$th smallest suffix of $\Text$.
  \end{bfdescription}
\end{framed}

\begin{theorem}[{\cite{SaPerfectEquiv}}]
  \label{th:prefix-select-suffix-array-equivalence}
  Let $\AlphabetSize,N\in\Z_{\geq2}$ satisfy $\AlphabetSize\leq N$.
  Consider the word RAM model with word size $w=c\log N$, where $c\geq2$
  is a constant. Consider the problem of indexing for prefix select queries
  from \cref{sec:prefix-select-problem-def} with
  $\ell=\lfloor\log_{\AlphabetSize}m\rfloor$, and the problem of indexing
  for suffix array queries defined above. We say that an input to the problem
  of indexing for prefix select queries has length at most $N$ if
  $m\ell\leq N$, and that an input to the problem of indexing for suffix array
  queries has length at most $N$ if $\Textlen\leq N$. Suppose that one of these
  problems admits, for every input of length at most $N$, a data structure
  with the following complexities:
  \begin{itemize}
  \item space usage $S(\AlphabetSize,N)$ bits,
  \item preprocessing time $P_t(\AlphabetSize,N)$,
  \item preprocessing space $P_s(\AlphabetSize,N)$ bits,
  \item query time $Q(\AlphabetSize,N)$.
  \end{itemize}
  Then there exists $N'=\Theta(N)$ with $N'\leq N$ such that the other
  problem admits, for every input of length at most $N'$, a data structure
  with the following complexities:
  \begin{itemize}
  \item space usage $\bigO(S(\AlphabetSize,N))$ bits,
  \item preprocessing time $\bigO(P_t(\AlphabetSize,N))$,
  \item preprocessing space $\bigO(P_s(\AlphabetSize,N))$ bits,
  \item query time $\bigO(Q(\AlphabetSize,N))$.
  \end{itemize}
\end{theorem}

\begin{theorem}\label{th:suffix-array-parameterized-tradeoffs}
  Let $\AlphabetSize,\Textlen\in\Z_{\geq2}$ satisfy
  $\AlphabetSize\leq\Textlen$. Consider the word RAM model with word size
  $w=c\log\Textlen$, where $c\geq2$ is a constant. Denote
  $K:=\max(2,\lceil\log_{\AlphabetSize}\Textlen\rceil)$. For every
  $B\in[2\dd K]$, there exist two data structures for the problem of indexing
  for suffix array queries over alphabet $[0\dd\AlphabetSize)$ defined above
  with the following complexities:
  \begin{itemize}
  \item A data structure with:
    \begin{itemize}
    \item space usage
      $\bigO(\Textlen\log\AlphabetSize\log_B K)$ bits,
    \item preprocessing time
      $\bigO(\Textlen
        \min(1,\log\AlphabetSize/\sqrt{\log\Textlen}))$,
    \item preprocessing space
      $\bigO(\Textlen\log\AlphabetSize\log_B K)$ bits,
    \item query time $\bigO(B\log_B K)$.
    \end{itemize}
  \item A data structure with:
    \begin{itemize}
    \item space usage
      $\bigO(B\Textlen\log\AlphabetSize\log_B K)$ bits,
    \item preprocessing time
      $\bigO(\Textlen
        \min(1,\log\AlphabetSize/\sqrt{\log\Textlen}))$,
    \item preprocessing space
      $\bigO(B\Textlen\log\AlphabetSize\log_B K)$ bits,
    \item query time $\bigO(\log_B K)$.
    \end{itemize}
  \end{itemize}
\end{theorem}
\begin{proof}
  Let $B\in[2\dd K]$ be arbitrary, and let
  $\Text\in[0\dd\AlphabetSize)^{\Textlen}$ be an arbitrary input text to the
  problem in the claim. We proceed in two steps.
  \begin{enumerate}

    \item \emph{Establish two intermediate exact-length constructions.}
      We apply \cref{th:prefix-select-suffix-array-equivalence} separately to
      the two prefix-select data structures from
      \cref{cor:prefix-select-tradeoffs-input-length}. The constant-factor
      loss in maximum input length in the equivalence is fixed. Hence, there
      is a fixed integer $C\geq1$, independent of $\AlphabetSize$, $B$, and
      $N$, such that applying the equivalence with length parameter $N$ gives
      bounds $N'_1,N'_2\geq N/C$ for the two applications. Let
      $N:=C\Textlen$, $\widehat w:=2\log N$, and
      $K_N:=\max(2,\lceil\log_{\AlphabetSize}N\rceil)$. Since
      $N\geq\Textlen$, it holds $K_N\geq K$, and hence
      $B\in[2\dd K_N]$.
      By \cref{cor:prefix-select-tradeoffs-input-length}, every prefix-select
      input of length at most $N$ admits, on a word RAM with word size
      $\widehat w$, the two data structures with the bounds stated in that
      corollary. Apply
      \cref{th:prefix-select-suffix-array-equivalence} with $c=2$ to each
      structure in the direction from prefix select to suffix array.
      The two conclusions hold for every text of length at most $N'_1$ and
      $N'_2$, respectively. Since $N'_1,N'_2\geq N/C=\Textlen$, both hold, in
      particular, for every text of length exactly $\Textlen$. Moreover,
      $N=\Theta(\Textlen)$, $\log N=\Theta(\log\Textlen)$, and
      $K_N=\Theta(K)$. Since $B\leq K$, it follows that
      $\log_B K_N=\bigO(\log_B K)$. We have therefore proved the following
      intermediate statement. For the fixed value of $B$, on a word RAM with
      word size $\widehat w=2\log(C\Textlen)$, given the packed representation
      $\PackedRepresentation{\widehat w}{\AlphabetSize}{\Text'}$ of any text
      $\Text'\in[0\dd\AlphabetSize)^{\Textlen}$, one can construct each of the
      following two data structures, both of which return
      $\SA{\Text'}[i]$ for every $i\in[1\dd\Textlen]$:
      \begin{itemize}
      \item A data structure with:
        \begin{itemize}
        \item space usage
          $\bigO(\Textlen\log\AlphabetSize\log_B K)$ bits,
        \item preprocessing time
          $\bigO(\Textlen
            \min(1,\log\AlphabetSize/\sqrt{\log\Textlen}))$,
        \item preprocessing space
          $\bigO(\Textlen\log\AlphabetSize\log_B K)$ bits,
        \item query time $\bigO(B\log_B K)$.
        \end{itemize}
      \item A data structure with:
        \begin{itemize}
        \item space usage
          $\bigO(B\Textlen\log\AlphabetSize\log_B K)$ bits,
        \item preprocessing time
          $\bigO(\Textlen
            \min(1,\log\AlphabetSize/\sqrt{\log\Textlen}))$,
        \item preprocessing space
          $\bigO(B\Textlen\log\AlphabetSize\log_B K)$ bits,
        \item query time $\bigO(\log_B K)$.
        \end{itemize}
      \end{itemize}

    \item \emph{Use the intermediate constructions on the word RAM from the
      claim.} The intermediate statement and the claim both concern texts over
      alphabet $[0\dd\AlphabetSize)$ of length exactly $\Textlen$. They differ
      only in the word size and the resulting packed representation of the
      input. The intermediate constructions use $\widehat w$-bit words and
      expect the text packed into such words, whereas the claim provides a
      word RAM with word size $w=c\log\Textlen$ and the preprocessing input
      $\PackedRepresentation{w}{\AlphabetSize}{\Text}$. Either $w$ or
      $\widehat w$ may be larger. Since $C$ and $c$ are fixed, it holds
      $\widehat w=2\log(C\Textlen)=\Theta(w)$. Hence, for
      $W:=\max(w,\widehat w)$, a constant number of $w$-bit words suffice to
      represent a $W$-bit word and simulate every operation on it in constant
      time. Compute $\lceil\log\AlphabetSize\rceil$ by repeated doubling. This
      takes $\bigO(\log\AlphabetSize)$ time, which is bounded by
      $\bigO(\Textlen\log\AlphabetSize/\log\Textlen)$ because
      $\Textlen/\log\Textlen\geq1$. For either of the two intermediate
      constructions, use this simulation to apply
      \cref{pr:packed-representation-word-size-conversion}, with length
      parameter $\Textlen$, and obtain
      $\PackedRepresentation{\widehat w}{\AlphabetSize}{\Text}$. The
      invocation is valid because $\AlphabetSize\leq\Textlen$,
      $w=c\log\Textlen>\log\Textlen$,
      $\widehat w=2\log(C\Textlen)>\log\Textlen$, and
      $\widehat w=\Theta(w)$. Applying
      \cref{pr:packed-representation-word-size-conversion} takes
      $\bigO(\Textlen/\log_{\AlphabetSize}\Textlen)
        =\bigO(\Textlen\log\AlphabetSize/\log\Textlen)$ time. This is upper
      bounded both by $\bigO(\Textlen)$ and by
      $\bigO(\Textlen\log\AlphabetSize/\sqrt{\log\Textlen})$. The two packed
      representations of $\Text$ and the working space used by
      \cref{pr:packed-representation-word-size-conversion} occupy
      $\bigO(\Textlen\log\AlphabetSize)$ bits. Then run the corresponding
      preprocessing algorithm. Representing every
      $\widehat w$-bit word by a constant number of $w$-bit words and
      simulating each operation changes the space usage, preprocessing time,
      preprocessing space, and query time of either structure by only constant
      factors. Since $\log_B K\geq1$, the representation conversion satisfies
      both preprocessing-space bounds. Thus, after the conversion,
      implementing either data structure by this word simulation satisfies
      all bounds in the claim.
  \end{enumerate}
  Since $B$ and $\Text$ were arbitrary, the claim follows.
\end{proof}

\begin{corollary}\label{th:suffix-array-tradeoffs}
  Let $\epsilon\in(0,1)$ be any constant. Let
  $\AlphabetSize,\Textlen\in\Z_{\geq2}$ satisfy
  $\AlphabetSize\leq\Textlen$. Consider the word RAM model with word size
  $w=c\log\Textlen$, where $c\geq2$ is a constant. There exist three data
  structures for the problem of indexing for suffix array queries over
  alphabet $[0\dd\AlphabetSize)$ defined above with the following
  complexities:
  \begin{itemize}
  \item A data structure with:
    \begin{itemize}
    \item space usage
      $\bigO(\Textlen\log\AlphabetSize
        (1+\log\log_{\AlphabetSize}\Textlen))$ bits,
    \item preprocessing time
      $\bigO(\Textlen
        \min(1,\log\AlphabetSize/\sqrt{\log\Textlen}))$,
    \item preprocessing space
      $\bigO(\Textlen\log\AlphabetSize
        (1+\log\log_{\AlphabetSize}\Textlen))$ bits,
    \item query time
      $\bigO(1+\log\log_{\AlphabetSize}\Textlen)$.
    \end{itemize}
  \item A data structure with:
    \begin{itemize}
    \item space usage $\bigO(\Textlen\log\AlphabetSize)$ bits,
    \item preprocessing time
      $\bigO(\Textlen
        \min(1,\log\AlphabetSize/\sqrt{\log\Textlen}))$,
    \item preprocessing space
      $\bigO(\Textlen\log\AlphabetSize)$ bits,
    \item query time
      $\bigO(\log^{\epsilon}_{\AlphabetSize}\Textlen)$.
    \end{itemize}
  \item A data structure with:
    \begin{itemize}
    \item space usage
      $\bigO(\Textlen\log\AlphabetSize
        \log^{\epsilon}_{\AlphabetSize}\Textlen)$ bits,
    \item preprocessing time
      $\bigO(\Textlen
        \min(1,\log\AlphabetSize/\sqrt{\log\Textlen}))$,
    \item preprocessing space
      $\bigO(\Textlen\log\AlphabetSize
        \log^{\epsilon}_{\AlphabetSize}\Textlen)$ bits,
    \item query time $\bigO(1)$.
    \end{itemize}
  \end{itemize}
  For constant $\AlphabetSize$, the space/query-time pairs become
  $(\bigO(\Textlen\log\log\Textlen),\bigO(\log\log\Textlen))$,
  $(\bigO(\Textlen),\bigO(\log^{\epsilon}\Textlen))$, and
  $(\bigO(\Textlen\log^{\epsilon}\Textlen),\bigO(1))$, respectively. In
  all three cases, the preprocessing time is
  $\bigO(\Textlen/\sqrt{\log\Textlen})$.
\end{corollary}
\begin{proof}
  Denote $K:=\max(2,\lceil\log_{\AlphabetSize}\Textlen\rceil)$, as in
  \cref{th:suffix-array-parameterized-tradeoffs}. Applying that theorem
  with $B=2$ gives the first data structure because
  $\log_2K=\Theta(1+\log\log_{\AlphabetSize}\Textlen)$.

  Denote $B:=\lceil K^{\epsilon}\rceil$. Then $B\in[2\dd K]$,
  $B\leq2K^{\epsilon}$, and $\log_B K\leq1/\epsilon$. The first alternative
  in \cref{th:suffix-array-parameterized-tradeoffs} gives the second data
  structure, and its second alternative gives the third. Since
  $K=\Theta(\log_{\AlphabetSize}\Textlen)$ and $\epsilon$ is constant, the
  stated bounds follow.
\end{proof}

\section{Prefix Special Rank Queries}\label{sec:prefix-special-rank}

\subsection{Problem Definition}\label{sec:prefix-special-rank-problem-def}

\begin{framed}
  \noindent
  \probname{Indexing for Prefix Rank Queries over Alphabet $[0 \dd \AlphabetSize)$}
  \begin{bfdescription}
  \item[Input:]
    Let $\AlphabetSize,m\in\Z_{\geq2}$ and $\ell\in\Z_{\geq1}$ satisfy
    $\AlphabetSize^\ell\leq m$. Consider the word RAM model with word size
    $w$ such that $w\geq\log\AlphabetSize$. The input to the problem is the
    packed sequence representation
    $\PackedSeqRepresentation{w}{\AlphabetSize}{W}$
    (\cref{def:packed-sequence-representation}) of a sequence
    $W[1\dd m]\in([0\dd\AlphabetSize)^\ell)^m$.
  \item[Output:]
    A data structure that, given the length $|X|$, the packed representation
    $\PackedRepresentation{w}{\AlphabetSize}{X}$ of any string
    $X\in[0\dd\AlphabetSize)^{\leq\ell}$
    (\cref{def:packed-representation}), and any $j\in[0\dd m]$, returns
    $\PrefixRank{W}{j}{X}$ (\cref{def:prefix-range-queries}), i.e., the number
    of indices $i\in[1\dd j]$ such that $X$ is a prefix of $W[i]$.
  \end{bfdescription}
\end{framed}

\begin{framed}
  \noindent
  \probname{Indexing for Prefix Special Rank Queries over Alphabet $[0 \dd \AlphabetSize)$}
  \begin{bfdescription}
  \item[Input:]
    Let $\AlphabetSize,m\in\Z_{\geq2}$ and $\ell\in\Z_{\geq1}$ satisfy
    $\AlphabetSize^\ell\leq m$. Consider the word RAM model with word size
    $w$ such that $w\geq\log\AlphabetSize$. The input to the problem is the
    packed sequence representation
    $\PackedSeqRepresentation{w}{\AlphabetSize}{W}$
    (\cref{def:packed-sequence-representation}) of a sequence
    $W[1\dd m]\in([0\dd\AlphabetSize)^\ell)^m$.
  \item[Output:]
    A data structure that, given any $j\in[1\dd m]$ and any
    $p\in[0\dd\ell]$, returns $\PrefixSpecialRank{W}{j}{p}$
    (\cref{def:prefix-range-queries}), i.e., the number of indices
    $i\in[1\dd j]$ such that $W[j][1\dd p]$ is a prefix of $W[i]$.
  \end{bfdescription}
\end{framed}

\subsection{Prefix Special Rank for Large Alphabets}
\label{sec:prefix-special-rank-large-alphabets}

\begin{proposition}\label{pr:prefix-special-rank-large-alphabets}
  Let $\AlphabetSize,m\in\Z_{\geq2}$ and $\ell\in\Z_{\geq1}$ satisfy
  $\AlphabetSize^\ell\leq m$. Consider the word RAM model with word size
  $w=c\log m$, where $c\geq2$ is a constant. There exists a data structure
  for the problem of indexing for prefix special rank queries over alphabet
  $[0\dd\AlphabetSize)$
  (see \cref{sec:prefix-special-rank-problem-def}) that achieves the following
  complexities:
  \begin{itemize}
  \item space usage
    $\bigO(m\ell(\log\AlphabetSize+\log\log m))$ bits,
  \item preprocessing time
    \[
      \bigO\left(
        m\min\left\{
          \ell,
          \frac{\ell(\log\AlphabetSize+\log\log m)}{\sqrt{\log m}},
          \frac{\ell^2(\log\AlphabetSize+\log\log m)^2}{\log m}
        \right\}
      \right),
    \]
  \item preprocessing space
    $\bigO(m\ell(\log\AlphabetSize+\log\log m))$ bits,
  \item query time $\bigO(1)$.
  \end{itemize}
\end{proposition}
\begin{proof}
  If $\log\log m<9$, construct and store all answers directly, and hence the
  claim follows immediately. Thus, assume that $\log\log m\geq9$. Denote
  $b:=\lceil\log m\rceil$ and $B:=b^2$.

  For $X\in[0\dd\AlphabetSize)^{\leq\ell}$, let
  $f_X:=\PrefixRank{W}{m}{X}$ (see \cref{def:prefix-range-queries}) and let $(p_{X,r})_{r \in [1 \dd f_{X}]}$ denote
  an increasing sequence of positions of the strings in $W$ having prefix $X$, i.e.,
  such that, $p_{X,r} = \PrefixSelect{W}{r}{X}$ (see \cref{def:prefix-range-queries}), where $r \in [1 \dd f_{X}]$.

  Let $R[1 \dd m] \in ([0 \dd B)^{\ell})^{m}$ be a sequence of $m$
  strings of length $\ell$ over alphabet $[0 \dd B)$ defined such that,
  for every $j \in [1 \dd m]$ and every $p \in [1 \dd \ell]$, it holds
  \[
    R[j][p] = \Big(\PrefixSpecialRank{W}{j}{p} - 1\Big) \bmod B.
  \]
  In other words, $R[j][p]$ is the local
  rank (decreased by one to fit in $[0 \dd B)$) of the $r$th occurrence
  of $X = W[j][1 \dd p]$ as a prefix in $W$ (where $r = \PrefixSpecialRank{W}{j}{p}$)
  inside its \emph{bucket}, where by a \emph{bucket} of string $X$ (or \emph{$X$-bucket})
  we mean a collection of positions $\{p_{X,r} : r \in (s \dd t]\}$ for some
  $s,t \in [0 \dd f_{X}]$ such that $s = (k-1)B$ and $t = \min(f_{X},kB)$ holds
  for some $k \in [1 \dd \lceil f_{X}/B \rceil]$. Informally, buckets of string $X$
  are size-$B$ groups of its occurrences as prefixes of strings in $W$
  (except possibly the last group, which may have fewer than $B$ elements)
  obtained by greedy left-to-right grouping, so that the first $X$-bucket
  is the set $\{p_{X,1}, \ldots, p_{X,B}\}$, the second $X$-bucket is the set
  $\{p_{X,B+1}, \ldots, p_{X,2B}\}$, etc.

  We call $X \in [0 \dd \AlphabetSize)^{\leq \ell}$ a \emph{frequent prefix}
  if it holds $f_{X} > B$, i.e., occurrences of $X$ as a prefix in $W$
  partition into at least two buckets. Otherwise, $X$ is an \emph{rare prefix}.
  The \emph{prefix code} of the $k$th bucket for string
  $X \in [0 \dd \AlphabetSize)^{\leq \ell}$, denoted $c_{X,k}$,
  is the longest common prefix, truncated at length $b-1$,
  of the length-$b$ binary encodings of the zero-based smallest and largest positions
  in the $k$th bucket of $X$. Formally, for any $k \in [1 \dd \lceil f_{X} / B \rceil]$,
  it holds $c_{X,k} := C[1 \dd \min(|C|,b-1)]$, where
  $s = (k-1)B$ and
  $t = \min(f_{X}, kB)$
  are the boundaries of the $k$th bucket of $X$,
  $p_{L} = p_{X,s+1}$ and
  $p_{R} = p_{X,t}$ are the leftmost and rightmost positions in the bucket,
  $C_L = \AlphabetMap{m}{2}{p_{L}-1}$ and
  $C_R = \AlphabetMap{m}{2}{p_{R}-1}$ (see \cref{def:alphabet-map})
  are the length-$b$ binary encodings of $p_{L}-1$ and $p_{R}-1$, and
  $C = C_{L}[1 \dd \lcp{C_{L}}{C_{R}}]$ is the common prefix of
  $C_{L}$ and $C_{R}$ (see \cref{sec:prelim-basic}).
  Observe that:
  \begin{itemize}
  \item Letting $\{p_{X,r} : r \in (s \dd t]\}$ be the elements of the $k$th $X$-bucket
    (i.e., $s = (k-1)B$ and $t = \min(f_{X}, kB)$),
    for every $r \in (s \dd t]$, the string $\AlphabetMap{m}{2}{p_{X,r}-1}$ has $c_{X,k}$
    as a prefix.
  \item For every $k_1, k_2 \in [1 \dd \lceil f_{X}/B \rceil]$,
    $k_1 \neq k_2$ implies $c_{X,k_1} \neq c_{X,k_2}$.
  \end{itemize}
  Indeed, every binary prefix identifies a consecutive interval of zero-based
  positions. Since the bucket endpoints have prefix $c_{X,k}$, every position
  between them also has prefix $c_{X,k}$. If $|c_{X,k}|<b-1$, the two
  endpoints of the bucket lie in different halves of the interval identified
  by $c_{X,k}$. Thus, two disjoint buckets cannot have the same such code.
  A code of length $b-1$ identifies only two positions. Since
  $B=b^2>2$ and only the last bucket may contain fewer than $B$ positions, at
  most one bucket has such a code. Thus, the bucket codes are pairwise
  distinct.
  By $L[1 \dd m] \in ([0 \dd b)^{\ell})^{m}$ we then denote
  a sequence of $m$ length-$\ell$ strings over alphabet $[0 \dd b)$ defined as
  follows. For every $j \in [1 \dd m]$ and $p \in [1 \dd \ell]$, let
  $X$ denote $W[j][1 \dd p]$. It holds
  $L[j][p] = 0$ if $X$ is a rare prefix, and $L[j][p] = |c_{X,k}|$
  (where $c_{X,k}$ is the prefix code of the $X$-bucket containing position $j$,
  i.e., $r = \PrefixSpecialRank{W}{j}{p}$ and $k = \lceil r/B \rceil$) otherwise.

  Finally, for every frequent prefix
  $X \in [0 \dd \AlphabetSize)^{\leq \ell}$, let $\mathcal D_X$ denote a
  static dictionary that maps
  $\BasicInt{2}{c_{X,k}}$ to $k$ for every
  $k \in [1 \dd \lceil f_X / B\rceil]$.
  The encodings include the string lengths and are
  therefore unambiguous (\cref{def:basic-int}).

  \DSComponents
  The data structure consists of the following components:
  \begin{enumerate}

  \item The packed representation
    $\PackedSeqRepresentation{w}{\AlphabetSize}{W}$ of the input sequence
    $W[1\dd m]$, together with the values $b$, $B$,
    $\lceil\log\AlphabetSize\rceil$, $\lceil\log b\rceil$, and
    $\lceil\log B\rceil$. It uses $\bigO(m\ell\log\AlphabetSize)$ bits.

  \item The data structure from
    \cref{pr:basic-int-encoding} with parameters $m$ and $\AlphabetSize$. It
    uses $\bigO(\sqrt m\log m)
    =\bigO(m\ell\log\AlphabetSize)$ bits.

  \item A prefix-frequency array
    $A_{\rm pfreq}[0\dd2\AlphabetSize^\ell)$. It satisfies
    $A_{\rm pfreq}[\BasicInt{\AlphabetSize}{X}]=f_X$ for every
    $X\in[0\dd\AlphabetSize)^{\leq\ell}$. All remaining entries are zero.
    By \cref{pr:packed-prefix-frequencies}\eqref{eq:packed-prefix-frequencies-universe}, it uses
    $\bigO(\AlphabetSize^\ell\log m)
    =\bigO(m\ell\log\AlphabetSize)$ bits.

  \item The packed representation
    $\PackedSeqRepresentation{w}{b}{L}$
    of the sequence $L[1 \dd m]$. It uses $\bigO(m\ell\log\log m)$ bits.

  \item The packed representation
    $\PackedSeqRepresentation{w}{B}{R}$ of $R[1 \dd m]$. It uses
    $\bigO(m\ell\log\log m)$ bits.

  \item An array
    $A_{\rm dict}[0\dd2\AlphabetSize^\ell)$ and the dictionaries
    $\mathcal D_X$ for all frequent prefixes $X$. For every such $X$, the entry
    at $\BasicInt{\AlphabetSize}{X}$ stores a pointer to $\mathcal D_X$. All
    remaining entries are null. By \cref{th:static-dictionaries}, every
    dictionary lookup takes $\bigO(1)$ time. Every frequent prefix $X$ has
    $\lceil f_X/B\rceil \leq 2f_X/B
    =\bigO(f_X/(\log m)^2)$ buckets. At every depth, the frequencies $f_X$
    sum to $m$. Hence, the dictionaries contain
    $\bigO(m\ell/(\log m)^2)$ key--value pairs and occupy
    $\bigO(m\ell/\log m)$ bits. The array uses
    $\bigO(\AlphabetSize^\ell\log m)
    =\bigO(m\ell\log\AlphabetSize)$ bits. Every dictionary key
    is smaller than $2^b<2m$, and every associated value is at most $m$.
    Thus, each key and value fits in one word.
  \end{enumerate}

  In total, the data structure uses
  $\bigO(m\ell(\log\AlphabetSize+\log\log m))$ bits.

  \DSQueries
  Given any $j \in [1 \dd m]$ and any $p \in [0 \dd \ell]$, we
  compute $\PrefixSpecialRank{W}{j}{p}$ as follows:
  \begin{enumerate}

  \item If $p=0$, return $j$. Otherwise, use
    \cref{pr:packed-representation}\eqref{pr:packed-representation-substring}
    with length parameter $m\ell$, to obtain the packed representation
    $\PackedRepresentation{w}{\AlphabetSize}{X}$ of $X:=W[j][1\dd p]$.
    The invocation is valid because
    $\AlphabetSize\leq m\leq m\ell<m^2$ and
    $w\geq2\log m>\log(m\ell)$. It takes constant time because
    $p\log\AlphabetSize\leq\ell\log\AlphabetSize\leq\log m$. Using
    \cref{pr:basic-int-encoding}, we then compute
    $x:=\BasicInt{\AlphabetSize}{X}$, and retrieve $f_X:=A_{\rm pfreq}[x]$.

  \item Retrieve $d:=L[j][p]$ and $\delta:=R[j][p]$ from their packed sequence
    representations using
    \cref{pr:packed-representation}\eqref{pr:packed-representation-access}.
    If $f_X\leq B$, we return $\delta+1$ and conclude the query algorithm.

  \item Compute $c:=\lfloor(j-1)/2^{b-d}\rfloor$ in $\bigO(1)$ time.
    Observe that, denoting
    $C:=\AlphabetMap{m}{2}{j-1}[1\dd d]$, it holds
    $c=\Val{2}{C}$, since $\AlphabetMap{m}{2}{j-1}$ is the length-$b$
    binary encoding of $j-1$. Let $k$ denote the number of the $X$-bucket
    containing position $j$. By the definition of $L$ and the fact that every
    position in a bucket has its bucket code as a prefix,
    $d=|c_{X,k}|$ and $C=c_{X,k}$. It remains to compute
    $k$. By the definition of $\mathcal D_X$, this is the value mapped from
    the key $\BasicInt{2}{c_{X,k}}=\BasicInt{2}{C}$. By
    \cref{def:basic-int} and $|C|=d$, it holds
    $\BasicInt{2}{C}=2^{|C|}+\Val{2}{C}=2^d+c$, which can also be computed
    in constant time. Retrieve the pointer to $\mathcal D_X$ stored at
    $A_{\rm dict}[x]$, compute $k:=\mathcal D_X[2^d+c]$, and return
    $(k-1)\cdot B+\delta+1$.
  \end{enumerate}

  Every step takes constant time. If $p=0$, the returned value $j$ is
  correct. Hence, assume that $p>0$. Denote $X=W[j][1\dd p]$ and let
  $r:=\PrefixSpecialRank{W}{j}{p}$, so that $j=p_{X,r}$. The query obtains
  $f_X$ from $A_{\rm pfreq}$ and hence determines whether $X$ is a frequent
  or rare prefix. If $X$ is rare, the query returns
  $R[j][p]+1 = r$. Suppose that $X$ is a frequent prefix
  and denote $k:=\lceil r/B \rceil$. The dictionary computation above shows
  that the query
  retrieves $k$ and returns
  $(k-1) \cdot B + R[j][p] + 1 = r$.

  \DSConstruction
  We construct the six components in six steps.
  \begin{enumerate}

    \item \emph{Store the input and construct the parameters.} Retain
      $\PackedSeqRepresentation{w}{\AlphabetSize}{W}$ and compute $b$ and
      $B$. Compute $\lceil\log\AlphabetSize\rceil$, $\lceil\log b\rceil$,
      and $\lceil\log B\rceil$ by repeated doubling. This
      takes $\bigO(\log m)$ time and uses $\bigO(\log m)$ working bits, within
      the claimed bounds.

    \item \emph{Construct the string-encoding structure.} Apply
      \cref{pr:basic-int-encoding} with parameters $(m,\AlphabetSize)$ and
      retain the resulting data structure. This takes $\bigO(\sqrt m)$ time
      and uses $\bigO(\sqrt m\log m)$ working bits, within the claimed
      bounds.

    \item \emph{Construct the prefix-frequency array.} Apply
      \cref{pr:packed-prefix-frequencies}, with parameters $\AlphabetSize$,
      $m$, and $\ell$, to the packed sequence representation
      $\PackedSeqRepresentation{w}{\AlphabetSize}{W}$ and retain its output as
      $A_{\rm pfreq}$. This takes
      $\bigO(m\ell\log\AlphabetSize/\log m)$ time and uses
      $\bigO(m\ell\log\AlphabetSize)$ working bits, within the claimed
      bounds.

    \item \emph{Construct the packed sequence representation of $L$.} Proceed
      in five substeps.
      \begin{enumerate}

        \item\label{step:large-alphabet-prefix-special-rank-temporary-select-l}
          Construct a temporary prefix-select structure on $W$. If
          $\ell\geq2$, apply the small-space tradeoff from
          \cref{th:prefix-select-tradeoffs}, with parameters
          $\AlphabetSize$, $m$, and $\ell$, its tradeoff parameter equal to
          $\ell$, and supplied packed sequence representation
          $\PackedSeqRepresentation{w}{\AlphabetSize}{W}$. It uses
          $\bigO(m\ell\log\AlphabetSize)$ bits and has query time
          $\bigO(\ell)$. Its preprocessing time is
          \[
            \bigO\left(
              m\min\left\{
                \ell,
                \frac{\ell\log\AlphabetSize}{\sqrt{\log m}},
                \frac{(\ell\log\AlphabetSize)^2}{\log m}
              \right\}
            \right).
          \]
          If $\ell=1$, apply
          \cref{pr:small-space-prefix-select-baseline}, with parameters
          $\AlphabetSize$, $m$, and $\ell$, to the packed sequence representation
          $\PackedSeqRepresentation{w}{\AlphabetSize}{W}$. Its query time is
          constant, and its preprocessing time is bounded by the same
          expression.

        \item For every $d\in[0\dd b)$, directly construct the packed representation
          $\PackedRepresentation{w}{b}{(d)}$ and apply
          \cref{pr:packed-string-repetition}, with target length $B$ and length
          upper bound $m$, to construct the packed representation of $d^B$
          over alphabet $[0\dd b)$.
          Store pointers to these representations in a temporary array
          $A_{\rm const}[0\dd b)$. These invocations are valid because
          $b,B\leq m$ and $w\geq2\log m>\log m$.

        \item Initialize a temporary array $A_{\rm pay}[0\dd2\AlphabetSize^\ell)$
          with null pointers and set $a:=\AlphabetSize$. For every
          $p\in[1\dd\ell]$ in increasing order, scan $x\in[0\dd a)$ while
          maintaining $a=\AlphabetSize^p$. For every scanned $x$, let $X$
          denote the length-$p$ string satisfying
          $\Val{\AlphabetSize}{X}=x$. Set
          $z:=a+x=\BasicInt{\AlphabetSize}{X}$ and retrieve
          $f_X:=A_{\rm pfreq}[z]$. If
          $f_X=0$ or $f_X>B$, continue to the next value of $x$. Let
          $P_X^L[1\dd f_X]\in[0\dd b)^{f_X}$ denote the string satisfying
          $P_X^L[r]=0$ for every $r\in[1\dd f_X]$. By the definition
          of $L$, these entries equal $L[p_{X,r}][p]$. Use the substring
          operation from
          \cref{pr:packed-representation}\eqref{pr:packed-representation-substring},
          with length parameter $m$, to extract the length-$f_X$ prefix of
          $0^B$ stored in $A_{\rm const}[0]$. Store a pointer to the resulting packed
          representation of $P_X^L$ in $A_{\rm pay}[z]$. The
          invocation is valid because $f_X\leq B\leq m$, $b\leq m$, and
          $w\geq2\log m>\log m$.
          After completing the scan at a depth $p<\ell$, set
          $a:=\AlphabetSize\cdot a$.

        \item Set $a:=\AlphabetSize$. For every $p\in[1\dd\ell]$ in
          increasing order, scan $x\in[0\dd a)$ while maintaining
          $a=\AlphabetSize^p$ and the packed length-$p$
          base-$\AlphabetSize$ representation $X$ of $x$. At each depth $p$,
          set $X:=0^p$ and advance it by adding one in base
          $\AlphabetSize$ and propagating carries. The total number of changed
          characters at depth $p$ is $\bigO(\AlphabetSize^p)$. Set
          $z:=a+x=\BasicInt{\AlphabetSize}{X}$ and retrieve
          $f_X:=A_{\rm pfreq}[z]$. If $f_X\leq B$, continue to the next value
          of $x$. Let $P_X^L[1\dd f_X]\in[0\dd b)^{f_X}$ denote the string
          satisfying $P_X^L[r]=|c_{X,\lceil r/B\rceil}|$ for every
          $r\in[1\dd f_X]$.
          By the definition of $L$, these entries equal $L[p_{X,r}][p]$.
          Proceed in two substeps.
          \begin{enumerate}

            \item Initialize temporary arrays $A_{\rm ptr}$ and
              $A_{\rm size}$, both indexed by
              $[1\dd\lceil f_X/B\rceil]$. For every
              $k\in[1\dd\lceil f_X/B\rceil]$, set
              $q:=\min\{B,f_X-(k-1)B\}$. Query the temporary prefix-select
              structure to compute
              $p_L:=\PrefixSelect{W}{(k-1)B+1}{X}$ and
              $p_R:=\PrefixSelect{W}{(k-1)B+q}{X}$. Compare
              $\AlphabetMap{m}{2}{p_L-1}$ and
              $\AlphabetMap{m}{2}{p_R-1}$ from left to right using shifts
              until their first mismatch or until $b-1$ characters have been
              examined. The number of equal leading bits examined is
              $|c_{X,k}|$, and the scan takes $\bigO(b)$ time. Use the
              substring operation from
              \cref{pr:packed-representation}\eqref{pr:packed-representation-substring},
              with length parameter $m$, to extract the length-$q$ prefix of
              $|c_{X,k}|^B$ stored in $A_{\rm const}[|c_{X,k}|]$. Store its pointer in
              $A_{\rm ptr}[k]$ and set $A_{\rm size}[k]:=q$.

            \item Apply the concatenation operation from
              \cref{pr:packed-representation}\eqref{pr:packed-representation-concat},
              with length parameter $m$, to the representations in
              $A_{\rm ptr}$ and their lengths in $A_{\rm size}$. This constructs
              $\PackedRepresentation{w}{b}{P_X^L}$. Store a pointer to it in
              $A_{\rm pay}[z]$.
          \end{enumerate}
          Both substeps are valid because the supplied substring lengths are
          positive and sum to $f_X\leq m$, while $b\leq m$ and
          $w\geq2\log m>\log m$.
          After completing the scan at a depth $p<\ell$, set
          $a:=\AlphabetSize\cdot a$.

        \item Apply \cref{pr:packed-prefix-path-payload-gathering}, with
          parameters $\AlphabetSize$, $m$, $\ell$, and payload alphabet $b$,
          to the stored packed representation of $W$ and $A_{\rm pay}$. For
          every nonempty $X$
          and every $r\in[1\dd f_X]$, the payload $P_X^L[r]$ equals
          $L[\PrefixSelect{W}{r}{X}][|X|]$. The algorithm therefore constructs
          $\PackedSeqRepresentation{w}{b}{L}$. Retain this representation.
          The application is valid because
          $\lceil\log b\rceil\leq\sqrt{\log m}$.
      \end{enumerate}

      Enumerating all $X\in[0\dd\AlphabetSize)^p$ for
      $p\in[1\dd\ell]$ and constructing the packed representations of the
      strings $P_X^L$ takes $\bigO(\AlphabetSize^\ell)$ time plus
      $\bigO(m\ell\log\log m/\log m)$ time apart from the prefix-select
      queries. There are
      $\bigO(m\ell/(\log m)^2)$ buckets of frequent prefixes. Hence, the
      queries take $\bigO(m\ell^2/(\log m)^2)$ time, and computing all prefix
      codes takes $\bigO(m\ell/\log m)$ time. The applications of
      \cref{pr:packed-string-repetition} add only
      $(\log m)^{\bigO(1)}$ time. The construction times
      $\bigO(\AlphabetSize^\ell)$,
      $\bigO(m\ell\log\log m/\log m)$,
      $\bigO(m\ell^2/(\log m)^2)$, and
      $\bigO(m\ell/\log m)$ are bounded by each preprocessing-time bound in the claim,
      using
      $\AlphabetSize^\ell
        =\bigO(m\ell\log\AlphabetSize/\log m)$.
      Constructing the temporary prefix-select structure and applying
      \cref{pr:packed-prefix-path-payload-gathering} satisfy the same bounds.
      The packed input, parameter values, string-encoding structure,
      $A_{\rm pfreq}$, temporary prefix-select structure, and packed
      representations of the strings $P_X^L$, $A_{\rm pay}$, the packed
      sequence representation of $L$, and the working space used by
      \cref{pr:packed-prefix-path-payload-gathering} occupy
      $\bigO(m\ell(\log\AlphabetSize+\log\log m))$ bits. The temporary arrays
      $A_{\rm const}$, $A_{\rm ptr}$, and $A_{\rm size}$ fit the same bound.
      Thus, this step has the claimed preprocessing-space bound. Of the
      objects constructed in this step, only the packed sequence
      representation of $L$ remains stored.

    \item \emph{Construct the packed sequence representation of $R$.} Proceed
      in three substeps.
      \begin{enumerate}

        \item Let $S[1\dd B]\in[0\dd B)^B$ denote the sequence satisfying
          $S[t]=t-1$ for every $t\in[1\dd B]$. Construct the packed representation
          $\PackedRepresentation{w}{B}{S}$ using the initialization and update
          operations from \cref{pr:packed-representation}, and keep it until
          the packed representation of $T$ has been constructed. Let
          $T[1\dd m]\in[0\dd B)^m$ denote $S^\infty[1\dd m]$. Apply
          \cref{pr:packed-string-repetition}, with target length $m$ and length
          upper bound $m$, to the packed representation of $S$ to construct
          the packed representation $\PackedRepresentation{w}{B}{T}$,
          which remains stored until the end of
          Step~\ref{step:large-alphabet-prefix-special-rank-gather-r}. The
          invocation is valid
          because $B\leq m$ and $w\geq2\log m>\log m$.

        \item Initialize a temporary array
          $A_{\rm pay}[0\dd2\AlphabetSize^\ell)$ with null pointers and set
          $a:=\AlphabetSize$. For every $p\in[1\dd\ell]$ in increasing order,
          scan $x\in[0\dd a)$ while maintaining $a=\AlphabetSize^p$. For every
          scanned $x$, let $X$ denote the length-$p$ string satisfying
          $\Val{\AlphabetSize}{X}=x$. Set
          $z:=a+x=\BasicInt{\AlphabetSize}{X}$ and retrieve
          $f_X:=A_{\rm pfreq}[z]$. If
          $f_X=0$, continue to the next value of $x$. Let
          $P_X^R[1\dd f_X]\in[0\dd B)^{f_X}$ denote the string satisfying
          $P_X^R[r]=(r-1)\bmod B$ for every $r\in[1\dd f_X]$.
          By the definition of $R$, these entries equal $R[p_{X,r}][p]$.
          Thus, $P_X^R$ is the length-$f_X$ prefix of $T$. Extract its packed
          representation using
          \cref{pr:packed-representation}\eqref{pr:packed-representation-substring},
          with length parameter $m$, and store a pointer to it in $A_{\rm pay}[z]$.
          The construction is valid because $f_X\leq m$, $B\leq m$, and
          $w\geq2\log m>\log m$.
          After completing the scan at a depth $p<\ell$, set
          $a:=\AlphabetSize\cdot a$.

        \item\label{step:large-alphabet-prefix-special-rank-gather-r}
          Apply \cref{pr:packed-prefix-path-payload-gathering}, with parameters
          $\AlphabetSize$, $m$, $\ell$, and payload alphabet $B$, to the stored
          packed representation of $W$ and $A_{\rm pay}$. For every nonempty $X$
          and every $r\in[1\dd f_X]$, the payload $P_X^R[r]$ equals
          $R[\PrefixSelect{W}{r}{X}][|X|]$. The algorithm therefore constructs
          $\PackedSeqRepresentation{w}{B}{R}$. Retain this representation.
          The application is valid because
          $\lceil\log B\rceil\leq\sqrt{\log m}$.
      \end{enumerate}

      Constructing $S$ and $T$, enumerating all
      $X\in[0\dd\AlphabetSize)^p$ for $p\in[1\dd\ell]$, and constructing the
      packed representations of the strings $P_X^R$ takes
      $\bigO((\log m)^2+\AlphabetSize^\ell
        +m\ell\log\log m/\log m)$ time. Together with the application of
      \cref{pr:packed-prefix-path-payload-gathering}, this is bounded by each of
      the three preprocessing-time bounds in the claim. The packed input, the
      string-encoding structure, $A_{\rm pfreq}$, the packed sequence
      representation of $L$, the packed representations of $S$, $T$, and the
      strings $P_X^R$, $A_{\rm pay}$, the
      packed sequence representation of $R$, and the working space used by the
      application of
      \cref{pr:packed-prefix-path-payload-gathering} use
      $\bigO(m\ell(\log\AlphabetSize+\log\log m))$ bits. Thus, this step has
      the claimed preprocessing-space bound. Of the objects constructed in
      this step, only the packed sequence representation of $R$ remains
      stored.

    \item \emph{Construct $A_{\rm dict}$ and the static dictionaries.} Proceed
      in two substeps.
      \begin{enumerate}

        \item Construct a new temporary prefix-select structure by repeating
          Step~\ref{step:large-alphabet-prefix-special-rank-temporary-select-l},
          with parameters $\AlphabetSize$, $m$, and $\ell$ and supplied packed
          sequence representation
          $\PackedSeqRepresentation{w}{\AlphabetSize}{W}$. Its query time is
          $\bigO(\ell)$.

        \item Initialize $A_{\rm dict}[0\dd2\AlphabetSize^\ell)$ with null
          pointers and set $a:=1$. For every $p\in[0\dd\ell]$ in increasing
          order, scan $x\in[0\dd a)$ while maintaining
          $a=\AlphabetSize^p$ and the packed length-$p$
          base-$\AlphabetSize$ representation $X$ of $x$. At each depth $p$,
          set $X:=0^p$ and advance it by adding one in base
          $\AlphabetSize$ and propagating carries. Set
          $z:=a+x=\BasicInt{\AlphabetSize}{X}$ and retrieve
          $f_X:=A_{\rm pfreq}[z]$. If $f_X\leq B$, continue to the next value
          of $x$. Proceed in two substeps.
          \begin{enumerate}

            \item Initialize a temporary array $A_X$ with
              $\lceil f_X/B\rceil$ entries. For every
              $k\in[1\dd\lceil f_X/B\rceil]$, set
              $q:=\min\{B,f_X-(k-1)B\}$. Query the temporary prefix-select
              structure to compute
              $p_L:=\PrefixSelect{W}{(k-1)B+1}{X}$ and
              $p_R:=\PrefixSelect{W}{(k-1)B+q}{X}$. Compare
              $\AlphabetMap{m}{2}{p_L-1}$ and
              $\AlphabetMap{m}{2}{p_R-1}$ from left to right using shifts
              until their first mismatch or until $b-1$ characters have been
              examined. The number of equal leading bits examined is
              $|c_{X,k}|$, and the scan takes $\bigO(b)$ time. Compute
              $\BasicInt{2}{c_{X,k}}
                =2^{|c_{X,k}|}
                  +\lfloor(p_L-1)/2^{b-|c_{X,k}|}\rfloor$
              in constant time and set
              $A_X[k]:=(\BasicInt{2}{c_{X,k}},k)$.

            \item Apply \cref{th:static-dictionaries} to the pairs in $A_X$
              to construct $\mathcal D_X$, and store a pointer to
              $\mathcal D_X$ in $A_{\rm dict}[z]$. The application is valid
              because distinct buckets have distinct prefix codes, and
              \cref{def:basic-int} maps these codes to distinct integers.
              Moreover, $f_X>B$ implies $|A_X|=\lceil f_X/B\rceil\geq2$.
              It also holds $|A_X|\leq m$, and hence
              $w\geq2\log m\geq\log|A_X|$.
              Every dictionary key is smaller than $2^b<2m$, and every bucket
              index belongs to $[1\dd m]$, so the key and bucket index each fit
              in one word.
          \end{enumerate}
          After completing the scan at a depth $p<\ell$, set
          $a:=\AlphabetSize\cdot a$.
      \end{enumerate}

      Initializing $A_{\rm dict}$ and enumerating all
      $X\in[0\dd\AlphabetSize)^p$ for $p\in[0\dd\ell]$ take
      $\bigO(\AlphabetSize^\ell)$ time.
      There are $\bigO(m\ell/(\log m)^2)$ buckets of frequent prefixes, so
      the queries to the temporary prefix-select structure take
      $\bigO(m\ell^2/(\log m)^2)$ time. Since $\log\log m\geq9$ and every
      dictionary has between $2$ and $m$ pairs, applying
      \cref{th:static-dictionaries} to all dictionaries takes
      \[
        \bigO(m\ell(\log\log m)^2/(\log m)^2)
      \]
      time.
      Computing all prefix codes takes $\bigO(m\ell/\log m)$ time.
      Constructing the temporary prefix-select structure satisfies the
      claimed preprocessing-time bound by
      \cref{th:prefix-select-tradeoffs,pr:small-space-prefix-select-baseline}.
      Using
      $\AlphabetSize^\ell
        =\bigO(m\ell\log\AlphabetSize/\log m)$, initializing $A_{\rm dict}$,
      enumerating prefixes, answering the prefix-select queries, computing the
      codes, and constructing the dictionaries also take at most the claimed
      preprocessing time. The
      temporary prefix-select structure uses
      $\bigO(m\ell\log\AlphabetSize)$ bits.
      All dictionaries $\mathcal D_X$ together use
      $\bigO(m\ell/\log m)$ bits. Since the arrays $A_X$ are constructed one
      at a time, they use at most $\bigO(m/\log m)$ bits at any time.
      Constructing one dictionary has a peak space usage of
      $\bigO(m(\log\log m)^2/\log m)=\bigO(m)$ bits by
      \cref{th:static-dictionaries}, and the dictionaries are constructed
      sequentially.
      The array $A_{\rm dict}$ uses
      $\bigO(m\ell\log\AlphabetSize)$ bits. Together with the packed input,
      the string-encoding structure, $A_{\rm pfreq}$, and the packed sequence
      representations of $L$ and $R$, they fit the claimed
      preprocessing-space bound and complete the construction.
      \qedhere
  \end{enumerate}
\end{proof}

\subsection{Prefix Rank on Small Universes}\label{sec:prefix-special-rank-short-strings}

\begin{proposition}\label{pr:short-string-general-prefix-rank}
  Let $\AlphabetSize,m\in\Z_{\geq2}$ and $\ell\in\Z_{\geq1}$ satisfy
  $\AlphabetSize^\ell\leq\sqrt{\lceil\log m\rceil}$. Consider the word RAM
  model with word size $w=c\log m$, where $c\geq2$ is a constant. There
  exists a data structure for the problem of indexing for prefix rank queries
  over alphabet $[0\dd\AlphabetSize)$
  (see \cref{sec:prefix-special-rank-problem-def}) that achieves the following
  complexities:
  \begin{itemize}
  \item space usage $\bigO(m\ell\log\AlphabetSize)$ bits,
  \item preprocessing time
    $\bigO(m(\ell\log\AlphabetSize)^2/\log m)$,
  \item preprocessing space $\bigO(m\ell\log\AlphabetSize)$ bits,
  \item query time $\bigO(1)$.
  \end{itemize}
\end{proposition}
\begin{proof}

  If $m<64$, we construct and store all answers, and the claim
  follows. Assume $m\geq64$.

  Denote $a:=\lceil\log\AlphabetSize\rceil$,
  $M:=\AlphabetSize^\ell$, and $n_b:=\lceil m/M\rceil$. Let
  $A_{\rm pow}[0\dd\ell]$ denote the array satisfying
  $A_{\rm pow}[d]=\AlphabetSize^d$ for every $d\in[0\dd\ell]$.
  Let $W_1,W_2,\ldots,W_{n_b}$ denote the partition of $W$ into consecutive
  blocks, each of length $M$ except possibly the last one. For every
  $X\in[0\dd\AlphabetSize)^{\leq\ell}$ and $t\in[1\dd n_b]$, denote
  $f_{X,t}:=\PrefixRank{W_t}{|W_t|}{X}$
  (see \cref{def:prefix-range-queries}). For every such $X$, denote
  $F_X:=\sum_{t=1}^{n_b}f_{X,t}$. Thus,
  $F_X=\PrefixRank{W}{m}{X}$. Let
  $A_{\rm freq}[0\dd2M)$ denote the array satisfying
  $A_{\rm freq}[\BasicInt{\AlphabetSize}{X}]=F_X$
  (see \cref{def:basic-int}) for every
  $X\in[0\dd\AlphabetSize)^{\leq\ell}$. All remaining entries are zero.
  For every $X\in[0\dd\AlphabetSize)^{\leq\ell}$, denote
  $s_X:=\AlphabetSize^{\ell-|X|}$, and let $B_X$ denote the bitvector
  \[
    B_X=
      \one^{f_{X,1}}\zero^{s_X}
      \one^{f_{X,2}}\zero^{s_X}\cdots
      \one^{f_{X,n_b}}\zero^{s_X}.
  \]

  We next define the lookup table used within one block. Since
  $a\leq2\AlphabetSize/3$ and
  $\ell\leq\AlphabetSize^{\ell-1}$, we have $\ell a\leq2M/3$.
  Together with the hypothesis, this gives
  $M\ell a\leq2M^2/3\leq2\lceil\log m\rceil/3<\log m$.
  For a sequence
  $C\in([0\dd\AlphabetSize)^\ell)^u$ with $u\in[0\dd M]$, let
  \[
    \operatorname{enc}_M(C)
      :=\sum_{i=1}^{u}\sum_{h=1}^{\ell}
        C[i][h]\cdot2^{((i-1)\ell+h-1)a}.
  \]
  This is the integer represented by the $M\ell a$-bit fixed-width encoding
  of $C\mathbin\odot(0^\ell)^{M-u}$. Let
  \[
    L_{\rm srank}[0\dd2^{M\ell a})[0\dd M+1)[0\dd2M)[0\dd M+1)
  \]
  denote a table whose entries belong to $[0\dd M]$. It satisfies
  \[
    L_{\rm srank}[\operatorname{enc}_M(C),u,
      \BasicInt{\AlphabetSize}{X},r]
      =
      |\{i\in[1\dd r]:X\text{ is a prefix of }C[i]\}|
  \]
  for every such $C$, every
  $X\in[0\dd\AlphabetSize)^{\leq\ell}$, and every $r\in[0\dd u]$.
  All remaining entries of $L_{\rm srank}$ are zero. The explicit length $u$
  makes the definition independent of the zero padding in
  $\operatorname{enc}_M(C)$.

  \DSComponents
  The data structure consists of the following components:
  \begin{enumerate}

    \item The packed sequence representation
      $\PackedSeqRepresentation{w}{\AlphabetSize}{W}$
      (see \cref{def:packed-sequence-representation}), the values $a$, $M$,
      and $n_b$, and the array $A_{\rm pow}[0\dd\ell]$. The packed sequence
      representation uses $\bigO(m\ell\log\AlphabetSize)$ bits. The values
      $a,M,n_b$ and the array $A_{\rm pow}$ use $\bigO(\ell\log m)$ bits.
      Thus, this component uses
      $\bigO(m\ell\log\AlphabetSize)$ bits in total.

    \item The string-encoding structure from
      \cref{pr:basic-int-encoding} with parameters $m$ and $\AlphabetSize$.
      By \cref{rm:space}, it uses $\bigO(\sqrt m\log m)
      =\bigO(m\ell\log\AlphabetSize)$ bits.

    \item The lookup table $L_{\rm srank}$. It is stored as an array of
      one-word entries, so every lookup takes $\bigO(1)$ time. Since
      $M\ell a\leq2\lceil\log m\rceil/3$ and
      $M\leq\sqrt{\lceil\log m\rceil}$, it uses
      $2^{2\lceil\log m\rceil/3}(\log m)^{\bigO(1)}
      =o(m/\log m)$ bits.

    \item The array $A_{\rm freq}[0\dd2M)$. It uses
      $\bigO(M\log m)=o(m)$ bits.

    \item The data structures from \cref{th:bin-rank-select} for the
      bitvectors $B_X$, for every
      $X\in[0\dd\AlphabetSize)^{\leq\ell}$, together with an array
      $A_{\rm sel}[0\dd2M)$. Its entry at
      $\BasicInt{\AlphabetSize}{X}$ stores a pointer to the data structure
      for $B_X$. All remaining entries are zero. Denote
      $N_B:=\sum_{X\in[0\dd\AlphabetSize)^{\leq\ell}}|B_X|$. Every string
      in $W$ has exactly one prefix of each length in $[0\dd\ell]$, and hence
      $\sum_{X\in[0\dd\AlphabetSize)^{\leq\ell}}F_X=m(\ell+1)$. Moreover,
      \[
        N_B
          =m(\ell+1)+n_b\sum_{p=0}^{\ell}
            \AlphabetSize^p\AlphabetSize^{\ell-p}
          =(m+n_bM)(\ell+1)
          =\Theta(m\ell).
      \]
      By \cref{th:bin-rank-select}, the returned data structures use
      $\bigO(N_B)=\bigO(m\ell)$ bits in total. The array $A_{\rm sel}$ uses
      $\bigO(M\log m)=o(m)$ bits. Thus, this component uses
      $\bigO(m\ell\log\AlphabetSize)$ bits.
  \end{enumerate}

  In total, the data structure uses
  $\bigO(m\ell\log\AlphabetSize)$ bits.

  \DSQueries
  Let $X\in[0\dd\AlphabetSize)^{\leq\ell}$ and $j\in[0\dd m]$. Given the
  length $|X|$, the packed representation
  $\PackedRepresentation{w}{\AlphabetSize}{X}$
  (see \cref{def:packed-representation}), and the position $j$, we compute
  $\PrefixRank{W}{j}{X}$ as follows:
  \begin{enumerate}

    \item If $j=0$, return zero. Otherwise, use
      \cref{pr:basic-int-encoding} to compute
      $x:=\BasicInt{\AlphabetSize}{X}$ and retrieve
      $s_X:=A_{\rm pow}[\ell-|X|]$. This step takes $\bigO(1)$ time.

    \item Compute $q:=\floor{(j-1)/M}+1$,
      $r:=j-(q-1)M$, and
      $u:=\min\{M,m-(q-1)M\}=|W_q|$.
      Apply
      \cref{pr:packed-representation}\eqref{pr:packed-representation-substring},
      with alphabet size $\AlphabetSize$ and length parameter $m\ell$, to
      $\PackedSeqRepresentation{w}{\AlphabetSize}{W}$. The starting position
      is $(q-1)M\ell+1$ and the substring length is $u\ell$. The result is
      $\PackedSeqRepresentation{w}{\AlphabetSize}{W_q}$. The invocation is
      valid because $\AlphabetSize\leq m\leq m\ell<m^2$ and
      $w\geq2\log m>\log(m\ell)$. Moreover,
      $u\ell a\leq M\ell a<\log m$, so the representation occupies one word,
      its unused bits are zero, and its value is
      $\operatorname{enc}_M(W_q)$. Retrieve
      \[
        r_1:=L_{\rm srank}[\operatorname{enc}_M(W_q),u,x,r]
        \quad\text{and}\quad
        f_{X,q}:=L_{\rm srank}[\operatorname{enc}_M(W_q),u,x,u].
      \]
      This step takes $\bigO(1)$ time.

    \item Using the data structure from \cref{th:bin-rank-select} pointed to
      by $A_{\rm sel}[x]$, compute
      \[
        z:=\Select{B_X}{qs_X}{\zero}
        \qquad\text{and}\qquad
        r_0:=z-qs_X-f_{X,q}.
      \]
      Return $r_0+r_1$. This step takes $\bigO(1)$ time.
  \end{enumerate}

  To prove correctness, observe that the $qs_X$th zero of $B_X$ is the last
  zero in the factor
  $\one^{f_{X,q}}\zero^{s_X}$ associated with $W_q$. Consequently,
  \[
    z=qs_X+\sum_{t=1}^{q}f_{X,t},
    \qquad\text{so}\qquad
    r_0=\sum_{t=1}^{q-1}f_{X,t}.
  \]
  The value $r_1$ counts the strings with prefix $X$ among the first $r$
  strings of $W_q$. Since $j=(q-1)M+r$, the value $r_0+r_1$ is
  $\PrefixRank{W}{j}{X}$. Every query step takes $\bigO(1)$ time, so the
  query takes $\bigO(1)$ time in total.

  \DSConstruction
  Given the packed sequence representation
  $\PackedSeqRepresentation{w}{\AlphabetSize}{W}$
  (see \cref{def:packed-sequence-representation}), we construct the five
  components in five steps.
  \begin{enumerate}

    \item \emph{Store the input and construct the parameters.} Retain
      $\PackedSeqRepresentation{w}{\AlphabetSize}{W}$. Compute
      $a=\lceil\log\AlphabetSize\rceil$ by repeated doubling. Allocate
      $A_{\rm pow}[0\dd\ell]$, set $A_{\rm pow}[0]:=1$, and, for every
      $d\in[1\dd\ell]$ in increasing order, set
      $A_{\rm pow}[d]:=\AlphabetSize\cdot A_{\rm pow}[d-1]$. Finally, set
      $M:=A_{\rm pow}[\ell]$ and $n_b:=\lceil m/M\rceil$. This step takes
      $\bigO(\log\AlphabetSize+\ell)$ time and uses $\bigO(\log m)$
      temporary bits. The packed sequence representation of $W$, the values
      $a,M,n_b$, and $A_{\rm pow}$ occupy
      $\bigO(m\ell\log\AlphabetSize)$ bits.

    \item \emph{Construct the string-encoding structure.} Apply
      \cref{pr:basic-int-encoding} with parameters $m$ and $\AlphabetSize$,
      and retain the resulting data structure. The invocation is valid because
      $\AlphabetSize\leq M\leq m$ and $w\geq2\log m>1+\log m$. It takes
      $\bigO(\sqrt m)$ time and, by \cref{rm:space}, the invocation has peak
      space usage $\bigO(\sqrt m\log m)$ bits. The constructed data structure
      occupies the same number of bits.

    \item \emph{Construct the lookup table.} Allocate $L_{\rm srank}$ and
      initialize all its entries to zero. Enumerate
      $z\in[0\dd2^{M\ell a})$ and $u\in[0\dd M]$. Inspect the $M\ell$
      fixed-width fields of $z$. If its first $u\ell$ fields encode the
      concatenation of a sequence
      $C\in([0\dd\AlphabetSize)^\ell)^u$ and all its remaining fields are
      zero, enumerate $p\in[0\dd\ell]$ and
      $y\in[0\dd\AlphabetSize^p)$. Let $X\in[0\dd\AlphabetSize)^p$ denote
      the string satisfying $y=\Val{\AlphabetSize}{X}$, and set
      $x:=A_{\rm pow}[p]+y=\BasicInt{\AlphabetSize}{X}$. For every
      $r\in[1\dd u]$ in increasing order, set
      $L_{\rm srank}[z,u,x,r]:=L_{\rm srank}[z,u,x,r-1]+1$ if $X$ is a
      prefix of $C[r]$, and set
      $L_{\rm srank}[z,u,x,r]:=L_{\rm srank}[z,u,x,r-1]$ otherwise. The field
      tests, base-$\AlphabetSize$ decoding, prefix tests, and assignments take
      $(\log m)^{\bigO(1)}$ time for each choice of $z,u,p,y,r$. The number of entries,
      including those that remain zero, is
      $2^{M\ell a}(M+1)^2(2M)$, and every one-word entry uses
      $\bigO(\log m)$ bits.
      Consequently, constructing
      $L_{\rm srank}$ takes
      $2^{2\lceil\log m\rceil/3}(\log m)^{\bigO(1)}=o(m/\log m)$ time. The
      table occupies
      $2^{2\lceil\log m\rceil/3}(\log m)^{\bigO(1)}=o(m/\log m)$ bits.

    \item \emph{Construct the prefix-frequency array.} Apply
      \cref{pr:packed-prefix-frequencies}, with parameters
      $\AlphabetSize,m,\ell$, to
      $\PackedSeqRepresentation{w}{\AlphabetSize}{W}$, and retain its output
      as $A_{\rm freq}[0\dd2M)$. Thus,
      $A_{\rm freq}[\BasicInt{\AlphabetSize}{X}]=F_X$ for every
      $X\in[0\dd\AlphabetSize)^{\leq\ell}$. The invocation takes
      $\bigO(m\ell\log\AlphabetSize/\log m)$ time and has peak space usage
      $\bigO(m\ell\log\AlphabetSize)$ bits. The constructed array occupies
      $\bigO(M\log m)=o(m)$ bits.

    \item \emph{Construct the select structures.} Proceed in two substeps.
      \begin{enumerate}

        \item \emph{Construct the packed representations of $B_X$.}
          The values $M,n_b$, the blocks $W_t$, the quantities
          $f_{X,t},F_X,s_X$, the bitvectors $B_X$, and the arrays
          $A_{\rm pow},A_{\rm freq}$ are defined as in the proof of
          \cref{pr:large-space-prefix-select-baseline}. We construct the
          packed representations of all bitvectors $B_X$ as in the packed
          construction in
          Step~\ref{large-space-baseline-packed-construction} of that proof,
          stopping before the bitvectors are augmented with select support.
          Its hypotheses hold because
          $M\leq\sqrt{\lceil\log m\rceil}\leq m$, $m\geq64$, and the word-size
          assumption is the same. The construction initializes a temporary
          array $A_{\rm bit}[0\dd2M)$ such that, for every
          $X\in[0\dd\AlphabetSize)^{\leq\ell}$,
          $A_{\rm bit}[\BasicInt{\AlphabetSize}{X}]$ points to
          $\PackedRepresentation{w}{2}{B_X}$. It takes
          $\bigO(m(\ell\log\AlphabetSize)^2/\log m)$ time and has peak space
          usage $\bigO(m\ell\log\AlphabetSize)$ bits. The packed
          representations of the bitvectors $B_X$ and $A_{\rm bit}$ occupy
          $\bigO(m\ell\log\AlphabetSize)$ bits.

        \item \emph{Construct select support for the bitvectors.} Initialize
          $A_{\rm sel}[0\dd2M)$ with zeros. Apply
          \cref{th:bin-rank-select} jointly to the bitvectors pointed to by
          $A_{\rm bit}$. For every $p\in[0\dd\ell]$ and
          $y\in[0\dd\AlphabetSize^p)$, supply the packed representation
          pointed to by $A_{\rm bit}[A_{\rm pow}[p]+y]$ and its length
          $A_{\rm freq}[A_{\rm pow}[p]+y]
          +n_bA_{\rm pow}[\ell-p]$.
          Store a pointer to the returned data structure in
          $A_{\rm sel}[A_{\rm pow}[p]+y]$. The bitvectors have total length
          $N_B<3m(\ell+1)<m^2$, and hence
          $w\geq2\log m>\log(3m(\ell+1))$. Thus, $3m(\ell+1)$ is a valid
          common length bound in \cref{th:bin-rank-select}. Moreover, the
          number of bitvectors is
          $\sum_{p=0}^{\ell}\AlphabetSize^p<2M$. The construction therefore
          takes $\bigO(M+m\ell/\log m)$ time. By \cref{rm:space}, it uses
          $\bigO(M\log m+m\ell)$ working bits. The packed representations
          $\PackedRepresentation{w}{2}{B_X}$ and the arrays $A_{\rm bit}$ and
          $A_{\rm sel}$ use $\bigO(m\ell\log\AlphabetSize)$ bits. Thus, this
          substep has peak space usage $\bigO(m\ell\log\AlphabetSize)$ bits.
          The constructed select structures and $A_{\rm sel}$ occupy
          $\bigO(m\ell\log\AlphabetSize)$ bits. The temporary array
          $A_{\rm bit}$ is no longer needed.
      \end{enumerate}

      Constructing the packed representations of the bitvectors takes
      $\bigO(m(\ell\log\AlphabetSize)^2/\log m)$ time. Adding select support
      takes $\bigO(M+m\ell/\log m)$ time, which is bounded by the same
      expression because $M=\bigO(m/\log m)$ and
      $\ell\log\AlphabetSize\geq1$. Thus, this step takes
      $\bigO(m(\ell\log\AlphabetSize)^2/\log m)$ time and has peak space usage
      $\bigO(m\ell\log\AlphabetSize)$ bits. The select structures and
      $A_{\rm sel}$ that remain stored occupy
      $\bigO(m\ell\log\AlphabetSize)$ bits.
  \end{enumerate}

  Constructing $a,M,n_b,A_{\rm pow}$ takes
  $\bigO(\log\AlphabetSize+\ell)$ time. Applying
  \cref{pr:basic-int-encoding} takes $\bigO(\sqrt m)$ time, constructing
  $L_{\rm srank}$ takes $o(m/\log m)$ time, and applying
  \cref{pr:packed-prefix-frequencies} takes
  $\bigO(m\ell\log\AlphabetSize/\log m)$ time. The sum of these costs is
  $\bigO(\log\AlphabetSize+\ell+\sqrt m+m/\log m
  +m\ell\log\AlphabetSize/\log m)$, which is
  $\bigO(m(\ell\log\AlphabetSize)^2/\log m)$ because
  $\sqrt m=\bigO(m/\log m)$,
  $\log\AlphabetSize+\ell=\bigO(m/\log m)$, and
  $\ell\log\AlphabetSize\geq1$. Together with the construction of the select
  structures, the total preprocessing time is
  $\bigO(m(\ell\log\AlphabetSize)^2/\log m)$. The space bounds for the
  components and the five construction steps give total preprocessing-space
  usage $\bigO(m\ell\log\AlphabetSize)$ bits.
\end{proof}

\subsection{The Layering Lemma}
\label{sec:prefix-special-rank-self-reduction}

\begin{lemma}[A two-layer composition lemma]
  \label{lem:prefix-special-rank-final-layer-composition}
  Let $\AlphabetSize,m\in\Z_{\geq2}$,
  $\ell\in\Z_{\geq2}$, and $\lambda\in[1\dd\ell)$ satisfy
  $\AlphabetSize^\ell\leq m$. Denote
  $k:=\AlphabetSize^\lambda$ and $h:=\lfloor\ell/\lambda\rfloor$, and consider the word
  RAM model with word size $w=c\log m$, where $c\geq2$ is a constant.
  Assume the existence of the following two data structures:
  \begin{itemize}
  \item a prefix-special-rank data structure for $m$ strings of length
    $\lambda$ over $[0\dd\AlphabetSize)$, with space usage $S_1$ bits,
    preprocessing time $P_{t,1}$, preprocessing space $P_{s,1}$ bits, and
    query time $Q_1$,
  \item a prefix-special-rank data structure for $m$ strings of length $h$
    over $[0\dd k)$, with space usage $S_2$ bits, preprocessing time
    $P_{t,2}$, preprocessing space $P_{s,2}$ bits, and query time $Q_2$.
  \end{itemize}
  Then there exists a data structure that, given
  $\PackedSeqRepresentation{w}{\AlphabetSize}{W}$
  (\cref{def:packed-sequence-representation}) for a sequence
  $W[1\dd m]$ of $m$ strings of length $\ell$ over
  $[0\dd\AlphabetSize)$, answers prefix-special-rank queries
  (\cref{def:prefix-range-queries}) on $W$ and has the following
  complexities:
  \begin{itemize}
  \item space usage
    $\bigO(\lceil\ell/\lambda\rceil\cdot S_1+S_2)$ bits,
  \item preprocessing time
    \[
      \bigO\!\left(
        m\min\!\left\{
          \ell,
          \frac{\ell\log\AlphabetSize}{\sqrt{\log m}},
          \frac{(\ell\log\AlphabetSize)^2}{\log m}
        \right\}
        +\left\lceil\frac{\ell}{\lambda}\right\rceil P_{t,1}+P_{t,2}
      \right),
    \]
  \item preprocessing space
    $\bigO(\lceil\ell/\lambda\rceil\cdot S_1+S_2
    +\max\{P_{s,1},P_{s,2}\})$ bits,
  \item query time $\bigO(Q_1+Q_2)$.
  \end{itemize}
\end{lemma}
\begin{proof}

  If $m<64$, construct and store an array $A_{\rm ans}$. For every
  $j\in[1\dd m]$ and $p\in[0\dd\ell]$, its entry $A_{\rm ans}[j][p]$ stores
  $\PrefixSpecialRank{W}{j}{p}$.
  All parameters and input sizes are then bounded by a constant, and hence the
  claim follows immediately. Thus, assume that $m\geq64$. The monotonicity of
  $x/\log x$ and $\AlphabetSize^\ell\leq m$ give
  $\AlphabetSize^\ell\log m=\bigO(m\ell\log\AlphabetSize)$.
  Denote
  \[
    T_0:=m\min\!\left\{
      \ell,
      \frac{\ell\log\AlphabetSize}{\sqrt{\log m}},
      \frac{(\ell\log\AlphabetSize)^2}{\log m}
    \right\}.
  \]

  Let $F_X$, $(X_{p,j})_{j\in[1\dd\AlphabetSize^p]}$, $W_{X,d}$,
  $W_{p,d}$, $b_X$, $e_X$, $L_p^{\rm beg}$, $L_p^{\rm end}$,
  $L_{p,d}^{\rm rank}$, $V_p$, and $U$ be defined as in the proof of
  \cref{lem:prefix-select-final-layer-composition}.

  For every $p\in[0\dd\ell)$, $X\in[0\dd\AlphabetSize)^p$, and
  $d\in[1\dd\ell-p]$, let $D_{X,d}\in[0\dd\AlphabetSize^d)^{F_X}$ denote
  the string satisfying
  $D_{X,d}[j]=\Val{\AlphabetSize}{W_{X,d}[j]}$ for every $j\in[1\dd F_X]$
  (see \cref{def:val}). Let
  $D_{p,d}\in[0\dd\AlphabetSize^d)^m$ denote the concatenation of the strings
  $D_{X,d}$ in lexicographic order of $X$. Thus,
  $D_{X,d}=D_{p,d}(b_X\dd e_X]$. By the definition of
  $L_{p,d}^{\rm rank}$,
  \[
    L_{p,d}^{\rm rank}[\Val{\AlphabetSize}{XY}]
      =\Rank{D_{p,d}}{b_X}{\Val{\AlphabetSize}{Y}}
      =\PrefixRank{W_{p,d}}{b_X}{Y}
  \]
  for every $X\in[0\dd\AlphabetSize)^p$ and
  $Y\in[0\dd\AlphabetSize)^d$. The rank notation is from
  \cref{def:rank-select}.

  \DSComponents
  The data structure consists of the following components:
  \begin{enumerate}

  \item The values $\lambda,h,k$ and $\lceil\log\AlphabetSize\rceil$, the array
    $A_{\rm pow}[0\dd\ell]$, the data structure from
    \cref{pr:val-encoding} with parameters $m$ and $\AlphabetSize$, and
    $\PackedSeqRepresentation{w}{\AlphabetSize}{W}$.
    The array satisfies $A_{\rm pow}[d]=\AlphabetSize^d$ for every
    $d\in[0\dd\ell]$. These objects use
    $\bigO(m\ell\log\AlphabetSize+\sqrt m\log m+\ell\log m)$ bits.

  \item The arrays $L_p^{\rm beg}$ and $L_p^{\rm end}$ for every
    $p\in[0\dd\ell]$, together with $L_{p,d}^{\rm rank}$ for every multiple
    $p<\ell$ of $\lambda$ and every $d\in[1\dd\min\{\lambda,\ell-p\}]$, all stored in
    plain form. The boundary arrays have $\bigO(\AlphabetSize^\ell)$ entries
    altogether. The pairs
    $(p,d)$ used by the rank arrays contain every value
    $p+d\in[1\dd\ell]$ exactly once, so the rank arrays also have
    $\bigO(\AlphabetSize^\ell)$ entries altogether. Every entry belongs to
    $[0\dd m]$ and fits in one word. Thus, this component uses
    $\bigO(\AlphabetSize^\ell\log m)
      =\bigO(m\ell\log\AlphabetSize)$ bits.

  \item An array $A_{\rm srank}[0\dd\lceil\ell/\lambda\rceil)$, copies of the
    first prefix-special-rank data structure from the claim constructed for
    the sequences $V_p$, and one copy of the second prefix-special-rank data
    structure from the claim constructed for $U$.
    For every multiple $p<\ell$ of $\lambda$, the entry
    $A_{\rm srank}[p/\lambda]$ stores a pointer to the prefix-special-rank
    structure for $V_p$. This component uses
    $\bigO(\lceil\ell/\lambda\rceil\cdot S_1+S_2)$ bits.
  \end{enumerate}

  By \cref{th:prefix-query-space-lower-bounds-exact}, exact prefix special
  rank on $m$ strings of length $\lambda$ over $[0\dd\AlphabetSize)$ requires
  $S_1=\Omega(m\lambda\log\AlphabetSize)$ bits.
  Since $\lambda\lceil\ell/\lambda\rceil\geq\ell$, the term
  $\lceil\ell/\lambda\rceil S_1$ is $\Omega(m\ell\log\AlphabetSize)$. For
  $m\geq64$, the values $\lambda,h,k,\lceil\log\AlphabetSize\rceil$, the arrays
  $A_{\rm pow}$, $L_p^{\rm beg}$, $L_p^{\rm end}$, and
  $L_{p,d}^{\rm rank}$, the string-encoding structure, and the packed input
  use $\bigO(m\ell\log\AlphabetSize)$ bits. This space is dominated by
  $\bigO(\lceil\ell/\lambda\rceil S_1)$. Hence, in total, the data structure uses
  $\bigO(\lceil\ell/\lambda\rceil S_1+S_2)$ bits.

  \DSQueries
  Given $j\in[1\dd m]$ and $p\in[0\dd\ell]$, compute
  $\PrefixSpecialRank{W}{j}{p}$ as follows:
  \begin{enumerate}

  \item Compute $q:=\lfloor p/\lambda\rfloor$ and $d:=p-q\lambda$. If $p=0$, return
    $j$. If $d=0$, use the stored prefix-special-rank structure for $U$ to
    return $\PrefixSpecialRank{U}{j}{q}$. This step takes $\bigO(Q_2)$ time.

  \item Set $\hat p:=q\lambda$ and let $Z:=W[j][1\dd p]$. Use
    \cref{pr:packed-representation}\eqref{pr:packed-representation-substring}
    with alphabet size $\AlphabetSize$, length parameter $m\ell$, and source
    length $m\ell$ on the packed sequence representation
    $\PackedSeqRepresentation{w}{\AlphabetSize}{W}$. For
    $Z$, supply starting position $(j-1)\ell+1$ and substring length $p$.
    This invocation returns the packed representation of $Z$. Use
    \cref{pr:val-encoding} with length $p$ and the packed representation of
    $Z$ to compute $x:=\Val{\AlphabetSize}{Z}$, and compute
    $x_{\rm pref}:=\lfloor x/A_{\rm pow}[d]\rfloor$.
    Retrieve $b:=L_{\hat p}^{\rm beg}[x_{\rm pref}]$ and
    $r_{\rm base}:=L_{\hat p,d}^{\rm rank}[x]$.
    These operations take $\bigO(1)$ time. The substring invocation is valid
    because $\AlphabetSize\leq m\leq m\ell$ and
    $w\geq2\log m>\log(m\ell)$. The invocation of
    \cref{pr:val-encoding} is valid because
    $p\leq\ell\leq\lfloor\log_{\AlphabetSize}m\rfloor$ and
    $w\geq1+\log m$.

  \item Use the stored prefix-special-rank structure for $U$ to compute
    $t:=\PrefixSpecialRank{U}{j}{q}$. Use the prefix-special-rank structure
    pointed to by $A_{\rm srank}[q]$ to compute
    $s:=\PrefixSpecialRank{V_{\hat p}}{b+t}{d}$. Return $s-r_{\rm base}$.
    This step takes $\bigO(Q_1+Q_2)$ time.
  \end{enumerate}

  Every operation outside the stored prefix-special-rank structures takes
  constant time. Moreover,
  $\lceil\log\AlphabetSize\rceil
  \leq2\log\AlphabetSize$, and hence each extracted string occupies at most
  $2\ell\log\AlphabetSize\leq2\log m\leq w$ bits. If $p=0$, the returned
  value $j$ is correct. If $d=0$, the length-$q$ prefixes of $U[i]$ and
  $U[j]$ are equal exactly when the length-$p$ prefixes of $W[i]$ and $W[j]$
  are equal, so the query to the stored prefix-special-rank structure
  constructed for $U$ returns the required answer.

  Suppose that $d>0$, and let $X:=W[j][1\dd\hat p]$ and
  $Y:=W[j][\hat p+1\dd p]$. The value
  $x_{\rm pref}$ equals $\Val{\AlphabetSize}{X}$, and hence $b=b_X$. The value
  $t$ is the position of
  $W[j][\hat p+1\dd\hat p+\min\{\lambda,\ell-\hat p\}]$ within
  $W_{X,\min\{\lambda,\ell-\hat p\}}$, because equality of length-$q$
  prefixes in $U$ is equivalent to equality of length-$\hat p$ prefixes in
  $W$. Consequently,
  \[
    V_{\hat p}[b+t]
      =W[j][\hat p+1\dd\hat p+\min\{\lambda,\ell-\hat p\}]
        \zero^{\lambda-\min\{\lambda,\ell-\hat p\}}.
  \]
  Since $d\leq\min\{\lambda,\ell-\hat p\}$, the length-$d$ prefix of every string
  in $V_{\hat p}$ equals the corresponding string in $W_{\hat p,d}$. Thus,
  the length-$d$ prefix of $V_{\hat p}[b+t]$ is $Y$. Hence, the local prefix
  special rank $s$ counts the occurrences of $Y$ in
  $W_{\hat p,d}[1\dd b+t]$, whereas
  $r_{\rm base}$ counts its occurrences in $W_{\hat p,d}[1\dd b]$. Their
  difference counts exactly the indices $i\leq j$ satisfying
  $W[i][1\dd\hat p]=X$ and $W[i][\hat p+1\dd p]=Y$.
  These equalities are equivalent to $W[i][1\dd p]=W[j][1\dd p]$. This proves
  correctness. The algorithm makes at most one query to the structure for $U$
  and one query to the structure for $V_{\hat p}$. Its query time is therefore
  $\bigO(Q_1+Q_2)$.

  \DSConstruction
  Given $\PackedSeqRepresentation{w}{\AlphabetSize}{W}$, construct the data
  structure as follows:
  \begin{enumerate}

  \item\label{step:prefix-special-rank-layering-parameters} \emph{Construct the
    parameters, power array, and string-encoding structure}: The construction
    of $h$, $k$, $\lceil\log\AlphabetSize\rceil$, $A_{\rm pow}$, and the
    string-encoding structure proceeds as in
    Step~\ref{step:prefix-select-layering-parameters} in the construction of
    \cref{lem:prefix-select-final-layer-composition}. Retain $\lambda$, $h$, $k$,
    $\lceil\log\AlphabetSize\rceil$, $A_{\rm pow}$, the string-encoding
    structure, and the packed input. This step takes $\bigO(T_0)$ time and
    has peak space usage
    $\bigO(m\ell\log\AlphabetSize)$ bits. The values
    $\lambda,h,k,\lceil\log\AlphabetSize\rceil$, $A_{\rm pow}$, the
    string-encoding structure, and the packed input use
    $\bigO(m\ell\log\AlphabetSize)$ bits.

  \item\label{step:prefix-special-rank-layering-lookup-arrays} \emph{Construct
    the boundary and rank arrays}: Apply \cref{pr:packed-prefix-frequencies},
    with parameters $\AlphabetSize$, $m$, and $\ell$, to
    $\PackedSeqRepresentation{w}{\AlphabetSize}{W}$ and retain its output as
    $A_{\rm pfreq}[0\dd2\AlphabetSize^\ell)$, indexed using the basic-integer
    encoding from \cref{def:basic-int}. Thus,
    $A_{\rm pfreq}[\BasicInt{\AlphabetSize}{X}]=F_X$ for every
    $X\in[0\dd\AlphabetSize)^{\leq\ell}$. The construction of
    $L_p^{\rm beg}$, $L_p^{\rm end}$, and $L_{p,d}^{\rm rank}$ proceeds as in
    Steps~\ref{step:prefix-select-layering-boundary-arrays}
    and~\ref{step:prefix-select-layering-rank-arrays} in the construction of
    \cref{lem:prefix-select-final-layer-composition}. Discard
    $A_{\rm pfreq}$ after the scans. Applying
    \cref{pr:packed-prefix-frequencies} takes
    $\bigO(m\ell\log\AlphabetSize/\log m)$ time. The two scans write
    $\bigO(\AlphabetSize^\ell)$ entries. Since
    $\AlphabetSize^\ell
    =\bigO(m\ell\log\AlphabetSize/\log m)$, this step takes
    $\bigO(m\ell\log\AlphabetSize/\log m)=\bigO(T_0)$ time and has peak
    space usage $\bigO(m\ell\log\AlphabetSize)$ bits. The arrays
    $L_p^{\rm beg}$, $L_p^{\rm end}$, and $L_{p,d}^{\rm rank}$ use
    $\bigO(m\ell\log\AlphabetSize)$ bits.

  \item\label{step:prefix-special-rank-layering-structures} \emph{Construct the
    local and coarse data structures}: Construct the data structure from
    \cref{pr:generalized-wavelet-tree-construction} for
    $\PackedSeqRepresentation{w}{\AlphabetSize}{W}$ with parameters
    $\AlphabetSize$, $m$, and $\ell$, initialize
    $A_{\rm srank}[0\dd\lceil\ell/\lambda\rceil)$ with null pointers, and proceed in
    two substeps.
    \begin{enumerate}

    \item\label{step:prefix-special-rank-layering-local-structures} For every
      multiple $p<\ell$ of $\lambda$, set $d:=\min\{\lambda,\ell-p\}$. The construction
      of $\PackedSeqRepresentation{w}{\AlphabetSize}{W_{p,d}}$ and
      $\PackedSeqRepresentation{w}{\AlphabetSize}{V_p}$ proceeds as in
      Steps~\ref{step:prefix-select-layering-local-concatenation}
      and~\ref{step:prefix-select-layering-local-encoding} in the construction
      of \cref{lem:prefix-select-final-layer-composition}. When constructing
      $W_{p,d}$, retain the generalized-wavelet-tree query output as
      $\mathcal W_{p,d}$ and allocate the temporary arrays $A_{\rm ptr}$ and
      $A_{\rm len}$ used for concatenation. The inequality
      $\AlphabetSize^\lambda\leq\AlphabetSize^\ell\leq m$ verifies the input-size
      hypothesis for the first prefix-special-rank data structure from the
      claim. Apply its preprocessing algorithm, with parameters
      $\AlphabetSize$, $m$, and $\lambda$, to the packed sequence representation
      $\PackedSeqRepresentation{w}{\AlphabetSize}{V_p}$, and store a pointer
      to the output in $A_{\rm srank}[p/\lambda]$. Discard
      $\mathcal W_{p,d}$, $A_{\rm ptr}$, $A_{\rm len}$, and the packed
      representation of $W_{p,d}$ before processing the next value of $p$.
      After all values of $p$ have been processed, discard the
      generalized-wavelet-tree data structure.

    \item\label{step:prefix-special-rank-layering-coarse-structure} Let
      $W^{\rm pref}[1\dd m]$ and $W^{\rm suf}[1\dd m]$ denote the sequences
      satisfying $W^{\rm pref}[j]=W[j][1\dd h\lambda]$ and
      $W^{\rm suf}[j]=W[j][h\lambda+1\dd\ell]$ for every $j\in[1\dd m]$.
      The construction of their packed representations and
      $\PackedSeqRepresentation{w}{k}{U}$ proceeds as in
      Steps~\ref{step:prefix-select-layering-coarse-partition}
      and~\ref{step:prefix-select-layering-coarse-encoding} in the construction
      of \cref{lem:prefix-select-final-layer-composition}. After the partition
      operation, retain the packed input and the packed representation of
      $W^{\rm pref}$, and discard the packed representation of $W^{\rm suf}$.
      Since
      $k^h=\AlphabetSize^{\lambda h}\leq\AlphabetSize^\ell\leq m$, the
      prefix-special-rank data structure from the claim is applicable to $U$.
      Apply its preprocessing algorithm, with parameters $k$, $m$, and $h$,
      to the packed sequence representation
      $\PackedSeqRepresentation{w}{k}{U}$, and retain the resulting data
      structure. Discard the packed representation of $W^{\rm pref}$.
    \end{enumerate}

    The time and space analysis is the same as for
    Step~\ref{step:prefix-select-layering-structures} in the construction of
    \cref{lem:prefix-select-final-layer-composition}. Thus, the
    generalized-wavelet-tree data structure and, subsequently, the packed
    representations of the sequences $V_p$ and $U$ are constructed in
    $\bigO(T_0)$ total time using
    $\bigO(m\ell\log\AlphabetSize)$ additional bits.
    Retaining the packed input does not change this space bound. Adding the
    time used by Steps~\ref{step:prefix-special-rank-layering-parameters}
    and~\ref{step:prefix-special-rank-layering-lookup-arrays} and by the
    invocations of the two prefix-special-rank preprocessing algorithms gives
    total preprocessing time
    $\bigO(T_0+\lceil\ell/\lambda\rceil P_{t,1}+P_{t,2})$.
    The packed input, $A_{\rm pow}$, the string-encoding structure, the
    boundary and rank arrays, $A_{\rm srank}$, the prefix-special-rank
    structures for the sequences $V_p$, and the prefix-special-rank structure
    for $U$ use
    $\bigO(\lceil\ell/\lambda\rceil S_1+S_2)$ bits. All invocations of the
    prefix-special-rank preprocessing algorithms are sequential, so their
    preprocessing spaces contribute at most
    $\max\{P_{s,1},P_{s,2}\}$ bits. Thus, the peak space usage is
    $\bigO(\lceil\ell/\lambda\rceil S_1+S_2
    +\max\{P_{s,1},P_{s,2}\})$ bits.
    \qedhere
  \end{enumerate}
\end{proof}

\subsection{Putting Everything Together}\label{sec:optimal-prefix-special-rank}

\begin{theorem}\label{th:optimal-prefix-special-rank}
  Let $m,\AlphabetSize\in\Z_{\geq2}$ and $\ell\in\Z_{\geq1}$ satisfy
  $\AlphabetSize^\ell\leq m$. Consider the word RAM model with word size
  $w=c\log m$, where $c\geq2$ is a constant. There exists a data structure
  for the problem of indexing for prefix special rank queries over alphabet
  $[0\dd\AlphabetSize)$
  (see \cref{sec:prefix-special-rank-problem-def}) that achieves the following
  complexities:
  \begin{itemize}
  \item space usage $\bigO(m\ell\log\AlphabetSize)$ bits,
  \item preprocessing time
    \[
      \bigO\left(
        m\min\left\{
          \ell,
          \frac{\ell\log\AlphabetSize}{\sqrt{\log m}},
          \frac{(\ell\log\AlphabetSize)^2}{\log m}
        \right\}
      \right),
    \]
  \item preprocessing space $\bigO(m\ell\log\AlphabetSize)$ bits,
  \item query time $\bigO(1)$.
  \end{itemize}
\end{theorem}
\begin{proof}
  If $m<2^{17}$, construct and store an array $A_{\rm ans}$. For every
  $j\in[1\dd m]$ and $p\in[0\dd\ell]$, its entry $A_{\rm ans}[j][p]$ stores
  $\PrefixSpecialRank{W}{j}{p}$.
  The claim follows immediately in this case. Thus, assume that
  $m\geq2^{17}$. Denote
  \[
    T_0:=
    m\min\left\{
      \ell,
      \frac{\ell\log\AlphabetSize}{\sqrt{\log m}},
      \frac{(\ell\log\AlphabetSize)^2}{\log m}
    \right\}.
  \]
  The assumption $\AlphabetSize^\ell\leq m$ implies
  $\ell\log\AlphabetSize\leq\log m$.

  Denote
  $b:=\max\{1,\lfloor\log\log m/(2\log\AlphabetSize)\rfloor\}$.
  This choice satisfies
  $b\log\AlphabetSize=\Omega(\log\log m)$. Indeed, if
  $\log\log m/(2\log\AlphabetSize)\geq2$, then its floor is at least
  $\log\log m/(4\log\AlphabetSize)$. Otherwise,
  $\log\AlphabetSize>\tfrac14\log\log m$ and $b\geq1$.
  Moreover, if $b>1$, then
  $b\log\AlphabetSize\leq\tfrac12\log\log m$ and
  $\AlphabetSize^b\leq\sqrt{\log m}$.

  We distinguish three cases.
  \begin{enumerate}
  \item \emph{Suppose that $b=1$.}
    The definition of $b$ gives
    $\log\AlphabetSize+\log\log m=\bigO(\log\AlphabetSize)$. Applying
    \cref{pr:prefix-special-rank-large-alphabets}, with parameters
    $\AlphabetSize$, $m$, and $\ell$, to the packed sequence representation
    $\PackedSeqRepresentation{w}{\AlphabetSize}{W}$ gives space usage and
    preprocessing space $\bigO(m\ell\log\AlphabetSize)$ bits,
    preprocessing time $\bigO(T_0)$, and constant query time.

  \item \emph{Suppose that $b>1$ and $b\geq\ell$.}
    It follows that $\AlphabetSize^\ell\leq\AlphabetSize^b
    \leq\sqrt{\log m}\leq\sqrt{\lceil\log m\rceil}$.
    Compute and store $a:=\lceil\log\AlphabetSize\rceil$ by repeated doubling.
    Apply \cref{pr:short-string-general-prefix-rank}, with parameters
    $\AlphabetSize$, $m$, and $\ell$, to the packed sequence representation
    $\PackedSeqRepresentation{w}{\AlphabetSize}{W}$. Retain the resulting
    prefix-rank structure and the supplied packed representation. Given a
    prefix-special-rank query $(j,p)$, return $j$ if $p=0$. Otherwise, let
    $P:=W[j][1\dd p]$. Apply
    \cref{pr:packed-representation}\eqref{pr:packed-representation-substring}
    to the packed sequence representation
    $\PackedSeqRepresentation{w}{\AlphabetSize}{W}$, with alphabet size
    $\AlphabetSize$, length parameter $m\ell$, source length $m\ell$, starting
    position $(j-1)\ell+1$, and substring length $p$,
    to construct
    $\PackedRepresentation{w}{\AlphabetSize}{P}$. This invocation is valid
    because $\AlphabetSize\leq m\leq m\ell<m^2\leq2^w$ and the requested
    interval is contained in $[1\dd m\ell]$. It takes constant time because
    $p\log\AlphabetSize\leq\ell\log\AlphabetSize\leq\log m$.
    Supply position $j$, pattern length $p$, and
    $\PackedRepresentation{w}{\AlphabetSize}{P}$ to the prefix-rank structure
    constructed for $W$, and return $\PrefixRank{W}{j}{P}$. This value equals
    $\PrefixSpecialRank{W}{j}{p}$, so the query takes constant time.
    The inequalities $\ell\log\AlphabetSize\leq b\log\AlphabetSize
    \leq\tfrac12\log\log m<\sqrt{\log m}$ imply
    $(\ell\log\AlphabetSize)^2/\log m
    \leq\ell\log\AlphabetSize/\sqrt{\log m}$ and
    $(\ell\log\AlphabetSize)^2/\log m\leq1\leq\ell$.
    Thus, $T_0=m(\ell\log\AlphabetSize)^2/\log m$, and the
    preprocessing-time bound from
    \cref{pr:short-string-general-prefix-rank} is $\bigO(T_0)$. Computing $a$
    takes $\bigO(\log\AlphabetSize)=\bigO(T_0)$ time. Its space
    usage and preprocessing space, including the stored packed input, are
    $\bigO(m\ell\log\AlphabetSize)$ bits.

  \item \emph{Suppose that $2\leq b<\ell$.}
    Denote $k:=\AlphabetSize^b$ and $h:=\lfloor\ell/b\rfloor$.
    Since $\AlphabetSize^b\leq\sqrt{\lceil\log m\rceil}$,
    \cref{pr:short-string-general-prefix-rank} is applicable. As shown in the
    preceding case, its structure also answers prefix special rank queries
    within the same bounds. It therefore supplies the local prefix-special-rank
    structure required by
    \cref{lem:prefix-special-rank-final-layer-composition}, with
    $S_1=\bigO(mb\log\AlphabetSize)$ bits of space, preprocessing time
    $P_{t,1}=\bigO(m(b\log\AlphabetSize)^2/\log m)$, preprocessing space
    $P_{s,1}=\bigO(mb\log\AlphabetSize)$ bits, and query time
    $Q_1=\bigO(1)$.
    Moreover,
    $k^h=\AlphabetSize^{bh}\leq\AlphabetSize^\ell\leq m$, so
    \cref{pr:prefix-special-rank-large-alphabets} supplies the coarse
    prefix-special-rank structure required by the lemma. Since
    $\log k=b\log\AlphabetSize$ and
    $\log\log m=\bigO(b\log\AlphabetSize)$, it holds
    $h(\log k+\log\log m)=\bigO(\ell\log\AlphabetSize)$.
    Moreover, $h\leq\ell$, and
    $h(\log k+\log\log m)=\bigO(\ell\log\AlphabetSize)$ gives
    $h^2(\log k+\log\log m)^2
    =\bigO((\ell\log\AlphabetSize)^2)$.
    Therefore, the three preprocessing-time expressions from
    \cref{pr:prefix-special-rank-large-alphabets} are bounded by
    $\bigO(m\ell)$,
    $\bigO(m\ell\log\AlphabetSize/\sqrt{\log m})$, and
    $\bigO(m(\ell\log\AlphabetSize)^2/\log m)$, respectively. Thus,
    \cref{pr:prefix-special-rank-large-alphabets} supplies a coarse structure
    with space $S_2=\bigO(m\ell\log\AlphabetSize)$ bits, preprocessing time
    $P_{t,2}=\bigO(T_0)$, preprocessing space
    $P_{s,2}=\bigO(m\ell\log\AlphabetSize)$ bits, and query time
    $Q_2=\bigO(1)$.
    It remains to bound the total contribution of the local structures.
    Since $b<\ell$, we have $\lceil\ell/b\rceil\leq2\ell/b$, and therefore
    \[
      \left\lceil\frac{\ell}{b}\right\rceil S_1
      =\bigO(m\ell\log\AlphabetSize),\qquad
      \left\lceil\frac{\ell}{b}\right\rceil P_{t,1}
      =
      \bigO\left(
        \frac{m\ell b(\log\AlphabetSize)^2}{\log m}
      \right).
    \]
    We next bound $\lceil\ell/b\rceil P_{t,1}$ by each expression in the
    preprocessing-time bound of the theorem. The inequality
    $b(\log\AlphabetSize)^2\leq(b\log\AlphabetSize)^2\leq\log m$
    gives the bound $\bigO(m\ell)$. Moreover,
    $b\log\AlphabetSize\leq\sqrt{\log m}$ gives the bound
    $\bigO(m\ell\log\AlphabetSize/\sqrt{\log m})$. Finally, $b<\ell$ gives
    the bound $\bigO(m(\ell\log\AlphabetSize)^2/\log m)$.
    Thus, $\lceil\ell/b\rceil P_{t,1}=\bigO(T_0)$.
    Applying \cref{lem:prefix-special-rank-final-layer-composition} with
    $\lambda:=b$ now gives space usage
    $\bigO(\lceil\ell/b\rceil S_1+S_2)
    =\bigO(m\ell\log\AlphabetSize)$ bits, preprocessing time
    $\bigO(T_0+\lceil\ell/b\rceil P_{t,1}+P_{t,2})=\bigO(T_0)$,
    preprocessing space
    \[
      \bigO\left(
        \left\lceil\frac{\ell}{b}\right\rceil S_1+S_2+
        \max\{P_{s,1},P_{s,2}\}
      \right)
      =\bigO(m\ell\log\AlphabetSize)
    \]
    bits, and query time $\bigO(Q_1+Q_2)=\bigO(1)$.
  \end{enumerate}

  The integer $b$ can be computed without real arithmetic or overflowing a
  word. Compute $u:=\lfloor\log m\rfloor$. If
  $\AlphabetSize>\lfloor u/\AlphabetSize\rfloor$, set $\beta:=0$.
  Otherwise, compute $s:=\AlphabetSize^2$, initialize $z:=1$ and $\beta:=0$,
  and repeatedly set $z:=sz$ and increase $\beta$ while
  $z\leq\lfloor u/s\rfloor$. Every product formed is at most $u$, and on
  termination $\beta$ is the largest nonnegative integer satisfying
  $\AlphabetSize^{2\beta}\leq\lfloor\log m\rfloor$. Since
  $\AlphabetSize^{2\beta}$ is integral, this is equivalent to
  $\AlphabetSize^{2\beta}\leq\log m$, so $\beta$ equals the floor in the
  definition of $b$. We then set $b:=\max\{1,\beta\}$. This takes
  $\bigO(\log m)$ time and uses $\bigO(\log m)$ bits. Since $m\geq2^{17}$
  and $\ell\log\AlphabetSize\geq1$, the $\bigO(\log m)$ time and
  $\bigO(\log m)$ bits are bounded by
  $\bigO(T_0)$ time and $\bigO(m\ell\log\AlphabetSize)$ bits, respectively.
  The three cases above therefore prove the theorem.
\end{proof}

The following corollary gives a version of
\cref{th:optimal-prefix-special-rank} in which the word size is based on an
upper bound $N$ on $m\ell$, and all four bounds are stated in terms of $N$.

\begin{corollary}\label{cor:optimal-prefix-special-rank-input-length}
  Let $\AlphabetSize,N\in\Z_{\geq2}$ satisfy $\AlphabetSize\leq N$.
  Consider the word RAM model with word size $w=2\log N$. Every input to the
  problem of indexing for prefix special rank queries from
  \cref{sec:prefix-special-rank-problem-def} whose parameters satisfy
  \[
    m\geq\AlphabetSize,\qquad
    \ell=\lfloor\log_{\AlphabetSize}m\rfloor,\qquad
    m\ell\leq N
  \]
  admits a data structure with the following complexities:
  \begin{itemize}
  \item space usage $\bigO(N\log\AlphabetSize)$ bits,
  \item preprocessing time
    $\bigO(N\min(1,\log\AlphabetSize/\sqrt{\log N}))$,
  \item preprocessing space $\bigO(N\log\AlphabetSize)$ bits,
  \item query time $\bigO(1)$.
  \end{itemize}
\end{corollary}
\begin{proof}
  If $N<2^{16}$, construct and store all answers directly, and hence the claim
  follows immediately. Thus, assume that $N\geq2^{16}$. Let $W$ be the input
  sequence of an arbitrary prefix-special-rank input satisfying the three
  conditions in the claim. Since
  $m\geq\AlphabetSize$, it holds $\ell\geq1$. The definition of $\ell$ also
  gives $\AlphabetSize^\ell\leq m$. The word size in
  \cref{th:optimal-prefix-special-rank} is a constant multiple of $\log m$,
  whereas the word size in this corollary is $2\log N$. If
  $m\geq\sqrt N$, then $2\log N=c_m\log m$ for a constant $c_m\in[2,4]$, as
  required by that theorem. This need not hold if $m<\sqrt N$, so we use a
  direct construction in that case.
  \begin{enumerate}

    \item \emph{Suppose that $m<\sqrt N$.}
      The data structure is an array
      $A_{\rm ans}[1\dd m][0\dd\ell]$ in plain form. Its entries belong to
      $[1\dd m]$, and it satisfies
      \[
        A_{\rm ans}[j][p]=\PrefixSpecialRank{W}{j}{p}
        \qquad
        \text{for every }j\in[1\dd m]\text{ and }p\in[0\dd\ell].
      \]
      We construct this array as follows. Compute
      $\lceil\log\AlphabetSize\rceil$ by repeated doubling. This takes
      $\bigO(\log\AlphabetSize)$ time and $\bigO(\log N)$ bits, which are
      dominated by the bounds below. Allocate an array
      $F[0\dd2\AlphabetSize^\ell)
        \in[0\dd m]^{2\AlphabetSize^\ell}$ of counters initialized to zero,
      and allocate $A_{\rm ans}$. Scan $j\in[1\dd m]$ in increasing order.
      For every $j$, set $x:=1$, increase $F[x]$ by one, and set
      $A_{\rm ans}[j][0]:=F[x]$. It holds
      $A_{\rm ans}[j][0]=j=\PrefixSpecialRank{W}{j}{0}$. Then scan
      $p\in[1\dd\ell]$ in increasing order. For every $p$, apply
      \cref{pr:packed-representation}\eqref{pr:packed-representation-access}
      to the packed sequence representation
      $\PackedSeqRepresentation{w}{\AlphabetSize}{W}$ with alphabet size
      $\AlphabetSize$, length parameter $N$, source length $m\ell$, and
      position $(j-1)\ell+p$ to obtain $c:=W[j][p]$. The application is valid
      because $\AlphabetSize\leq N$, $m\ell\leq N$, and
      $w=2\log N>\log N$. Set $x:=\AlphabetSize x+c$, increase $F[x]$ by one,
      and set $A_{\rm ans}[j][p]:=F[x]$.
      By \cref{def:basic-int}, the value of $x$ after the update in the
      iteration for $p$ is
      $\BasicInt{\AlphabetSize}{W[j][1\dd p]}$. Consequently, $F[x]$ is the
      number of indices $i\in[1\dd j]$ for which $W[i][1\dd p]$ equals
      $W[j][1\dd p]$. Thus, the constructed array satisfies its defining
      equation, and every query is answered by one access to $A_{\rm ans}$.
      The array $F$ has $2\AlphabetSize^\ell\leq2m$ entries. Initializing it
      and constructing $A_{\rm ans}$ takes $\bigO(m\ell)$ time. The input,
      $F$, and $A_{\rm ans}$ use $\bigO(m\ell\log N)$ bits altogether during
      preprocessing. After preprocessing, $A_{\rm ans}$ satisfies the same
      bound. We have $\ell\leq\log m\leq\log N$. Since $N\geq2^{16}$, it holds
      $(\log N)^2\leq\sqrt N$ and
      $\sqrt N(\log N)^{3/2}\leq N$. Therefore, the space usage and
      preprocessing space are $\bigO(N)$ bits and hence
      $\bigO(N\log\AlphabetSize)$ bits, while the preprocessing time is
      \[
        \bigO(m\ell)
          =\bigO(\sqrt N\log N)
          =\bigO(N/\sqrt{\log N}).
      \]
      Since $\log\AlphabetSize\geq1$, this time is bounded by the required
      preprocessing time.

    \item \emph{Suppose that $m\geq\sqrt N$.}
      Since $\ell\geq1$ and $m\ell\leq N$, it holds $m\leq N$. Hence,
      $\log m\in[\tfrac12\log N,\log N]$ and
      $w=c_m\log m$ for a constant $c_m\in[2,4]$. We may therefore apply
      \cref{th:optimal-prefix-special-rank}. Its space usage and preprocessing
      space are
      $\bigO(m\ell\log\AlphabetSize)
        =\bigO(N\log\AlphabetSize)$ bits, and its query time is constant. Its
      preprocessing time is upper bounded both by $m\ell\leq N$ and by
      \[
        \frac{m\ell\log\AlphabetSize}{\sqrt{\log m}}
          =\bigO\left(
            \frac{N\log\AlphabetSize}{\sqrt{\log N}}
          \right).
      \]
      Thus, it is bounded by the preprocessing time in the claim.
      \qedhere
  \end{enumerate}
\end{proof}

\subsection{Consequences}\label{sec:prefix-special-rank-consequences}

\begin{framed}
  \noindent
  \probname{Indexing for Inverse Suffix Array Queries over Alphabet $[0 \dd \AlphabetSize)$}
  \begin{bfdescription}
  \item[Input:]
    Let $\AlphabetSize,\Textlen\in\Z_{\geq2}$ satisfy
    $\AlphabetSize\leq\Textlen$. Consider the word RAM model with word size
    $w$ such that $w\geq\log\AlphabetSize$. The input is the packed
    representation $\PackedRepresentation{w}{\AlphabetSize}{\Text}$
    (\cref{def:packed-representation}) of a text
    $\Text\in[0\dd\AlphabetSize)^{\Textlen}$.
  \item[Output:]
    A data structure that, given $j\in[1\dd\Textlen]$, returns
    $\ISA{\Text}[j]$ (\cref{def:inverse-suffix-array}), i.e., the
    lexicographic rank of the suffix $\Text[j\dd\Textlen]$ among all suffixes
    of $\Text$.
  \end{bfdescription}
\end{framed}

\begin{theorem}[{\cite{SaPerfectEquiv}}]
  \label{th:prefix-special-rank-inverse-suffix-array-equivalence}
  Let $\AlphabetSize,N\in\Z_{\geq2}$ satisfy $\AlphabetSize\leq N$.
  Consider the word RAM model with word size $w=c\log N$, where $c\geq2$
  is a constant. Consider the problem of indexing for prefix special rank
  queries from \cref{sec:prefix-special-rank-problem-def} with
  $\ell=\lfloor\log_{\AlphabetSize}m\rfloor$, and the problem of indexing for
  inverse suffix array queries defined above. We say that an input to the
  problem of indexing for prefix special rank queries has length at most $N$
  if $m\ell\leq N$, and an input to the problem of indexing for inverse suffix
  array queries has length at most $N$ if $\Textlen\leq N$. Suppose that one
  of these problems admits, for every input of length at most $N$, a data
  structure with the following complexities:
  \begin{itemize}
  \item space usage $S(\AlphabetSize,N)$ bits,
  \item preprocessing time $P_t(\AlphabetSize,N)$,
  \item preprocessing space $P_s(\AlphabetSize,N)$ bits,
  \item query time $Q(\AlphabetSize,N)$.
  \end{itemize}
  Then there exists $N'=\Theta(N)$ with $N'\leq N$ such that the other
  problem admits, for every input of length at most $N'$, a data structure
  with the following complexities:
  \begin{itemize}
  \item space usage $\bigO(S(\AlphabetSize,N))$ bits,
  \item preprocessing time $\bigO(P_t(\AlphabetSize,N))$,
  \item preprocessing space $\bigO(P_s(\AlphabetSize,N))$ bits,
  \item query time $\bigO(Q(\AlphabetSize,N))$.
  \end{itemize}
\end{theorem}

\begin{theorem}\label{th:inverse-suffix-array-optimal}
  Let $\AlphabetSize,\Textlen\in\Z_{\geq2}$ satisfy
  $\AlphabetSize\leq\Textlen$. Consider the word RAM model with word size
  $w=c\log\Textlen$, where $c\geq2$ is a constant. There exists a data
  structure for the problem of indexing for inverse suffix array queries over
  alphabet $[0\dd\AlphabetSize)$ defined above with the following
  complexities:
  \begin{itemize}
  \item space usage $\bigO(\Textlen\log\AlphabetSize)$ bits,
  \item preprocessing time
    $\bigO(\Textlen
      \min(1,\log\AlphabetSize/\sqrt{\log\Textlen}))$,
  \item preprocessing space $\bigO(\Textlen\log\AlphabetSize)$ bits,
  \item query time $\bigO(1)$.
  \end{itemize}
  For $\AlphabetSize=2$, the space usage and preprocessing space are
  $\bigO(\Textlen)$ bits, and the preprocessing time is
  $\bigO(\Textlen/\sqrt{\log\Textlen})$.
\end{theorem}
\begin{proof}
  Let $\Text\in[0\dd\AlphabetSize)^{\Textlen}$ be an arbitrary input text to
  the problem in the claim. We proceed in two steps.
  \begin{enumerate}

    \item \emph{Establish an intermediate exact-length construction.}
      The constant-factor loss in maximum input length in
      \cref{th:prefix-special-rank-inverse-suffix-array-equivalence} is fixed.
      Hence, there is a fixed integer $C\geq1$, independent of
      $\AlphabetSize$ and $N$, such that applying the theorem with length
      parameter $N$ yields a bound $N'\geq N/C$. Let
      $N:=C\Textlen$ and $\widehat w:=2\log N$. By
      \cref{cor:optimal-prefix-special-rank-input-length}, every
      prefix-special-rank input of length at most $N$ admits, on a word RAM
      with word size $\widehat w$, a data structure with the bounds stated in
      that corollary. Apply
      \cref{th:prefix-special-rank-inverse-suffix-array-equivalence} with
      $c=2$ in the direction from prefix special rank to inverse suffix array.
      Its conclusion holds for every text of length at most $N'$, where
      $N'\geq N/C=\Textlen$, and hence, in particular, for every text of
      length exactly $\Textlen$. Since $C$ is fixed, it holds
      $N=\Theta(\Textlen)$ and $\log N=\Theta(\log\Textlen)$. We have
      therefore proved the following intermediate statement. On a word RAM
      with word size
      $\widehat w=2\log(C\Textlen)$, given the packed representation
      $\PackedRepresentation{\widehat w}{\AlphabetSize}{\Text'}$ of any text
      $\Text'\in[0\dd\AlphabetSize)^{\Textlen}$, one can construct a data
      structure that returns $\ISA{\Text'}[j]$ for every
      $j\in[1\dd\Textlen]$ with the following complexities:
      \begin{itemize}
      \item space usage $\bigO(\Textlen\log\AlphabetSize)$ bits,
      \item preprocessing time
        $\bigO(\Textlen
          \min(1,\log\AlphabetSize/\sqrt{\log\Textlen}))$,
      \item preprocessing space
        $\bigO(\Textlen\log\AlphabetSize)$ bits,
      \item query time $\bigO(1)$.
      \end{itemize}

    \item \emph{Use the intermediate construction on the word RAM from the
      claim.} The intermediate statement and the claim both concern texts over
      alphabet $[0\dd\AlphabetSize)$ of length exactly $\Textlen$. They differ
      only in the word size and the resulting packed representation of the
      input. The intermediate construction uses $\widehat w$-bit words and
      expects the text packed into such words, whereas in the claim the text
      is supplied as $\PackedRepresentation{w}{\AlphabetSize}{\Text}$ on a
      word RAM with word size $w=c\log\Textlen$. Either $w$ or
      $\widehat w$ may be larger. Since $C$ and $c$ are fixed, it holds
      $\widehat w=2\log(C\Textlen)=\Theta(w)$. Hence, for
      $W:=\max(w,\widehat w)$, a constant number of $w$-bit words suffice to
      represent a $W$-bit word and simulate every operation on it in constant
      time. Compute $\lceil\log\AlphabetSize\rceil$ by repeated doubling. This
      takes $\bigO(\log\AlphabetSize)$ time, which is bounded by
      $\bigO(\Textlen\log\AlphabetSize/\log\Textlen)$ because
      $\Textlen/\log\Textlen\geq1$. Using this simulation, apply
      \cref{pr:packed-representation-word-size-conversion}, with length
      parameter $\Textlen$, to construct
      $\PackedRepresentation{\widehat w}{\AlphabetSize}{\Text}$. The
      invocation is valid because
      $\AlphabetSize\leq\Textlen$,
      $w=c\log\Textlen>\log\Textlen$,
      $\widehat w=2\log(C\Textlen)>\log\Textlen$, and
      $\widehat w=\Theta(w)$. Applying
      \cref{pr:packed-representation-word-size-conversion} takes
      $\bigO(\Textlen/\log_{\AlphabetSize}\Textlen)
        =\bigO(\Textlen\log\AlphabetSize/\log\Textlen)$ time. This is upper
      bounded both by $\bigO(\Textlen)$ and by
      $\bigO(\Textlen\log\AlphabetSize/\sqrt{\log\Textlen})$. The two packed
      representations of $\Text$ and the working space used by
      \cref{pr:packed-representation-word-size-conversion} occupy
      $\bigO(\Textlen\log\AlphabetSize)$ bits. Then run the intermediate
      preprocessing algorithm on $\Text$, and denote the resulting
      inverse-suffix-array structure by $\mathcal D$. Representing every
      $\widehat w$-bit word by a constant number of $w$-bit words and
      simulating each operation changes the space usage, preprocessing time,
      preprocessing space, and query time of $\mathcal D$ by only constant
      factors. Thus, the conversion and the implementation of $\mathcal D$
      using this word simulation satisfy all bounds in the claim.
  \end{enumerate}
  Since $\Text$ was arbitrary, the claim follows.
\end{proof}

\section{Conditional Lower Bounds for Preprocessing Time}
  \label{sec:conditional-lower-bounds}

The data structures from
\cref{th:suffix-array-parameterized-tradeoffs,th:inverse-suffix-array-optimal}
have preprocessing time
$\bigO(\Textlen/\sqrt{\log\Textlen})$ over the binary alphabet. We provide
conditional evidence that this bound cannot be improved. The evidence is based
on the difficulty of Dictionary Matching. We begin by recalling the known
reduction to inverse suffix array queries.

\subsection{Dictionary Matching}
  \label{sec:conditional-lower-bounds-dictionary-matching}

\begin{framed}
  \noindent
  \probname{Dictionary Matching}
  \begin{bfdescription}
  \item[Input:]
    Let $\Textlen\in\Z_{\geq2}$ and $k,m\in\Z_{\geq1}$. The input consists of
    the packed representation (\cref{def:packed-representation}) of a text
    $\Text\in\BinaryAlphabet^{\Textlen}$ and the packed representations of a
    collection
    $\mathcal D=\{\Pat_1,\ldots,\Pat_k\}\subseteq\BinaryAlphabet^m$ of
    nonempty patterns, where $k=\Theta(\Textlen/\log\Textlen)$ and
    $m=\Theta(\log\Textlen)$.
  \item[Output:]
    Return \texttt{YES} if
    $\OccTwo{\Pat_i}{\Text}\neq\emptyset$ for some $i\in[1\dd k]$, and return
    \texttt{NO} otherwise (\cref{def:occ}).
  \end{bfdescription}
\end{framed}

Throughout this section, the word size is $w=c\log\Textlen$, where $c\geq2$
is a constant. The text and all patterns contain $\Theta(\Textlen)$ bits
altogether, and hence the input occupies
$\Theta(\Textlen/\log\Textlen)$ words. Our conditional lower bounds use the
premise that no algorithm solves every instance of this problem
in $o(\Textlen/\sqrt{\log\Textlen})$ time. Equivalently, if
$s=\Theta(\Textlen/\log\Textlen)$ denotes the number of input words, the
excluded running time is $o(s\sqrt{\log s})$.

\subsection{Inverse Suffix Array Queries}
  \label{sec:conditional-lower-bounds-inverse-suffix-array}

\begin{theorem}[{\cite{hierarchy}}]
  \label{th:dictionary-matching-to-batched-inverse-suffix-array}
  Given any instance $(\Text,\mathcal D)$ of Dictionary Matching from
  \cref{sec:conditional-lower-bounds-dictionary-matching}, there is an
  algorithm that takes
  $\bigO(\Textlen/\log\Textlen)$ time and constructs the packed representation
  (\cref{def:packed-representation}) of a binary text
  $S\in\BinaryAlphabet^N$, where $N=\Theta(\Textlen)$, and positions
  $a_i,b_i\in[1\dd N]$ for every $i\in[1\dd k]$, such that
  \[
    \OccTwo{\Pat_i}{\Text}\neq\emptyset
      \quad\Longleftrightarrow\quad
    \ISA{S}[a_i]+1<\ISA{S}[b_i].
  \]
\end{theorem}

The two sequences of positions form a batch of
$2k=\Theta(N/\log N)$ inverse suffix array queries. This yields the following
consequence for preprocessing an inverse suffix array data structure.

\begin{corollary}
  \label{cor:dictionary-matching-via-inverse-suffix-array}
  Suppose that an algorithm, given the packed representation
  $\PackedRepresentation{w}{2}{S}$
  (\cref{def:packed-representation}) of any binary text $S$ of length $N$,
  constructs in $P(N)$ time a data structure that, given
  $j\in[1\dd N]$, returns $\ISA{S}[j]$ in $Q(N)$ time. Then every Dictionary
  Matching instance from
  \cref{sec:conditional-lower-bounds-dictionary-matching} can be solved in
  \[
    \bigO\left(
      \frac{\Textlen}{\log\Textlen}
      +P(N)
      +\frac{\Textlen}{\log\Textlen}Q(N)
    \right)
  \]
  time for some $N=\Theta(\Textlen)$.
\end{corollary}
\begin{proof}
  Apply
  \cref{th:dictionary-matching-to-batched-inverse-suffix-array} to construct
  $S$ and the positions $a_i,b_i$. Construct the data structure for $S$, query
  all $2k$ positions, and check whether
  $\ISA{S}[a_i]+1<\ISA{S}[b_i]$ for some $i\in[1\dd k]$. Constructing $S$ and
  the positions and checking the inequalities takes
  $\bigO(\Textlen/\log\Textlen)$ time. Preprocessing $S$ takes $P(N)$ time,
  and the $2k=\Theta(\Textlen/\log\Textlen)$ queries take
  $\bigO((\Textlen/\log\Textlen)Q(N))$ time.
\end{proof}

\begin{corollary}
  \label{cor:inverse-suffix-array-conditional-preprocessing-lower-bound}
  Suppose that no algorithm solves every Dictionary Matching
  instance from
  \cref{sec:conditional-lower-bounds-dictionary-matching} in
  $o(\Textlen/\sqrt{\log\Textlen})$ time. Let $P(N)$ and $Q(N)$ denote the
  preprocessing time and query time of any data structure for
  inverse suffix array queries over a binary text of length $N$. No such data
  structure can simultaneously satisfy $P(N)=o(N/\sqrt{\log N})$ and
  $Q(N)=o(\sqrt{\log N})$. In particular, the binary-alphabet data structure
  from \cref{th:inverse-suffix-array-optimal} answers inverse suffix array
  queries in $\bigO(1)$ time, and its $\bigO(N/\sqrt{\log N})$ preprocessing
  time cannot be improved to $o(N/\sqrt{\log N})$.
\end{corollary}
\begin{proof}
  Suppose that the bounds on $P(N)$ and $Q(N)$ in the claim hold. By
  \cref{cor:dictionary-matching-via-inverse-suffix-array}, Dictionary Matching
  on a text of length $\Textlen$ can be solved in
  $\bigO(\Textlen/\log\Textlen+P(N)
  +(\Textlen/\log\Textlen)Q(N))$ time for some
  $N=\Theta(\Textlen)$. Each of $\Textlen/\log\Textlen$, $P(N)$, and
  $(\Textlen/\log\Textlen)Q(N)$ is
  $o(\Textlen/\sqrt{\log\Textlen})$, contradicting the premise. Since the
  data structure in \cref{th:inverse-suffix-array-optimal} has query time
  $\bigO(1)=o(\sqrt{\log N})$, its preprocessing time cannot be
  $o(N/\sqrt{\log N})$.
\end{proof}

\subsection{Suffix Array Queries}
  \label{sec:conditional-lower-bounds-suffix-array}

We next give a reduction that uses suffix array queries on the Dictionary
Matching text itself.

\begin{theorem}
  \label{th:dictionary-matching-to-suffix-array}
  Given any instance $(\Text,\mathcal D)$ of Dictionary Matching from
  \cref{sec:conditional-lower-bounds-dictionary-matching}, there is an
  algorithm that decides whether the instance is positive using
  $\bigO(\Textlen(1+\log\log\Textlen)/\log\Textlen)$ suffix array queries on
  $\Text$. In addition to the time spent answering these queries, the
  algorithm takes
  $\bigO(\Textlen(1+\log\log\Textlen)/\log\Textlen)$ time. The queries can be
  divided into
  $\bigO(1+\log\log\Textlen)$ rounds, each containing
  $\bigO(\Textlen/\log\Textlen)$ queries, such that the queries in a round
  depend only on the answers from the preceding rounds.
\end{theorem}
\begin{proof}
  If $m>\Textlen$, return \texttt{NO}. If $\Textlen<16$, compare every
  pattern character by character with every length-$m$ substring of $\Text$.
  This takes constant time. Thus, assume that $m\leq\Textlen$ and
  $\Textlen\geq16$. We proceed in three steps.
  \begin{enumerate}

    \item\label{step:dictionary-matching-sa-comparisons}
      \emph{Sort the patterns and implement lexicographic comparisons.}
      \begin{enumerate}

        \item Construct the value-encoding data structure from
          \cref{pr:val-encoding} with parameters $\Textlen$ and $2$. Set
          $g:=\floor{(\log\Textlen)/2}$ and
          $d:=\ceil{m/g}=\bigO(1)$. For every $r\in[1\dd d]$, set
          $a_r:=(r-1)g+1$ and $e_r:=\min(g,m-(r-1)g)$.
          For every $i\in[1\dd k]$ and $r\in[1\dd d]$, let
          $\Pat_{i,r}:=\Pat_i[a_r\dd a_r+e_r)$. Apply
          \cref{pr:packed-representation}\eqref{pr:packed-representation-substring},
          with length parameter
          $\Textlen$, supplied representation
          $\PackedRepresentation{w}{2}{\Pat_i}$, source length $m$, starting
          position $a_r$, and substring length $e_r$, to construct
          $\PackedRepresentation{w}{2}{\Pat_{i,r}}$. Apply
          \cref{pr:val-encoding} to the returned representation, with length
          $e_r$, to compute $y_{i,r}:=\Val{2}{\Pat_{i,r}}$.

        \item Sort the patterns by applying stable counting sort according to
          $y_{i,d},y_{i,d-1},\ldots,y_{i,1}$ in this order. Every value belongs
          to $[0\dd2^g)$, so one pass takes
          $\bigO(k+2^g)=\bigO(\Textlen/\log\Textlen)$ time. Since
          $d=\bigO(1)$, sorting all patterns takes
          $\bigO(\Textlen/\log\Textlen)$ time.

        \item To compare a pattern $\Pat_i$ with a suffix
          $\Text[p\dd\Textlen]$, set
          $e:=\min(m,\Textlen-p+1)$ and $d_e:=\ceil{e/g}$. For every
          $r\in[1\dd d_e]$, set $e'_r:=\min(g,e-(r-1)g)$. Apply
          \cref{pr:packed-representation}\eqref{pr:packed-representation-substring},
          with length parameter
          $\Textlen$, supplied representation
          $\PackedRepresentation{w}{2}{\Pat_i}$, source length $m$, starting
          position $(r-1)g+1$, and substring length $e'_r$. Apply
          \cref{pr:packed-representation}\eqref{pr:packed-representation-substring},
          with length parameter
          $\Textlen$, supplied representation
          $\PackedRepresentation{w}{2}{\Text}$, source length $\Textlen$,
          starting position $p+(r-1)g$, and substring length $e'_r$. Apply
          \cref{pr:val-encoding} with length $e'_r$ to the two returned
          representations and compare their integer values. The pieces in
          every comparison have equal lengths, so the first unequal pair has
          the same order as the two strings. If every pair is equal and
          $e<m$, then $\Text[p\dd\Textlen]$ is a proper prefix of $\Pat_i$.
          If every pair is equal and $e=m$, then $\Pat_i$ is a prefix of the
          suffix. These operations determine the lexicographic order and
          whether $\Pat_i$ is a prefix of the suffix in
          $\bigO(d_e)=\bigO(1)$ time.
      \end{enumerate}

      Every application of
      \cref{pr:packed-representation}\eqref{pr:packed-representation-substring}
      uses alphabet size $2$,
      length parameter $\Textlen$, and source length at most $\Textlen$. The
      assumptions hold because $2\leq\Textlen$ and $w>\log\Textlen$. Every
      piece has length at most $g\leq\floor{\log\Textlen}$, so each substring
      operation takes constant time. Every application of the value-encoding
      data structure is valid and takes constant time because the piece length
      is at most $\floor{\log\Textlen}$ and
      $w\geq2\log\Textlen\geq1+\log\Textlen$. Since $d=\bigO(1)$,
      constructing the value-encoding data structure, computing all values
      $y_{i,r}$, and sorting the patterns takes
      $\bigO(\sqrt\Textlen+d(k+2^g))
        =\bigO(\Textlen/\log\Textlen)$ time.

    \item\label{step:dictionary-matching-sa-intervals} \emph{Compute intervals containing
      $\RangeBegTwo{\Pat_i}{\Text}$.}
      \begin{enumerate}

        \item Set $h:=\ceil{\log\Textlen}$ and
          $q:=\floor{\Textlen/h}$. For every $t\in[1\dd q]$, query
          $\SA{\Text}[th]$. The sampled suffixes
          $\Text[\SA{\Text}[th]\dd\Textlen]$, for $t\in[1\dd q]$, occur in
          increasing lexicographic order.

        \item Merge the sampled suffixes with the sorted patterns using the
          comparisons from
          Step~\ref{step:dictionary-matching-sa-comparisons}. For every
          $i\in[1\dd k]$, compute $c_i$ as
          the number of sampled suffixes that are strictly smaller than
          $\Pat_i$, and set $L_i:=c_i h$ and
          $U_i:=\min((c_i+1)h,\Textlen+1)$. The computed values satisfy
          \[
            L_i\leq\RangeBegTwo{\Pat_i}{\Text}<U_i
              \qquad\text{and}\qquad U_i-L_i\leq h.
          \]
      \end{enumerate}

      The sampling uses $q=\bigO(\Textlen/\log\Textlen)$ suffix array
      queries. The merge takes
      $\bigO(k+q)=\bigO(\Textlen/\log\Textlen)$ additional time.

    \item\label{step:dictionary-matching-sa-occurrence}
      \emph{Determine whether a pattern occurs in $\Text$.}
      \begin{enumerate}

        \item For every $i\in[1\dd k]$, maintain the interval
          $(L_i\dd U_i]$. If $L_i>0$, the suffix at rank $L_i$ is smaller than
          $\Pat_i$. If $U_i\leq\Textlen$, the suffix at rank $U_i$ is not
          smaller than $\Pat_i$. The values $0$ and $\Textlen+1$ serve as
          boundaries when the corresponding suffix does not exist. While
          $U_i-L_i>1$, set $x_i:=\floor{(L_i+U_i)/2}$, query
          $\SA{\Text}[x_i]$, and compare the returned suffix with $\Pat_i$.
          If the suffix is smaller, set $L_i:=x_i$. Otherwise, set
          $U_i:=x_i$. Perform these searches in parallel, issuing at most one
          query for every pattern in each round. Since the initial interval
          has length at most $h$, after $\ceil{\log h}$ rounds it holds
          $U_i=1+\RangeBegTwo{\Pat_i}{\Text}$.

        \item For every $i\in[1\dd k]$ satisfying $U_i\leq\Textlen$, query
          $\SA{\Text}[U_i]$ and use the comparison from
          Step~\ref{step:dictionary-matching-sa-comparisons} to determine
          whether $\Pat_i$ is a prefix of the returned suffix. By
          \cref{def:occ}, at least one pattern passes this test if and only if
          the Dictionary Matching instance is positive.
      \end{enumerate}

      Including the final prefix tests, Step~\ref{step:dictionary-matching-sa-occurrence} uses
      $\bigO(1+\log h)=\bigO(1+\log\log\Textlen)$ rounds, each containing at
      most $k=\bigO(\Textlen/\log\Textlen)$ suffix array queries. The
      comparisons and updates in one round take $\bigO(k)$ time. Together
      with Steps~\ref{step:dictionary-matching-sa-comparisons}
      and~\ref{step:dictionary-matching-sa-intervals}, this proves the claimed bounds.\qedhere
  \end{enumerate}
\end{proof}

\begin{corollary}
  \label{cor:dictionary-matching-via-suffix-array}
  Suppose that an algorithm, given the packed representation
  $\PackedRepresentation{w}{2}{\Text}$
  (\cref{def:packed-representation}) of any binary text $\Text$ of length
  $\Textlen$, constructs in $P(\Textlen)$ time a data structure that, given
  $i\in[1\dd\Textlen]$, returns $\SA{\Text}[i]$ in $Q(\Textlen)$ time. Then
  every Dictionary Matching instance from
  \cref{sec:conditional-lower-bounds-dictionary-matching} can be solved in
  \[
    \bigO\left(
      P(\Textlen)
      +\frac{\Textlen(1+\log\log\Textlen)}{\log\Textlen}
        (Q(\Textlen)+1)
    \right)
  \]
  time.
\end{corollary}
\begin{proof}
  Construct the data structure for $\Text$ and apply
  \cref{th:dictionary-matching-to-suffix-array}. The preprocessing takes
  $P(\Textlen)$ time. The theorem uses
  $\bigO(\Textlen(1+\log\log\Textlen)/\log\Textlen)$ queries, which take
  $\bigO((\Textlen(1+\log\log\Textlen)/\log\Textlen)Q(\Textlen))$ time. The
  additional work in
  \cref{th:dictionary-matching-to-suffix-array} takes
  $\bigO(\Textlen(1+\log\log\Textlen)/\log\Textlen)$, proving the claim.
\end{proof}

\begin{corollary}
  \label{cor:suffix-array-conditional-lower-bound}
  Suppose that no algorithm solves every Dictionary Matching
  instance from
  \cref{sec:conditional-lower-bounds-dictionary-matching} in
  $o(\Textlen/\sqrt{\log\Textlen})$ time. Let $P(\Textlen)$ and
  $Q(\Textlen)$ denote the preprocessing time and query time of any
  data structure for suffix array queries over a binary text of
  length $\Textlen$. No such data structure can satisfy
  \begin{equation}
    \label{eq:suffix-array-dm-time}
    P(\Textlen)
      +\frac{\Textlen(1+\log\log\Textlen)}{\log\Textlen}
        (Q(\Textlen)+1)
      =o\left(\frac{\Textlen}{\sqrt{\log\Textlen}}\right).
  \end{equation}
  In particular, no such data structure can simultaneously have
  $P(\Textlen)=o(\Textlen/\sqrt{\log\Textlen})$ and query time
  $Q(\Textlen)=o(\sqrt{\log\Textlen}/(1+\log\log\Textlen))$.
\end{corollary}
\begin{proof}
  Suppose that a data structure satisfies the relation in
  \cref{cor:suffix-array-conditional-lower-bound}\eqref{eq:suffix-array-dm-time}.
  Then the running time in
  \cref{cor:dictionary-matching-via-suffix-array} is
  $o(\Textlen/\sqrt{\log\Textlen})$, contradicting the premise. The bounds on
  $P(\Textlen)$ and $Q(\Textlen)$ imply the relation in
  \cref{cor:suffix-array-conditional-lower-bound}\eqref{eq:suffix-array-dm-time}
  because
  $Q(\Textlen)+1
  =o(\sqrt{\log\Textlen}/(1+\log\log\Textlen))$.
\end{proof}

\begin{corollary}
  \label{cor:suffix-array-tradeoffs-conditional-preprocessing-lower-bound}
  Suppose that no algorithm solves every Dictionary Matching
  instance from
  \cref{sec:conditional-lower-bounds-dictionary-matching} in
  $o(\Textlen/\sqrt{\log\Textlen})$ time. Denote
  $K:=\max(2,\lceil\log_2\Textlen\rceil)$. For the binary alphabet, the
  following statements hold for the data structures from
  \cref{th:suffix-array-parameterized-tradeoffs}:
  \begin{itemize}
  \item For every choice $B=B(\Textlen)\in[2\dd K]$ satisfying
    $B=o(\sqrt K/\log K)$, no algorithm taking
    $o(\Textlen/\sqrt{\log\Textlen})$ time can construct the data structure
    with query time $\bigO(B\log_B K)$.
  \item For every choice $B=B(\Textlen)\in[2\dd K]$, no algorithm taking
    $o(\Textlen/\sqrt{\log\Textlen})$ time can construct the
    data structure with query time $\bigO(\log_B K)$.
  \end{itemize}
  Moreover, the following statements hold for the binary-alphabet data
  structures from \cref{th:suffix-array-tradeoffs}:
  \begin{itemize}
  \item No algorithm taking
    $o(\Textlen/\sqrt{\log\Textlen})$ time can construct the data structure
    with query time $\bigO(1+\log\log\Textlen)$.
  \item For every fixed $\epsilon\in(0,1/2)$, no algorithm
    taking $o(\Textlen/\sqrt{\log\Textlen})$ time can construct the data
    structure with query time $\bigO((\log\Textlen)^\epsilon)$.
  \item For every fixed $\epsilon\in(0,1)$, no algorithm taking
    $o(\Textlen/\sqrt{\log\Textlen})$ time can construct the data structure
    with query time $\bigO(1)$.
  \end{itemize}
\end{corollary}
\begin{proof}
  Since $K=\Theta(\log\Textlen)$, it holds
  $\sqrt{\log\Textlen}/(1+\log\log\Textlen)
  =\Theta(\sqrt K/\log K)$.
  If $B\log_B K=o(\sqrt K/\log K)$, then
  $B=o(\sqrt K/\log K)$ because $\log_B K\geq1$. Conversely, suppose that
  $B=o(\sqrt K/\log K)$. Then
  \[
    \frac{B\log_B K}{\sqrt K/\log K}
      =
      \begin{cases}
        \bigO((\log K)^2/K^{1/4}) & \text{if } B\leq K^{1/4},\\
        \bigO(B\log K/\sqrt K) & \text{if } B>K^{1/4}
      \end{cases}
      =o(1).
  \]
  Thus, $B\log_B K=o(\sqrt K/\log K)$ exactly when
  $B=o(\sqrt K/\log K)$. Moreover,
  $\log_B K\leq\log_2K=o(\sqrt K/\log K)$ for every
  $B\in[2\dd K]$. This proves both conclusions concerning
  \cref{th:suffix-array-parameterized-tradeoffs}.
  For \cref{th:suffix-array-tradeoffs},
  $1+\log\log\Textlen=o(\sqrt{\log\Textlen}/(1+\log\log\Textlen))$ and
  $1=o(\sqrt{\log\Textlen}/(1+\log\log\Textlen))$. For every fixed
  $\epsilon\in(0,1)$, it holds
  $(\log\Textlen)^\epsilon
  =o(\sqrt{\log\Textlen}/(1+\log\log\Textlen))$ exactly when
  $\epsilon<1/2$. Applying
  \cref{cor:suffix-array-conditional-lower-bound} proves the statements
  concerning \cref{th:suffix-array-tradeoffs}.
\end{proof}

\bibliographystyle{alphaurl}
\bibliography{paper}

\newcommand{\etalchar}[1]{$^{#1}$}
\begin{thebibliography}{BCKM20}

\bibitem[ABM08]{bwtbook}
Donald Adjeroh, Tim Bell, and Amar Mukherjee.
\newblock {\em The {B}urrows-{W}heeler Transform: Data Compression, Suffix
  Arrays, and Pattern Matching}.
\newblock Springer, Boston, MA, USA, 2008.
\newblock \href {https://doi.org/10.1007/978-0-387-78909-5}
  {\path{doi:10.1007/978-0-387-78909-5}}.

\bibitem[AKO04]{AbouelhodaKO04}
Mohamed~Ibrahim Abouelhoda, Stefan Kurtz, and Enno Ohlebusch.
\newblock Replacing suffix trees with enhanced suffix arrays.
\newblock {\em J. Discrete Algorithms}, 2(1):53--86, 2004.
\newblock \href {https://doi.org/10.1016/S1570-8667(03)00065-0}
  {\path{doi:10.1016/S1570-8667(03)00065-0}}.

\bibitem[BCKM20]{BelazzouguiCKM20}
Djamal Belazzougui, Fabio Cunial, Juha K{\"{a}}rkk{\"{a}}inen, and Veli
  M{\"{a}}kinen.
\newblock Linear-time string indexing and analysis in small space.
\newblock {\em {ACM} Trans. Algorithms}, 16(2):17:1--17:54, 2020.
\newblock \href {https://doi.org/10.1145/3381417} {\path{doi:10.1145/3381417}}.

\bibitem[Bel14]{Belazzougui14}
Djamal Belazzougui.
\newblock Linear time construction of compressed text indices in compact space.
\newblock In David~B. Shmoys, editor, {\em 46th Annual {ACM} Symposium on
  Theory of Computing, {STOC} 2014}, pages 148--193. {ACM}, 2014.
\newblock \href {https://doi.org/10.1145/2591796.2591885}
  {\path{doi:10.1145/2591796.2591885}}.

\bibitem[BGKS15]{WaveletSuffixTree}
Maxim Babenko, Pawe{\l} Gawrychowski, Tomasz Kociumaka, and Tatiana
  Starikovskaya.
\newblock Wavelet trees meet suffix trees.
\newblock In {\em SODA}, pages 572--591, 2015.
\newblock \href {https://doi.org/10.1137/1.9781611973730.39}
  {\path{doi:10.1137/1.9781611973730.39}}.

\bibitem[BHMP14]{BartonHMP14}
Carl Barton, Alice H{\'{e}}liou, Laurent Mouchard, and Solon~P. Pissis.
\newblock Linear-time computation of minimal absent words using suffix array.
\newblock {\em BMC Bioinformatics}, 15:388, 2014.
\newblock \href {https://doi.org/10.1186/S12859-014-0388-9}
  {\path{doi:10.1186/S12859-014-0388-9}}.

\bibitem[BN15]{BelazzouguiN15}
Djamal Belazzougui and Gonzalo Navarro.
\newblock Optimal lower and upper bounds for representing sequences.
\newblock {\em {ACM} Trans. Algorithms}, 11(4):31:1--31:21, 2015.
\newblock \href {https://doi.org/10.1145/2629339} {\path{doi:10.1145/2629339}}.

\bibitem[BP16]{BelazzouguiP16}
Djamal Belazzougui and Simon~J. Puglisi.
\newblock Range predecessor and {L}empel-{Z}iv parsing.
\newblock In Robert Krauthgamer, editor, {\em Proceedings of the Twenty-Seventh
  Annual {ACM-SIAM} Symposium on Discrete Algorithms, {SODA} 2016, Arlington,
  VA, USA, January 10-12, 2016}, pages 2053--2071. {SIAM}, 2016.
\newblock URL: \url{https://doi.org/10.1137/1.9781611974331.ch143}, \href
  {https://doi.org/10.1137/1.9781611974331.CH143}
  {\path{doi:10.1137/1.9781611974331.CH143}}.

\bibitem[BW94]{bwt}
Michael Burrows and David~J. Wheeler.
\newblock A block-sorting lossless data compression algorithm.
\newblock Technical Report 124, Digital Equipment Corporation, Palo Alto,
  California, 1994.
\newblock URL:
  \url{https://www.hpl.hp.com/techreports/Compaq-DEC/SRC-RR-124.pdf}.

\bibitem[CILP20]{Charalampopoulos20}
Panagiotis Charalampopoulos, Costas~S. Iliopoulos, Chang Liu, and Solon~P.
  Pissis.
\newblock Property suffix array with applications in indexing weighted
  sequences.
\newblock {\em {ACM} J. Exp. Algorithmics}, 25:1--16, 2020.
\newblock \href {https://doi.org/10.1145/3385898} {\path{doi:10.1145/3385898}}.

\bibitem[CIS08]{CrochemoreIS08}
Maxime Crochemore, Lucian Ilie, and William~F. Smyth.
\newblock A simple algorithm for computing the {L}empel {Z}iv factorization.
\newblock In {\em 2008 Data Compression Conference, {DCC} 2008}, pages
  482--488. {IEEE} Computer Society, 2008.
\newblock \href {https://doi.org/10.1109/DCC.2008.36}
  {\path{doi:10.1109/DCC.2008.36}}.

\bibitem[CKPR21]{Charalampopoulos21}
Panagiotis Charalampopoulos, Tomasz Kociumaka, Solon~P. Pissis, and Jakub
  Radoszewski.
\newblock Faster algorithms for longest common substring.
\newblock In Petra Mutzel, Rasmus Pagh, and Grzegorz Herman, editors, {\em 29th
  Annual European Symposium on Algorithms, {ESA} 2021}, volume 204 of {\em
  LIPIcs}, pages 30:1--30:17. Schloss Dagstuhl--Leibniz-Zentrum f{\"{u}}r
  Informatik, 2021.
\newblock \href {https://doi.org/10.4230/LIPIcs.ESA.2021.30}
  {\path{doi:10.4230/LIPIcs.ESA.2021.30}}.

\bibitem[Cla98]{Clark98}
David~R. Clark.
\newblock {\em Compact Pat Trees}.
\newblock PhD thesis, University of Waterloo, 1998.

\bibitem[CPZ20]{CaceresPZ20}
Manuel C{\'{a}}ceres, Simon~J. Puglisi, and Bella Zhukova.
\newblock Fast indexes for gapped pattern matching.
\newblock In Alexander Chatzigeorgiou, Riccardo Dondi, Herodotos Herodotou,
  Christos~A. Kapoutsis, Yannis Manolopoulos, George~A. Papadopoulos, and
  Florian Sikora, editors, {\em {SOFSEM} 2020: Theory and Practice of Computer
  Science - 46th International Conference on Current Trends in Theory and
  Practice of Informatics, {SOFSEM} 2020, Limassol, Cyprus, January 20-24,
  2020, Proceedings}, volume 12011 of {\em Lecture Notes in Computer Science},
  pages 493--504. Springer, 2020.
\newblock \href {https://doi.org/10.1007/978-3-030-38919-2\_40}
  {\path{doi:10.1007/978-3-030-38919-2\_40}}.

\bibitem[FFM00]{Farach-ColtonFM00}
Martin Farach{-}Colton, Paolo Ferragina, and S.~Muthukrishnan.
\newblock On the sorting-complexity of suffix tree construction.
\newblock {\em J. {ACM}}, 47(6):987--1011, 2000.
\newblock \href {https://doi.org/10.1145/355541.355547}
  {\path{doi:10.1145/355541.355547}}.

\bibitem[FM00]{FerraginaM00}
Paolo Ferragina and Giovanni Manzini.
\newblock Opportunistic data structures with applications.
\newblock In {\em 41st Annual Symposium on Foundations of Computer Science,
  {FOCS} 2000}, pages 390--398. {IEEE} Computer Society, 2000.
\newblock \href {https://doi.org/10.1109/SFCS.2000.892127}
  {\path{doi:10.1109/SFCS.2000.892127}}.

\bibitem[FM05]{FerraginaM05}
Paolo Ferragina and Giovanni Manzini.
\newblock Indexing compressed text.
\newblock {\em Journal of the ACM}, 52(4):552--581, 2005.
\newblock \href {https://doi.org/10.1145/1082036.1082039}
  {\path{doi:10.1145/1082036.1082039}}.

\bibitem[GB13]{GotoB13}
Keisuke Goto and Hideo Bannai.
\newblock Simpler and faster {L}empel-{Z}iv factorization.
\newblock In Ali Bilgin, Michael~W. Marcellin, Joan Serra{-}Sagrist{\`{a}}, and
  James~A. Storer, editors, {\em 2013 Data Compression Conference, {DCC} 2013},
  pages 133--142. {IEEE}, 2013.
\newblock \href {https://doi.org/10.1109/DCC.2013.21}
  {\path{doi:10.1109/DCC.2013.21}}.

\bibitem[GGV03]{wt}
Roberto Grossi, Ankur Gupta, and Jeffrey~Scott Vitter.
\newblock High-order entropy-compressed text indexes.
\newblock In {\em 14th Annual {ACM-SIAM} Symposium on Discrete Algorithms,
  {SODA} 2003}, pages 841--850. {ACM/SIAM}, 2003.
\newblock URL: \url{http://dl.acm.org/citation.cfm?id=644108.644250}.

\bibitem[GHN20]{Gao0N20}
Younan Gao, Meng He, and Yakov Nekrich.
\newblock Fast preprocessing for optimal orthogonal range reporting and range
  successor with applications to text indexing.
\newblock In {\em 28th Annual European Symposium on Algorithms, {ESA} 2020},
  volume 173 of {\em LIPIcs}, pages 54:1--54:18. Schloss Dagstuhl -
  Leibniz-Zentrum f{\"{u}}r Informatik, 2020.
\newblock \href {https://doi.org/10.4230/LIPICS.ESA.2020.54}
  {\path{doi:10.4230/LIPICS.ESA.2020.54}}.

\bibitem[GK12]{GonnellaK12}
Giorgio Gonnella and Stefan Kurtz.
\newblock Readjoiner: a fast and memory efficient string graph-based sequence
  assembler.
\newblock {\em {BMC} Bioinform.}, 13:82, 2012.
\newblock \href {https://doi.org/10.1186/1471-2105-13-82}
  {\path{doi:10.1186/1471-2105-13-82}}.

\bibitem[GNP20]{Gagie2020}
Travis Gagie, Gonzalo Navarro, and Nicola Prezza.
\newblock Fully functional suffix trees and optimal text searching in
  {BWT}-runs bounded space.
\newblock {\em Journal of the ACM}, 67(1):2:1--2:54, 2020.
\newblock \href {https://doi.org/10.1145/3375890} {\path{doi:10.1145/3375890}}.

\bibitem[Gro11]{Grossi11}
Roberto Grossi.
\newblock A quick tour on suffix arrays and compressed suffix arrays.
\newblock {\em Theor. Comput. Sci.}, 412(27):2964--2973, 2011.
\newblock \href {https://doi.org/10.1016/J.TCS.2010.12.036}
  {\path{doi:10.1016/J.TCS.2010.12.036}}.

\bibitem[GV00]{GrossiV00}
Roberto Grossi and Jeffrey~Scott Vitter.
\newblock Compressed suffix arrays and suffix trees with applications to text
  indexing and string matching (extended abstract).
\newblock In F.~Frances Yao and Eugene~M. Luks, editors, {\em 32nd Annual {ACM}
  Symposium on Theory of Computing, {STOC} 2000}, pages 397--406. {ACM}, 2000.
\newblock \href {https://doi.org/10.1145/335305.335351}
  {\path{doi:10.1145/335305.335351}}.

\bibitem[GV05]{GrossiV05}
Roberto Grossi and Jeffrey~Scott Vitter.
\newblock Compressed suffix arrays and suffix trees with applications to text
  indexing and string matching.
\newblock {\em SIAM Journal on Computing}, 35(2):378--407, 2005.
\newblock \href {https://doi.org/10.1137/S0097539702402354}
  {\path{doi:10.1137/S0097539702402354}}.

\bibitem[Hag98]{Hagerup98}
Torben Hagerup.
\newblock Sorting and searching on the word {RAM}.
\newblock In Michel Morvan, Christoph Meinel, and Daniel Krob, editors, {\em
  15th Annual Symposium on Theoretical Aspects of Computer Science, {STACS}
  1998}, volume 1373 of {\em LNCS}, pages 366--398. Springer, 1998.
\newblock \href {https://doi.org/10.1007/BFb0028575}
  {\path{doi:10.1007/BFb0028575}}.

\bibitem[HSS09]{HonSS03}
Wing{-}Kai Hon, Kunihiko Sadakane, and Wing{-}Kin Sung.
\newblock Breaking a time-and-space barrier in constructing full-text indices.
\newblock {\em SIAM Journal on Computing}, 38(6):2162--2178, 2009.
\newblock \href {https://doi.org/10.1137/070685373}
  {\path{doi:10.1137/070685373}}.

\bibitem[IFI11]{ilie2011hitec}
Lucian Ilie, Farideh Fazayeli, and Silvana Ilie.
\newblock Hitec: accurate error correction in high-throughput sequencing data.
\newblock {\em Bioinformatics}, 27(3):295--302, 2011.
\newblock \href {https://doi.org/10.1093/bioinformatics/btq653}
  {\path{doi:10.1093/bioinformatics/btq653}}.

\bibitem[IS11]{IlieS11}
Lucian Ilie and William~F. Smyth.
\newblock Minimum unique substrings and maximum repeats.
\newblock {\em Fundamenta Informaticae}, 110(1-4):183--195, 2011.
\newblock \href {https://doi.org/10.3233/FI-2011-536}
  {\path{doi:10.3233/FI-2011-536}}.

\bibitem[Jac89]{Jac89}
Guy Jacobson.
\newblock Space-efficient static trees and graphs.
\newblock In {\em FOCS}, pages 549--554, 1989.
\newblock \href {https://doi.org/10.1109/SFCS.1989.63533}
  {\path{doi:10.1109/SFCS.1989.63533}}.

\bibitem[KA03]{KoA03}
Pang Ko and Srinivas Aluru.
\newblock Space efficient linear time construction of suffix arrays.
\newblock In Ricardo~A. Baeza{-}Yates, Edgar Ch{\'{a}}vez, and Maxime
  Crochemore, editors, {\em Combinatorial Pattern Matching, 14th Annual
  Symposium, {CPM} 2003, Morelia, Michoc{\'{a}}n, Mexico, June 25-27, 2003,
  Proceedings}, volume 2676 of {\em Lecture Notes in Computer Science}, pages
  200--210. Springer, 2003.
\newblock \href {https://doi.org/10.1007/3-540-44888-8\_15}
  {\path{doi:10.1007/3-540-44888-8\_15}}.

\bibitem[KK19]{sss}
Dominik Kempa and Tomasz Kociumaka.
\newblock String synchronizing sets: Sublinear-time {BWT} construction and
  optimal {LCE} data structure.
\newblock In Moses Charikar and Edith Cohen, editors, {\em 51st Annual {ACM}
  {SIGACT} Symposium on Theory of Computing, {STOC} 2019}, pages 756--767.
  {ACM}, 2019.
\newblock \href {https://doi.org/10.1145/3313276.3316368}
  {\path{doi:10.1145/3313276.3316368}}.

\bibitem[KK23]{breaking}
Dominik Kempa and Tomasz Kociumaka.
\newblock Breaking the ${O(n)}$-barrier in the construction of compressed
  suffix arrays and suffix trees.
\newblock In Nikhil Bansal and Viswanath Nagarajan, editors, {\em 34th Annual
  {ACM-SIAM} Symposium on Discrete Algorithms, SODA 2023}, pages 5122--5202.
  {SIAM}, 2023.
\newblock \href {https://doi.org/10.1137/1.9781611977554.ch187}
  {\path{doi:10.1137/1.9781611977554.ch187}}.

\bibitem[KK25]{hierarchy}
Dominik Kempa and Tomasz Kociumaka.
\newblock On the hardness hierarchy for the ${O}(n\sqrt{\log n})$ complexity in
  the word {RAM}.
\newblock In {\em 57th Annual {ACM} Symposium on Theory of Computing, {STOC}
  2025}, pages 290--300. {ACM}, 2025.
\newblock \href {https://doi.org/10.1145/3717823.3718291}
  {\path{doi:10.1145/3717823.3718291}}.

\bibitem[KK26a]{SaPerfectEquiv}
Dominik Kempa and Tomasz Kociumaka.
\newblock Cell-probe lower bounds and complexity-preserving reductions for
  suffix array queries, 2026.
\newblock Posted simultaneously to arXiv on August 19, 2026.

\bibitem[KK26b]{PrefixEquiv}
Dominik Kempa and Tomasz Kociumaka.
\newblock Explaining the inherent tradeoffs for suffix array functionality:
  Equivalences between string problems and prefix range queries.
\newblock In Kasper~Green Larsen and Barna Saha, editors, {\em Proceedings of
  the 2026 Annual {ACM-SIAM} Symposium on Discrete Algorithms, {SODA} 2026,
  Vancouver, BC, Canada, January 11-14, 2026}, pages 1841--1858. {SIAM}, 2026.
\newblock \href {https://doi.org/10.1137/1.9781611978971.66}
  {\path{doi:10.1137/1.9781611978971.66}}.

\bibitem[KPD{\etalchar{+}}04]{kurtz2004versatile}
Stefan Kurtz, Adam Phillippy, Arthur~L. Delcher, Michael Smoot, Martin Shumway,
  Corina Antonescu, and Steven~L. Salzberg.
\newblock Versatile and open software for comparing large genomes.
\newblock {\em Genome biology}, 5:1--9, 2004.
\newblock \href {https://doi.org/10.1186/gb-2004-5-2-r12}
  {\path{doi:10.1186/gb-2004-5-2-r12}}.

\bibitem[KSB06]{KarkkainenSB06}
Juha K{\"{a}}rkk{\"{a}}inen, Peter Sanders, and Stefan Burkhardt.
\newblock Linear work suffix array construction.
\newblock {\em J. {ACM}}, 53(6):918--936, 2006.
\newblock \href {https://doi.org/10.1145/1217856.1217858}
  {\path{doi:10.1145/1217856.1217858}}.

\bibitem[LD09]{bwa}
Heng Li and Richard Durbin.
\newblock Fast and accurate short read alignment with {B}urrows-{W}heeler
  transform.
\newblock {\em Bioinformatics}, 25(14):1754--1760, 2009.
\newblock \href {https://doi.org/10.1093/bioinformatics/btp324}
  {\path{doi:10.1093/bioinformatics/btp324}}.

\bibitem[LS12]{bowtie2}
Ben Langmead and Steven~L. Salzberg.
\newblock Fast gapped-read alignment with {B}owtie 2.
\newblock {\em Nature methods}, 9(4):357, 2012.
\newblock \href {https://doi.org/10.1038/nmeth.1923}
  {\path{doi:10.1038/nmeth.1923}}.

\bibitem[MBCT23]{MBCT2023}
Veli M{\"{a}}kinen, Djamal Belazzougui, Fabio Cunial, and Alexandru~I. Tomescu.
\newblock {\em Genome-Scale Algorithm Design: Bioinformatics in the Era of
  High-Throughput Sequencing (2nd edition)}.
\newblock Cambridge University Press, 2023.

\bibitem[MM90]{ManberM90}
Udi Manber and Gene Myers.
\newblock Suffix arrays: {A} new method for on-line string searches.
\newblock In David~S. Johnson, editor, {\em 1st Annual {ACM-SIAM} Symposium on
  Discrete Algorithms, {SODA} 1990}, pages 319--327. {SIAM}, 1990.
\newblock URL: \url{http://dl.acm.org/citation.cfm?id=320176.320218}.

\bibitem[MNN17]{MunroNN17}
J.~Ian Munro, Gonzalo Navarro, and Yakov Nekrich.
\newblock Space-efficient construction of compressed indexes in deterministic
  linear time.
\newblock In Philip~N. Klein, editor, {\em 28th Annual {ACM-SIAM} Symposium on
  Discrete Algorithms, {SODA} 2017}, pages 408--424. {SIAM}, 2017.
\newblock \href {https://doi.org/10.1137/1.9781611974782.26}
  {\path{doi:10.1137/1.9781611974782.26}}.

\bibitem[MNN20]{MunroNN20}
J.~Ian Munro, Gonzalo Navarro, and Yakov Nekrich.
\newblock Text indexing and searching in sublinear time.
\newblock In Inge~Li G{\o}rtz and Oren Weimann, editors, {\em 31st Annual
  Symposium on Combinatorial Pattern Matching, {CPM} 2020, June 17-19, 2020,
  Copenhagen, Denmark}, volume 161 of {\em LIPIcs}, pages 24:1--24:15. Schloss
  Dagstuhl - Leibniz-Zentrum f{\"{u}}r Informatik, 2020.
\newblock \href {https://doi.org/10.4230/LIPICS.CPM.2020.24}
  {\path{doi:10.4230/LIPICS.CPM.2020.24}}.

\bibitem[MNV16]{MunroNV16}
J.~Ian Munro, Yakov Nekrich, and Jeffrey~Scott Vitter.
\newblock Fast construction of wavelet trees.
\newblock {\em Theor. Comput. Sci.}, 638:91--97, 2016.
\newblock \href {https://doi.org/10.1016/j.tcs.2015.11.011}
  {\path{doi:10.1016/j.tcs.2015.11.011}}.

\bibitem[MSST20]{MatsudaSST20}
Kotaro Matsuda, Kunihiko Sadakane, Tatiana Starikovskaya, and Masakazu
  Tateshita.
\newblock Compressed orthogonal search on suffix arrays with applications to
  range {LCP}.
\newblock In Inge~Li G{\o}rtz and Oren Weimann, editors, {\em 31st Annual
  Symposium on Combinatorial Pattern Matching, {CPM} 2020, June 17-19, 2020,
  Copenhagen, Denmark}, volume 161 of {\em LIPIcs}, pages 23:1--23:13. Schloss
  Dagstuhl - Leibniz-Zentrum f{\"{u}}r Informatik, 2020.
\newblock \href {https://doi.org/10.4230/LIPICS.CPM.2020.23}
  {\path{doi:10.4230/LIPICS.CPM.2020.23}}.

\bibitem[Nav14]{Navarro14}
Gonzalo Navarro.
\newblock Wavelet trees for all.
\newblock {\em Journal of Discrete Algorithms}, 25:2--20, 2014.
\newblock \href {https://doi.org/10.1016/j.jda.2013.07.004}
  {\path{doi:10.1016/j.jda.2013.07.004}}.

\bibitem[Nav16]{navarrobook}
Gonzalo Navarro.
\newblock {\em Compact data structures: A practical approach}.
\newblock Cambridge University Press, Cambridge, UK, 2016.
\newblock \href {https://doi.org/10.1017/cbo9781316588284}
  {\path{doi:10.1017/cbo9781316588284}}.

\bibitem[Nav21]{NavarroIndexes}
Gonzalo Navarro.
\newblock Indexing highly repetitive string collections, part {II}: Compressed
  indexes.
\newblock {\em ACM Computing Surveys}, 54(2):26:1--26:32, 2021.
\newblock \href {https://doi.org/10.1145/3432999} {\path{doi:10.1145/3432999}}.

\bibitem[NM07]{NavarroM07}
Gonzalo Navarro and Veli M{\"{a}}kinen.
\newblock Compressed full-text indexes.
\newblock {\em ACM Computing Surveys}, 39(1):2:1--2:61, 2007.
\newblock \href {https://doi.org/10.1145/1216370.1216372}
  {\path{doi:10.1145/1216370.1216372}}.

\bibitem[NT21]{NishimotoT21}
Takaaki Nishimoto and Yasuo Tabei.
\newblock Optimal-time queries on {BWT}-runs compressed indexes.
\newblock In Nikhil Bansal, Emanuela Merelli, and James Worrell, editors, {\em
  48th International Colloquium on Automata, Languages, and Programming,
  {ICALP} 2021}, volume 198 of {\em LIPIcs}, pages 101:1--101:15. Schloss
  Dagstuhl--Leibniz-Zentrum f{\"{u}}r Informatik, 2021.
\newblock \href {https://doi.org/10.4230/LIPICS.ICALP.2021.101}
  {\path{doi:10.4230/LIPICS.ICALP.2021.101}}.

\bibitem[OG11]{OhlebuschG11}
Enno Ohlebusch and Simon Gog.
\newblock {L}empel-{Z}iv factorization revisited.
\newblock In Raffaele Giancarlo and Giovanni Manzini, editors, {\em 22nd Annual
  Symposium on Combinatorial Pattern Matching, {CPM} 2011}, volume 6661 of {\em
  LNCS}, pages 15--26. Springer, 2011.
\newblock \href {https://doi.org/10.1007/978-3-642-21458-5_4}
  {\path{doi:10.1007/978-3-642-21458-5_4}}.

\bibitem[Ohl13]{ennobook}
Enno Ohlebusch.
\newblock {\em Bioinformatics algorithms: Sequence analysis, genome
  rearrangements, and phylogenetic reconstruction}.
\newblock Oldenbusch Verlag, Ulm, Germany, 2013.

\bibitem[OS09]{OkanoharaS09}
Daisuke Okanohara and Kunihiko Sadakane.
\newblock A linear-time {B}urrows-{W}heeler {T}ransform using induced sorting.
\newblock In Jussi Karlgren, Jorma Tarhio, and Heikki Hyyr{\"{o}}, editors,
  {\em String Processing and Information Retrieval, 16th International
  Symposium, {SPIRE} 2009, Saariselk{\"{a}}, Finland, August 25-27, 2009,
  Proceedings}, volume 5721 of {\em Lecture Notes in Computer Science}, pages
  90--101. Springer, 2009.
\newblock \href {https://doi.org/10.1007/978-3-642-03784-9\_9}
  {\path{doi:10.1007/978-3-642-03784-9\_9}}.

\bibitem[PST07]{PuglisiST07}
Simon~J. Puglisi, William~F. Smyth, and Andrew Turpin.
\newblock A taxonomy of suffix array construction algorithms.
\newblock {\em {ACM} Comput. Surv.}, 39(2):4, 2007.
\newblock \href {https://doi.org/10.1145/1242471.1242472}
  {\path{doi:10.1145/1242471.1242472}}.

\bibitem[Rao02]{Rao02}
S.~Srinivasa Rao.
\newblock Time-space trade-offs for compressed suffix arrays.
\newblock {\em Inf. Process. Lett.}, 82(6):307--311, 2002.
\newblock \href {https://doi.org/10.1016/S0020-0190(01)00298-8}
  {\path{doi:10.1016/S0020-0190(01)00298-8}}.

\bibitem[Ru{\v{z}}08]{Ruzic08}
Milan Ru{\v{z}}i{\'c}.
\newblock Constructing efficient dictionaries in close to sorting time.
\newblock In Luca Aceto, Ivan Damg{\aa}rd, Leslie~Ann Goldberg, Magn{\'u}s~M.
  Halld{\'o}rsson, Anna Ing{\'o}lfsd{\'o}ttir, and Igor Walukiewicz, editors,
  {\em 35th International Colloquium on Automata, Languages and Programming,
  {ICALP} 2008}, volume 5125 of {\em LNCS}, pages 84--95. Springer, 2008.
\newblock \href {https://doi.org/10.1007/978-3-540-70575-8_8}
  {\path{doi:10.1007/978-3-540-70575-8_8}}.

\bibitem[Tha26]{ISA26}
Sharma~V. Thankachan.
\newblock Compressed inverse suffix arrays.
\newblock In {\em 67th {IEEE} Annual Symposium on Foundations of Computer
  Science, {FOCS} 2026}, 2026.
\newblock To appear.
\newblock \href {http://arxiv.org/abs/2607.17287} {\path{arXiv:2607.17287}}.

\end{thebibliography}

\end{document}